\documentclass[english, 9pt]{article}
\usepackage{jcappub}
\usepackage{lmodern}
\usepackage{graphicx}
\usepackage{esint}
\usepackage[toc,page]{appendix}
\usepackage{array}
\usepackage{longtable}
\usepackage{booktabs}
\usepackage{titlesec}
\usepackage{slashed}
\usepackage[utf8]{inputenc}
\usepackage{ifthen}
\usepackage{enumitem}
\usepackage{adjustbox}
\setcitestyle{square}

\begin{document}
\title{Machian Gravity: Confronting Galaxy Cluster Mass Profiles}
\author{Santanu Das}
\affiliation{Oridigm Inc., 43575 Mission Blvd \# 716, Fremont, CA 94539, USA}
\emailAdd{santanu@cosmocommunity.in}
\date{\today}

\abstract{
The general theory of relativity (GR) has excelled in explaining gravitational phenomena at the scale of the solar system with remarkable precision. However, when extended to the galactic or cosmological scale, it requires dark matter and dark energy to explain observations. In our previous article (\href{https://arxiv.org/abs/2308.04503}{arXiv:2308.04503}), we've formulated a gravity theory based in Mach's principle, known as Machian gravity. We demonstrated that the theory successfully explains galactic velocity profiles without requiring additional dark matter components. 
In previous studies, for a selected set of galaxy clusters, we also showed its ability to explain the velocity dispersion in the clusters without extra unseen matter components. This paper primarily explores the mass profiles of galaxy clusters. We test the Machian Gravity acceleration law on two distinct sets comprising galaxy clusters sourced from various studies. We fitted the dynamic mass profiles using the Machian gravity model. The outcomes of our study show good agreement between the theory and observational results.
}

\maketitle

\section{Introduction}

The longstanding puzzles of galaxy rotation curves and galaxy cluster masses have long challenged physicists, as they appear to demand some additional form of matter whose presence is inferred only through its gravitational effects. The most straightforward explanation is to posit an extra, invisible component of matter, commonly referred to as dark matter. In the standard picture, dark matter behaves as a cold, pressureless fluid that interacts only very weakly with baryonic matter. Over the years, numerous candidates for dark matter have been proposed. Nonetheless, none of these proposed dark matter particles has yet been detected. Dark matter, however, does not resolve all the observed phenomena. For example, it has been noted that every feature in a galaxy’s rotation curve is accompanied by a corresponding feature in its light distribution~\cite{Sancisi:2003xt}. If dark matter is not coupled with the baryon, it would be difficult to account for these correlations.

Galaxy clusters form in regions corresponding to peaks in the initial mass distribution. They are the most massive gravitationally bound structures and contain a significant fraction of mass content in the Universe. It is generally accepted that about 80–90\% of the matter in these clusters is dark matter, which interacts only via gravity. The galaxies we observe across the electromagnetic spectrum—together with their stellar populations and the hot intracluster gas emitting X-rays—constitute the remaining 10–20\% of the matter. A standard NFW fit for each cluster provides a good description. However, the NFW profile itself was originally derived from numerical simulations using the P3M code developed by Efstathiou et al.~\cite{navarro1997universal,efstathiou1985numerical}. The evolution of the dark matter fluid is computed through numerical simulations~\cite{davis1985evolution} based on Newtonian mechanics. If, however, the Newtonian equations of motion—validated in the solar system—do not straightforwardly apply on galactic scales, then alternative theories or models of gravity must be considered.

Over the past decades, several modified gravity frameworks have been proposed. Milgrom introduced Modified Newtonian Dynamics (MOND), which modifies Newtonian dynamics to account for galactic rotation curves \cite{Milgrim1983,Milgrim1983a,Milgrim1983b,Milgrom2011}. Subsequently, Bekenstein formulated the AQUAL theory, providing a Lagrangian-based mathematical foundation for MOND \cite{Bekenstein1984,Bekenstein2009,Milgrom1986}. 

Despite its success on galactic scales, the standard MOND prescription fails to reproduce the observed mass profiles of galaxy clusters and still requires an additional dark matter component. In particular, the MOND acceleration law does not adequately resolve the mass discrepancy in galaxy clusters, as demonstrated by multiple independent analyses across diverse data sets~\cite{sanders1998virial,ettori2019hydrostatic,li2023measuring,gerbal1992analysis}. Cosmological simulations carried out within the MOND framework also indicate an overabundance or higher densities of galaxy clusters compared to observations~\cite{angus2011abundance}. 

To address these shortcomings, Milgrom proposed the external field effect (EFE) hypothesis. Although the EFE can partially improve the modeling of certain galaxy clusters, the resulting fits are, in general, inferior to those obtained with the standard NFW dark matter halo profile~\cite{pointecouteau2005new}.

Moffat proposed the Modified Gravity (MoG) framework to bridge this gap without invoking dark matter \cite{Moffat2005,Brownstein2005,Brownstein2005a,Moffat2005a}, and argued that the theory can account for the mass of galaxy clusters. Although the fits to observational data are generally good, yielding notable phenomenological success, the requirement that the characteristic length scale be different for distinct galaxies and clusters appears ad hoc. Alternative approaches, including Tensor–Vector–Scalar (TeVeS) gravity \cite{Bekenstein2005}, Massive Gravity theories \cite{Dam1970,Zakharov1970,Babichev2010,Babichev2013}, and higher-dimensional models such as Induced Matter Theory \cite{Overduin1998,Ponce1993,Wesson1992,de2010schwarzschild,moraes2016cosmic}, have also been advanced in the literature.

Although several existing theoretical frameworks attempt to address the missing-mass problem without invoking additional dark matter components, they are predominantly constructed as phenomenological models of astrophysical observations rather than being derived from fundamental principles within general relativity. In our earlier work, we introduced a gravitational theory explicitly grounded in Mach’s principle, formulated as a metric theory in a five-dimensional spacetime~\cite{das2023aspects}. We have previously shown that this theory can accurately reproduce the rotation velocity profiles of spiral galaxies~\cite{das2023aaspects}, based on a detailed analysis of the SPARC galaxy rotation-curve dataset. In the present paper, I demonstrate that the same theoretical framework can also account for the mass profiles of galaxy clusters, using a large and statistically significant sample of clusters.

 %In this paper, we investigate how Machian Gravity can address the missing mass corresponding to galaxy clusters. 

The article has been organized as follows. In the second section, we have briefly described the Machian gravity theory in the limits of the galactic clusters. The third section discusses the details of the measurements involved in the galactic clusters. In the next section, we discuss various analyses and results.
We fit the Machian Gravity formula on two data sets. The first data set consists of 106 X-ray clusters compiled from XMM-Newton data~\cite{reiprich2003cosmological}. We have shown a parametric fit to the data using the same acceleration law, which was earlier used to explain the galactic velocity profile, can explain the velocity dispersion in the clusters. Here, we like to mention that the MOND-like acceleration law fails to explain the cluster mass and requires additional dark matter components~\cite{mcgaugh2015tale}. We then slightly modify the acceleration law based on the data from the first set and apply it to another set of data comprised of 44 clusters compiled from  Chandra X-ray telescope data~\cite{main2016relationship}. This modified acceleration law shows an excellent fit for the data.  %We fit the Machian gravity formula with 106 galaxy cluster mass distribution profiles. This highlights Machian gravity's ability to explain these profiles without additional dark matter. Notably, conventional MOND-based theories fail in this regard \cite{mcgaugh2015tale}.
The final section provides a comprehensive discussion and conclusion, underscoring the significance of Machian gravity.

\section{Machian gravity an overview}

Machian gravity is motivated by Mach’s principle, which connects inertia with the large-scale distribution of matter in the universe. A comprehensive review of Mach’s principle is provided in~\cite{Jammer}. 

At a more general level, the underlying idea can be summarized as follows. The inertia of a particle can only be characterized once an inertial reference frame has been specified. However, the identification of such an inertial frame is intrinsically problematic, and possibly impossible, because there is no external, absolute reference frame relative to which its acceleration can be unambiguously measured. To address this conceptual difficulty, Ernst Mach proposed that the inertial reference frame is determined by the motions and distribution of distant astronomical objects. 

This hypothesis implies that the distant matter in the universe plays a decisive role in fixing the inertial properties of local bodies, a viewpoint commonly referred to as Mach’s principle. Consequently, if two otherwise identical objects are placed at different locations in the universe, their inertial properties may differ as a function of the surrounding mass–energy distribution. In particular, two identical particles could, in principle, possess different inertial masses or experience distinct effective forces due to differences in their cosmological backgrounds.

A well-known illustrative case is Newton’s bucket experiment. Consider a bucket filled with water that is set into rotation. Due to the centripetal force, the free surface of the water assumes a concave shape. 

If, however, we adopt a coordinate system rigidly attached to the rotating bucket, then the water is at rest in this non-inertial frame. In that description, no centripetal force appears to act on the water, and one might naively expect the free surface to remain flat. Empirically, this is not observed. The resolution is that, in the non-inertial (rotating) frame, one must introduce fictitious (inertial) forces to restore Newton’s second law in its usual form. In this context, the centrifugal force is invoked to account for the curvature of the water surface. The physical origin of this inertial force, however, is not explained within classical mechanics alone.

Mach’s principle proposes a conceptual answer: in the rotating frame of the bucket, the remainder of the universe—distant stars, galaxies, and all other mass distributions—appears to rotate. Consequently, in Mach's principle it is hypothesized that the gravitational interaction arising from the rotating mass distribution of the universe generates the inertial forces experienced in a local reference frame, including the centrifugal effects observed in the bucket experiment.

In our previous work, we have shown that in a 5-dimensional coordinate frame, we can recover all these inertial forces from the background~\cite{das2023aspects}. In absence of any stress energy tensor, the field equation for the theory is given by $\widetilde{G}_{AB}=0$, where $A,B$ varies from $0$ to $4$, and the tilde represents that the quantities are in $5$ dimensional spacetime. This, after some rearrangements, can be written as $\widetilde{R}_{AB}=0$, where $\widetilde{R}_{AB}$ is the Ricci tensor. 
Let us assume that the perturbation in the metric due to the gravitational field is $\widetilde{\gamma}_{A B}$. For this weak field limit, only $\widetilde{R}_{00}=\widetilde{R}^C_{0 C 0}$ is the important term, as all the other components are of the order of $O(1/c)$ or less, where the term on the right-hand side is the Riemann tensor. 

\begin{equation}
\widetilde{R}_{0 A 0}^B=\partial_A \widetilde{\Gamma}_{00}^B-\partial_0 \widetilde{\Gamma}_{A 0}^B+\widetilde{\Gamma}_{A C}^B \widetilde{\Gamma}_{00}^C-\widetilde{\Gamma}_{0 C}^B \widetilde{\Gamma}_{A 0}^C
\end{equation}

\noindent The second term here is a time derivative, which vanishes for static fields. The third and fourth terms are of the form $(\widetilde{\Gamma})^2$, and since $\widetilde{\Gamma}$ is first-order in the metric perturbation, these contribute only at second order and can be neglected, giving 

\begin{equation}
\widetilde{R}_{00} = \widetilde{R}_{0 A 0}^A = \partial_A\left(\frac{1}{2} \widetilde{g}^{A C}\left(\partial_0 \widetilde{g}_{C 0}+\partial_0 \widetilde{g}_{0 C}-\partial_C \widetilde{g}_{00}\right)\right)  =-\frac{1}{2} \widetilde{g}^{A B} \partial_A \partial_B \widetilde{\gamma}_{00}
\end{equation}

\noindent For the static solution, the time derivative also vanishes, and the equation becomes

\begin{equation}
\partial_{\zeta}^{2}\tilde{\gamma}_{00}+\partial_{x}^{2}\tilde{\gamma}_{00}+\partial_{y}^{2}\tilde{\gamma}_{00}+\partial_{z}^{2}\tilde{\gamma}_{00}=0\,.\label{eq:laplace equation}
\end{equation}

\noindent Here $\zeta$ is the fifth coordinate. Under the assumption of spherical symmetry of the special part, it can be written as

\begin{equation}
\partial_{\zeta}^{2}(r\tilde{\gamma}_{00})+\partial_{r}^{2}(r\tilde{\gamma}_{00})=0\,.\label{eq:waveequation}
\end{equation}

\noindent Using `separation of variables' and considering $(r\tilde{\gamma}_{00})=R(r)\chi(\zeta)$, we get 

\begin{equation}
\frac{1}{R}\frac{\partial^{2}R}{\partial r^{2}}=-\frac{1}{\chi}\frac{\partial^{2}\chi}{\partial\zeta^{2}}=\lambda^{2}\,,
\end{equation}

\noindent where, $\lambda$ is a constant. %Putting  This gives
To relate it with Newtonian gravity, we can put  $\tilde{\gamma}_{00}=2\Phi$, where $\Phi$ is the Newtonian potential of the gravitational field~\cite{das2023aspects}. Few straightforward manipulations lead us to 

\begin{equation}
\Phi=\frac{GM}{r}\left[1+K\left(1-e^{-\lambda r}\right)\right]\,.\label{eq:potential}
\end{equation}

\noindent Here, M is the enclosed mass within radius $r$. $G$ is the Newton's gravitational constant. $\lambda$ and $K$ are the quantities that depend on the background mass distribution. They may depend on mass $M$ but are independent on $r$. Galactic velocity profiles show that $\lambda$
is of the order of few $\texttt{kpc}^{-1}$. When $r$ is small, $e^{-\lambda r}\sim1$, $\Phi$ takes the form of Newtonian potential i.e. $\Phi=\frac{GM}{r}$.
This gives the Newtonian gravitational equation at the solar system
scale. In the asymptotic limit of $r\rightarrow\infty$, the
exponential term goes to $0$. Hence, for large values of $r$, it becomes $(1+K)$ times that of Newtonian potential and can provide additional gravitational force in large gravitationally balanced systems, such as galaxies. In fact, a similar form of potential has previously been used 
by other groups to explain the galactic velocity profile \cite{Moffat2009,Moffat2005,Brownstein2005,Brownstein2005a,Moffat2005a}.

In our previous work cited as~\cite{das2023aspects,das2023aaspects}, we demonstrated that when a particle orbits in a circular path in this potential, the velocity curve flattens out when $\lambda r \in (0.4,2.5)$ and $K \in (11,19)$. This flattened-out velocity profile can be used to explain the rotational velocity profile of the outer edge of a galaxy. We used the equation $K=\left({\frac{M_c}{M}}\right)^\alpha-1$ to recover the baryonic Tully-Fisher relation for spiral galaxies. In this equation, $M_c$ represents a constant mass and $\alpha$ is some power law power. When we take $\alpha = \frac{1}{2}$, it gives the baryonic Tully-Fisher relation, given by $v^4\propto M$. This gives us the velocity in a circular orbit as 

\begin{equation}
v^2=\frac{GM}{r}\left[1+\left(\sqrt{\frac{M_c}{M}} - 1\right)\left(1-e^{-\lambda r}\left(1+\lambda r\right)\right)\right] \,.
\label{eq:velocityFinal}
\end{equation}

It's important to note that the term enclosed within the bracket relies on the background mass distribution. Otherwise, it would violate the symmetry argument. For instance, in a scenario with only two particles, the gravitational attraction between them would obey Newtonian laws, i.e. the resulting force should be symmetric about both masses. In this case, the term enclosed within the bracket would be unity. However, in the presence of a mass distribution, such as in galaxies or galaxy clusters, the term within the bracket varies depending on the specific mass distribution. We don't have an exact form of $K$ from our equations. Hence, we resort to using a power law with a power %$\alpha$, where 
$\alpha \approx 0.5$. %, which satisfies the baryonic Tully-Fisher relation for spiral galaxies.

%Here we should note that the term inside the bracket depends of the background mass distribution. Otherwise it will violate the symmetry argument, i.e. if there were only two particles, then the gravitation attraction between them would have been given by the Newtonian laws, such that the force between them is symmetric abbout both the masses. The term inside the bracket would have been unity. However when there is a mass distribution as in case of galaxy or galaxy cluster the term in the bracket will pop up depending on the mass distribution. We don't know the exact form of $K$. So we use a power law, where $\alpha \approx 0.5$ satisfy the baryonic Tully-Fisher relation for spiral galaxies.  

\section{A theoretical background of galaxy clusters}

Clusters of galaxies are believed to comprise three primary components. Galaxy clusters, as suggested by their name, have been identified as groupings of galaxies. Each cluster can contain roughly 100-1000 galaxies, constituting approximately $1\%$ of its mass. The expanse between these galaxies is not devoid of matter; instead, it contains substantial amounts of intracluster gas (ICG), comprising about $9\%$. The predominant portion of the overall gravitational mass in clusters is believed to exist in the form of dark matter, accounting for the remaining $90\%$ (based on calculations employing Newtonian Mechanics). Of course in modified gravity theories the dark matter components will not exist.

\subsection{Galactic cluster mass}

The density distribution of hot gas in a cluster has been well described by the King $\beta$-model~\cite{King1966,Brownstein2005a,Cavaliere1976}. 

\begin{equation}
\rho(r) = \rho_0 \left[ 1+\left(\frac{r}{r_c}\right)^2\right]^{-3\beta/2}\;,
\label{eq:Kingbeta}
\end{equation}

\noindent where $\rho_0$ is the central density and $r_c$ is a core radius.  Here $\beta$ is known as the shape parameter. %, and it denotes the ratio of the specific kinetic energies of the galaxies and the gas. 
By fitting this model to the mean radial profile of X-ray surface brightness in clusters, the quantity $\beta$ can be found. While the assumptions of the $\beta$-model may not be completely accurate, %be violated in detail. Nevertheless, 
the surface brightness profile derived from it represents the measured profile well within a wide range, which makes this model widely accepted. 

\noindent The mass profile of the cluster can be calculated by integrating the density profile, i.e. 

\begin{equation}
M_{b}(r)=4\pi\int_0^r\rho(r')r'^2dr'   \;.
\label{BaryonMass1}
\end{equation}

\noindent We can integrate it numerically to get the gas mass of the galaxy cluster. For large radii and small $\beta$ values we can get an approximate analytical folrmula for the gas mass as~\cite{King1966,Cavaliere1976} 

\begin{equation}
    M_b(r)\approx \frac{4\pi \rho_0 r_c^3}{-3\beta + 3}\left(\frac{r}{r_c}\right)^{-3\beta + 3}\qquad\qquad \frac{r}{r_c}\gg 1 , \beta<1\,.
    \label{MassBaryon}
\end{equation}
 
\noindent Here we should note that the above function is a divergent function for $\beta>1$. Therefore, we must have $\beta<1$. 

\subsection{Calculating $\beta$, $r_c$ and $\rho_0$}
Intracluster gas (ICG) is trapped and heated to $10^7 \-- 10^8$K in the cluster gravitational potential. Thermal bremsstrahlung emission from this ICG makes clusters luminous X-ray sources. The main constituents of the ICG are the Hydrogen and Helium, where a good approximation of the number densities is $n_{He} = n_H / 10$. Due to the high temperature the gas can be considered completely ionized and mean molecular weight of the gas including the electrons is given by

\begin{equation}
    \mu \approx \frac{1+2\sum\limits_{Z>1}w_Z Z}{2+\sum\limits_{Z>1}w_Z (Z+1)}\,,
\end{equation}

\noindent where $Z$ is the atomic number and $w_Z$ the the relative weight e.g., $w_2 =0.1$ and $w_Z=0$ for $Z>2$. This gives $\mu \approx 0.61$ and $\rho(r) \approx 1.17n_e m_p$. Therefore, the electron number density should also follow the relation 
\begin{equation}
    n_e(r) = n_{e0} \left[ 1+\left(\frac{r}{r_c}\right)^2\right]^{-3\beta/2}\;.
\end{equation}

For the X-ray surface brightness, i.e. the number of photons detected in a defined energy
range per unit time and per unit solid angle, one has
\begin{equation}
    S \propto \int_{-\infty}^{\infty} n_e^2 dl
    \qquad\implies \qquad
    S \propto  \int_{-\infty}^{\infty} \left[ 1+\left(\frac{r}{r_c}\right)^2\right]^{-3\beta} dl \;,
\end{equation}

\noindent where the integration in along the light of sight. This integral can be reduced to a form 

\begin{equation}
    S(R)  = S_0  \left[ 1+\left(\frac{R}{r_c}\right)^2\right]^{-3\beta+\frac{1}{2}} \;,
\end{equation}

\noindent Here $R$ denotes as the projected distance from the center of the cluster. This can be fitted with the observed surface brightness profile and measure the values for $S_0$, $r_c$ and $\beta$. Thereby $\rho_o$ can be calculated. The measured values of these for different clusters are taken from~\cite{reiprich2003cosmological}.

For calculating the total gas mass of a cluster, we also need the outer radius of the cluster. The above equation holds to a point where the cluster X-ray emission is lost in the background~\cite{sanders1994faber}. For parametric analysis, define the outer radius of the cluster as the radius where the density of the cluster is about 250 times the cosmological baryon density, and this radius is taken as $r_{250}$. So, from Eq.~\ref{BaryonMass1}, we can get the total baryonic gas mass of the galaxy cluster.

\paragraph{Dynamic mass of a galaxy}
The basic assumption that is taken for calculating the dynamic mass of a cluster is that the gas inside the cluster is in hydrostatic equilibrium, which gives

\begin{equation}
    \frac{dP(r)}{dr} = -\frac{\rho(r) G M_d(r) }{r^2}
\end{equation}

\noindent where $P(r)$ represents the gas pressure, $G$ is the gravitational constant. $M_d(r)$ is the gravitating mass of the cluster inside a radius $r$. With the ideal gas equation, 

\begin{equation}
    P=\frac{k_B}{\mu m_p} \rho T \,,
\end{equation}

\noindent this leads to 

\begin{equation}
    M_d(r) = \frac{k_B T r^2}{\mu m_p G}\left(\frac{1}{\rho (r)}\frac{d\rho (r)}{dr}+
    \frac{1}{T(r)}\frac{dT(r)}{dr}\right)\,.
    \label{DynamicMass1}
\end{equation}

\noindent where $T$ is the temperature of the cluster, $m_p$ is the mass of proton, $\mu=0.609$ is mean atomic weight and $k$ is the Boltzmann constant. From XMM-Newton it is also seen that the clusters are isotheomal at least up to half of the virial radius. In this analysis we consider the clusters to be completely isothermal. For isothermal clusters, we can take the temperature derivative to be $0$. Putting the value of $\rho(r)$ we get the dynamical mass of the cluster as

\begin{equation} 
M_d(r) = \frac{3\beta kT}{\mu m_pG}\left(\frac{r^3}{r^2+r_c^2}\right)\,.
\label{eq:M_N}
\end{equation}

\paragraph{Machian Gravity Formulation}

Under Machian gravity, the gravitational equation will change. As we have not considered any dark matter, the matter component will be given by the baryonic component only. Therefore, by putting the gravitational equation in the hydrostatic equilibrium equation and doing similar analysis, we get 

\begin{equation}
    M_{b}(r)\left\{1+K\left[1-\exp(-\lambda r)\left(1+\lambda r\right)\right]\right\} = \frac{3\beta kT}{\mu m_pG}\left(\frac{r^3}{r^2+r_c^2}\right)\,,
    \label{eq:Machiangrav}
\end{equation}

\noindent $K$ and $\lambda$ are the background dependent quantities. Earlier, for explaining the spiral galactic Tully-Fisher relation, we used $K = \left(\sqrt{\frac{M_c}{M_b(r)}}-1\right)$. However, it does not imply that galaxy clusters should also follow a similar relation because the mass distribution is different than spiral galaxies. Secondly, they are gravitationally bound systems, while galaxy clusters are supported by hydrostatic equilibrium.

%Clusters are the largest gravitating body in the universe. 
As $\lambda^{-1}$ is of the order of 10s to 100s of kpc, as can be seen for galaxy velocity profile study~\cite{das2023aaspects}, at the edge of the clusters, we can expect $\exp(-\lambda r)\rightarrow 0$. There the equation will take the form 
\begin{equation}
\label{eq:MachMassFormulaOuter}
        M_b(r_{out})(1+K) = \frac{3\beta kT}{\mu m_pG}\left(\frac{r_{out}^3}{r_{out}^2+r_c^2}\right)\,,
\end{equation}

\noindent where $r_{out}$ is the outer radius of the cluster.  

%where the cluster density is 250 times the cosmic baryon density. Also for $\frac{r_{out}}{r_c} \gg 1$ we can take the limit to get 
%\begin{equation}
%    M_b(r_{out})(1+K) \propto \beta T r_{out} 
%\end{equation}
%\noindent Therefore, we can use the data to calculate the form of $K$ for clusters. 

\begin{figure}
\centering
\includegraphics[trim=1.0cm 1.0cm 0.0cm 0.0cm, clip=true, width=0.8\columnwidth]{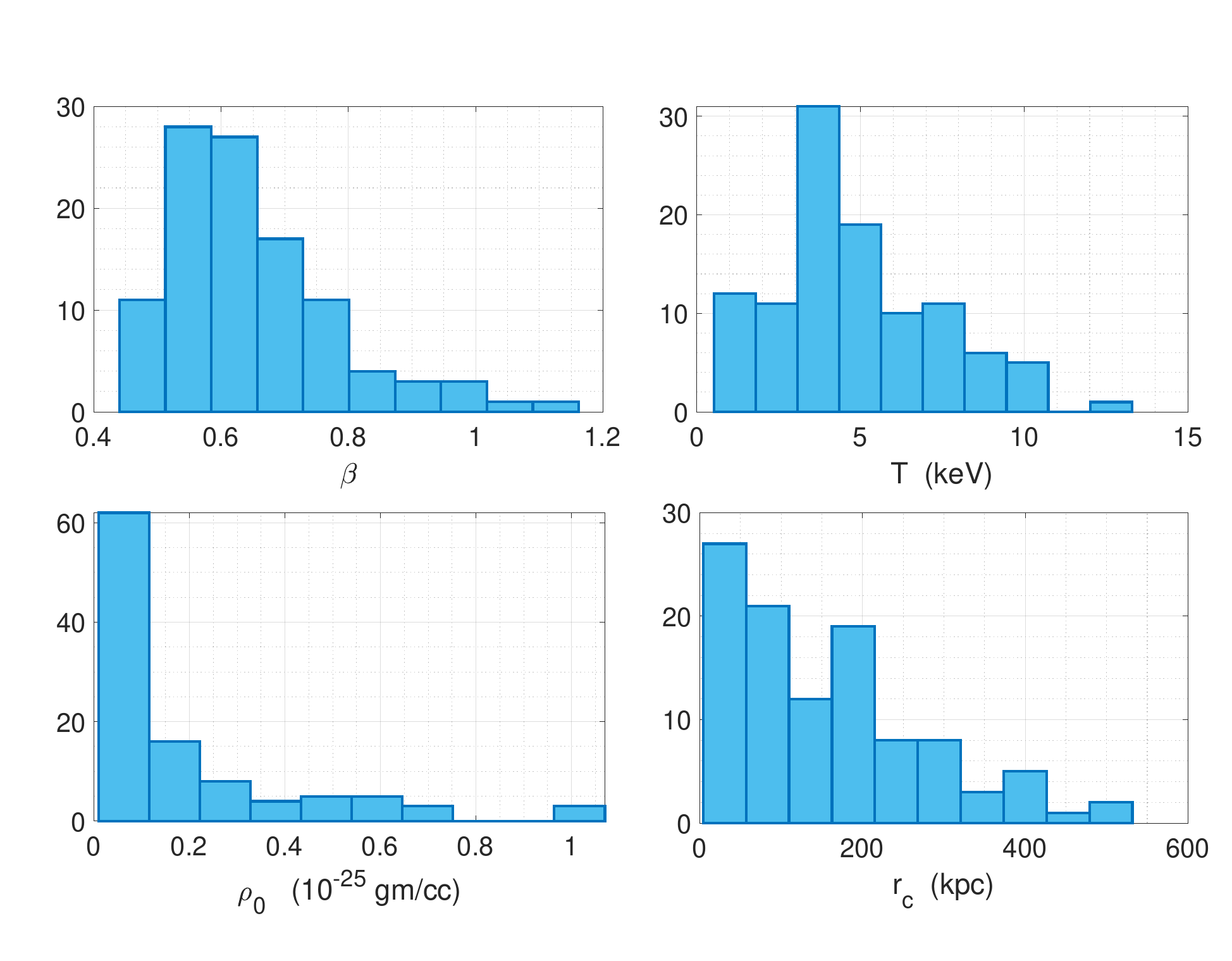}
\caption{\label{fig:fig1}The plot presents the distributions of several physical properties for the sample of 106 galaxy clusters. The distribution of $\beta$ is comparatively smooth, exhibiting a pronounced peak at approximately $\beta \approx 0.56$. In the upper-right panel, the distribution of the isothermal temperature appears relatively broad, with a maximum around $T \approx 4\,\mathrm{keV}$. Although the central gas density of the clusters spans a wide range, it remains below $0.1 \times 10^{-25}\,\mathrm{g\,cm^{-3}}$ for more than $60\%$ of the systems. Finally, the core radius $r_c$ also shows substantial cluster-to-cluster variation, indicating significant diversity in the spatial extent of the intracluster medium. }
\end{figure}

%//////////////////////////////

%\noindent As the background of the objects are changing  the background of the system will change significantly from the center to the edge and hence $K$ and $\lambda$ should also change with the radius i.e $r$. However, here for simplicity we use radius independent values $K$ and $\lambda$ . We fix $\lambda = 0.07kpc^{-1}$ for all the clusters and only change $K$.

%For testing our theory with the observational results we pick up $6$ galaxy clusters and match the dynamical mass profile  from our theory with the physical mass given by King's $\beta$-model. We fix the values of $K$ just by eye estimation and  not by any parameter estimation package. The plots shows that the dynamical mass calculated using our theory (shown in orange color)  matches well with the gas mass calculated using the King's $\beta$-model (show in yellow ocher). However, using Newtonian theory of gravity (cyan dotted lines) we need more dynamical masses than observed, i.e. we need the dark matter. Recently it is seen that   the MOND \cite{Milgrom1983,Sanders2002} also doesn't provide proper explanation for the galaxy cluster masses. Therefore, it can be said  that the new theory i.e. Machian Gravity (MG) is better then others alternate theories in all prospects. 

\section{Analysis and Results}

\subsection{Parametric Galaxy Cluster Sample of 106 Clusters from the Reiprich Catalogue~\cite{reiprich2003cosmological}}
We have analyzed our results with two sets of samples. The first set consists a total of 106 galaxy clusters taken from~\cite{reiprich2003cosmological}. The properties of the sample with their error bars are shown in the table~\ref{Table1}. For understanding the sample, in Fig.~\ref{fig:fig1} we have plotted the distribution of $\beta$, $T$, $\rho_0$, $r_c$ for all the clusters. We can see that the $\beta$ is roughly centered around $0.6483\pm 0.1349$. The temperature distribution of the clusters is roughly flat between $(0,10) {\rm keV}$ with a peak at about $4-5 {\rm keV}$. For most of the clusters, $\rho_0$ is less than $0.2 {\rm gm/cc}$ and $r_c$, i.e. the core radius of the cluster is mostly within $500{\rm kpc}$. 
The outer radius of the cluster, $r_{out}$, is taken as the radius where the density of the cluster is about 250 times the cosmological baryon density. In different places this radius is also denoted by  $r_{250}$.
We use Eq.~\ref{BaryonMass1} to calculate the baryon mass in the galaxy cluster. As $\beta$ centered around $0.65$, at $\frac{r}{r_c}\gg 1$, the baryonic mass go as $M_b(r) \sim r^{1.05}$ for most of the clusters according to the Eq.~\ref{MassBaryon}. The dynamic mass (calculated from Newtonian mechanics) is given by Eq.~\ref{DynamicMass1}, and for $r\gg r_c$, the dynamic mass $M_d(r) \sim r$ as given by Eq.~\ref{eq:M_N}.  Therefore, roughly, we can see that for $r\gg r_c$, the ratio of the dynamic mass and the baryonic mass becomes almost independent of $r$ for the mean radius. However, for different galaxy clusters, the ratio may depend on different powers of $r$. 
%Therefore, for $r\gg r_c$ we can have $M_d(r) \sim M_b(r)^\alpha$, where $\alpha$ can have quite a large range. No fixed form of power of $M_b(r)$ in $K$. However, these are just fitting formulae. There is no reason to believe that there should be some fixed power-law relation between $M_b(r)$ and $M_d(r)$. The error bar on $M_d(r)$ is obtained from the error bars on $T$, $\beta$, and $r_c$. 

\begin{figure}
\centering
\includegraphics[trim=0.0cm 0.0cm 1.0cm 0.0cm, clip=true, width=0.49\columnwidth]{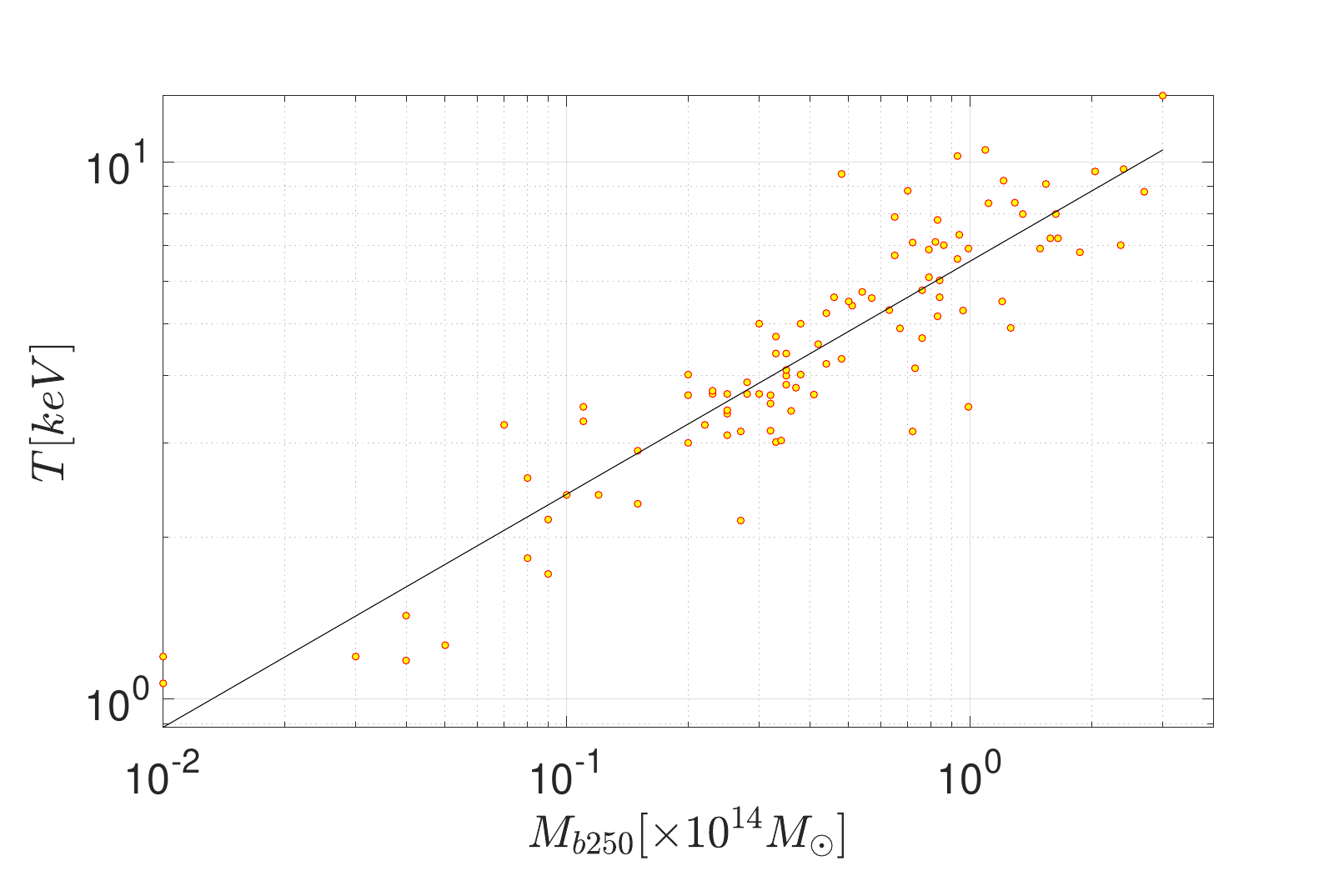}
\includegraphics[trim=0.0cm 0.0cm 1.0cm 0.0cm, clip=true, width=0.49\columnwidth]{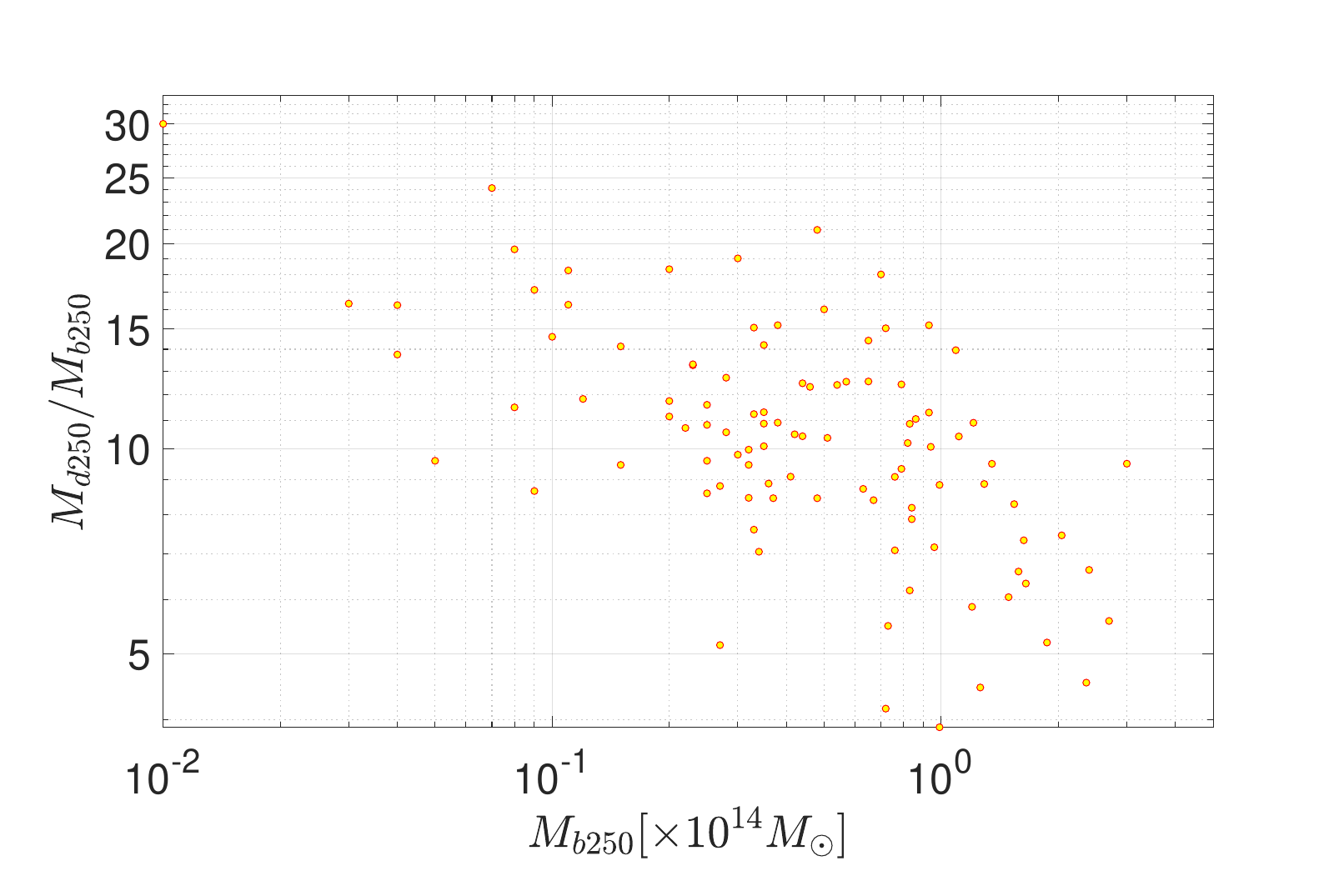}
\caption{\label{fig:ClusterProperties}The plot illustrates various characteristics of the clusters. On the left, the graph depicts the relationship between $T$ and $M_{b250}$. The black curve represents the best-fit regression line derived from the data points, indicating an approximately quadratic dependence, with $M_{b250} \propto T^2$. On the right, the plot demonstrates $M_{d250}/M_{b250}$ as a function of $M_{b250}$. It is exhibiting a slight negative correlation, the ratio remains relatively constant (about 10) across the range of $M_{b250}$ values. However, the number of data points is small and their dispersion is too large to support any definitive conclusions.}
\end{figure}

The total baryon mass and dynamic mass of the galaxy cluster upto radius $r_{250}$ is taken as $M_{b250}$ and $M_{d250}$. In Fig.~\ref{fig:ClusterProperties}, we have plotted the temperature of the 106 clusters as a function of $M_{b250}$.  The best-fit regression lines give 
\begin{equation}
 \log(T) = 0.43\log(M_{b250}) + 0.82  \,, 
\end{equation}
\noindent i.e. roughly $M_{b250} \sim T^2$ which is in agreement with previous analysis~\cite{sanders1994faber}. The ratio of the Newtonian dynamic mass, $M_{d250}$, and $M_{b250}$ is roughly constant, which is about $10$. However, there is a slight negative correlation between the ratio and the $\log(M_{b250})$. A regression fit to the total dynamic mass and the baryonic mass is given by

\begin{equation}
 \log(M_{d250}) = 0.7\log(M_{b250}) + 3.87   \,.
 \label{eq:Md250Mb250}
\end{equation}

\noindent From Eq.~\ref{eq:MachMassFormulaOuter} we can see that at the outer edge of a galaxy cluster $M_d \propto M_b (1+K)$. Therefore, the above relation shows that for clusters $K=\left(\frac{M_c}{M_b}\right)^{0.3}-1$ may provide a better fit than the square root relation, assuming $M_c$ is constant. However, the mass distribution of each galaxy cluster can be different. Therefore, there is no reason that $M_c$ can be constant. Therefore, we fit this first set of data with the same square root relation as used for explaining the galaxy velocity profiles. Our analysis shows that the relation fits the galaxy cluster mass profiles well.

\begin{figure}
\centering
\includegraphics[trim=0.0cm 0.0cm 1.0cm 0.0cm, clip=true, width=0.49\columnwidth]{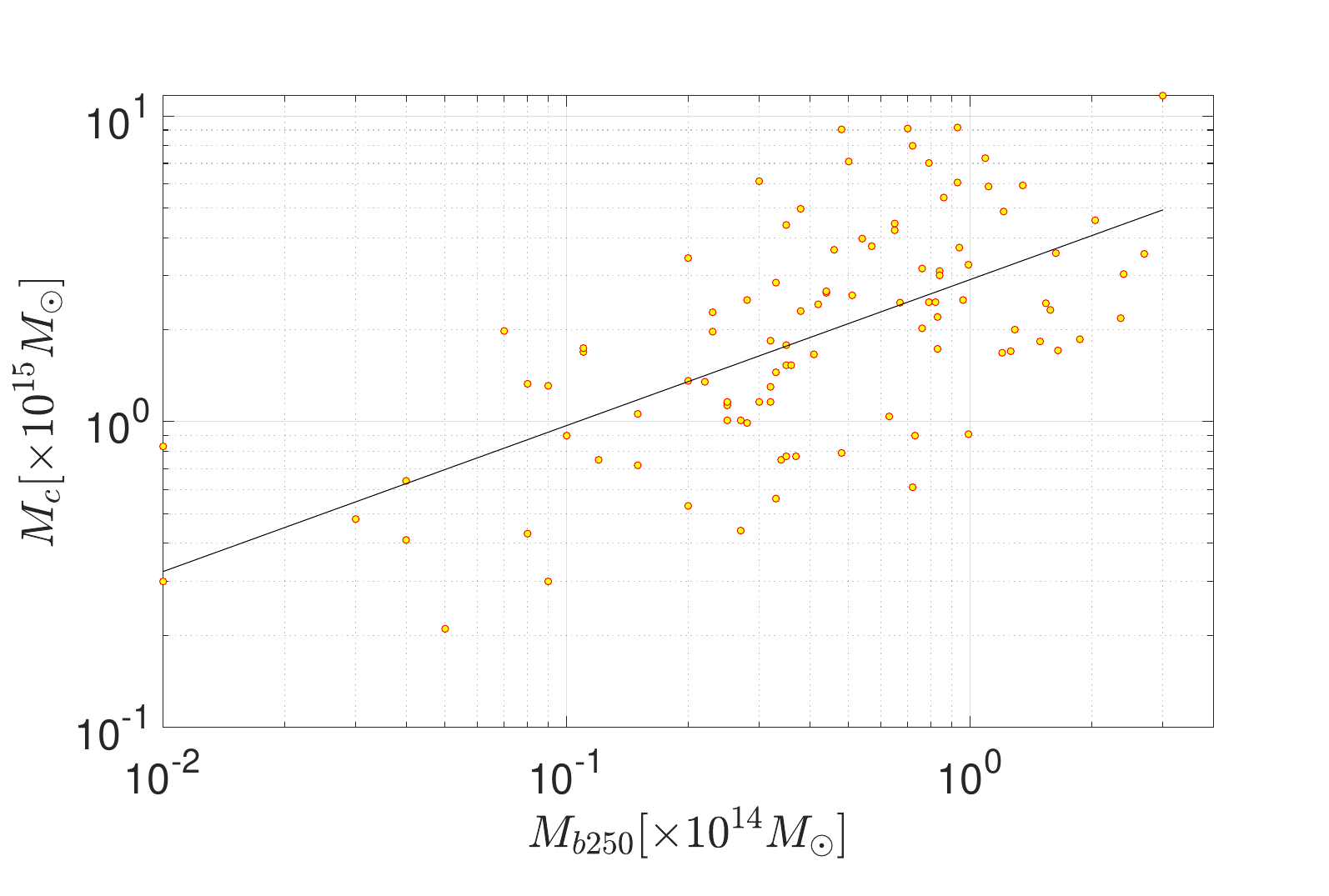}
\includegraphics[trim=0.0cm 0.0cm 1.0cm 0.0cm, clip=true, width=0.49\columnwidth]{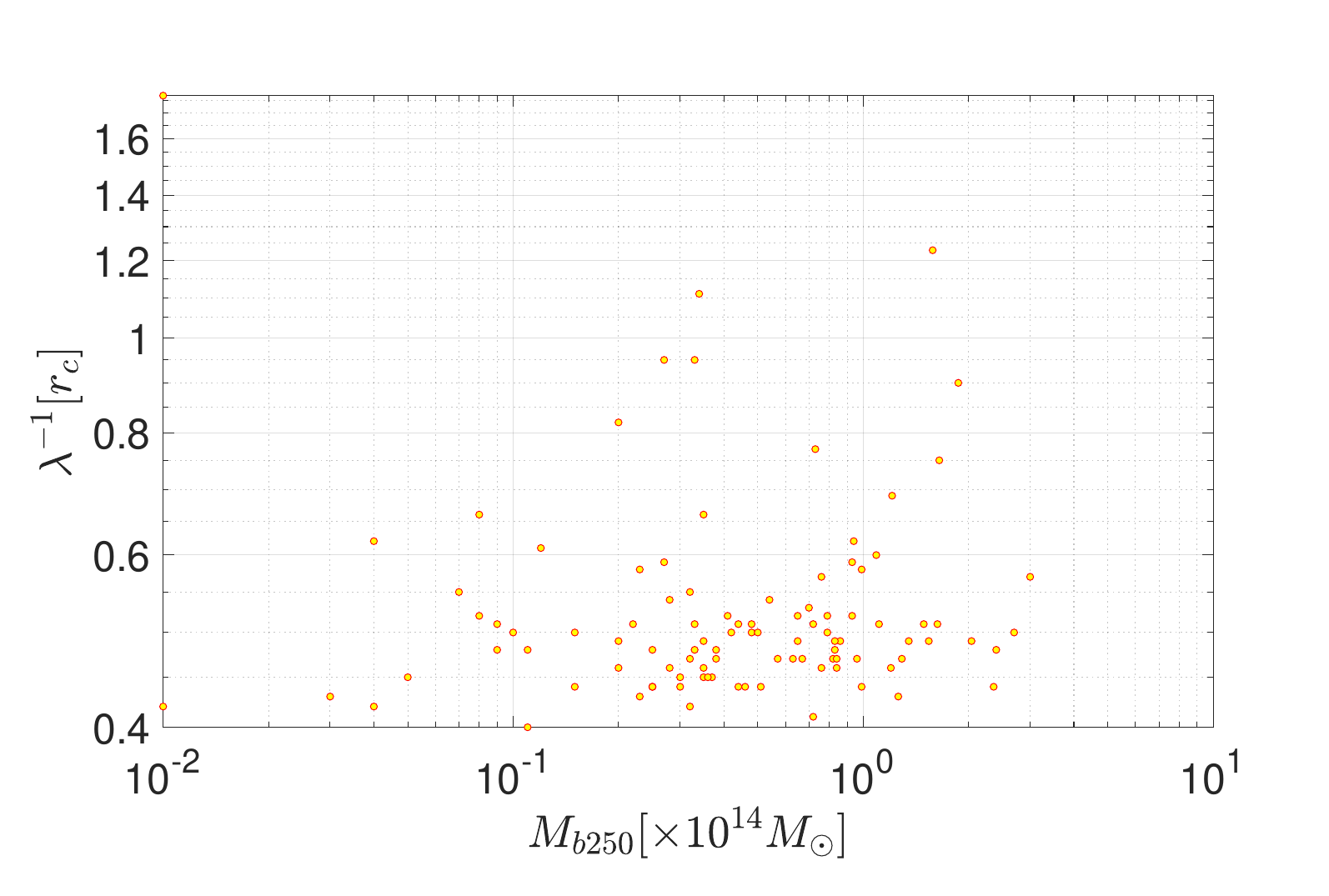}
\includegraphics[trim=0.0cm 0.0cm 1.0cm 0.0cm, clip=true, width=0.49\columnwidth]{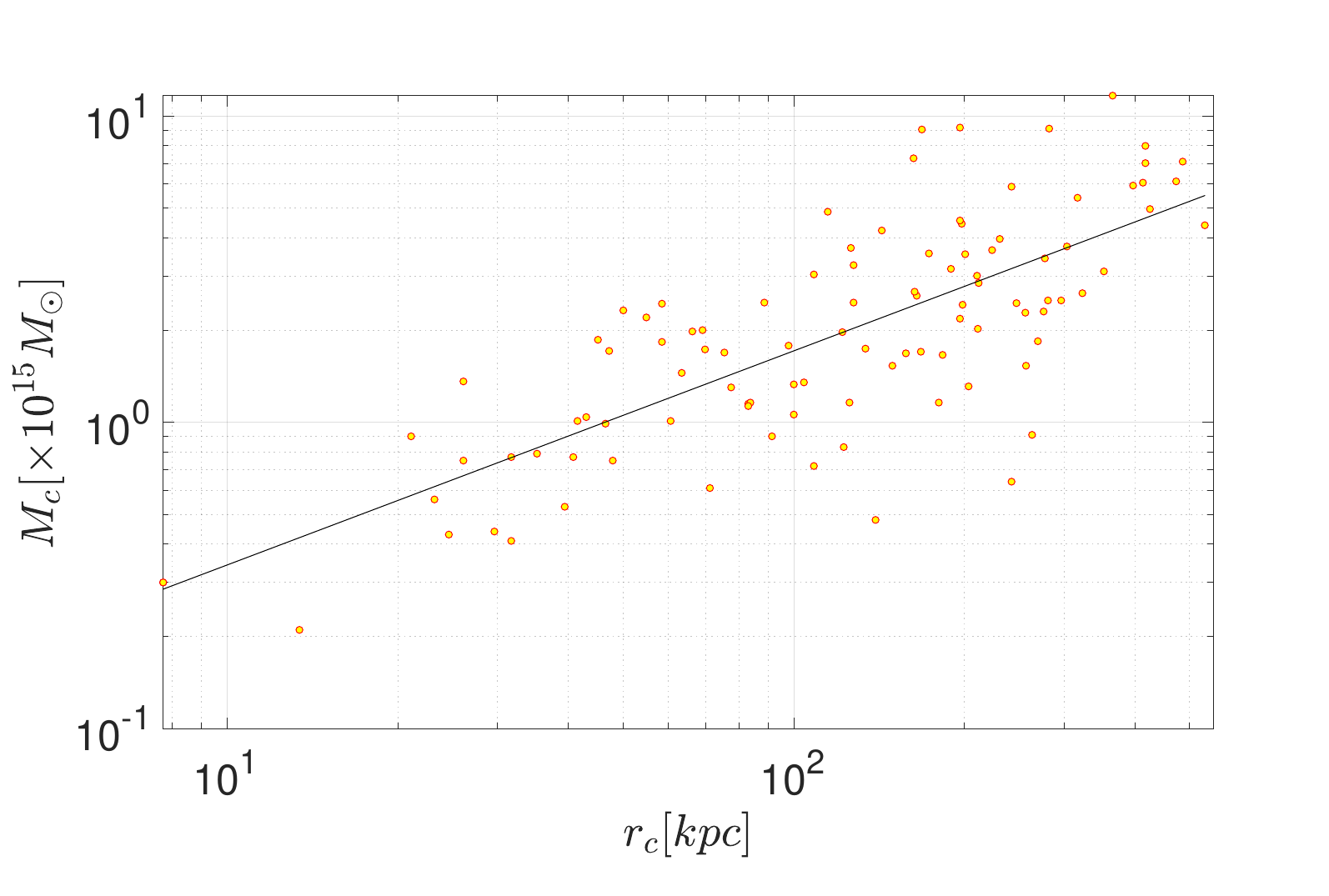}
\includegraphics[trim=0.0cm 0.0cm 1.0cm 0.0cm, clip=true, width=0.49\columnwidth]{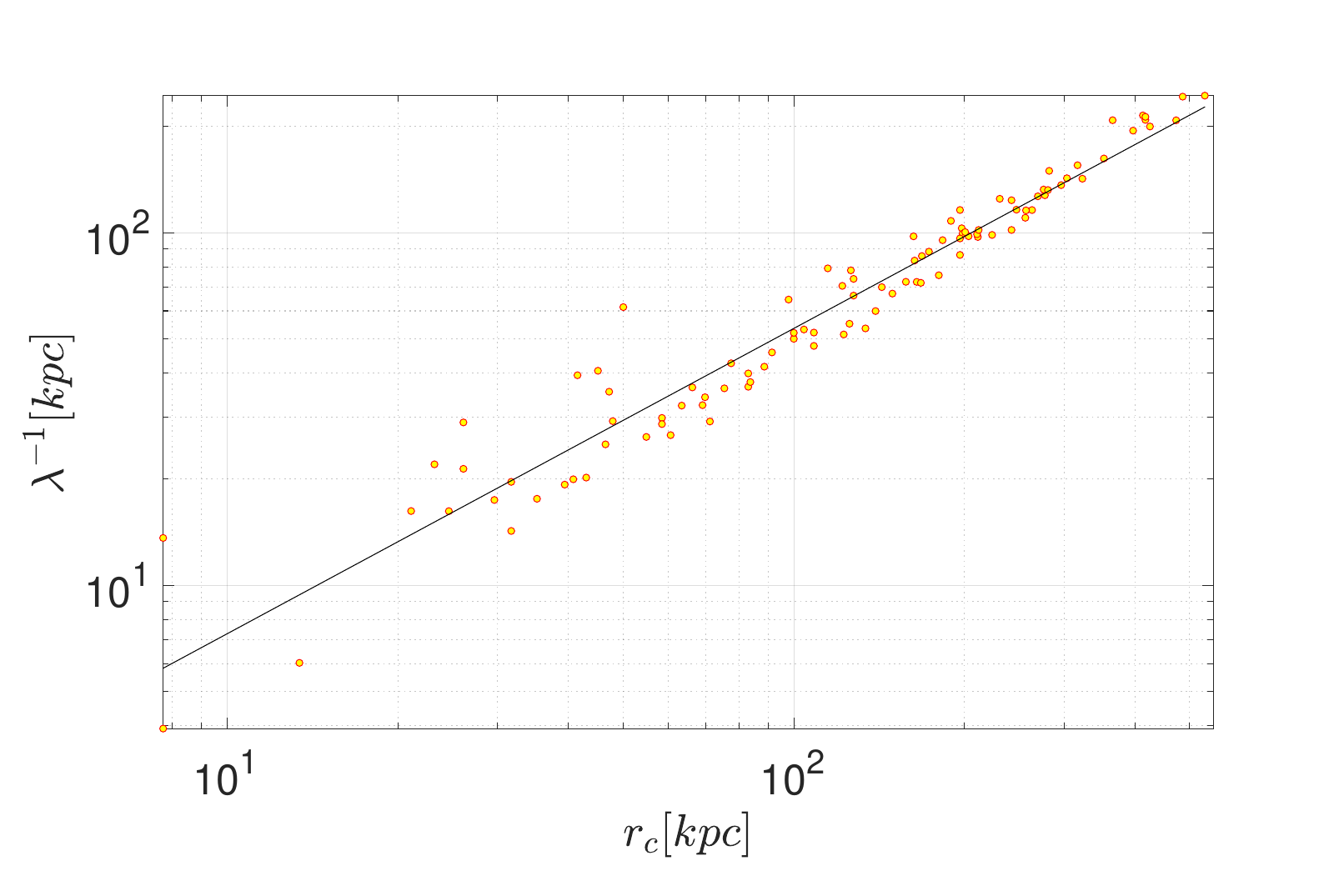}
\caption{\label{fig:ClusterProperties4val}
The figure presents $M_c$ and $\lambda^{-1}$ as functions of $M_{b250}$ and $r_c$. The results indicate that there is nearly no correlation between $\lambda^{-1}$ and $M_{b250}$. However, all other depicted relationships exhibit a robust positive correlation.
}
\end{figure}

%By optimizing the parameter $M_c$ and $\lambda^{-1}$ 
We fit the left-hand side of Eq.~\ref{eq:Machiangrav} against the Newtonian dynamic mass of the cluster given by Eq.~\ref{eq:M_N} and optimize $M_c$ and $\lambda^{-1}$ by minimizing the $\chi^2$ given by 

\begin{equation}
    \chi^2 = \sum_i\left[M_d(r_i) - M_b(r_i)\left[1+\left(\sqrt{\frac{M_c}{M_b(r_i)}} - 1\right)\left(1-e^{-\lambda r_i}\left(1+\lambda r_i\right)\right)\right]\right]^2 \Bigg/ \Delta M_d(r_i )^2\,.
\end{equation}

\noindent We divided the entire radius into 20 logarithmic spacing points. In each point, we calculate the dynamic and the baryonic mass. The upper and lower error bars of $M_d$ are calculated from the error bars of $\beta$ and $T$. We use the root mean square of these two error bars in the $\chi^2$ formula. Fig.~\ref{fig:ClusterProperties4val} shows how the value of best fit $M_c$ and $\lambda^{-1}$ varies as a function of $M_{b250}$ and $r_c$. We can see a very strong correlation between $M_c$ and $M_{b250}$. As $M_c$ depends on the mass distribution, intuitively, we may expect a correlation between $M_c$  and $M_{b250}$. From Fig~\ref{fig:ClusterProperties4val} we can see that they are positively correlated and the regression line between the logarithmic quantities is given by
%The relation is given by 

\begin{equation}
\log(M_c) = 0.4776\log(M_{b250}) + 0.4649\,.    
\label{eq:McMb}
\end{equation}

\noindent However, without a proper theory of gravity and without knowing how the background affects gravity, it's impossible to know the exact theoretical relation. This result also aligned with the previous result in Eq.~\ref{eq:Md250Mb250}, where we had seen that a better fit will be $K+1\propto M_{b250}^{-0.3}$, while %From 
Eq.~\ref{eq:McMb}, we get $K+1\propto M_{b250}^{(0.48-1)/2} = M_{b250}^{-0.26}$. 
%If we investigate the relation between $\lambda^{-1}$ (as a function of $r_c$) and $M_b$, then we can see 
Our results show that $\lambda^{-1}$ is almost independent on $M_b$. %Of course, in kpc we should expect some positive correlation.  with $M_b$.
We have also plotted $M_c$ and $\lambda^{-1}$ against the core radius of the clusters, $r_c$. Both of them show a positive correlation. The best-fit regression lines give
\begin{equation}
    \log(M_c) = 0.6988\log(r_c) - 1.1643\,, \qquad\qquad
    \log(\lambda^{-1}) = 0.8652\log(r_c) - 0.0019\,.    
\end{equation}

The core radius, $r_c$, of a cluster somehow gives its size and shape of the background matter distribution. Therefore, we can expect a strong relation between the $M_c$ and $r_c$. The relation of $r_c$ with $\lambda^{-1}$ is very close to linear. However, here one should note that $\lambda^{-1}$ are very small compared to the $r_{250}$. Therefore, as discussed before the exponential term containing the $\lambda^{-1}$ term will have almost no effect in the analysis. Therefore, the constraints on $\lambda^{-1}$ may not be very strong, and are mostly determined by the data-points at low radius. 

%Again, the exact reason is not known as we don't have an exact form of $K$ and $\lambda^{-1}$ from the theory. 

\begin{figure}
    \centering
    \includegraphics[width=0.32\columnwidth]{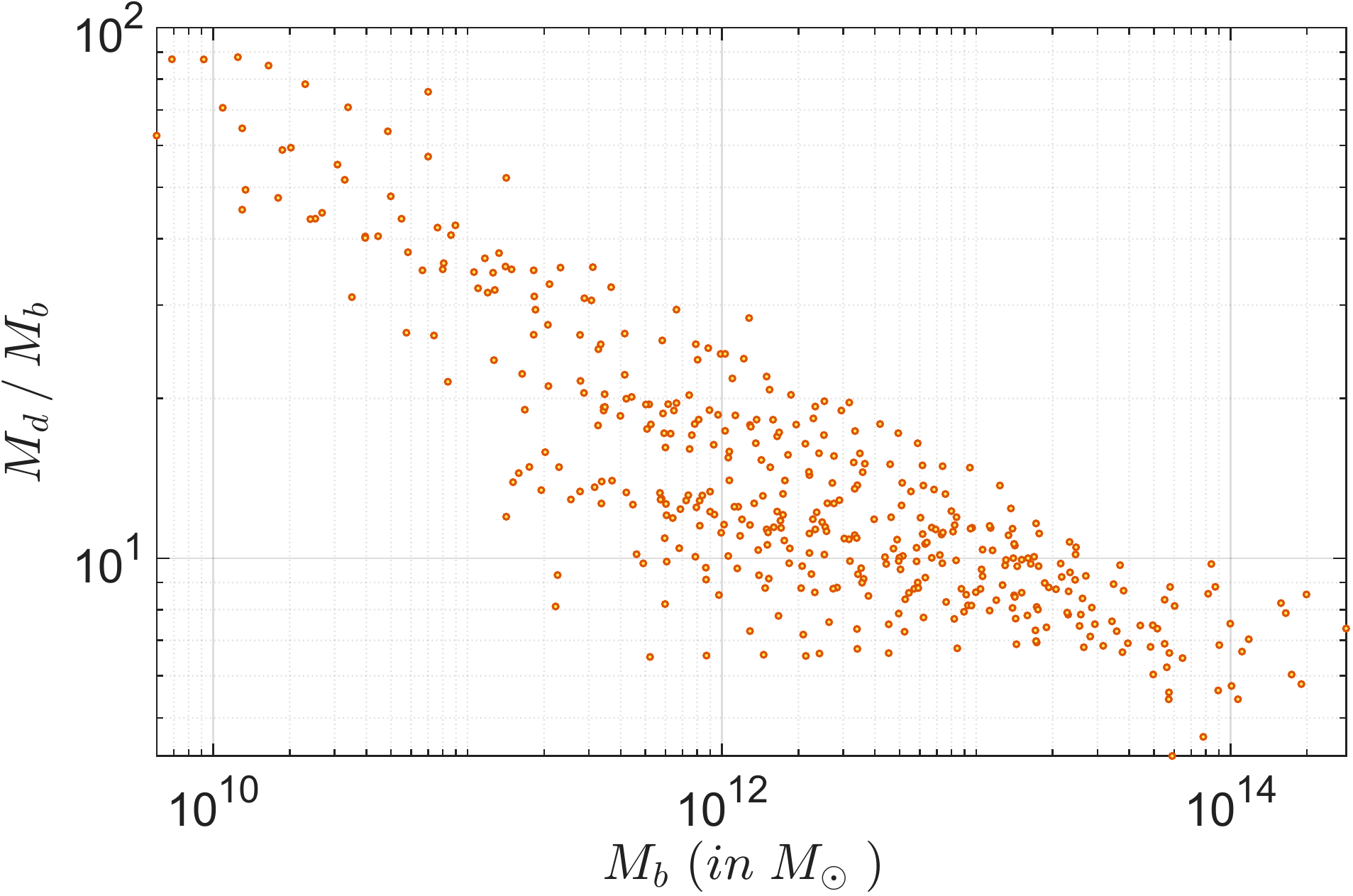}
    \includegraphics[width=0.32\columnwidth]{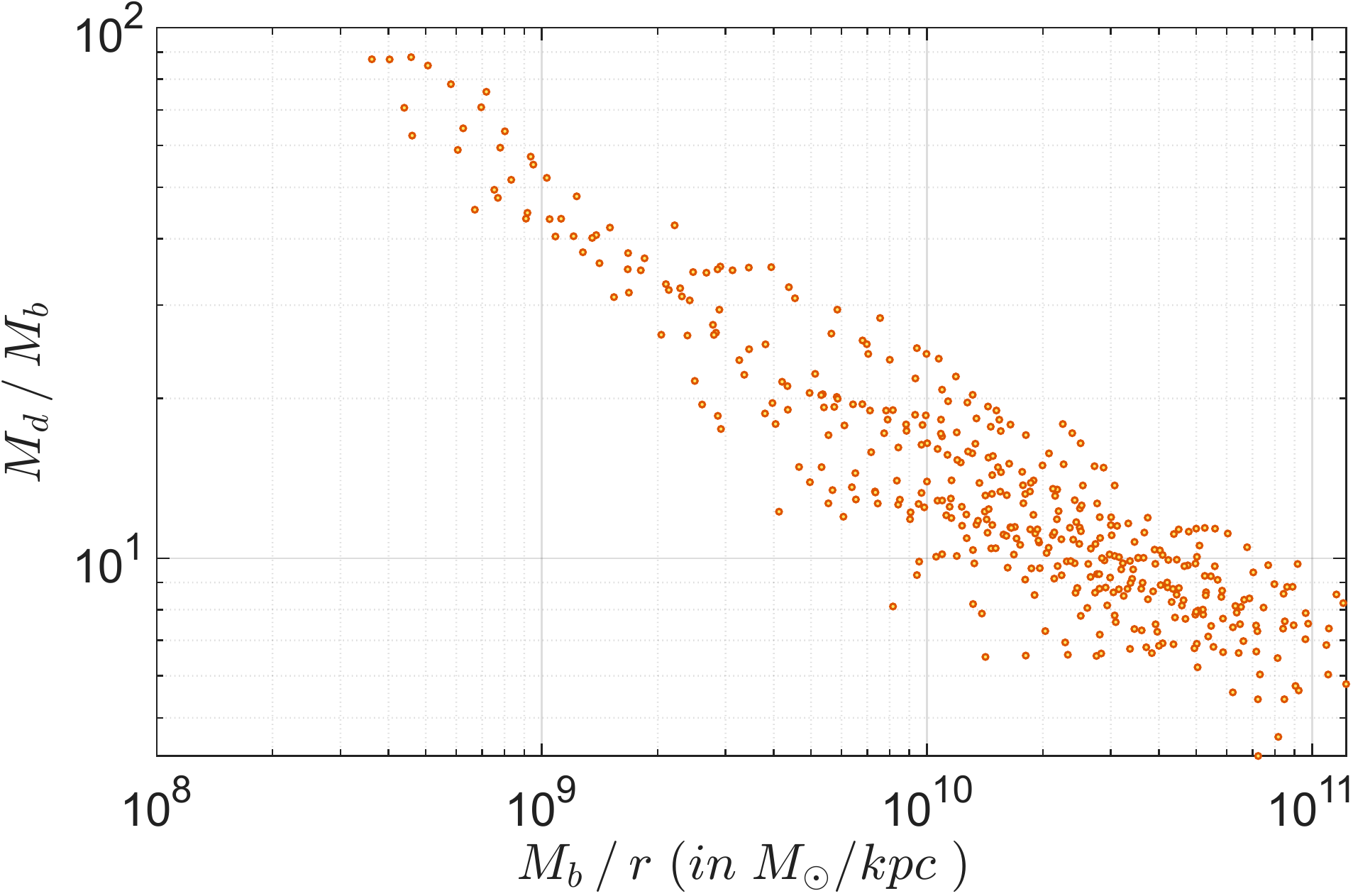}
    \includegraphics[width=0.32\columnwidth]{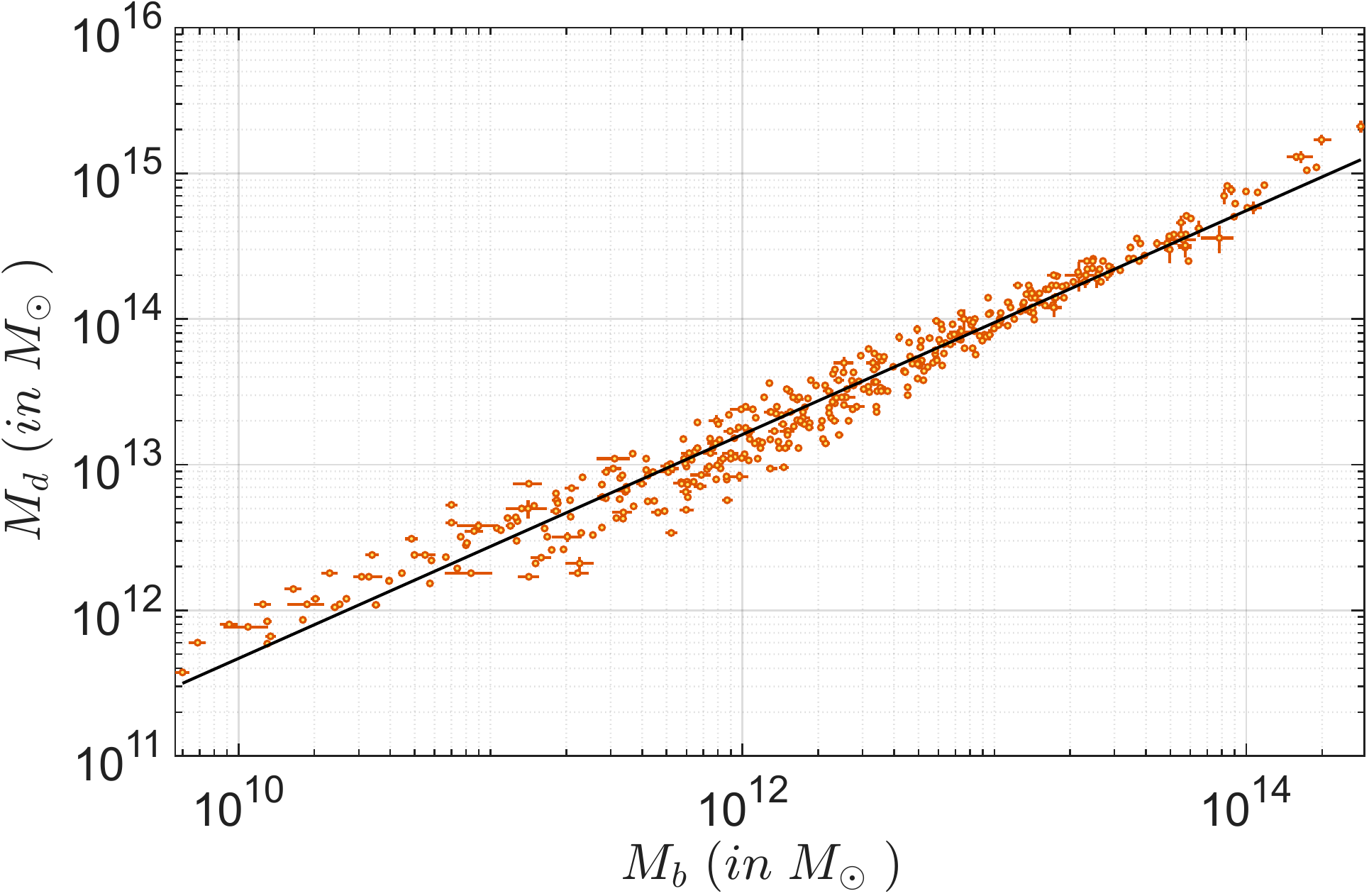}    
    \includegraphics[width=0.32\columnwidth]{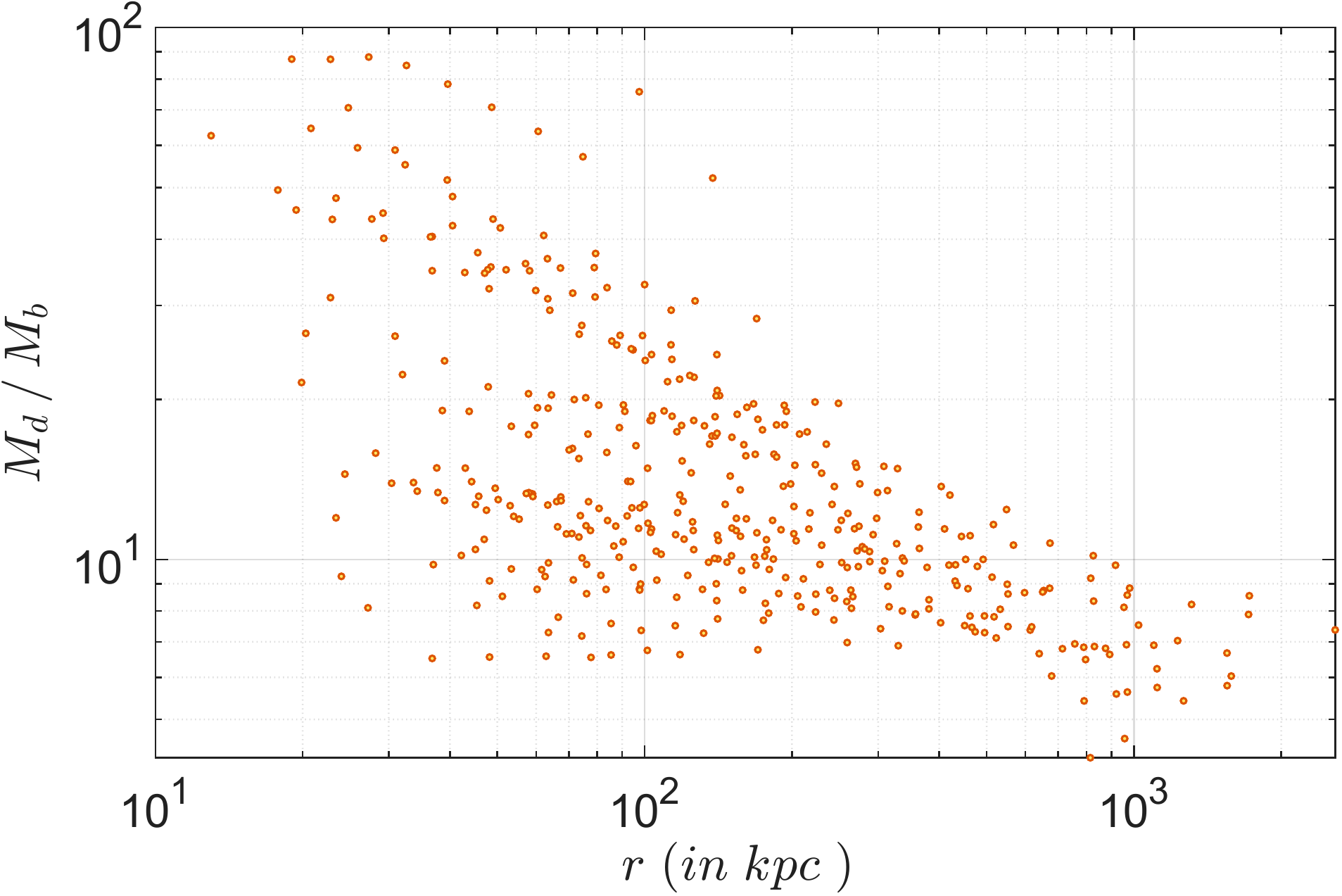}   
    \includegraphics[width=0.32\columnwidth]{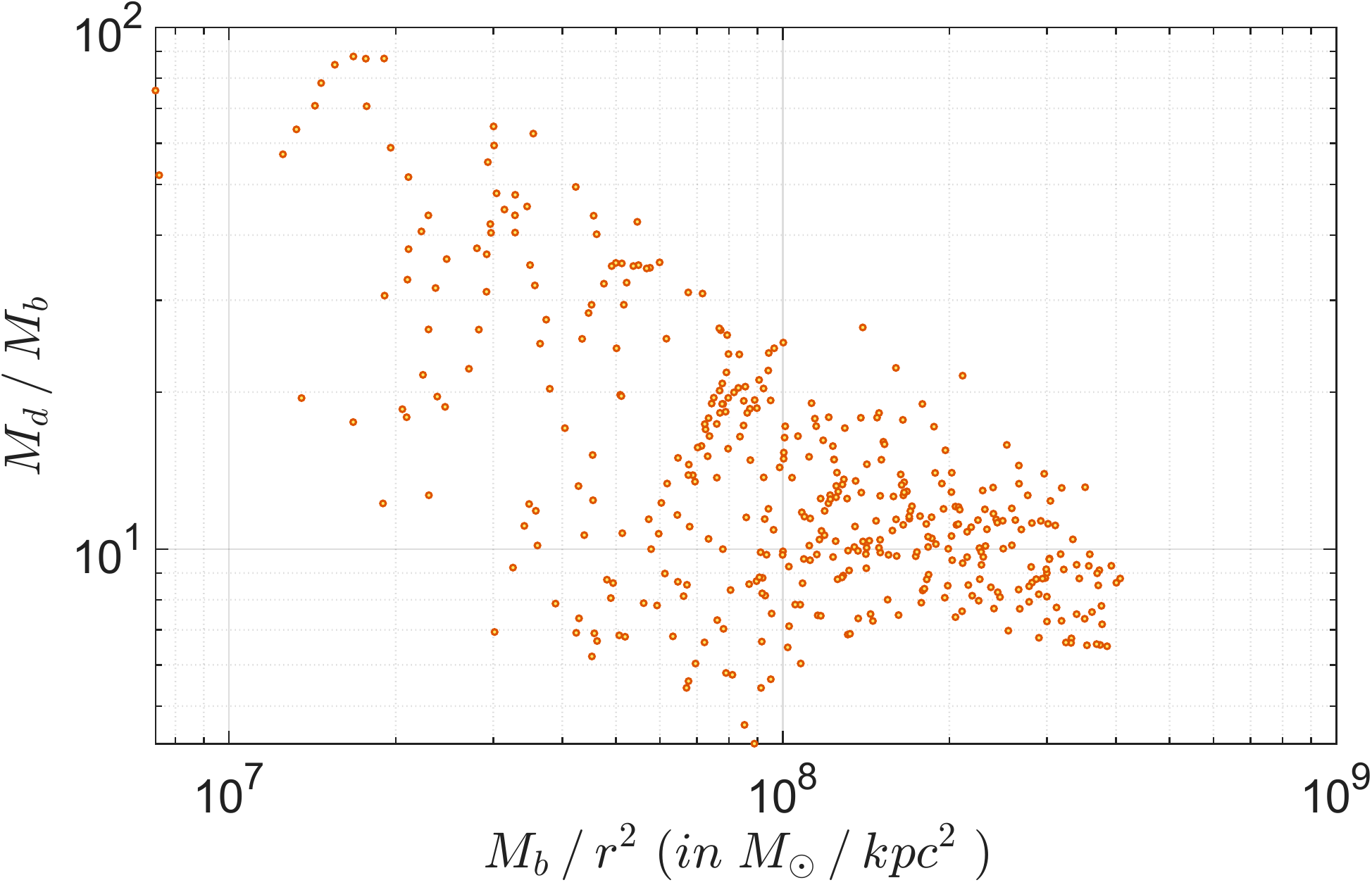}
    \includegraphics[width=0.32\columnwidth]{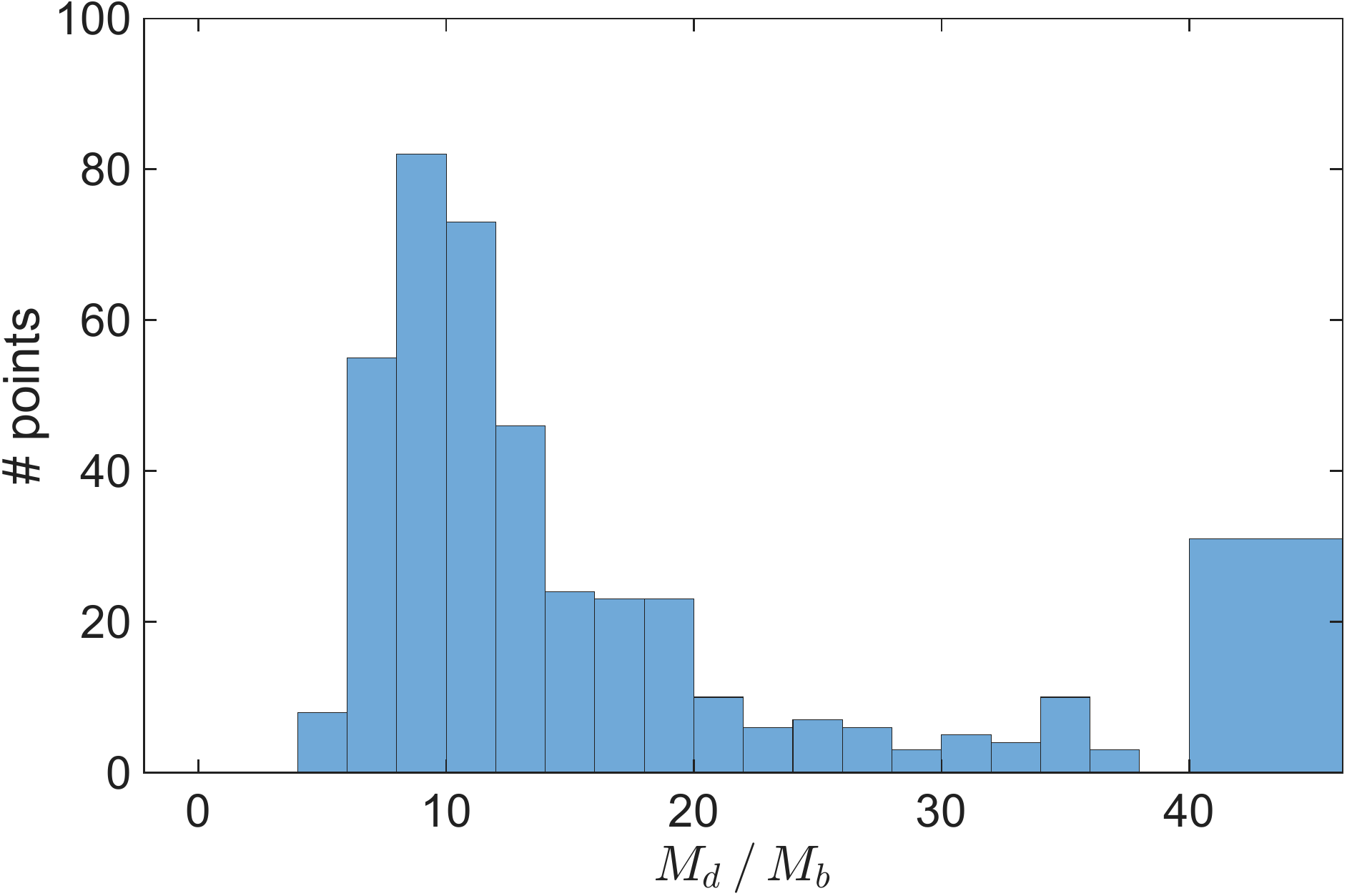}
    \caption{We analyze all data points from the HIFLUGCS galaxy cluster sample, comprising 419 measurements drawn from 45 clusters. When examining the ratio \(M_d/M_b\) as a function of \(M_b\), we find a substantial degree of scatter, although a clear negative trend is apparent. The scatter becomes even more pronounced when \(M_d/M_b\) is plotted against \(r\) or against \(M_b / r^2\). Despite this, all of these relations exhibit an overall negative correlation. In contrast, a markedly stronger correlation emerges when \(M_d/M_b\) is plotted against \(M_b / r\). Since there is no a priori reason to assume that any of these relations should be linear, we employ the Spearman rank correlation coefficient to quantify the strength and monotonicity of the associations. The rank correlation between \(M_d/M_b\) and \(M_b\) is \(-0.798\), indicating a strong negative correlation. The corresponding coefficient for \(M_d/M_b\) versus \(r\) is \(-0.604\). For \(M_d/M_b\) versus \(M_b / r\), the rank correlation reaches \(-0.874\), which is very strong. The rank correlation between \(M_d/M_b\) and \(M_b / r^2\) is \(-0.526\). If we seek the best-fitting monotonic relation of the form \(M_b / r^\alpha\), we find that the correlation between \(M_d/M_b\) and \(M_b / r^{0.9}\) is maximized at \(-0.874\), which is effectively identical to the correlation obtained with \(M_b / r\). The histogram of \(M_d/M_b\) values shows that, for the majority of data points, the ratio lies in the range \(\sim 5\)–\(15\).
 }
    \label{fig:MainPaperDataUnderstanding}
\end{figure}

In Fig.~\ref{fig:GalaxyCusterMass}, we present our results for all 106 clusters. The greenish-yellow curve denotes the baryonic mass of each cluster, computed using Eq.~\ref{BaryonMass1}, while the solid red curve represents the Newtonian dynamical mass obtained from Eq.~\ref{eq:M_N}. The dotted yellow-ochre curve corresponds to the left-hand side of Eq.~\ref{eq:Machiangrav}, i.e. the dynamical mass inferred from the Machian Gravity formalism. The comparison indicates that the different mass estimates are in good mutual agreement. We additionally display a fit obtained using fixed values of $M_c$ and $\lambda^{-1}$ as a black dotted curve. For this fit, we employ the median-averaged values of the relevant parameters, yielding $\lambda^{-1} = 0.4949\,r_c$ and $M_c = 2.759\times 10^{15} M_{\odot}$. Median averaging is used to mitigate the influence of outliers, which, if included, could substantially bias the inferred parameter values. Even with fixed $M_c$ and $\lambda^{-1}$, the resulting profiles still reproduce the Newtonian dynamical mass of the clusters to a high degree of accuracy. The full set of results from all analyses is summarized in Table~\ref{Table1}. A visual inspection further confirms that the agreement is generally very good.

It is, however, important to note that the best-fit Machian Gravity curves exhibit slightly different curvature compared to the Newtonian dynamical mass profiles. Specifically, at small radii the slope of the Machian Gravity curve is marginally steeper than that of the Newtonian dynamical mass, whereas at larger radii the slope becomes slightly shallower. This behavior reflects the curvature induced by the Machian Gravity term: to decrease (or increase) the overall curvature of the fitted profile, one must decrease (or increase) the exponent of the factor $\frac{M_c}{M_b}$ in the expression for $K$.

We further emphasize that, for this particular sample, we do not possess observational data points at several distinct radii. Instead, the cluster masses are inferred using analytic formulae under the assumption that each cluster is strictly isothermal. Consequently, actual observational data may deviate from the profiles predicted by the fitting formula. However, this level of accuracy should be able to provide a qualitative trend for our analysis.

Finally, we note that the parameters $M_c$ and $\lambda^{-1}$ are reasonably well constrained for all galaxy clusters. The parameter $\lambda^{-1}$ is effectively constrained only by data at small radii, because, as previously discussed, at large radii the contribution from the $\lambda^{-1}$ term becomes negligible.

\begin{figure}
    \centering
    \includegraphics[trim=6.0cm 0.0cm 7.0cm 0.0cm, clip=true, width=0.24\columnwidth]{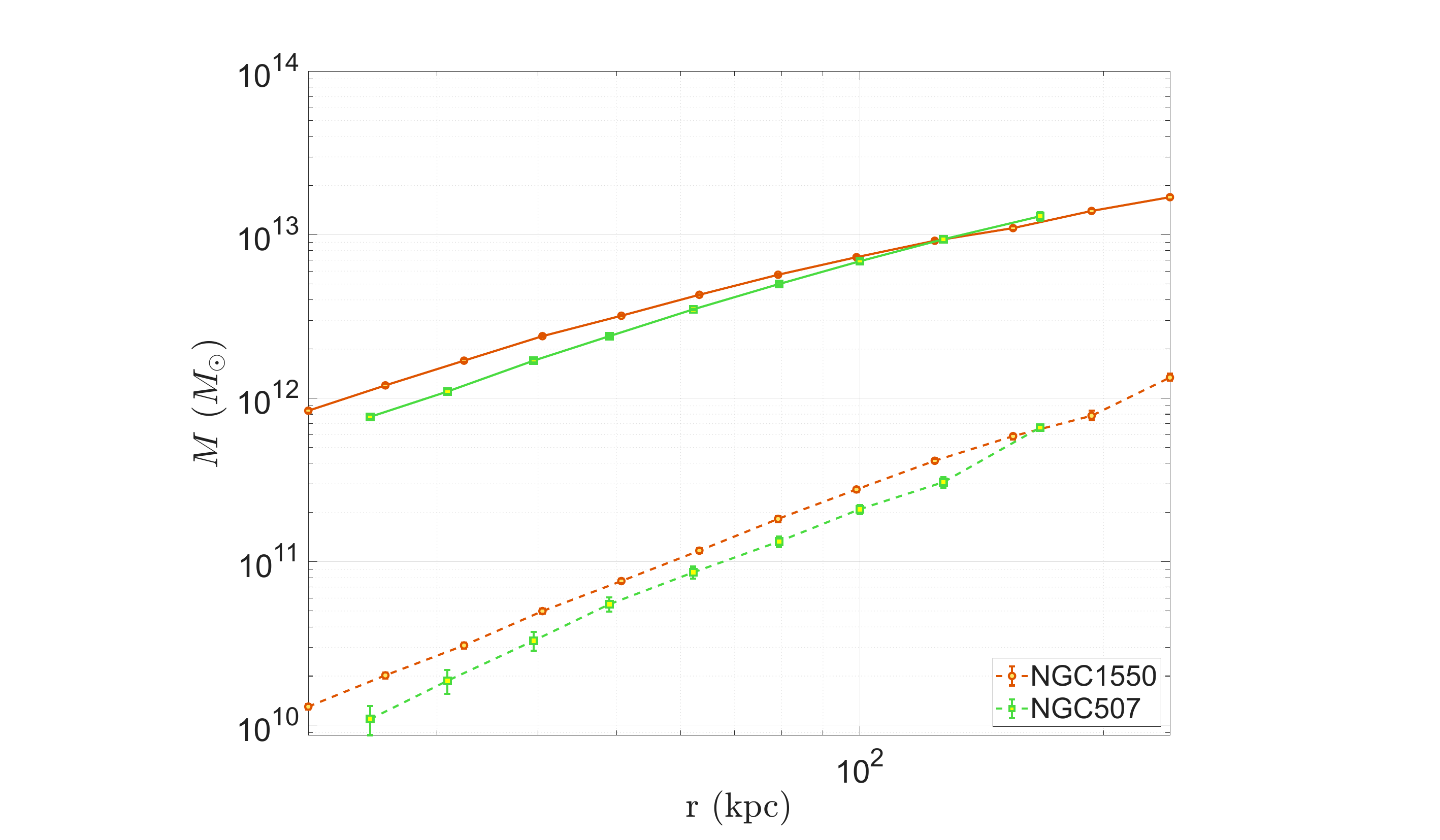}
    \includegraphics[trim=6.0cm 0.0cm 7.0cm 0.0cm, clip=true, width=0.24\columnwidth]{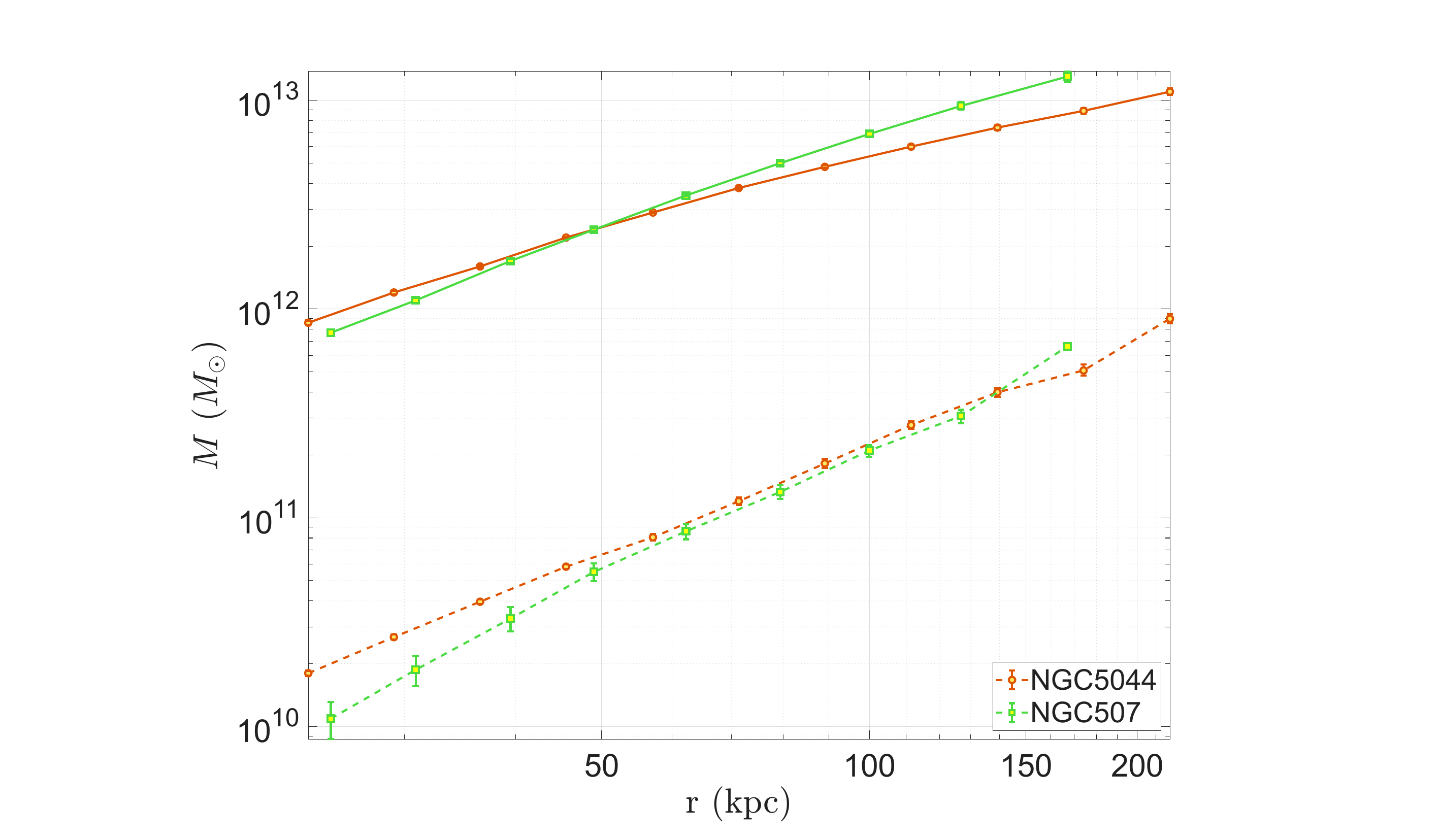}
    \includegraphics[trim=6.0cm 0.0cm 7.0cm 0.0cm, clip=true, width=0.24\columnwidth]{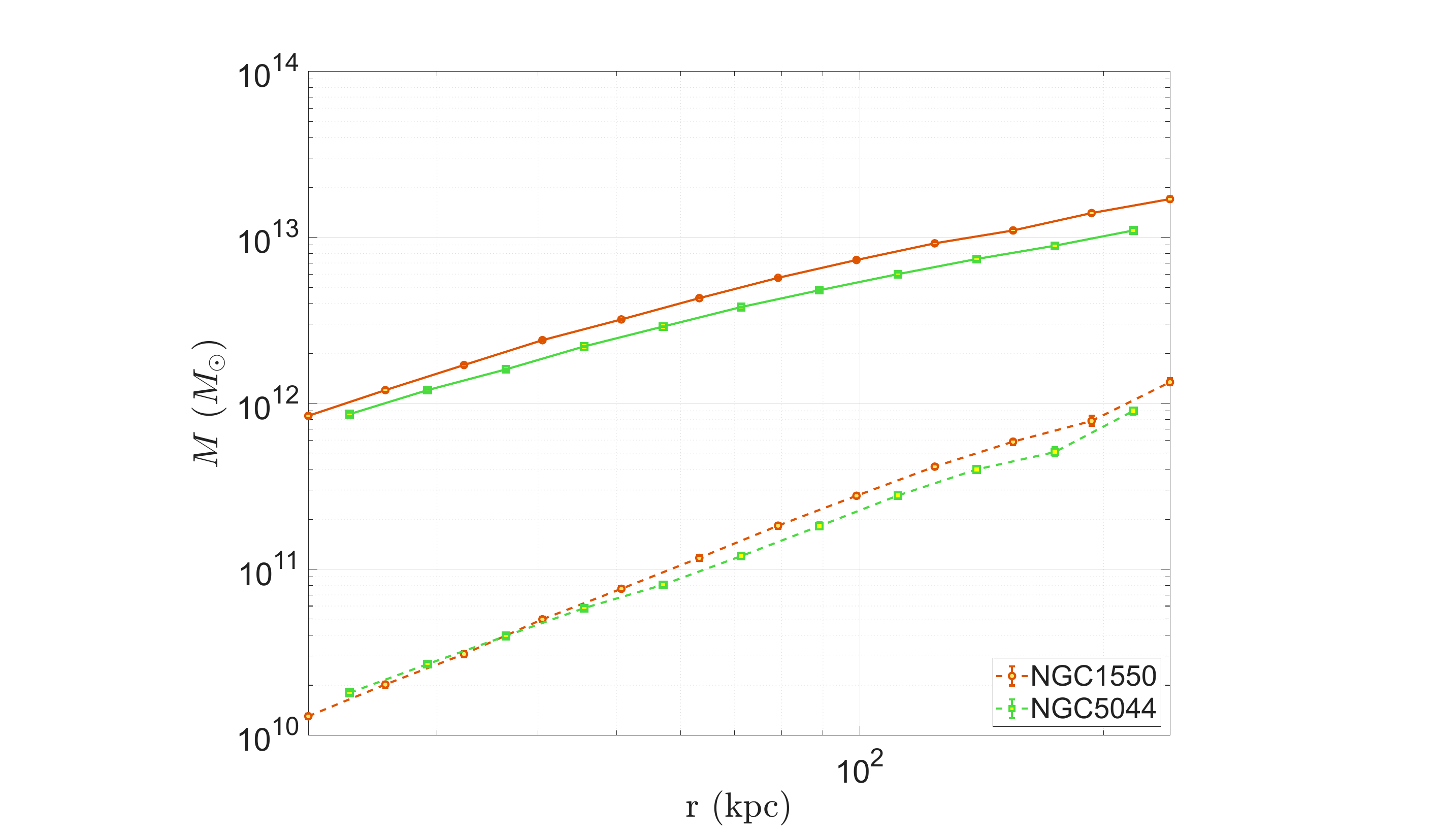}
    \includegraphics[trim=6.0cm 0.0cm 7.0cm 0.0cm, clip=true, width=0.24\columnwidth]{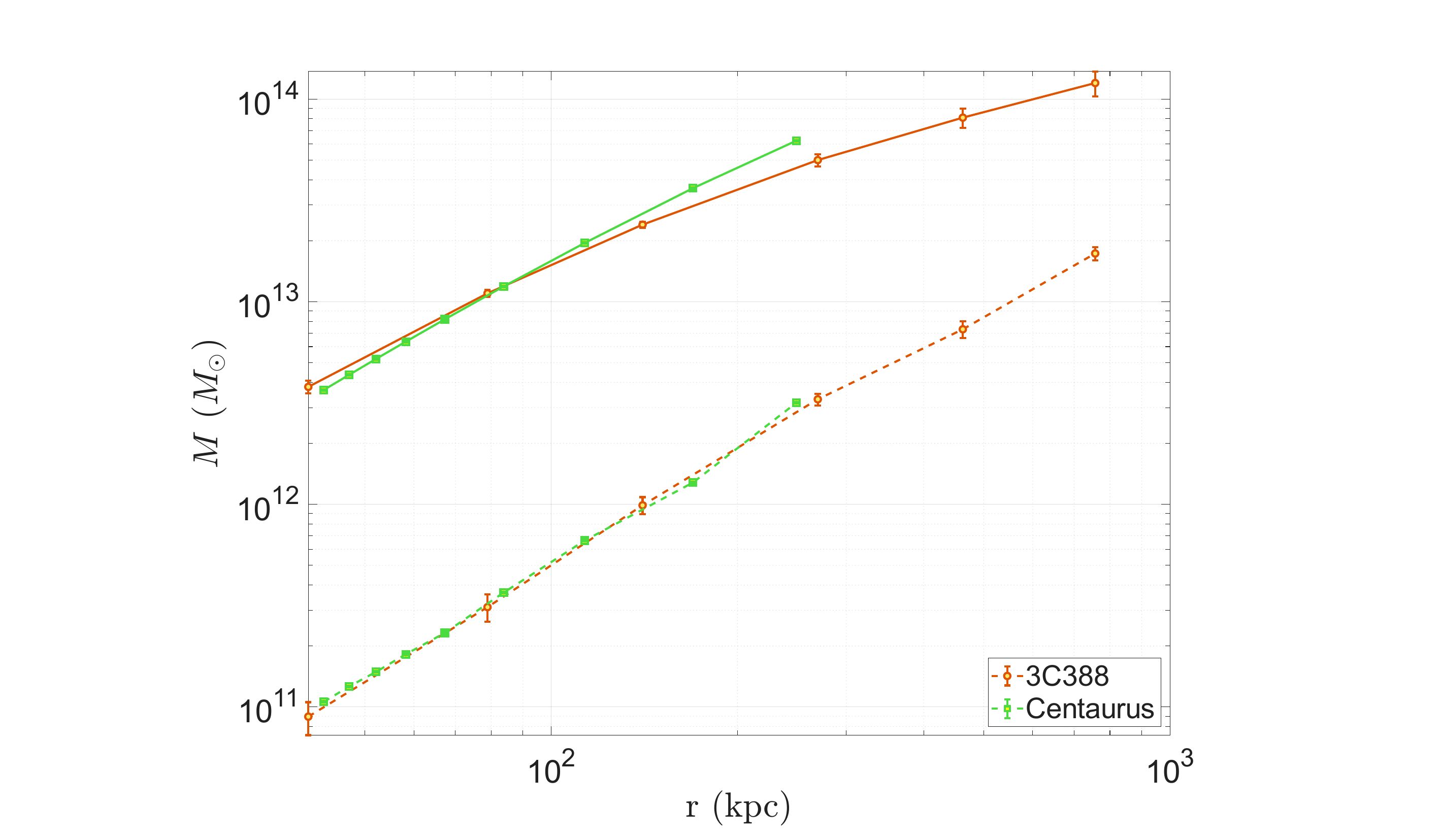}
    \includegraphics[trim=6.0cm 0.0cm 7.0cm 0.0cm, clip=true, width=0.24\columnwidth]{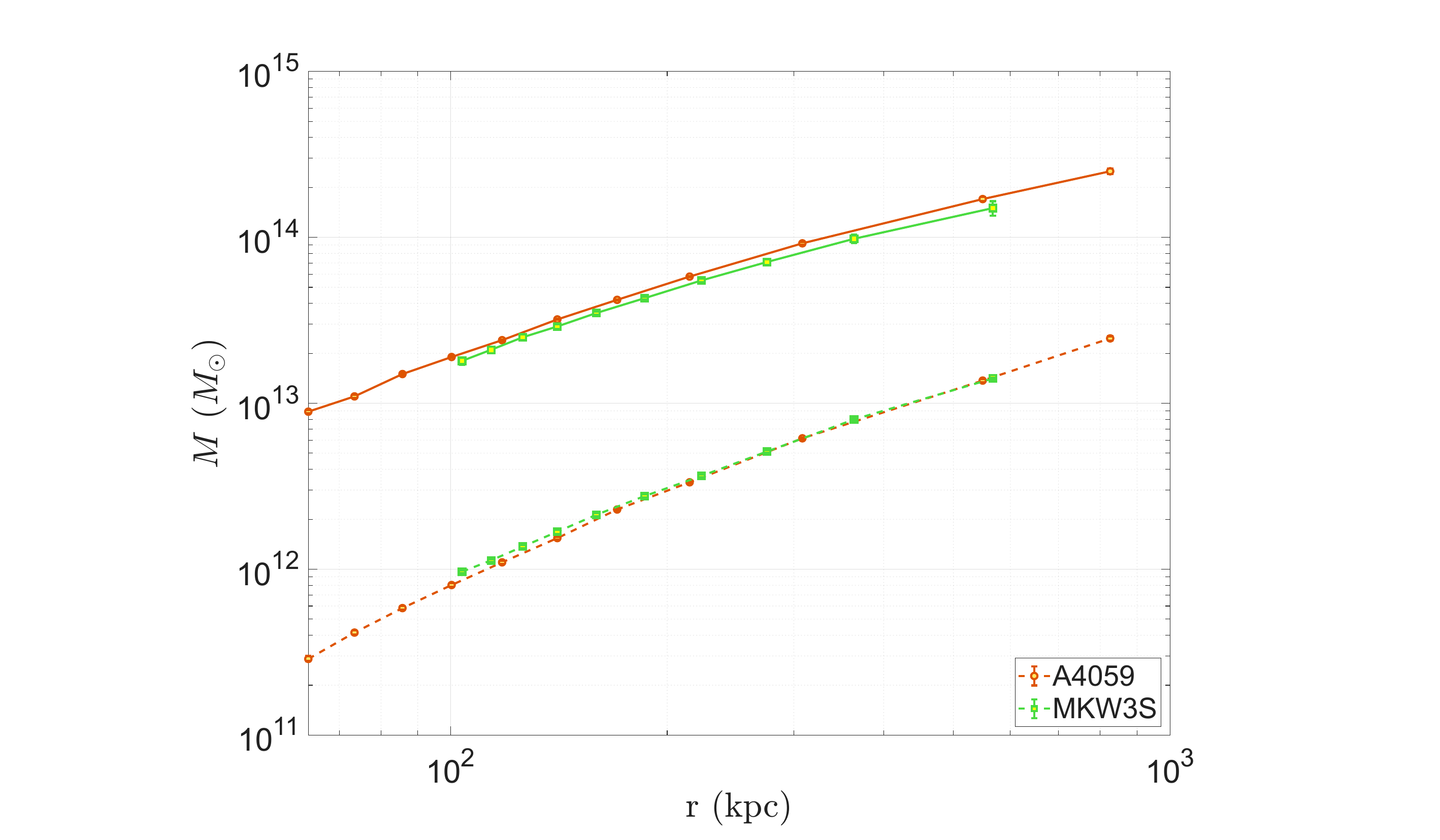}
    \includegraphics[trim=6.0cm 0.0cm 7.0cm 0.0cm, clip=true, width=0.24\columnwidth]{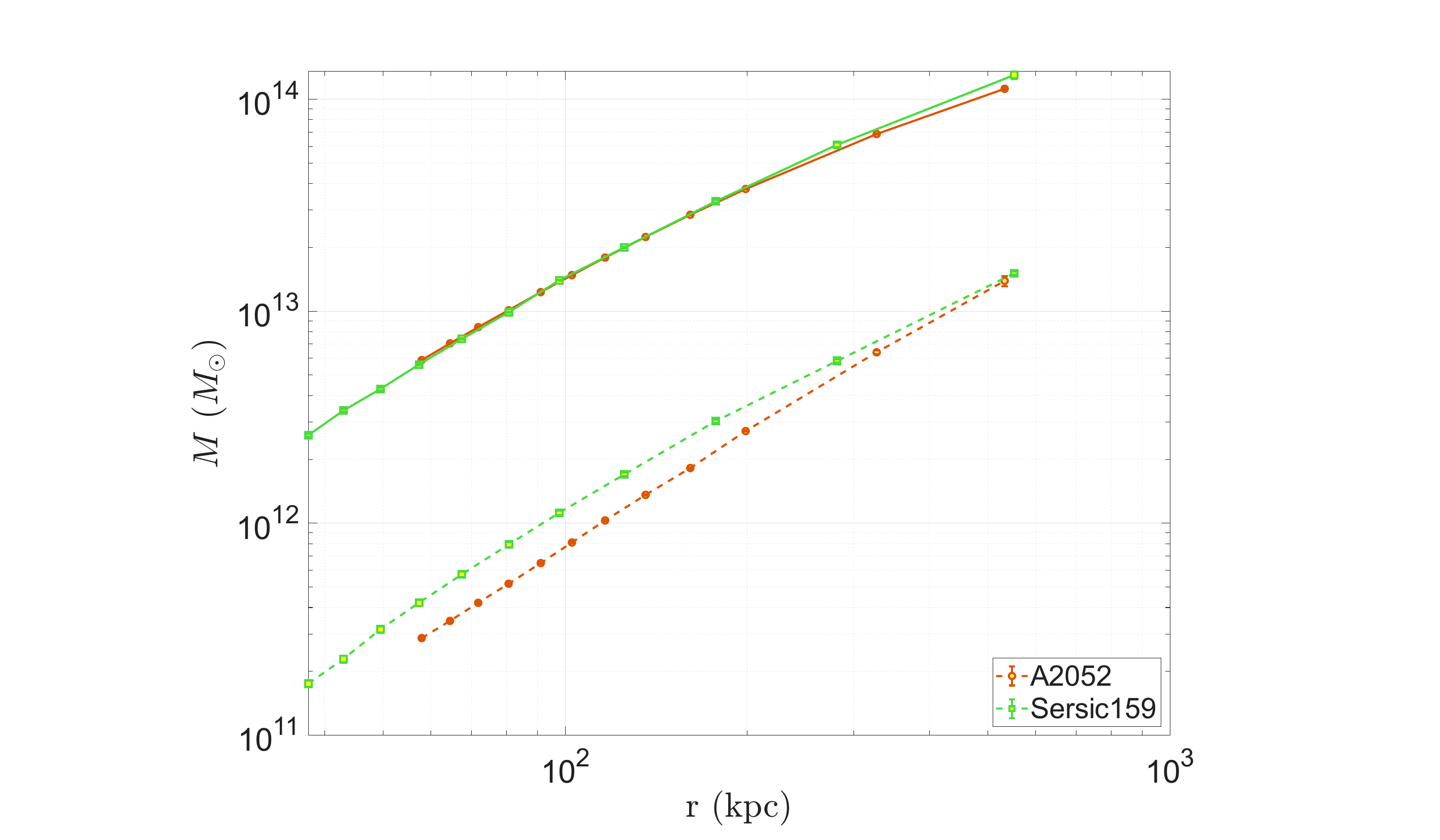}
    \includegraphics[trim=6.0cm 0.0cm 7.0cm 0.0cm, clip=true, width=0.24\columnwidth]{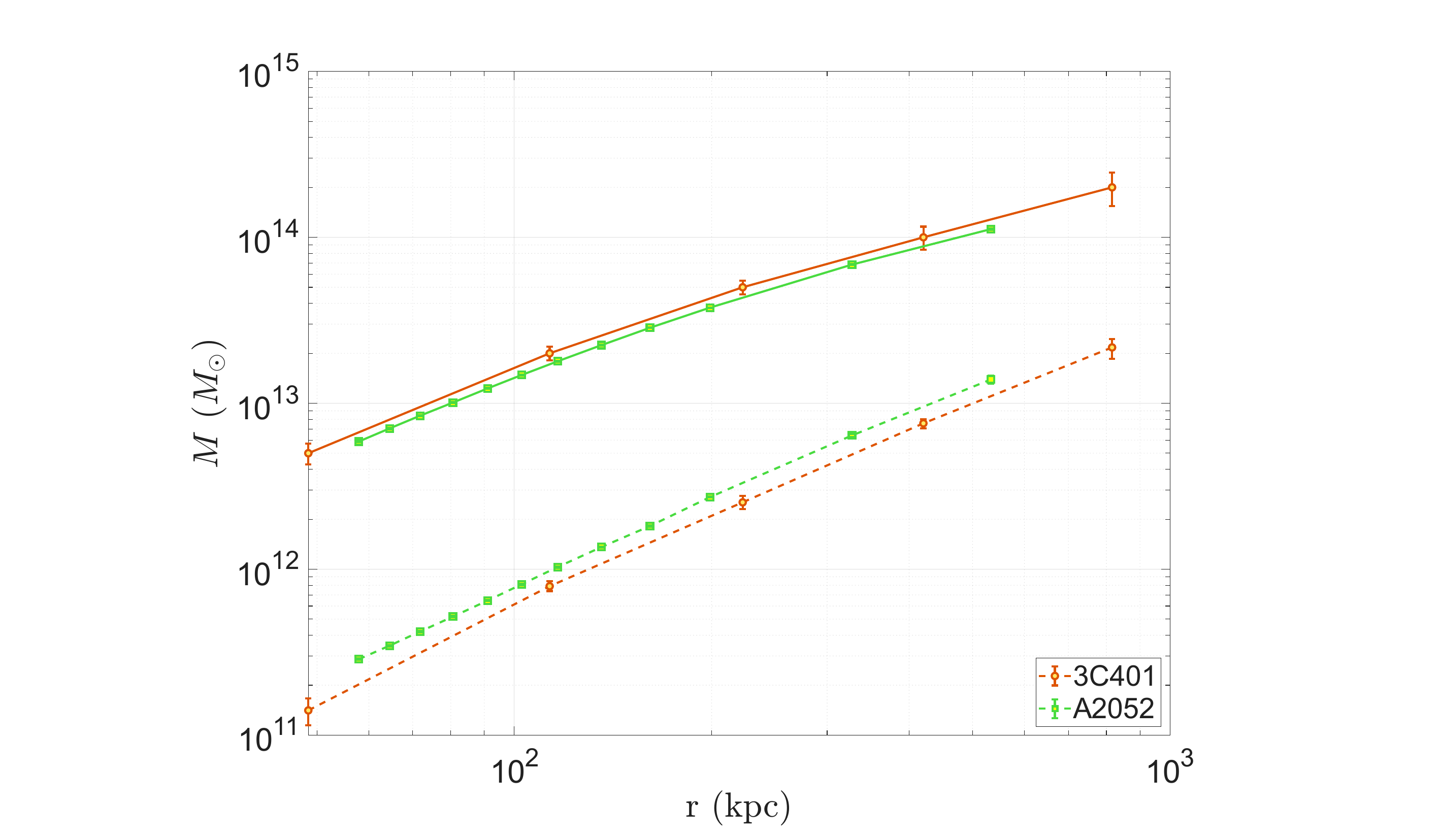}
    \includegraphics[trim=6.0cm 0.0cm 7.0cm 0.0cm, clip=true, width=0.24\columnwidth]{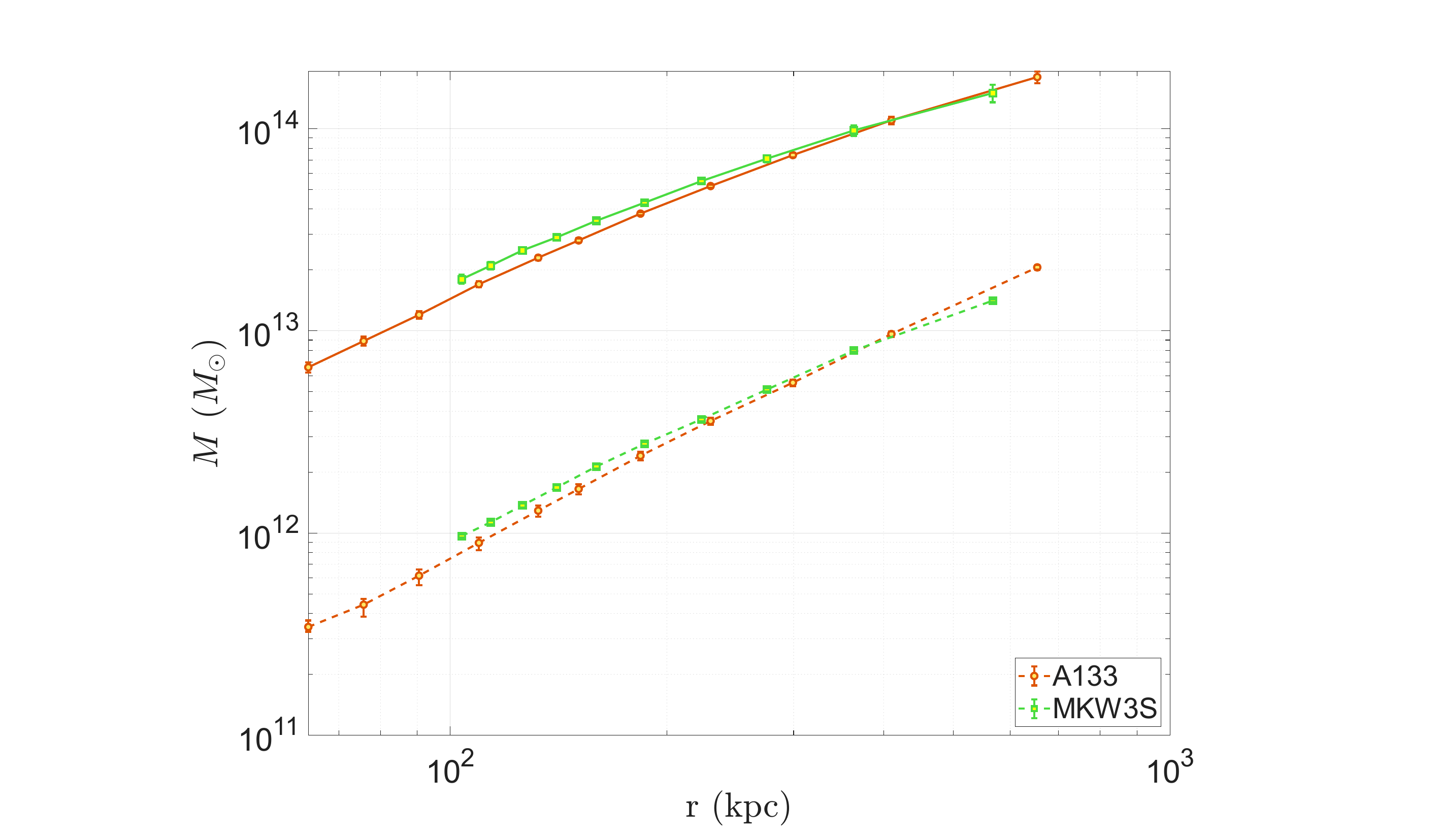}
    \includegraphics[trim=6.0cm 0.0cm 7.0cm 0.0cm, clip=true, width=0.24\columnwidth]{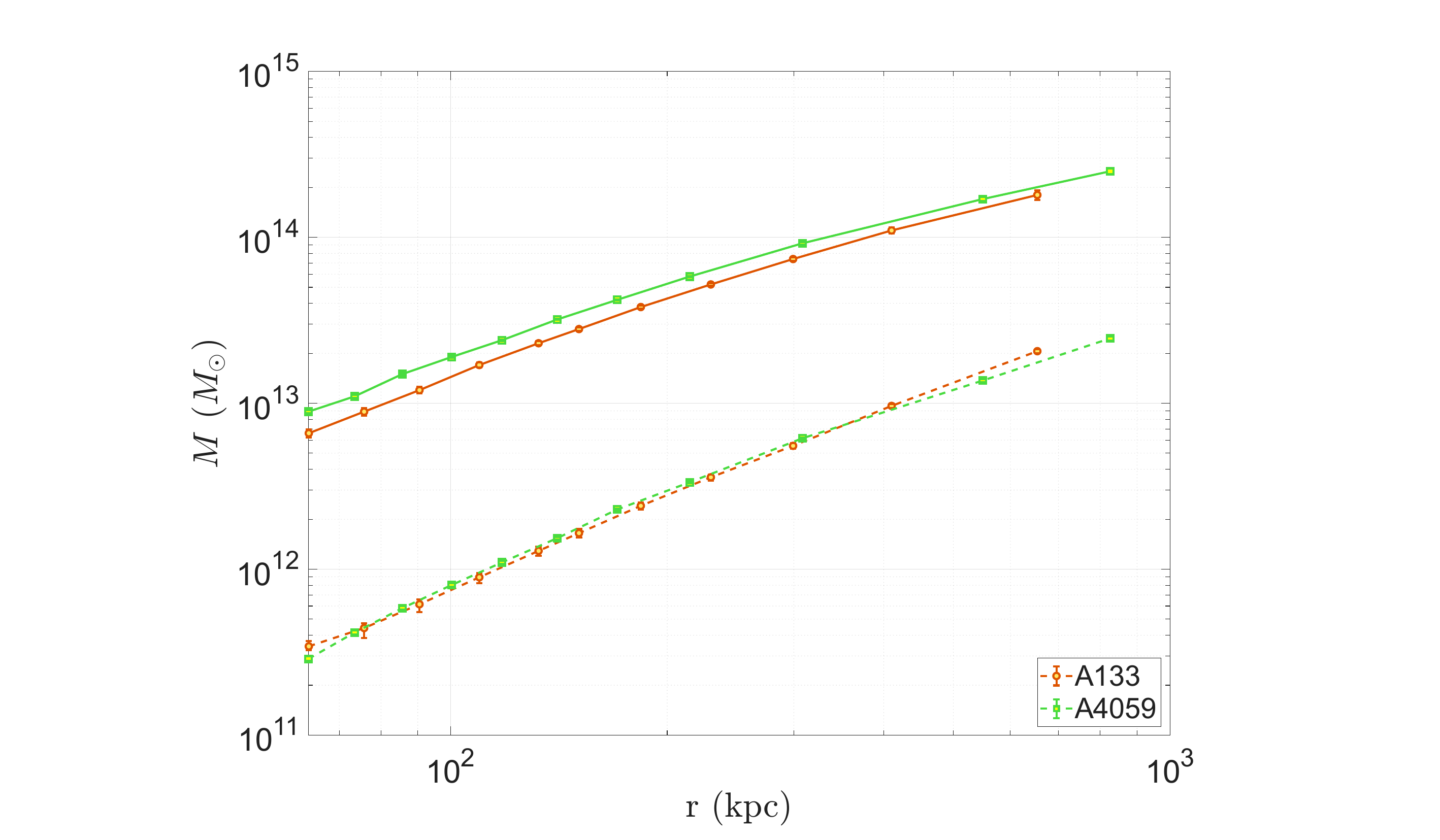}
    \includegraphics[trim=6.0cm 0.0cm 7.0cm 0.0cm, clip=true, width=0.24\columnwidth]{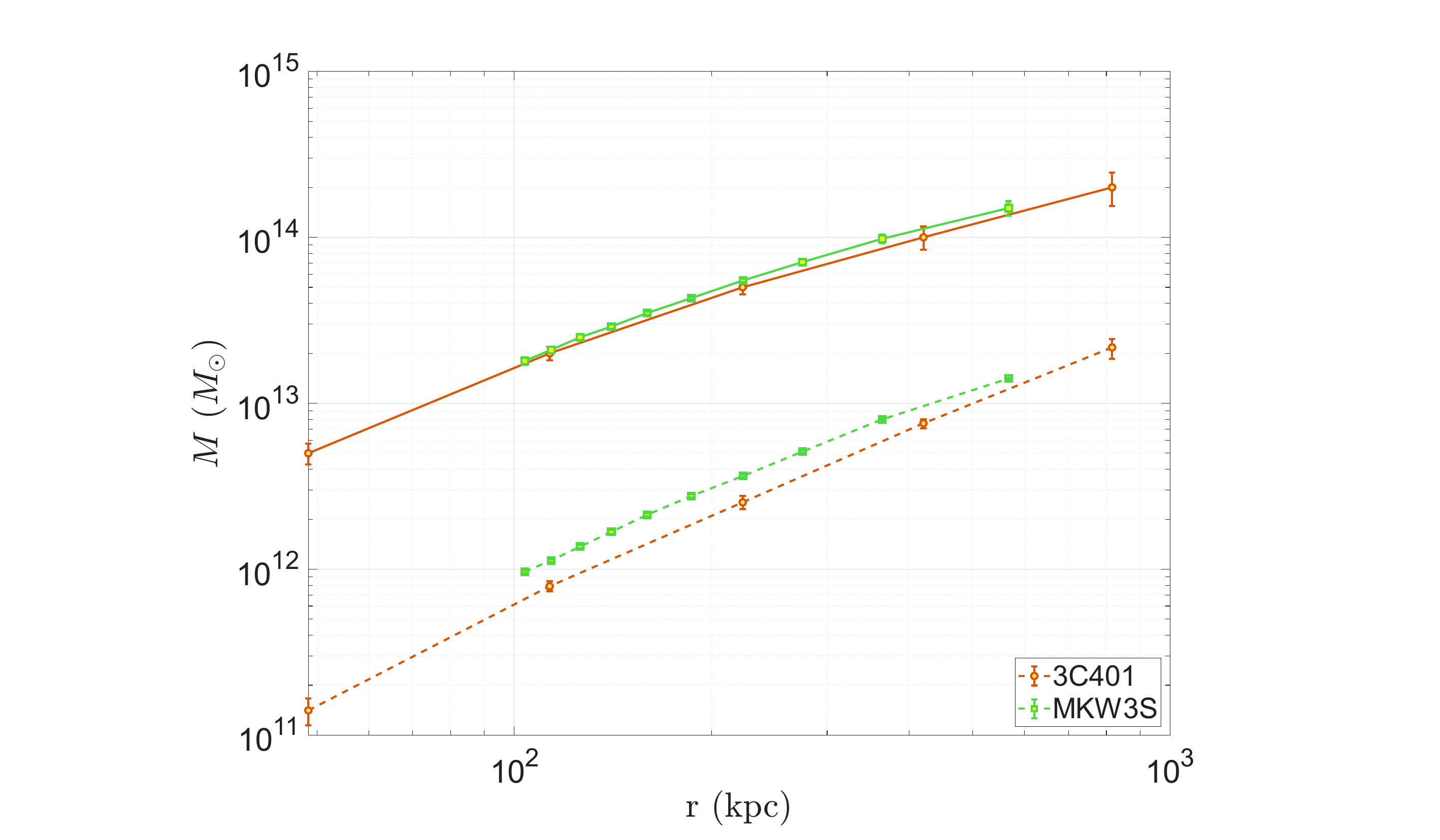}
    \includegraphics[trim=6.0cm 0.0cm 7.0cm 0.0cm, clip=true, width=0.24\columnwidth]{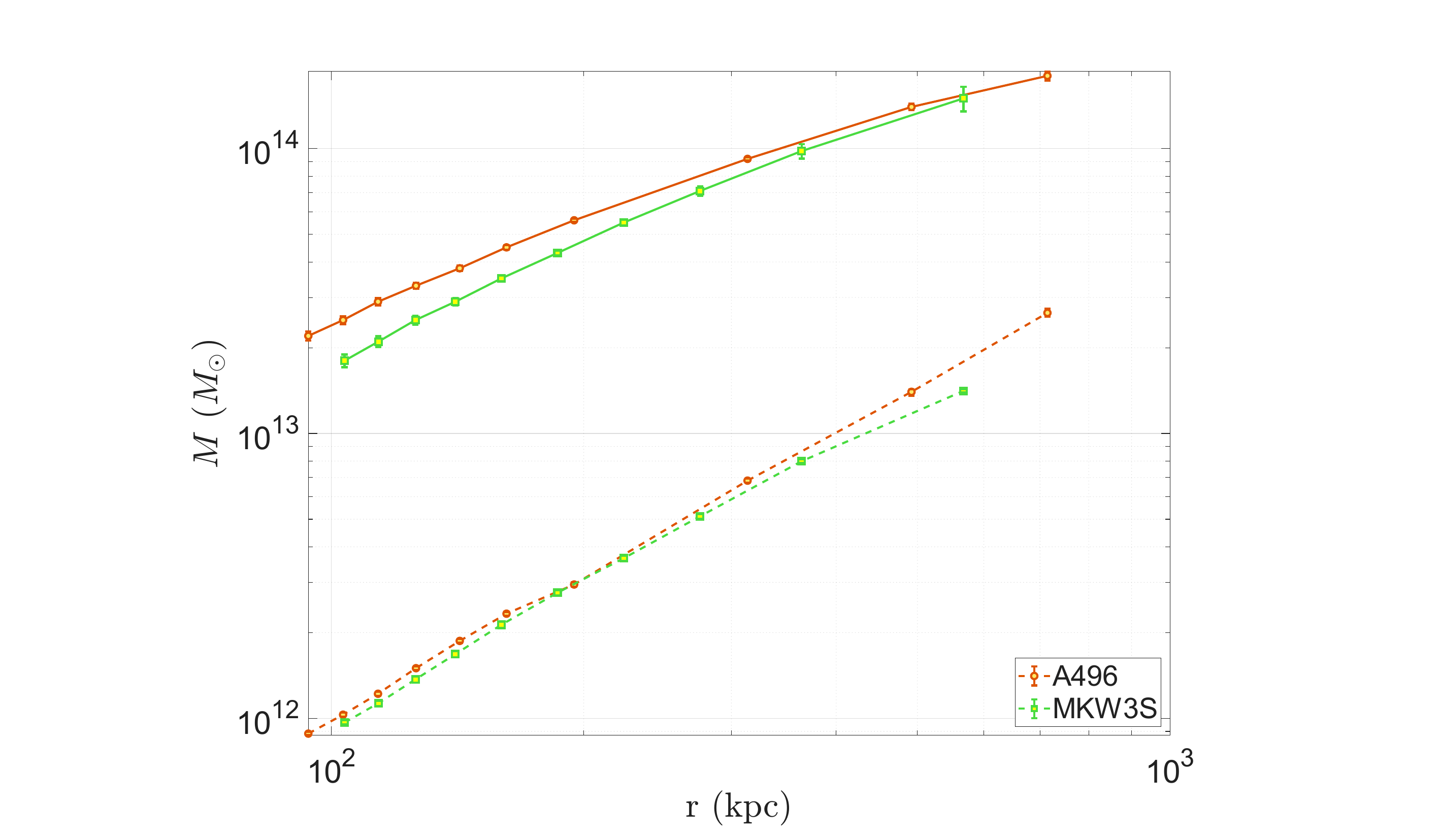}
    \includegraphics[trim=6.0cm 0.0cm 7.0cm 0.0cm, clip=true, width=0.24\columnwidth]{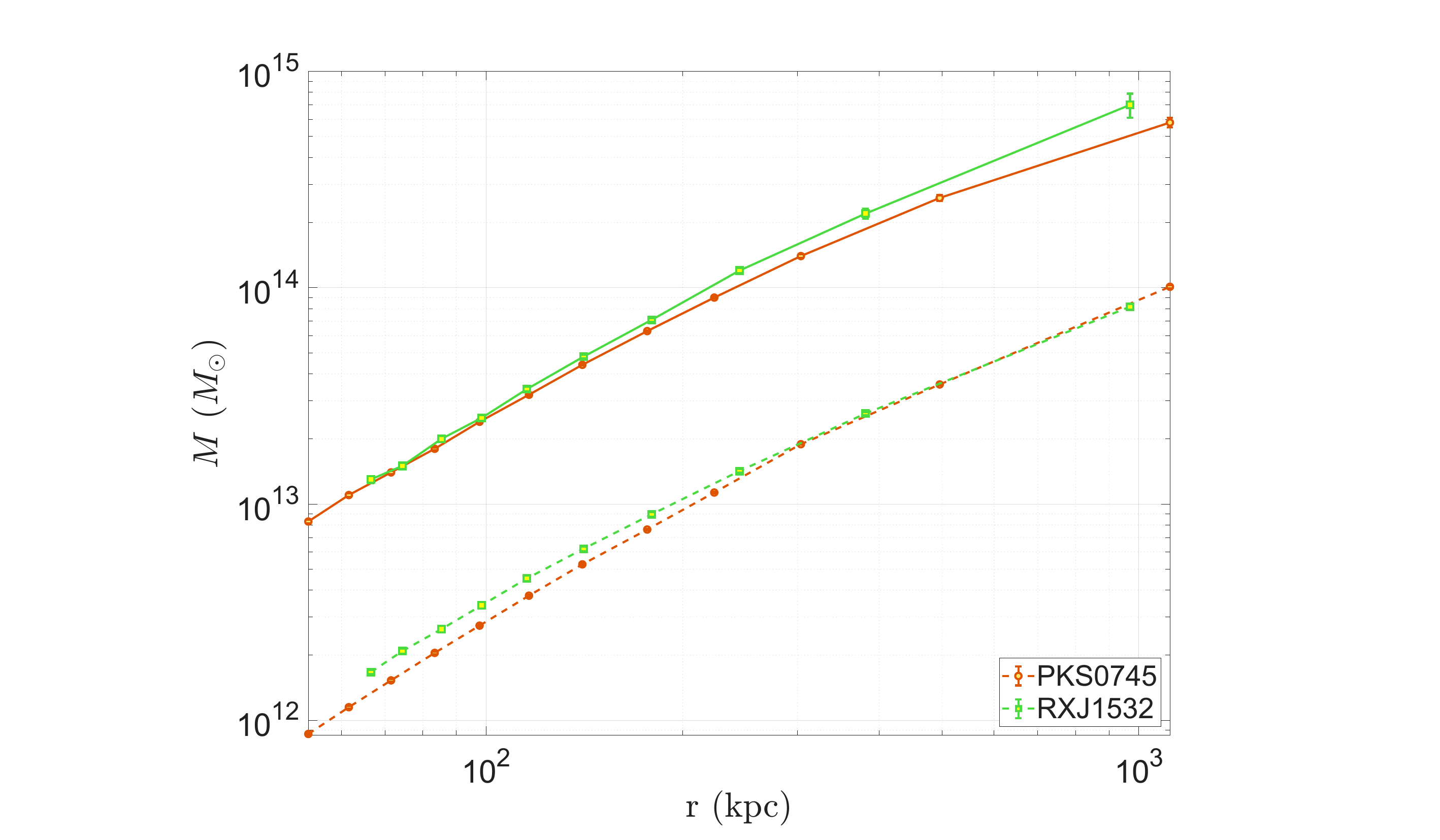}
    \caption{This figure shows 12 cluster pairs whose dynamic mass profiles coincide. We also observe that the baryonic mass distributions of each pair agree with one another. In a framework where dark matter is treated as a fully independent component, there is no obvious theoretical reason why agreement in one type of mass profile should imply agreement in the other. Thus, in these examples, Machian Gravity appears to offer a more satisfactory explanation than the standard dark matter paradigm, since in this theory both $M_c$ and $\lambda^{-1}$ depend on the underlying mass distribution.} 
    \label{fig:ClusterPair}
\end{figure}

\subsection{Testing Machian gravity curve on galaxy clusters from HIFLUGCS sample}

For our analysis, we use the data presented in Main et al.~\cite{main2016relationship}, where mass models are provided for 45 clusters from the HIFLUGCS sample~\cite{reiprich2002mass}. We employ their dataset to test our model. The full sample comprises 419 radial data points from these 45 galaxy clusters. In Fig.~\ref{fig:MainPaperDataUnderstanding} we present several diagnostic plots to characterize the data.

In the four panels on the left, we show the mass discrepancy ($M_d/M_b$) as a function of various parameters. The ratio $M_d/M_b$ exhibits a distinctly negative correlation with $M_b$, despite the plots appearing relatively scattered. The scatter increases when $M_d/M_b$ is plotted against $r$ or $M_b/r^2$, although a negative trend is still apparent in both cases. By contrast, a much stronger correlation is observed when $M_d/M_b$ is plotted against $M_b/r$.

Since there is no a priori reason to assume that the relationship should be linear, we quantify these trends using the Spearman rank correlation coefficient. The rank correlation between $M_d/M_b$ and $M_b$ is $-0.798$, indicating a strong negative correlation. The corresponding coefficient for $M_d/M_b$ and $r$ is $-0.604$. The correlation between $M_d/M_b$ and $M_b/r$ is even stronger, with a rank coefficient of $-0.874$. For $M_d/M_b$ versus $M_b/r^2$, the coefficient is $-0.526$. If we search for a power-law combination that maximizes the correlation, we find that the rank correlation between $M_d/M_b$ and $M_b/r^{0.9}$ is also $-0.875$, i.e., essentially identical to the correlation with $M_b/r$.

This relation is quite intriguing because, from the galactic radial velocity profiles, it is well established that the mass discrepancy depends on the acceleration, i.e. on $M_b/r^2$, rather than on $M_b/r$. Consequently, these findings are in substantial tension with those inferred from galactic rotation curves. This discrepancy is one of the primary reasons why MOND, in which the mass discrepancy is postulated to be a function of acceleration, has not succeeded in reproducing the mass profiles of galaxy clusters.

In the top-right plot we have plotted $M_b$ vs $M_d$, where we can see that the 
\begin{equation}
  \log (M_d) = 0.7686 \log (M_b) + 3.9840  \,.
\end{equation}
 
\noindent In the bottom-right panel, we present the histogram of the ratio $M_d / M_b$. The distribution indicates that, for the majority of data points, $M_d / M_b$ lies in the range $5 \text{--} 15$.

\begin{figure}
    \centering
    \includegraphics[width=0.40\columnwidth]{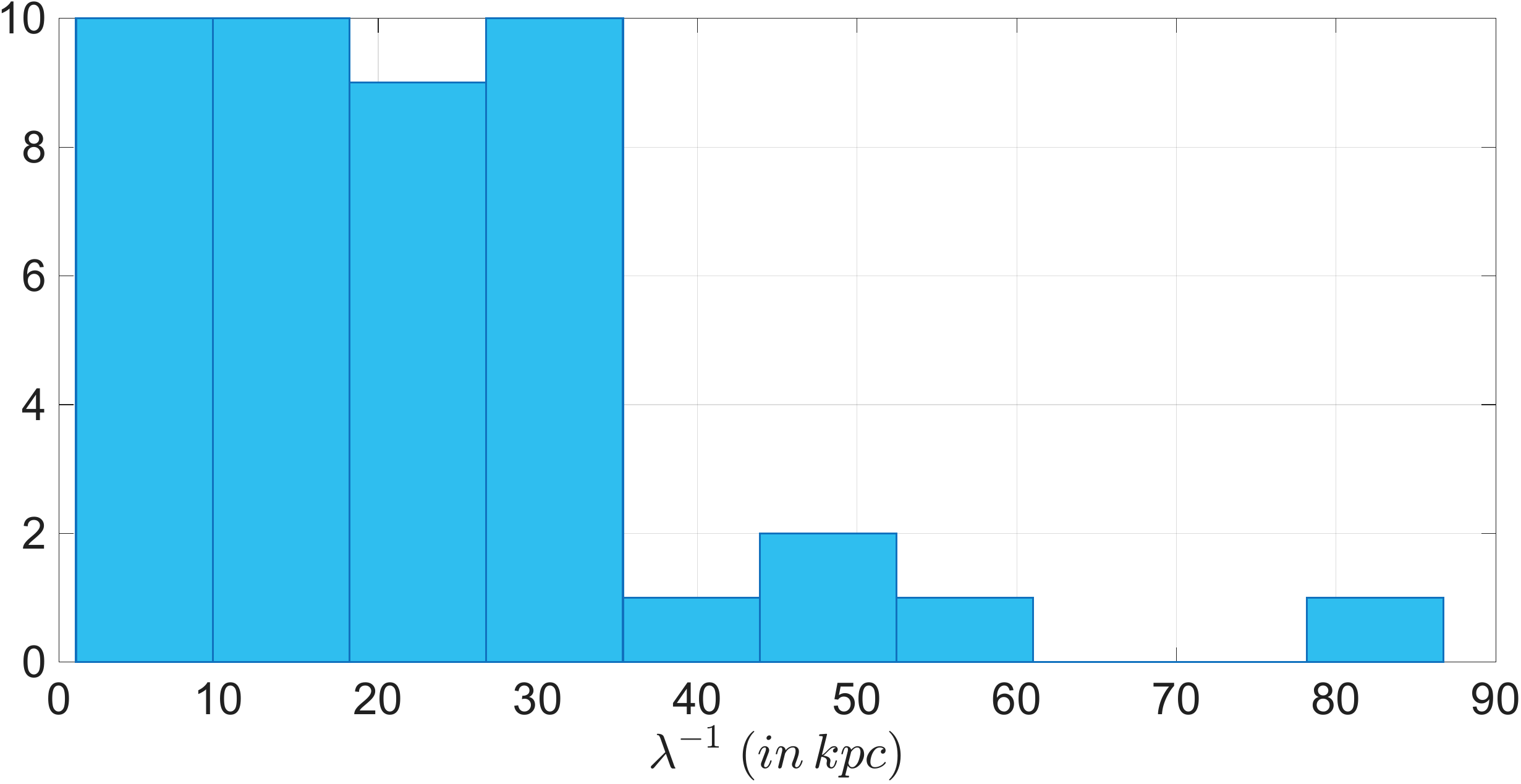}
    \includegraphics[width=0.40\columnwidth]{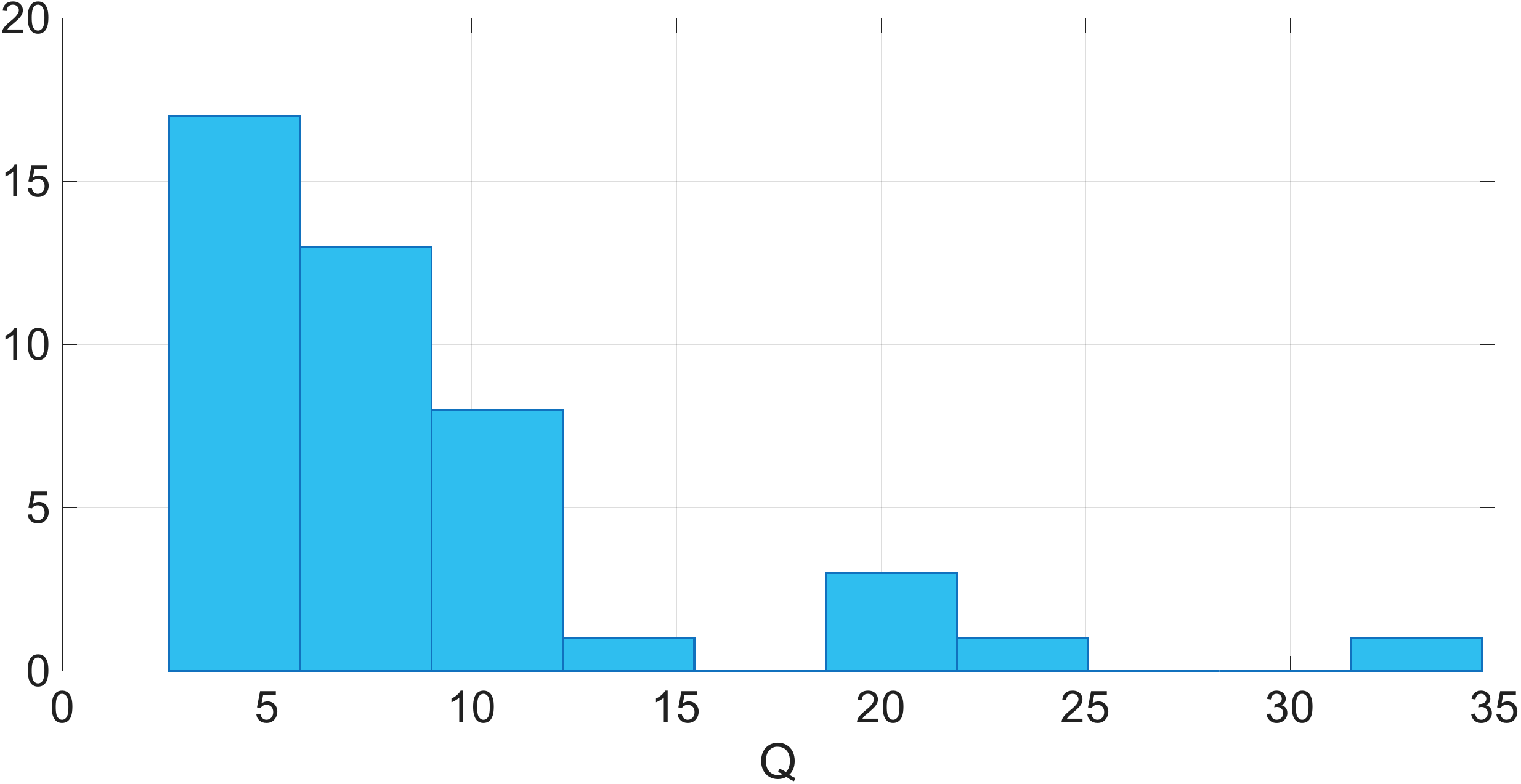}
    \caption{The plot shows the best-fit values of $\lambda^{-1}$ and $Q$ for the HIFLUGCS clusters, assuming the exponent in Eq.~\ref{eq:KQform} is fixed at $0.25$. In all best-fit models, $\lambda^{-1}$ is below 100 kpc, and in most cases it is even below 40 kpc. This means that, for any given cluster, $\lambda^{-1}$ is always much smaller than the radius of interest, so the exponential term has almost no impact. In addition, for most clusters the value of $Q = \left({\frac{M_c}{M_b}}\right)^{0.25}$ is typically in the range of about 10–15.} 
    \label{fig:linv_Q_25}
\end{figure}

By fitting a regression line to $M_d / M_b$ as a function of $M_b / r$, we obtain $\frac{M_d}{M_b} \sim \left(\frac{M_b}{r}\right)^{-0.48}$, which implies $M_d \sim M_b^{0.5}$,
in close analogy with the scaling inferred from galaxy velocity distribution relations. However, in this case we also find $M_d \sim r^{0.5}$, which differs from the corresponding galaxy scaling relations, where the dependence is approximately linear in $r$.

In Fig.~\ref{fig:ClusterPair}, 12 pairs of galaxy clusters are presented for which the dynamical masses of the clusters are comparable. The dotted red or green curves denote the baryonic mass distributions, while the solid curves correspond to the Newtonian dynamical mass profiles of these systems. It is evident that the baryonic mass distributions also exhibit mutually similar profiles within each pair. 

If one assumes that the dark matter distribution is completely independent of the baryonic mass distribution, it becomes difficult to account for the empirical fact that, whenever the baryonic mass profiles of two clusters are closely matched, their total (dynamical) mass profiles likewise coincide. This correlation implies that, for galaxy clusters, the dynamical mass distribution is functionally related to the baryonic mass distribution. Such behaviour cannot be naturally accommodated within the framework of the standard NFW dark matter halo profile.

MOND can reproduce some of these trends; however, it is well established that MOND alone fails to account for the full mass profiles of galaxy clusters. Extensions such as MOND with an external field effect (EFE) can improve the agreement for certain clusters. Nevertheless, explaining the present phenomenon within that framework would require assuming that the external field is effectively identical for both members of each cluster pair, which is not a particularly well‑justified or generic assumption.

It remains possible that external gravitational fields play some role. However, the observational evidence presented here indicates that the dynamical mass distribution must be primarily determined by the distribution of baryonic matter. In this context, Machian gravity provides a more promising framework for explaining the observed mass distributions of these galaxy cluster pairs.

\begin{figure}
\includegraphics[trim=0.0cm 0.0cm 0.0cm 0.0cm, clip=true, width=0.33\columnwidth]{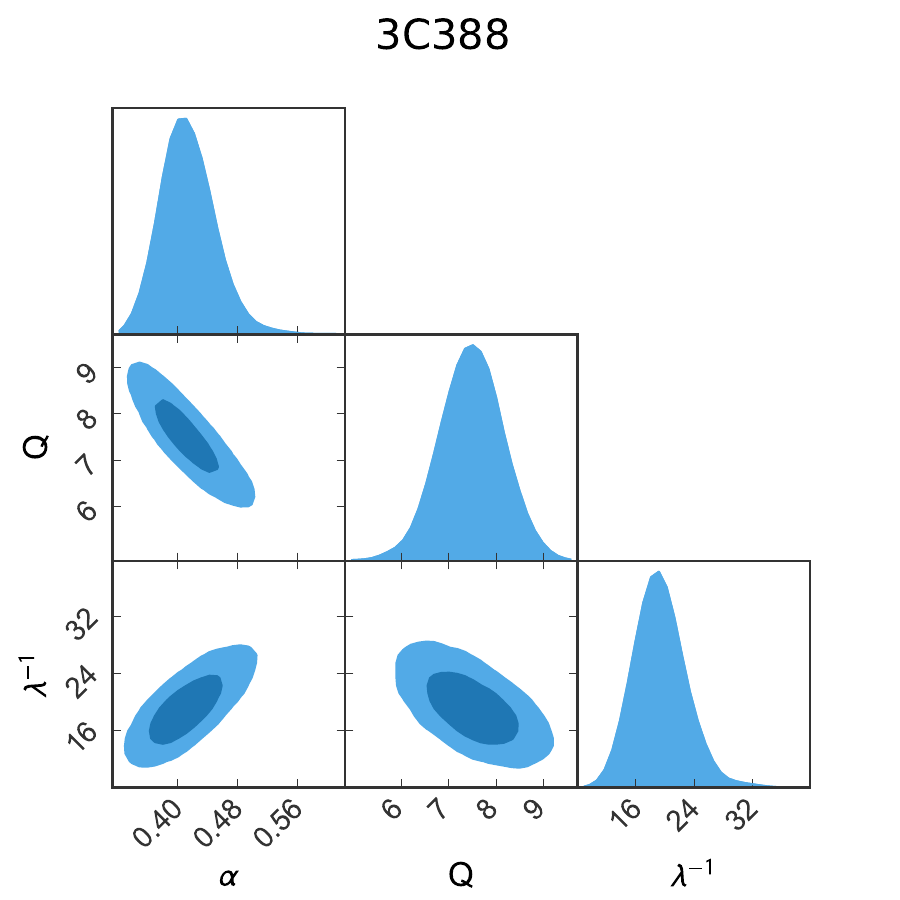}
\includegraphics[trim=0.0cm 0.0cm 0.0cm 0.0cm, clip=true, width=0.33\columnwidth]{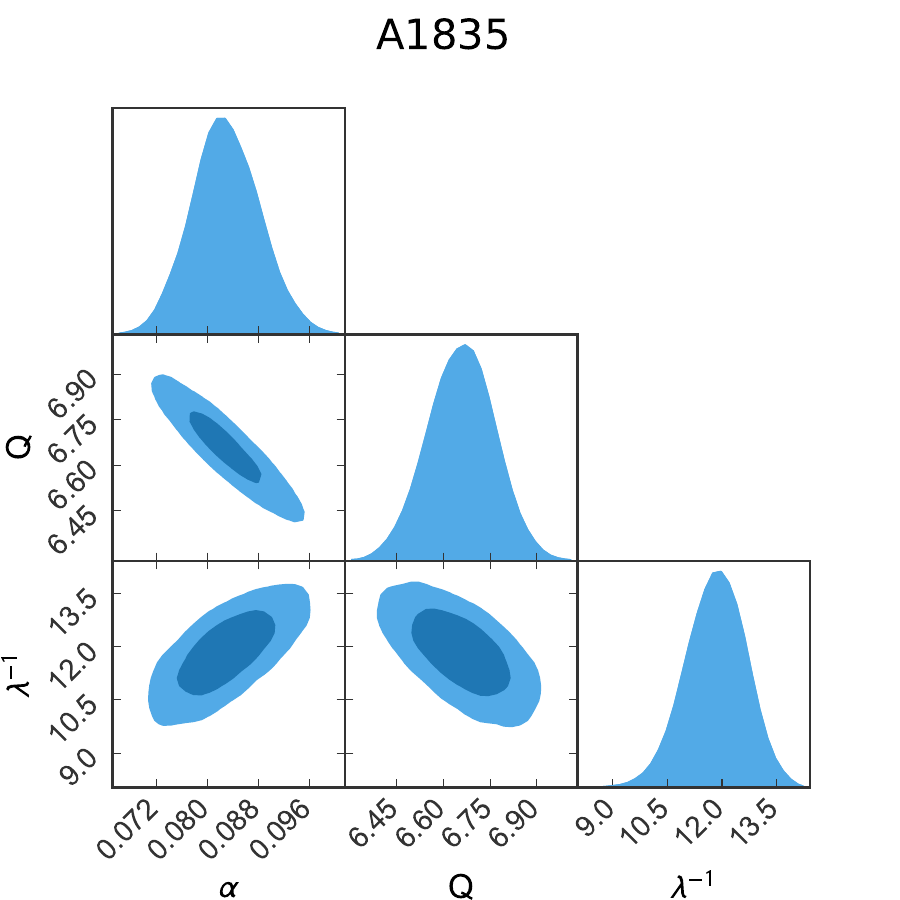}
\includegraphics[trim=0.0cm 0.0cm 0.0cm 0.0cm, clip=true, width=0.33\columnwidth]{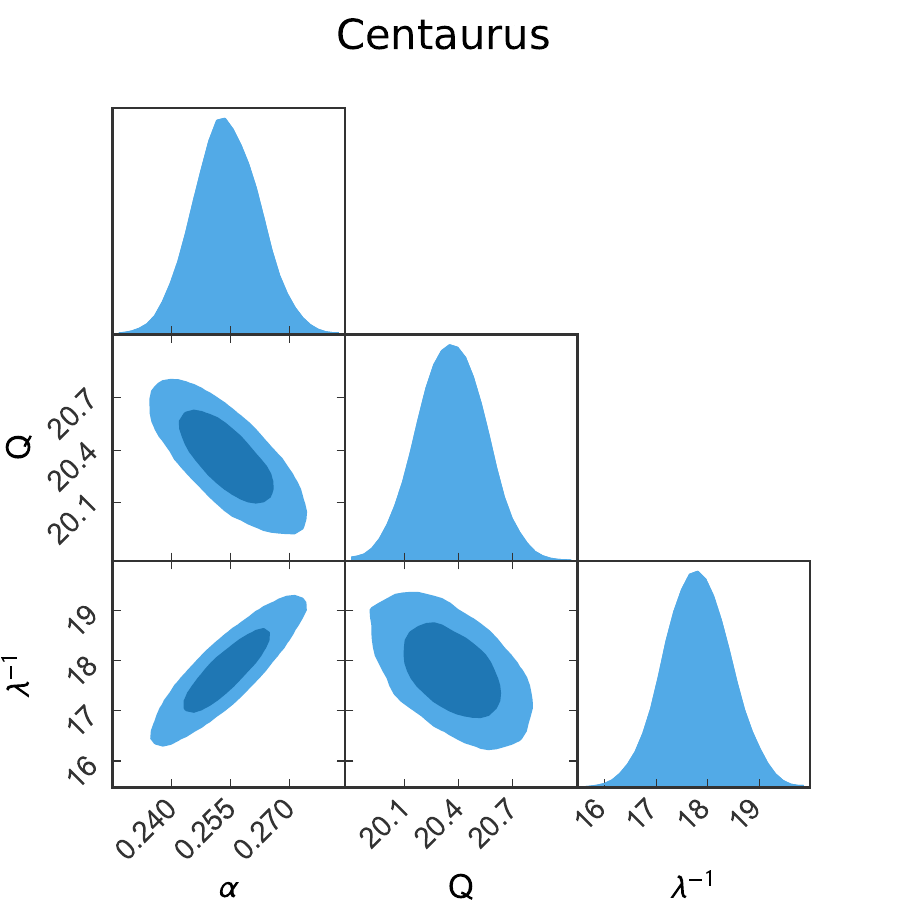}
\includegraphics[trim=0.0cm 0.0cm 0.0cm 0.0cm, clip=true, width=0.33\columnwidth]{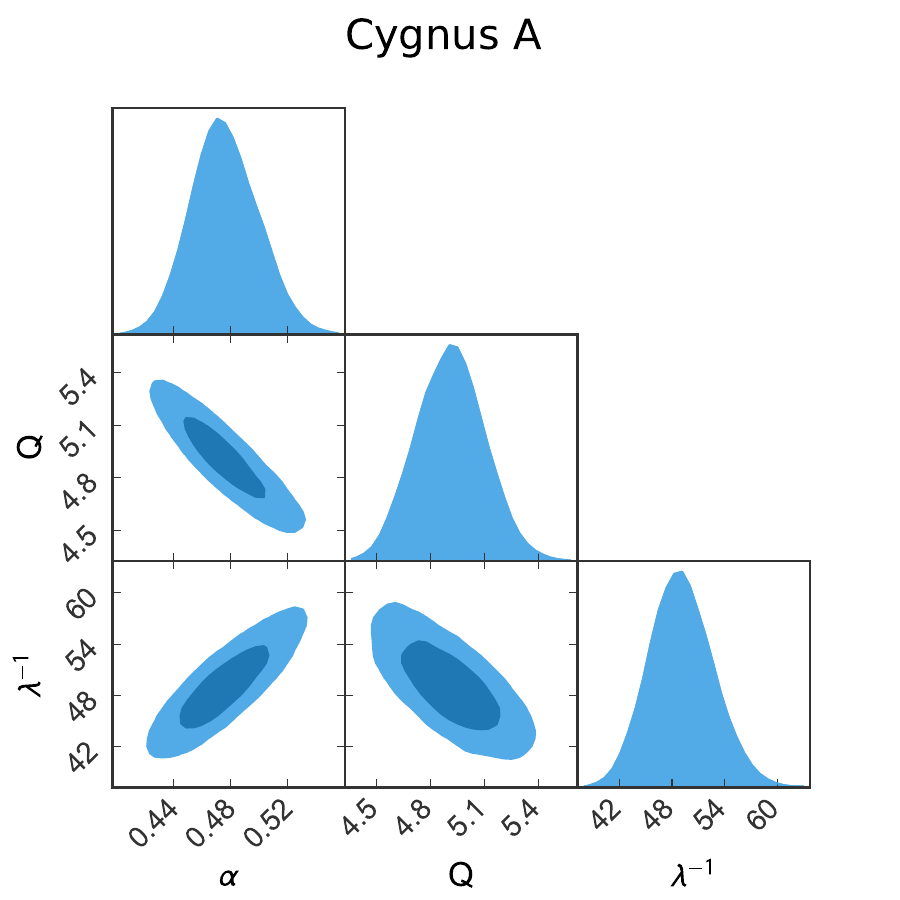}
\includegraphics[trim=0.0cm 0.0cm 0.0cm 0.0cm, clip=true, width=0.33\columnwidth]{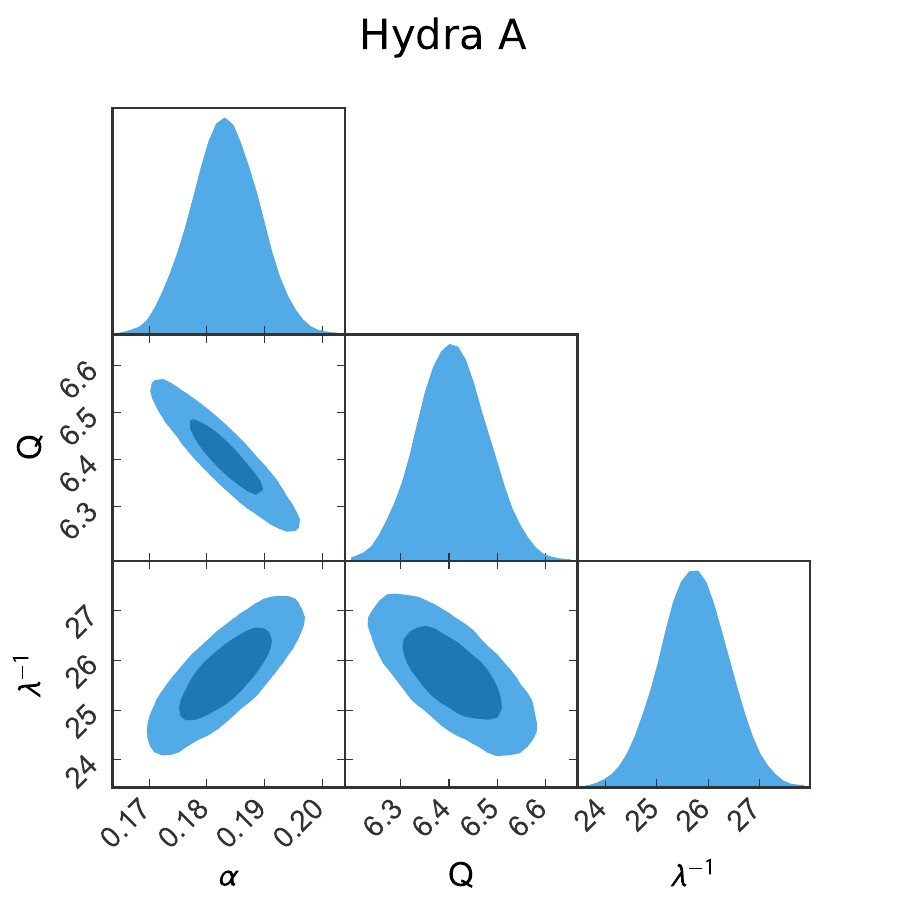}
\includegraphics[trim=0.0cm 0.0cm 0.0cm 0.0cm, clip=true, width=0.33\columnwidth]{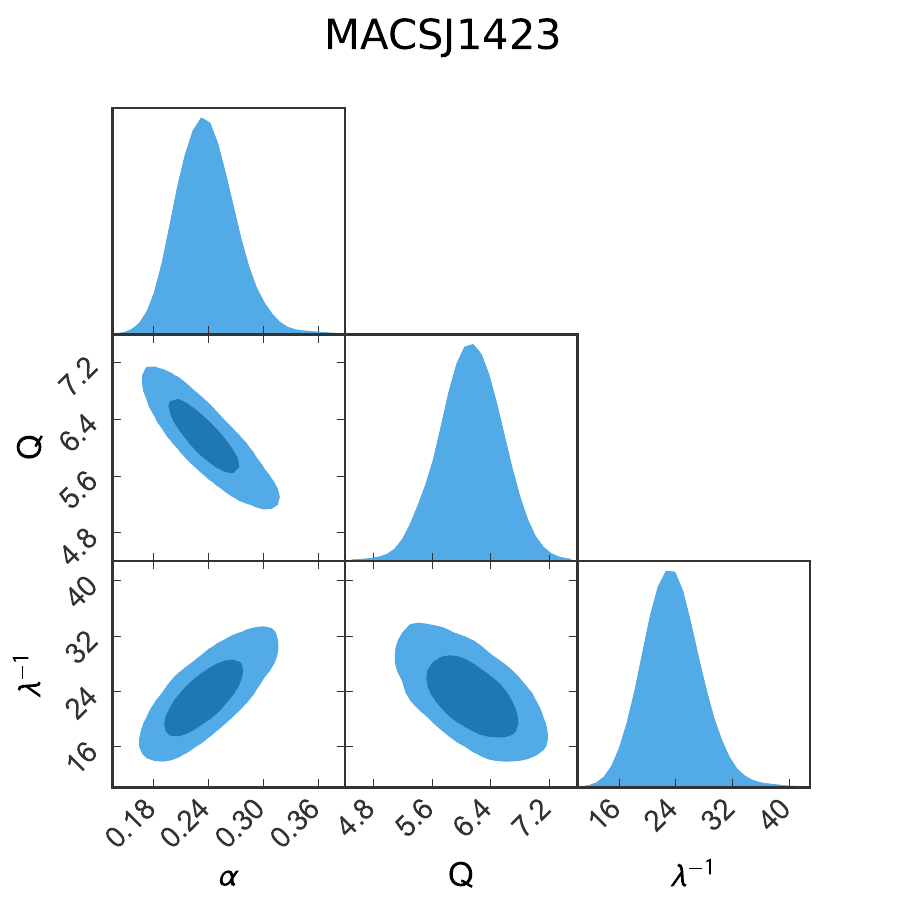}
\caption{\label{fig:CornerPLots}
The figure presents corner plots for six galaxy clusters, illustrating a three-parameter analysis across the HIFLUGCS sample. In every plot, a clear negative correlation is observed between $\lambda^{-1}$ and $\alpha$, while with respect to $Q$, other quantities, i.e. $\alpha$ and $\lambda^{-1}$ themselves exhibit a positive correlation. Corner plots for the remaining clusters are available at \href{https://machiangravity.github.io/galaxy_cluster_mass/corner_plots/}{corner plots}. With the exception of a few cases where the data are poorly constrained, all plots generally display the same overall behavior.}
\end{figure}

From our analysis of the preceding dataset, we found that the Machian gravity model reproduces the dynamical mass profile very accurately. However, the data suggest a power-law dependence with an exponent in the range \(0.25\text{--}0.3\), rather than a strict square-root relation. Moreover, in the plots shown in Fig.~\ref{fig:GalaxyCusterMass}, the Machian gravity model yields more strongly curved dynamical mass profiles for the clusters. This behaviour can be mitigated by decreasing the power of \(M_b\) in the relevant relation. Consequently, we relax the assumption of a fixed square-root dependence and investigate whether the quality of the fits can be improved by introducing a variable exponent, i.e.,
\begin{equation}
    K = \left(\frac{M_c}{M_b(r)}\right)^\alpha -1 \,.
    \label{eq:kformula}
\end{equation}

\noindent However, this type of parameterization is difficult to use because $\alpha$ and $M_c$ are strongly correlated and depending on the value of $\alpha$, the value of the $M_c$ can be several orders of magnitude different. Therefore, we just multiply and divide this fraction by the total mass of the cluster and redefine the $K$ as 

\begin{equation}
K = \left(\frac{M_c}{M_b(r_{out})}\right)^\alpha\left(\frac{M_b(r_{out})}{M_b(r)}\right)^\alpha -1
 = Q\left(\frac{M_b(r_{out})}{M_b(r)}\right)^\alpha -1
 \label{eq:KQform}
\end{equation}

\noindent where $Q = \left(\frac{M_c}{M_b(r_{out})}\right)^\alpha$. In this way, the factor $Q$ will get limited by the fraction of the mass discrepancy, and the power can be varied without affecting $Q$. So, the effective dynamic mass from the Machian gravity can be written as

\begin{equation}
    M_d(r) = M_b(r)\left[1+\left( Q\left(\frac{M_b(r_{out})}{M_b(r)}\right)^\alpha - 1\right)\left(1-e^{-\lambda r}\left(1+\lambda r\right)\right)\right]  \,.
    \label{MachMdMbrelation}
\end{equation}

\noindent Now, we can use MCMC to fit our model against the observations and constrain the $Q$, $\alpha$, and $\lambda^{-1}$.

\begin{figure}
\includegraphics[trim=0.0cm 0.0cm 0.0cm 0.0cm, clip=true, width=0.33\columnwidth]{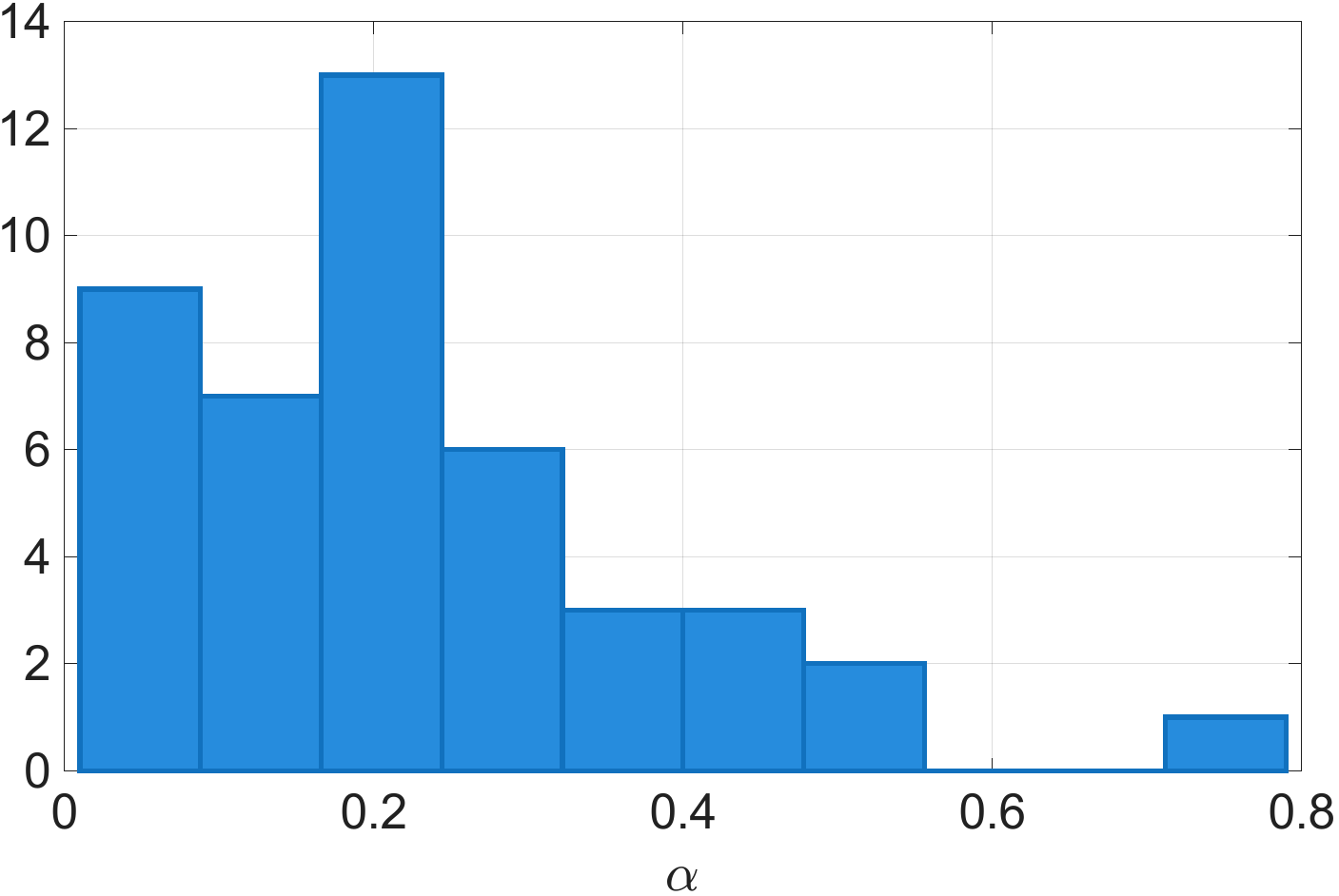}
\includegraphics[trim=0.0cm 0.0cm 0.0cm 0.0cm, clip=true, width=0.33\columnwidth]{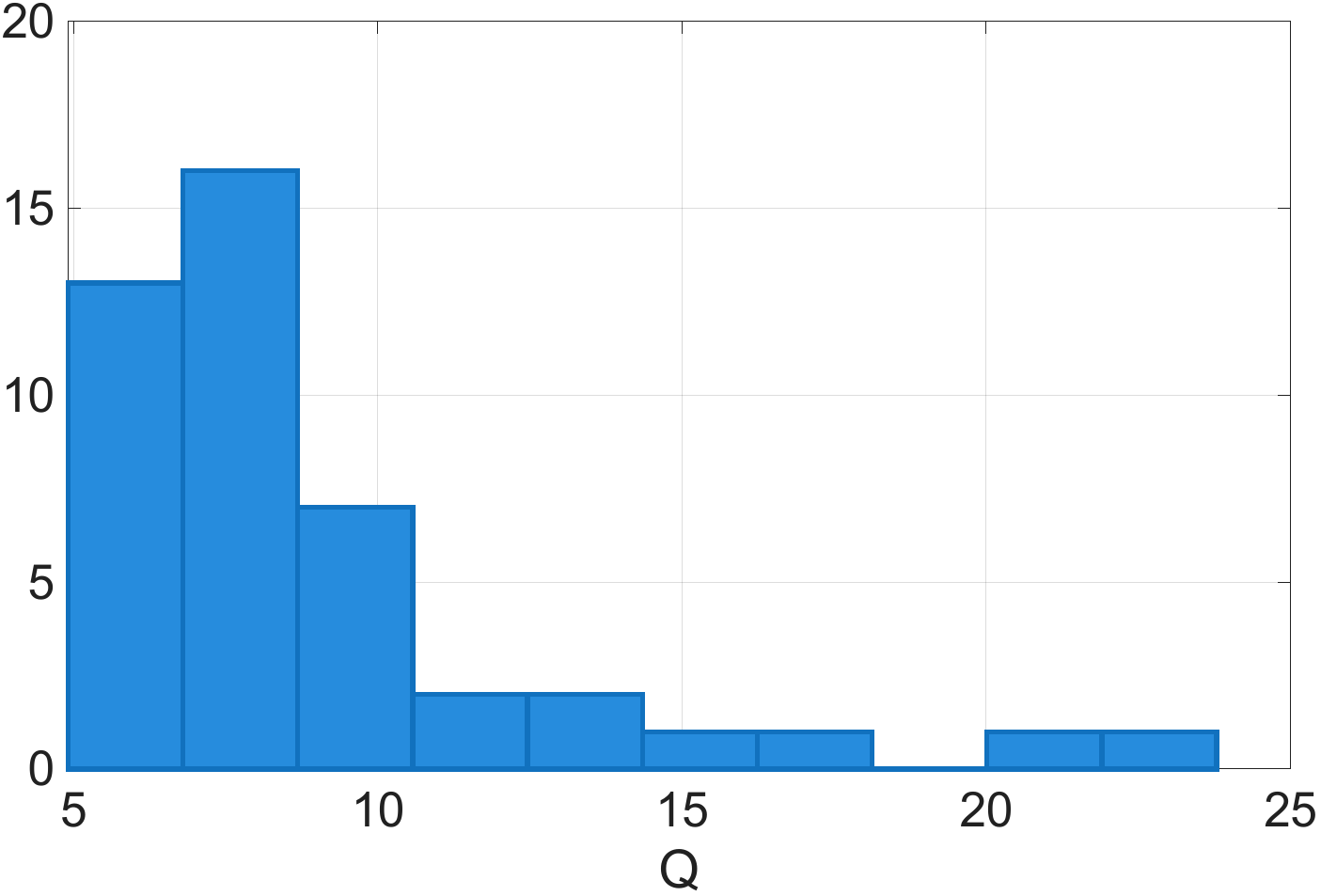}
\includegraphics[trim=0.0cm 0.0cm 0.0cm 0.0cm, clip=true, width=0.33\columnwidth]{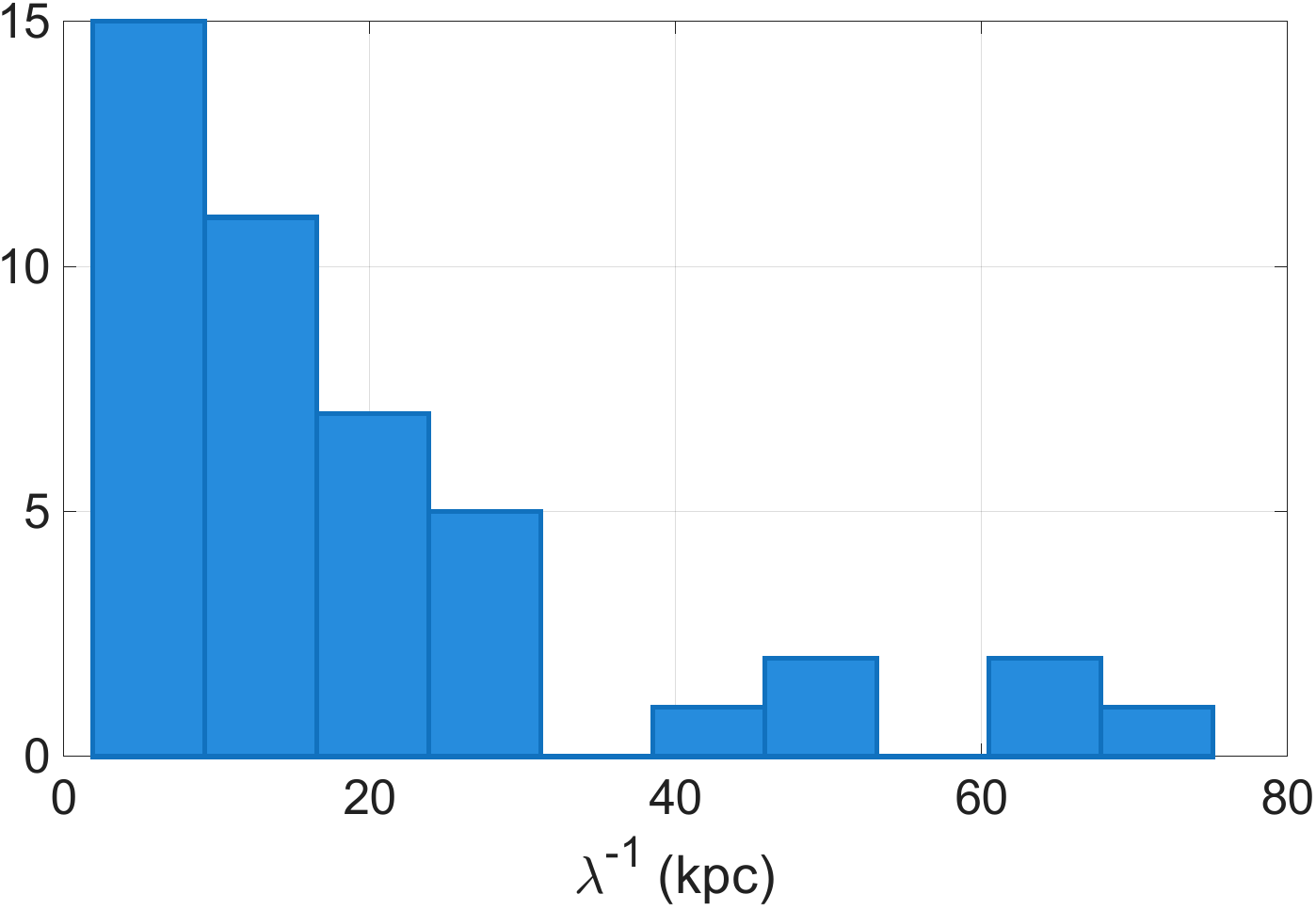}
\caption{\label{fig:threeparamproperties}
The figure presents the histograms of the best-fit values of $\alpha$, $Q$, and $\lambda^{-1}$ from the three-parameter MCMC analysis of the HIFLUGCS sample. With the exception of a single case, all values of $\alpha$ are below 0.5, with the distribution peaking around $\alpha = 0.25$. In particular, for some clusters we obtain $\alpha < 0.1$, implying that, for these systems, the dynamical mass is nearly proportional to the enclosed baryonic mass. For most clusters, $Q$ is found to be less than 12, and $\lambda^{-1}$ is below 30 in the majority of cases. This indicates that $\lambda^{-1}$ has only a very minor influence on the fit.}
\end{figure}

In Fig.~\ref{fig:CornerPLots}, we show the corner plots for three-parameter fits for six representative clusters. These plots indicate a strong negative correlation between $Q$ and each of the other two parameters, while $\lambda^{-1}$ and $\alpha$ are positively correlated. For clarity, we display corner plots only for a subset of six clusters; however, the qualitative behavior is similar across the full sample, although in some clusters the parameter constraints are comparatively weaker. (all plots are available \href{https://machiangravity.github.io/galaxy_cluster_mass/corner_plots/}{here}.)

The distributions of the parameters from the three-parameter model fits are presented in Fig.~\ref{fig:threeparamproperties}. We find that, except for one case, all clusters have $\alpha < 0.5$, whereas for galaxy fits $\alpha$ is expected to be close to $0.5$ in order to satisfy the Tully–Fisher relation~\cite{das2023aaspects}. The histogram of $\alpha$ instead peaks near $0.25$. The distributions of $Q$ and $\lambda^{-1}$ resemble those obtained for the three-parameter fit: most $Q$ values lie below 10, and most $\lambda^{-1}$ values are below 30 kpc.

We have also used a 2 parameter model where we set the exponent to $\alpha = 0.25$ and check how well it can explain the observations. In Fig.~\ref{fig:linv_Q_25}, we present the histograms of $\lambda^{-1}$ and $Q$ obtained for $\alpha = 0.25$ from the analysis of 45 clusters. In the majority of cases, $\lambda^{-1}$ is of the order of a few tens of kpc, which is extremely small compared to the typical sizes of galaxy clusters. Consequently, apart from a few data points at small cluster-centric radii, most of the remaining data points have a negligible impact on constraining $\lambda^{-1}$. The parameter $Q$ typically lies in the range $5 \text{–} 15$, with a few outliers exhibiting higher values.

In Fig.~\ref{MainPaper}, we present the fits obtained for all the clusters. The red curves with error bars represent the baryonic mass profiles of the galaxy clusters, while the blue points with error bars denote their dynamical mass profiles. The three-parameter fit is shown by the orange curve. The correspondence between the fitted curves and the observational data is remarkably good. The best-fit values of the fitting parameters are summarized in Table~\ref{table2}. %For all clusters, the inferred best-fit values of $\alpha$ are significantly smaller than $0.5$. 

The green curves in Fig.~\ref{fig:CornerPLots} display the best-fit models obtained when $\alpha$ is fixed to $\alpha = 0.25$. For the majority of clusters, the quality of the fit remains high even under this constraint, with only a few systems showing noticeable deviations, primarily in the vicinity of $r_{2500}$. This behavior suggests that the function $K$ in Eq.~\ref{eq:Machiangrav} depends sensitively on the mass distribution within a galaxy or galaxy cluster. Moreover, the functional form of $K$ appears to differ between galaxies and clusters. Although Machian gravity does not predict the exact analytic form of $K$, our results demonstrate that adopting the form given in Eq.~\ref{eq:kformula} yields an excellent fit to the available data.

\section{Discussion and Conclusion}
The Machian gravity framework introduces a modified acceleration law designed to account for gravitational phenomena on large scales~\cite{das2023aspects}. In our previous work, we demonstrated that this modified acceleration law successfully reproduces a wide variety of spiral galaxy rotation curves~\cite{das2023aaspects}. In the present study, we extend this analysis by applying the same acceleration law to a diverse array of X-ray–derived galaxy cluster mass profiles.

Although the MOND acceleration law provides a compelling phenomenological description of spiral galaxy rotation curves, it fails to adequately reproduce galaxy cluster mass profiles without invoking an additional dark matter component. For the HIFLUGCS galaxy cluster sample, we have shown that the mass discrepancy correlates with $M_b$ or $M_b/r$, whereas its correlation with $M_b/r^2$is significantly small, which is the primary assumption of the MOND dynamics. Furthermore, by comparing several pairs of galaxy clusters, we have demonstrated that whenever the dynamical mass profiles of the clusters are matched, their baryonic mass profiles also closely coincide. This behavior cannot be satisfactorily explained either by the standard NFW dark matter density profile or by MOND-like theories incorporating external field effects. In contrast, Machian gravity offers a more natural and consistent explanation for these empirical results.  

Our analysis further indicates a key result: on cluster scales, the effective gravitational behavior is consistent with an inverse-square law for the acceleration, in contradiction with the MOND prescription. Consequently, the cluster data provide strong support for Machian gravity over MOND.

Another noteworthy result is that the characteristic scale $\lambda^{-1}$ is found to be of the order of a few 10s of kpc. This value is comparable to what we obtained in the case of galaxies, which is a particularly intriguing outcome. Given that the physical sizes of galaxy clusters are order of magnitude larger than those of individual galaxies, one would naively expect that the value of $\lambda^{-1}$ inferred from any MCMC analysis for clusters should be correspondingly larger. However, our results do not align with this expectation. Moreover, $\lambda^{-1}$ does not exhibit any significant dependence on the $K$ term. We obtain consistent values across both cluster samples, and the inferred scale remains essentially insensitive to variations in the exponent of the $K$ term. In all cases, the order of magnitude of $\lambda^{-1}$ remains of the order of tens of kpc.

We also find that, although a square-root dependence of the parameter $K$ is sufficient to explain the Tully–Fisher relation in spiral galaxies and yields an acceptable fit to the cluster data even with fixed $M_c$ and $\lambda^{-1}$, the cluster analysis favors a slight modification of this scaling. In particular, adopting a power-law form for $K$ with an exponent of approximately $0.25$ significantly improves the quality of the fits. Since $K$ depends on the large-scale background mass distribution, a more fundamental theoretical treatment is required to determine its precise functional form. Developing such a theory, and thereby deriving the exact dependence of $K$, is left for future work.

\bibliographystyle{JHEP}
\bibliography{reference}

\providecommand{\href}[2]{#2}\begingroup\raggedright\begin{thebibliography}{10}

\bibitem{reiprich2003cosmological}
T.H.~Reiprich, \emph{Cosmological implications and physical properties of an
  x-ray flux-limited sample of galaxy clusters},  2003.

\bibitem{Sancisi:2003xt}
R.~Sancisi, \emph{{The Visible matter - Dark matter coupling}}, {\emph{IAU
  Symp.} {\bfseries 220} (2004) 233}
  [\href{https://arxiv.org/abs/astro-ph/0311348}{{\ttfamily
  astro-ph/0311348}}].

\bibitem{navarro1997universal}
J.F.~Navarro, C.S.~Frenk and S.D.~White, \emph{A universal density profile from
  hierarchical clustering}, {\emph{The Astrophysical Journal} {\bfseries 490}
  (1997) 493}.

\bibitem{efstathiou1985numerical}
G.~Efstathiou, M.~Davis, S.~White and C.~Frenk, \emph{Numerical techniques for
  large cosmological n-body simulations}, {\emph{Astrophysical Journal
  Supplement Series (ISSN 0067-0049), vol. 57, Feb. 1985, p. 241-260. Research
  supported by Cambridge University, Science and Engineering Research Council
  of England, and NASA.} {\bfseries 57} (1985) 241}.

\bibitem{davis1985evolution}
M.~Davis, G.~Efstathiou, C.S.~Frenk and S.D.~White, \emph{The evolution of
  large-scale structure in a universe dominated by cold dark matter},
  {\emph{Astrophysical Journal, Part 1 (ISSN 0004-637X), vol. 292, May 15,
  1985, p. 371-394. Research supported by the Science and Engineering Research
  Council of England and NASA.} {\bfseries 292} (1985) 371}.

\bibitem{Milgrim1983}
M.~Milgrom, \emph{{A Modification of the Newtonian dynamics as a possible
  alternative to the hidden mass hypothesis}},
  \href{https://doi.org/10.1086/161130}{\emph{Astrophys.J.} {\bfseries 270}
  (1983) 365}.

\bibitem{Milgrim1983a}
M.~Milgrom, \emph{{A Modification of the Newtonian dynamics: Implications for
  galaxies}}, \href{https://doi.org/10.1086/161131}{\emph{Astrophys.J.}
  {\bfseries 270} (1983) 371}.

\bibitem{Milgrim1983b}
M.~Milgrom, \emph{{A modification of the Newtonian dynamics: implications for
  galaxy systems}}, \href{https://doi.org/10.1086/161132}{\emph{Astrophys.J.}
  {\bfseries 270} (1983) 384}.

\bibitem{Milgrom2011}
M.~Milgrom, \emph{{MD or DM? Modified dynamics at low accelerations vs dark
  matter}}, {\emph{PoS} {\bfseries HRMS2010} (2010) 033}
  [\href{https://arxiv.org/abs/1101.5122}{{\ttfamily 1101.5122}}].

\bibitem{Bekenstein1984}
J.~Bekenstein and M.~Milgrom, \emph{{Does the missing mass problem signal the
  breakdown of Newtonian gravity?}},
  \href{https://doi.org/10.1086/162570}{\emph{Astrophys.J.} {\bfseries 286}
  (1984) 7}.

\bibitem{Bekenstein2009}
J.D.~Bekenstein, \emph{{Relativistic MOND as an alternative to the dark matter
  paradigm}},
  \href{https://doi.org/10.1016/j.nuclphysa.2009.05.122}{\emph{Nucl.Phys.}
  {\bfseries A827} (2009) 555C}
  [\href{https://arxiv.org/abs/0901.1524}{{\ttfamily 0901.1524}}].

\bibitem{Milgrom1986}
M.~Milgrom, \emph{{Solutions for the modified Newtonian dynamics field
  equation}}, \href{https://doi.org/10.1086/164021}{\emph{Astrophys.J.}
  {\bfseries 302} (1986) 617}.

\bibitem{sanders1998virial}
R.~Sanders, \emph{The virial discrepancy in clusters of galaxies in the context
  of modified newtonian dynamics}, {\emph{The Astrophysical Journal} {\bfseries
  512} (1998) L23}.

\bibitem{ettori2019hydrostatic}
S.~Ettori, V.~Ghirardini, D.~Eckert, E.~Pointecouteau, F.~Gastaldello,
  M.~Sereno et~al., \emph{Hydrostatic mass profiles in x-cop galaxy clusters},
  {\emph{Astronomy \& Astrophysics} {\bfseries 621} (2019) A39}.

\bibitem{li2023measuring}
P.~Li, Y.~Tian, M.P.~J{\'u}lio, M.S.~Pawlowski, F.~Lelli, S.S.~McGaugh et~al.,
  \emph{Measuring galaxy cluster mass profiles into the low-acceleration regime
  with galaxy kinematics}, {\emph{Astronomy \& Astrophysics} {\bfseries 677}
  (2023) A24}.

\bibitem{gerbal1992analysis}
D.~Gerbal, F.~Durret, M.~Lachieze-Rey and G.~Lima-Neto, \emph{Analysis of x-ray
  galaxy clusters in the framework of modified newtonian dynamics},
  {\emph{Astronomy and Astrophysics (ISSN 0004-6361), vol. 262, no. 2, p.
  395-400.} {\bfseries 262} (1992) 395}.

\bibitem{angus2011abundance}
G.W.~Angus and A.~Diaferio, \emph{The abundance of galaxy clusters in modified
  newtonian dynamics: cosmological simulations with massive neutrinos},
  {\emph{Monthly Notices of the Royal Astronomical Society} {\bfseries 417}
  (2011) 941}.

\bibitem{pointecouteau2005new}
E.~Pointecouteau and J.~Silk, \emph{New constraints on modified newtonian
  dynamics from galaxy clusters}, {\emph{Monthly Notices of the Royal
  Astronomical Society} {\bfseries 364} (2005) 654}.

\bibitem{Moffat2005}
J.~Moffat, \emph{{Scalar-tensor-vector gravity theory}},
  \href{https://doi.org/10.1088/1475-7516/2006/03/004}{\emph{JCAP} {\bfseries
  0603} (2006) 004} [\href{https://arxiv.org/abs/gr-qc/0506021}{{\ttfamily
  gr-qc/0506021}}].

\bibitem{Brownstein2005}
J.~Brownstein and J.~Moffat, \emph{{Galaxy rotation curves without non-baryonic
  dark matter}}, \href{https://doi.org/10.1086/498208}{\emph{Astrophys.J.}
  {\bfseries 636} (2006) 721}
  [\href{https://arxiv.org/abs/astro-ph/0506370}{{\ttfamily
  astro-ph/0506370}}].

\bibitem{Brownstein2005a}
J.~Brownstein and J.~Moffat, \emph{{Galaxy cluster masses without non-baryonic
  dark matter}},
  \href{https://doi.org/10.1111/j.1365-2966.2006.09996.x}{\emph{Mon.Not.Roy.Astron.Soc.}
  {\bfseries 367} (2006) 527}
  [\href{https://arxiv.org/abs/astro-ph/0507222}{{\ttfamily
  astro-ph/0507222}}].

\bibitem{Moffat2005a}
J.~Moffat and V.~Toth, \emph{{Modified Gravity: Cosmology without dark matter
  or Einstein's cosmological constant}},
  \href{https://arxiv.org/abs/0710.0364}{{\ttfamily 0710.0364}}.

\bibitem{Bekenstein2005}
J.D.~Bekenstein, \emph{{Relativistic gravitation theory for the MOND
  paradigm}}, \href{https://doi.org/10.1103/PhysRevD.70.083509,
  10.1103/PhysRevD.70.083509 10.1103/PhysRevD.71.069901,
  10.1103/PhysRevD.71.069901}{\emph{Phys.Rev.} {\bfseries D70} (2004) 083509}
  [\href{https://arxiv.org/abs/astro-ph/0403694}{{\ttfamily
  astro-ph/0403694}}].

\bibitem{Dam1970}
H.~van Dam and M.~Veltman, \emph{{Massive and massless Yang-Mills and
  gravitational fields}},
  \href{https://doi.org/10.1016/0550-3213(70)90416-5}{\emph{Nucl.Phys.}
  {\bfseries B22} (1970) 397}.

\bibitem{Zakharov1970}
V.~Zakharov, \emph{{Linearized gravitation theory and the graviton mass}},
  {\emph{JETP Lett.} {\bfseries 12} (1970) 312}.

\bibitem{Babichev2010}
E.~Babichev, C.~Deffayet and R.~Ziour, \emph{{The Recovery of General
  Relativity in massive gravity via the Vainshtein mechanism}},
  \href{https://doi.org/10.1103/PhysRevD.82.104008}{\emph{Phys.Rev.} {\bfseries
  D82} (2010) 104008} [\href{https://arxiv.org/abs/1007.4506}{{\ttfamily
  1007.4506}}].

\bibitem{Babichev2013}
E.~{Babichev} and M.~{Crisostomi}, \emph{{Restoring general relativity in
  massive bigravity theory}},
  \href{https://doi.org/10.1103/PhysRevD.88.084002}{\emph{Phys.Rev.} {\bfseries
  D88} (2013) 084002} [\href{https://arxiv.org/abs/1307.3640}{{\ttfamily
  1307.3640}}].

\bibitem{Overduin1998}
J.~Overduin and P.~Wesson, \emph{{Kaluza-Klein gravity}},
  \href{https://doi.org/10.1016/S0370-1573(96)00046-4}{\emph{Phys.Rept.}
  {\bfseries 283} (1997) 303}
  [\href{https://arxiv.org/abs/gr-qc/9805018}{{\ttfamily gr-qc/9805018}}].

\bibitem{Ponce1993}
J.~{Ponce de Leon} and P.S.~{Wesson}, \emph{{Exact solutions and the effective
  equation of state in Kaluza-Klein theory}},
  \href{https://doi.org/10.1063/1.530028}{\emph{Journal of Mathematical
  Physics} {\bfseries 34} (1993) 4080}.

\bibitem{Wesson1992}
P.S.~{Wesson} and J.P.~{de Leon}, \emph{{Kaluza-Klein equations, Einstein's
  equations, and an effective energy-momentum tensor.}},
  \href{https://doi.org/10.1063/1.529834}{\emph{Journal of Mathematical
  Physics} {\bfseries 33} (1992) 3883}.

\bibitem{de2010schwarzschild}
J.P.~de~Leon, \emph{Schwarzschild-like exteriors for stars in kaluza-klein
  gravity}, {\emph{arXiv preprint arXiv:1003.3151} (2010) }.

\bibitem{moraes2016cosmic}
P.~Moraes, \emph{Cosmic acceleration from varying masses in five dimensions},
  {\emph{International Journal of Modern Physics D} {\bfseries 25} (2016)
  1650009}.

\bibitem{das2023aspects}
S.~Das, \emph{Machian gravity: A mathematical formulation for mach's
  principle},  2023.

\bibitem{das2023aaspects}
S.~Das, \emph{Machian gravity: Modeling rotation curves and radial acceleration
  in the sparc galaxy sample},  2023.

\bibitem{mcgaugh2015tale}
S.S.~McGaugh, \emph{A tale of two paradigms: the mutual incommensurability of
  $\lambda$cdm and mond}, {\emph{Canadian Journal of Physics} {\bfseries 93}
  (2015) 250}.

\bibitem{main2016relationship}
R.~Main, B.~McNamara, P.~Nulsen, H.~Russell and A.~Vantyghem, \emph{A
  relationship between halo mass, cooling, agn heating, and the coevolution of
  massive black holes}, {\emph{Monthly Notices of the Royal Astronomical
  Society} (2016) stw2644}.

\bibitem{Jammer}
M.~Jammer, \emph{{Concepts of mass in contemporary physics and philosophy}},
  Princeton Univ. Press (2000).

\bibitem{Moffat2009}
J.~Moffat and V.~Toth, \emph{{Fundamental parameter-free solutions in modified
  gravity}},
  \href{https://doi.org/10.1088/0264-9381/26/8/085002}{\emph{Class.Quant.Grav.}
  {\bfseries 26} (2009) 085002}
  [\href{https://arxiv.org/abs/0712.1796}{{\ttfamily 0712.1796}}].

\bibitem{King1966}
I.R.~{King}, \emph{{The structure of star clusters. IV. Photoelectric surface
  photometry in nine globular clusters}},
  \href{https://doi.org/10.1086/109918}{\emph{Astronomical Journal} {\bfseries
  71} (1966) 276}.

\bibitem{Cavaliere1976}
A.~{Cavaliere} and R.~{Fusco-Femiano}, \emph{{X-rays from hot plasma in
  clusters of galaxies}}, {\emph{Astronomy and Astrophysics} {\bfseries 49}
  (1976) 137}.

\bibitem{sanders1994faber}
R.~Sanders, \emph{A faber-jackson relation for clusters of galaxies:
  Implications for modified dynamics}, {\emph{Astronomy and Astrophysics}
  {\bfseries 284} (1994) L31}.

\bibitem{reiprich2002mass}
T.H.~Reiprich and H.~Boehringer, \emph{The mass function of an x-ray
  flux-limited sample of galaxy clusters}, {\emph{The Astrophysical Journal}
  {\bfseries 567} (2002) 716}.

\end{thebibliography}\endgroup

\begin{figure}
    \centering
    \includegraphics[trim=2.0cm 2.0cm 3.0cm 3.0cm, clip=true, width=0.32\columnwidth]{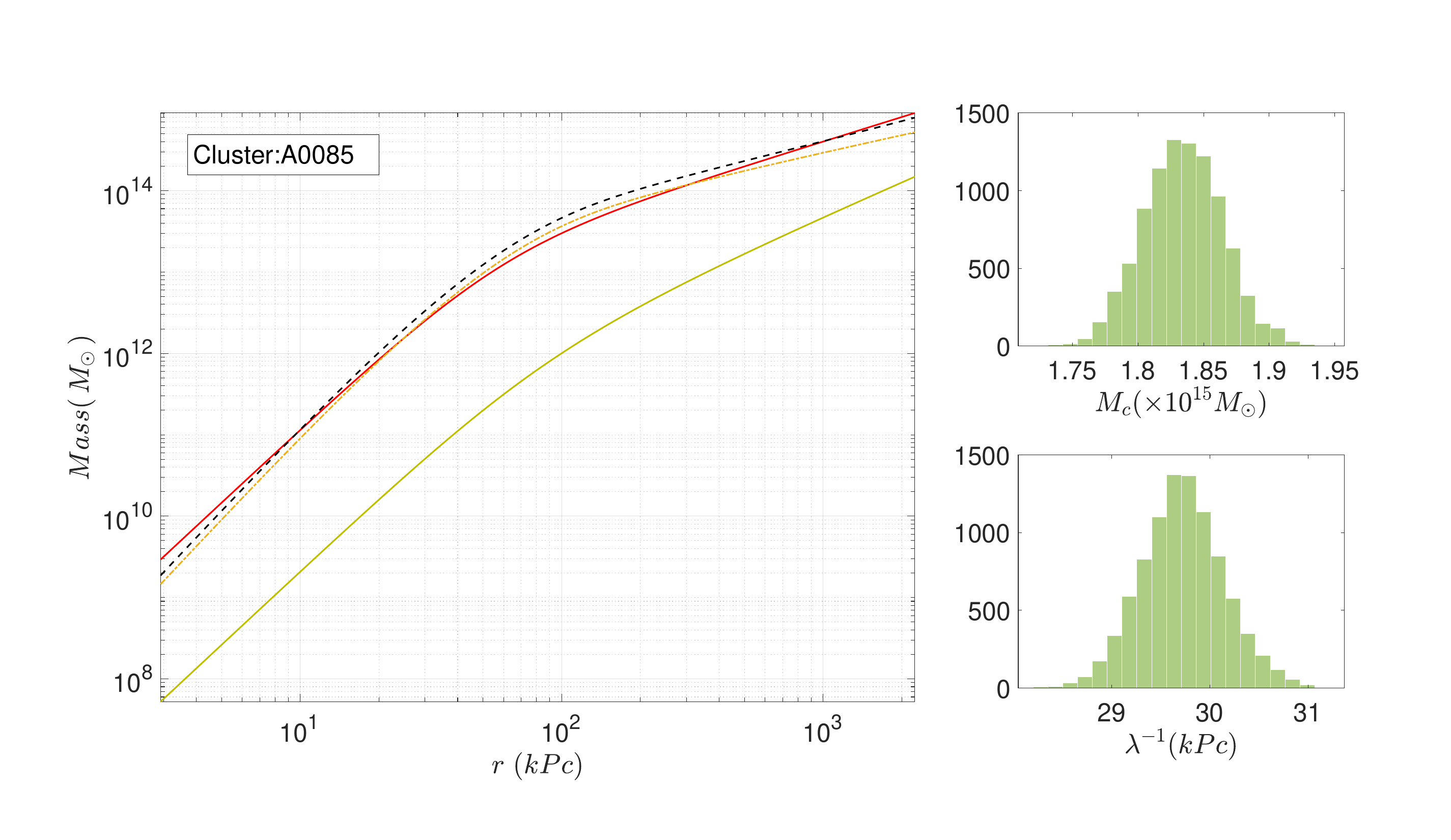}
    \includegraphics[trim=2.0cm 2.0cm 3.0cm 3.0cm, clip=true, width=0.32\columnwidth]{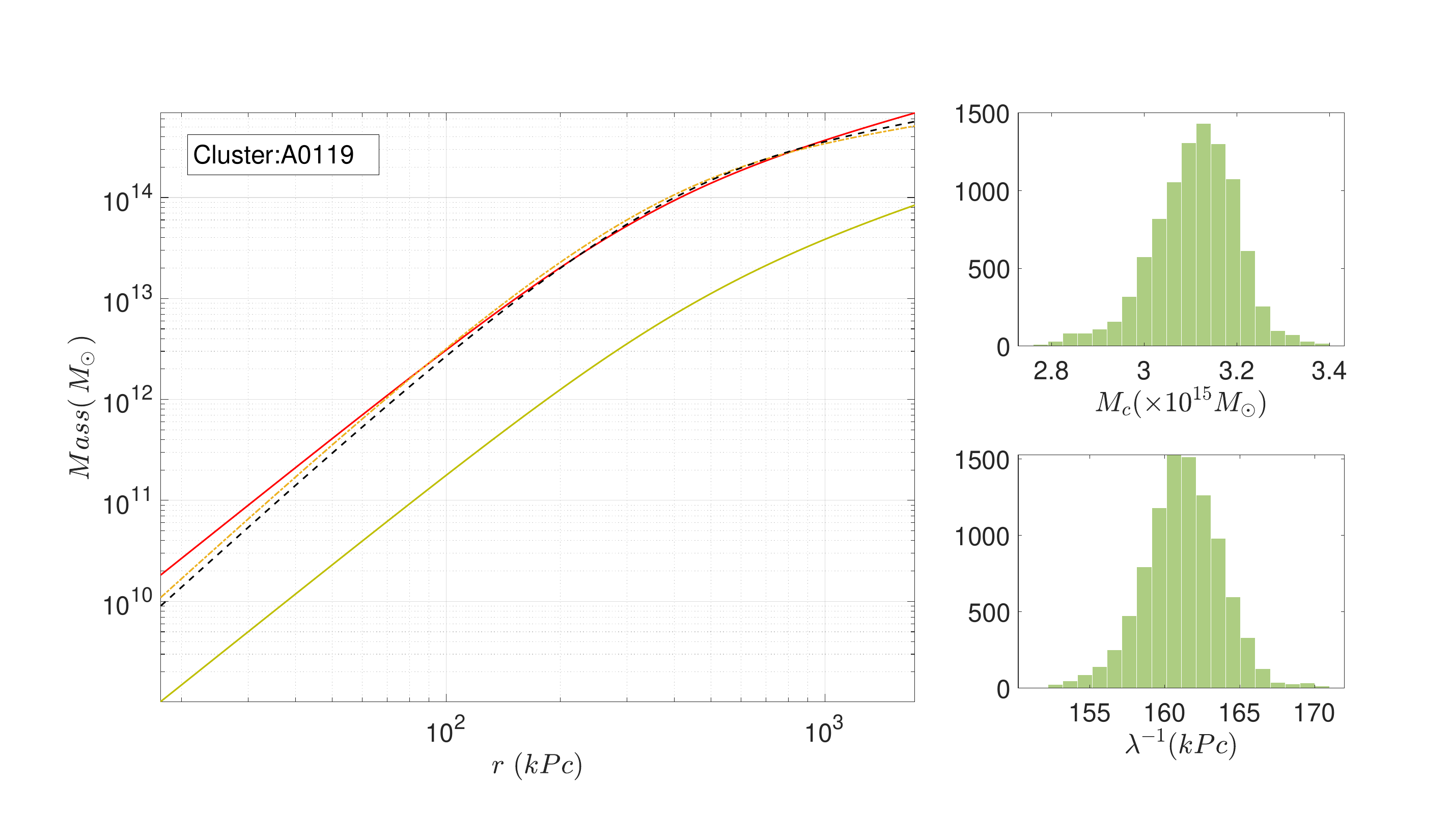}
    \includegraphics[trim=2.0cm 2.0cm 3.0cm 3.0cm, clip=true, width=0.32\columnwidth]{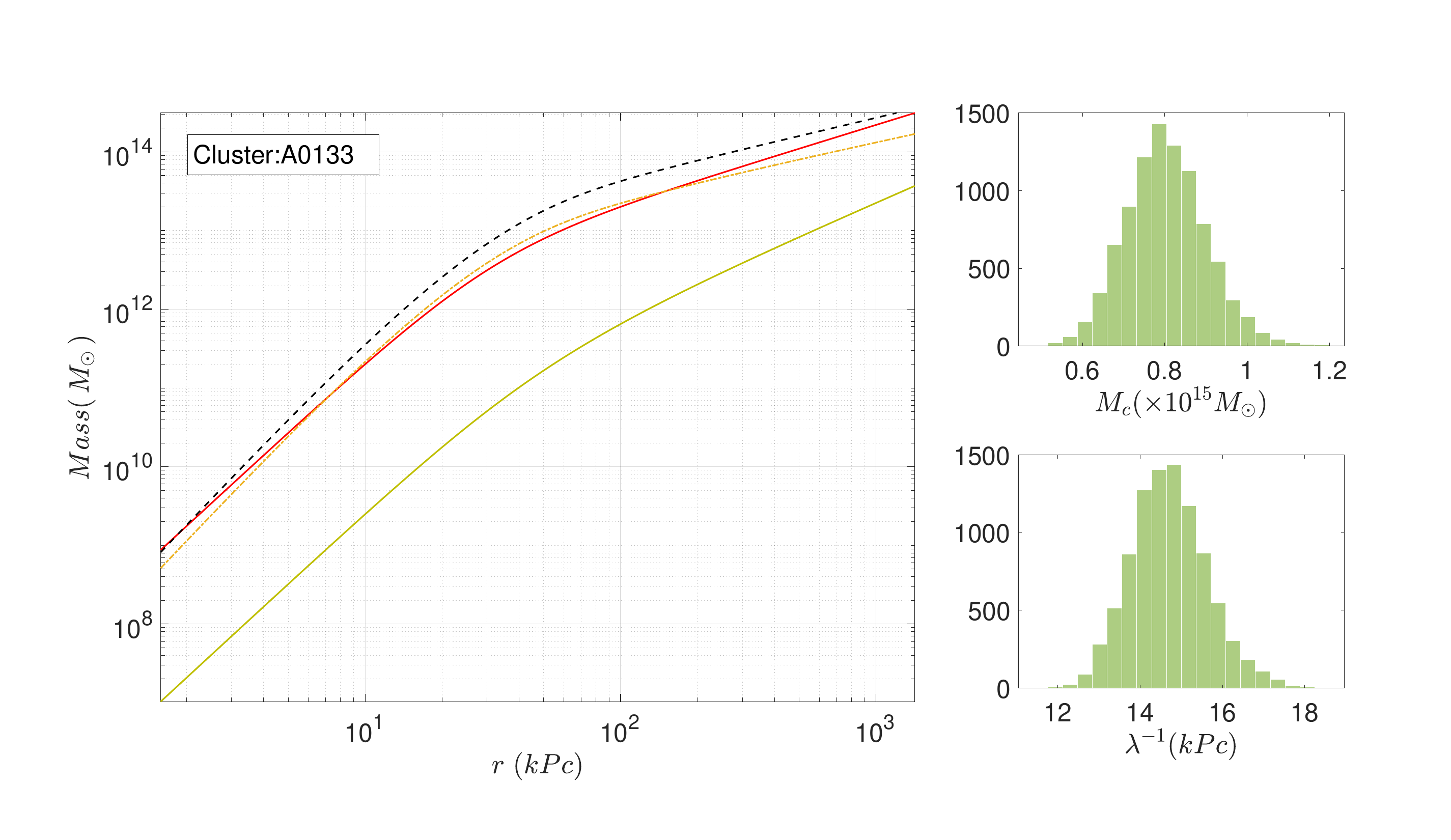}
    \includegraphics[trim=2.0cm 2.0cm 3.0cm 3.0cm, clip=true, width=0.32\columnwidth]{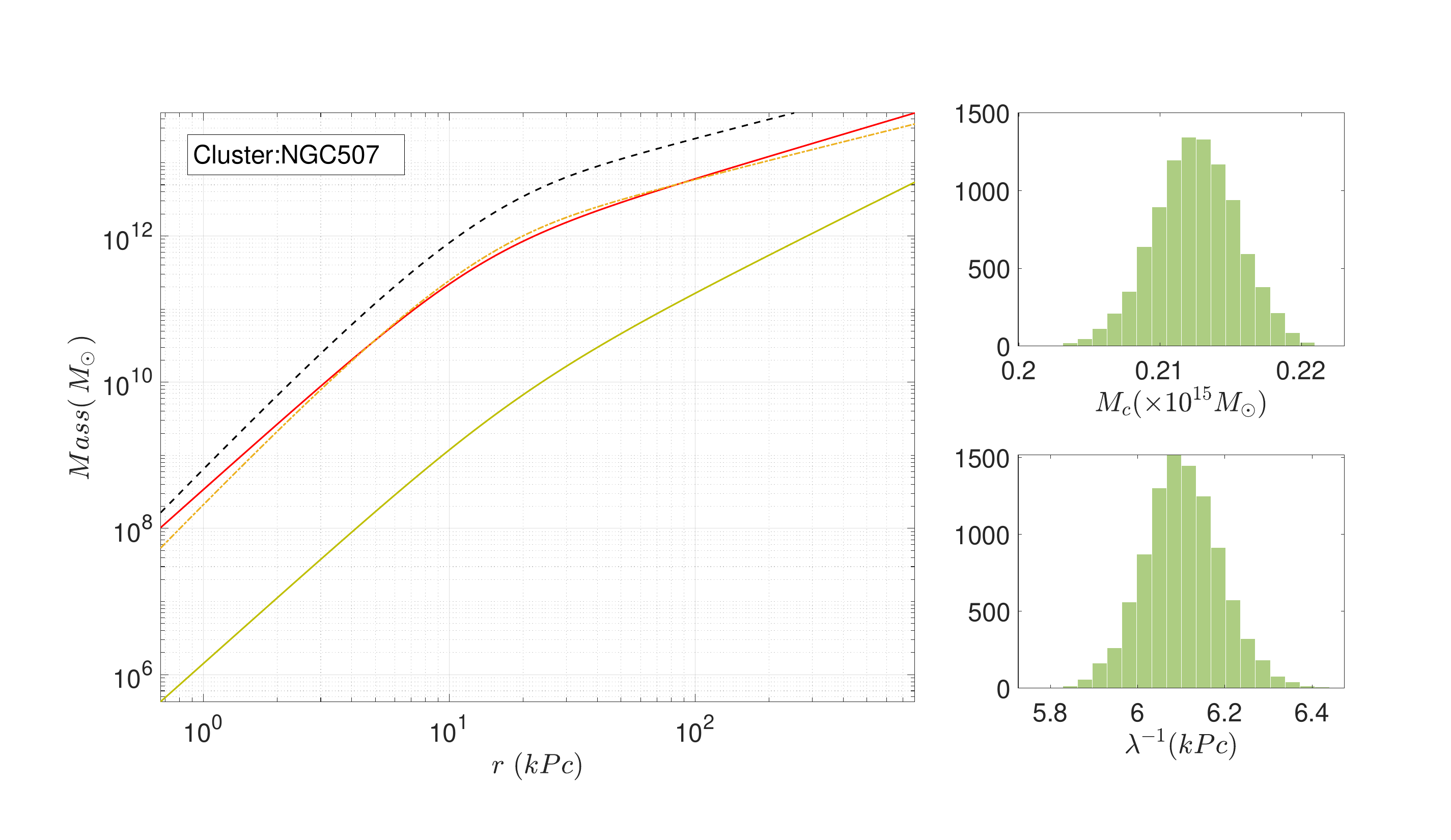}
    \includegraphics[trim=2.0cm 2.0cm 3.0cm 3.0cm, clip=true, width=0.32\columnwidth]{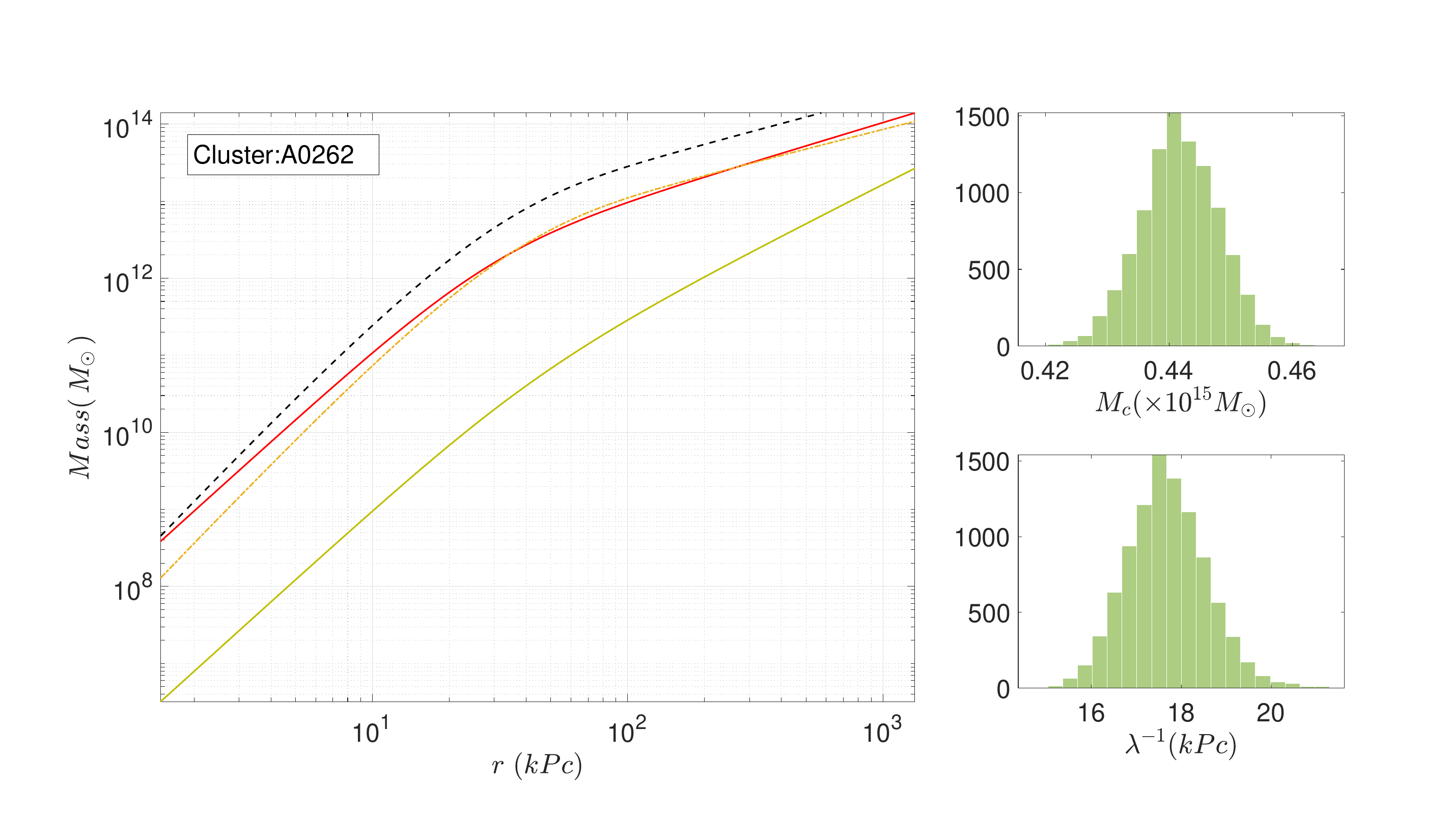}
    \includegraphics[trim=2.0cm 2.0cm 3.0cm 3.0cm, clip=true, width=0.32\columnwidth]{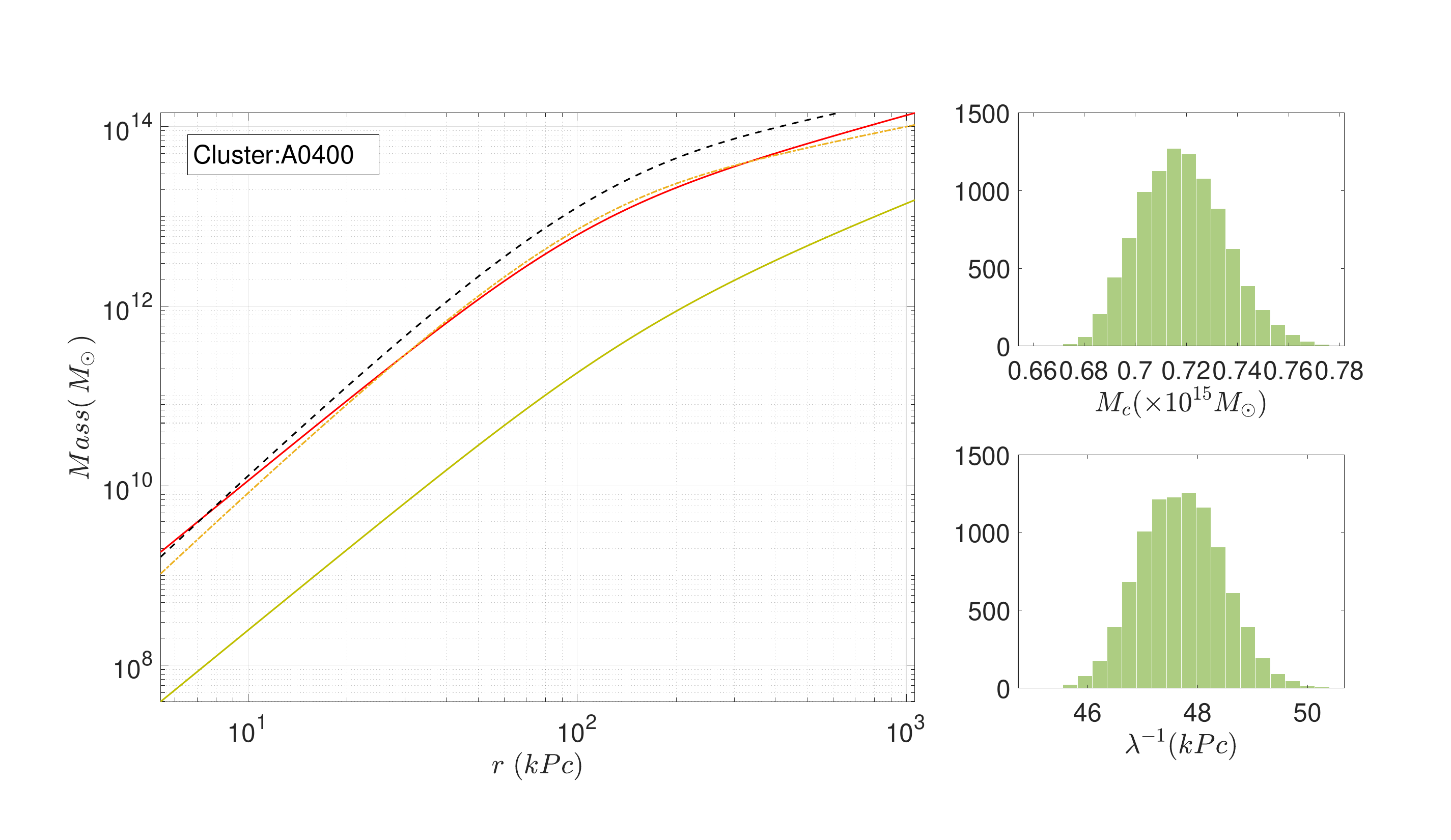}
    \includegraphics[trim=2.0cm 2.0cm 3.0cm 3.0cm, clip=true, width=0.32\columnwidth]{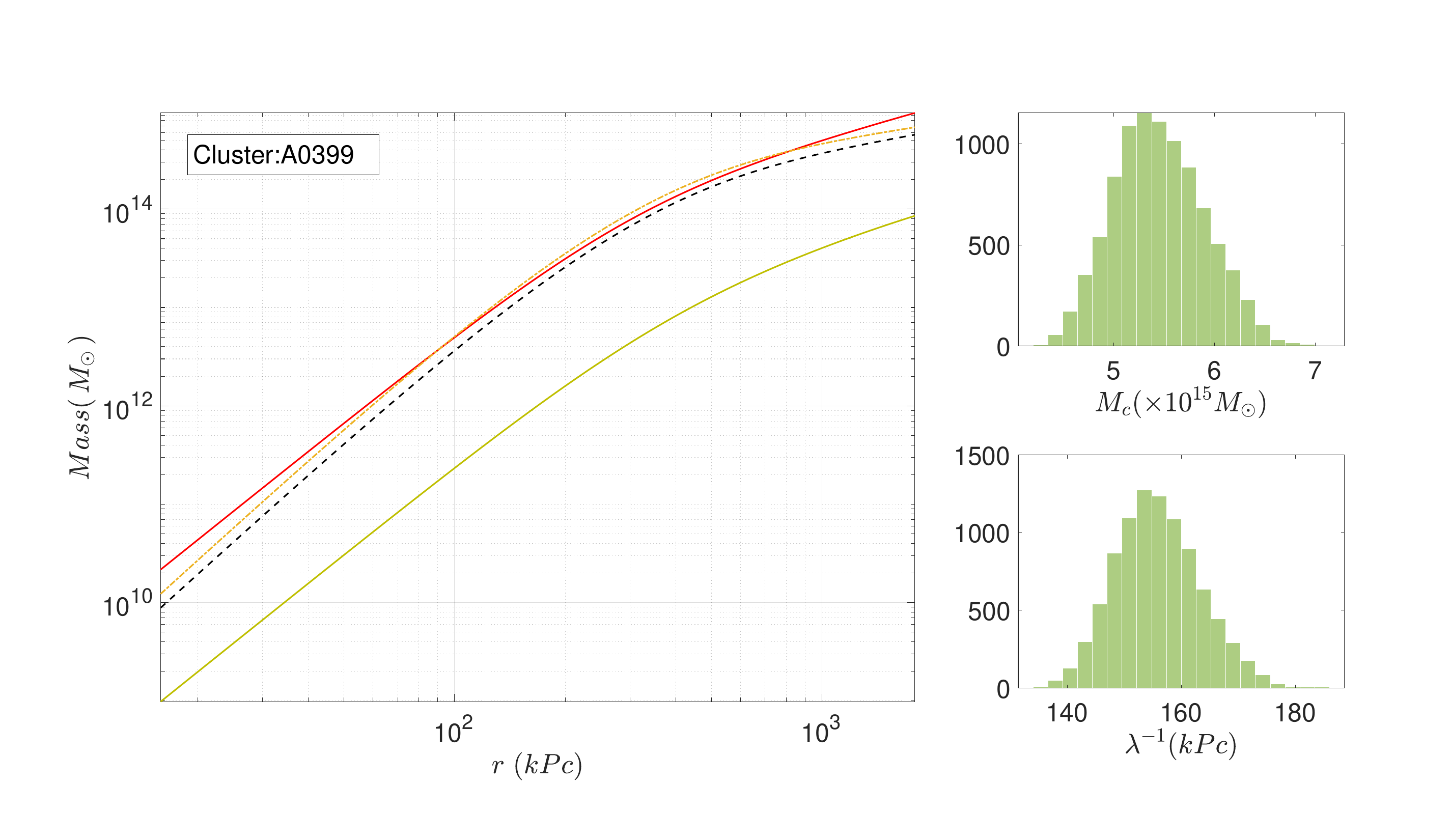}
    \includegraphics[trim=2.0cm 2.0cm 3.0cm 3.0cm, clip=true, width=0.32\columnwidth]{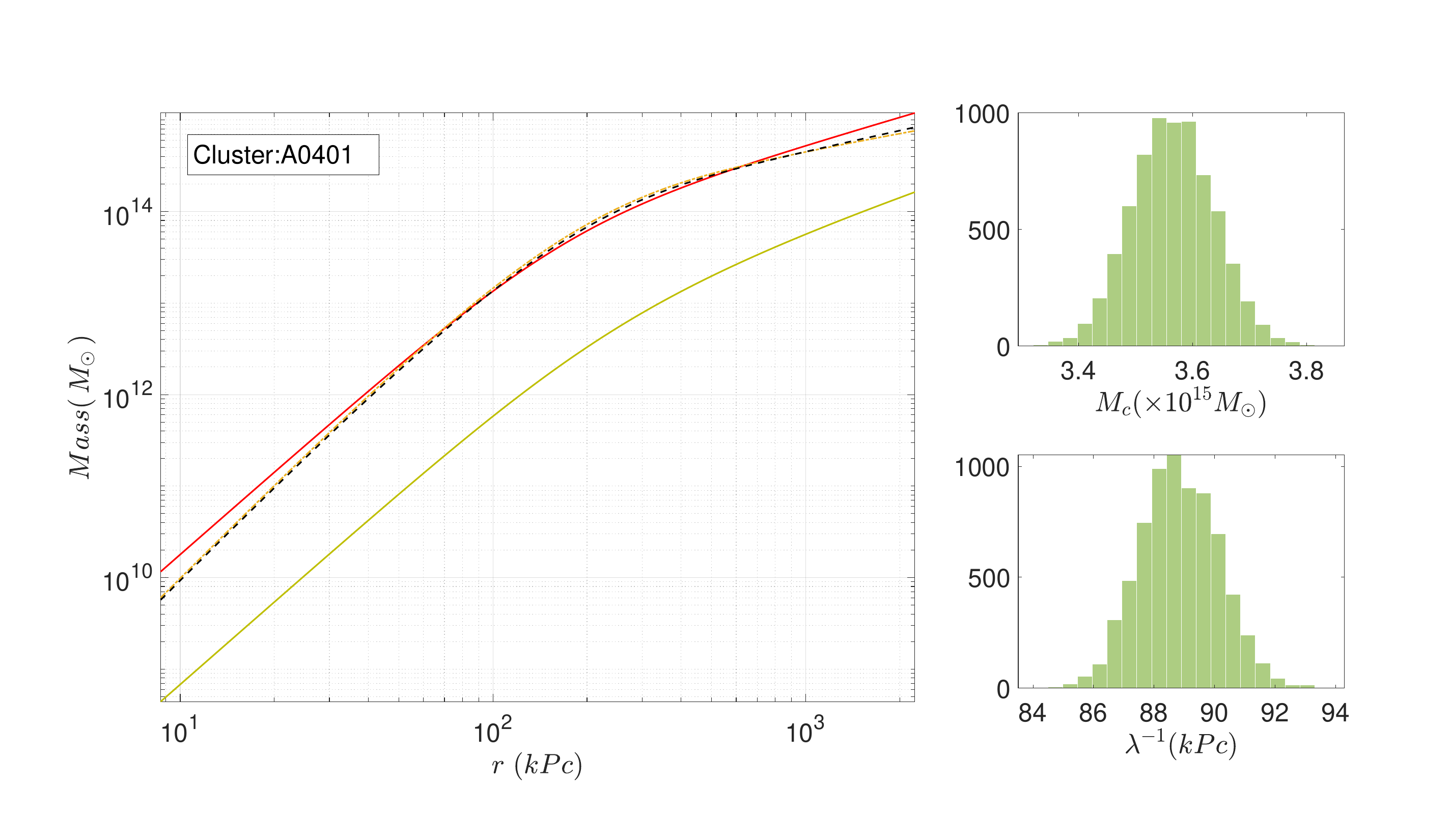}
    \includegraphics[trim=2.0cm 2.0cm 3.0cm 3.0cm, clip=true, width=0.32\columnwidth]{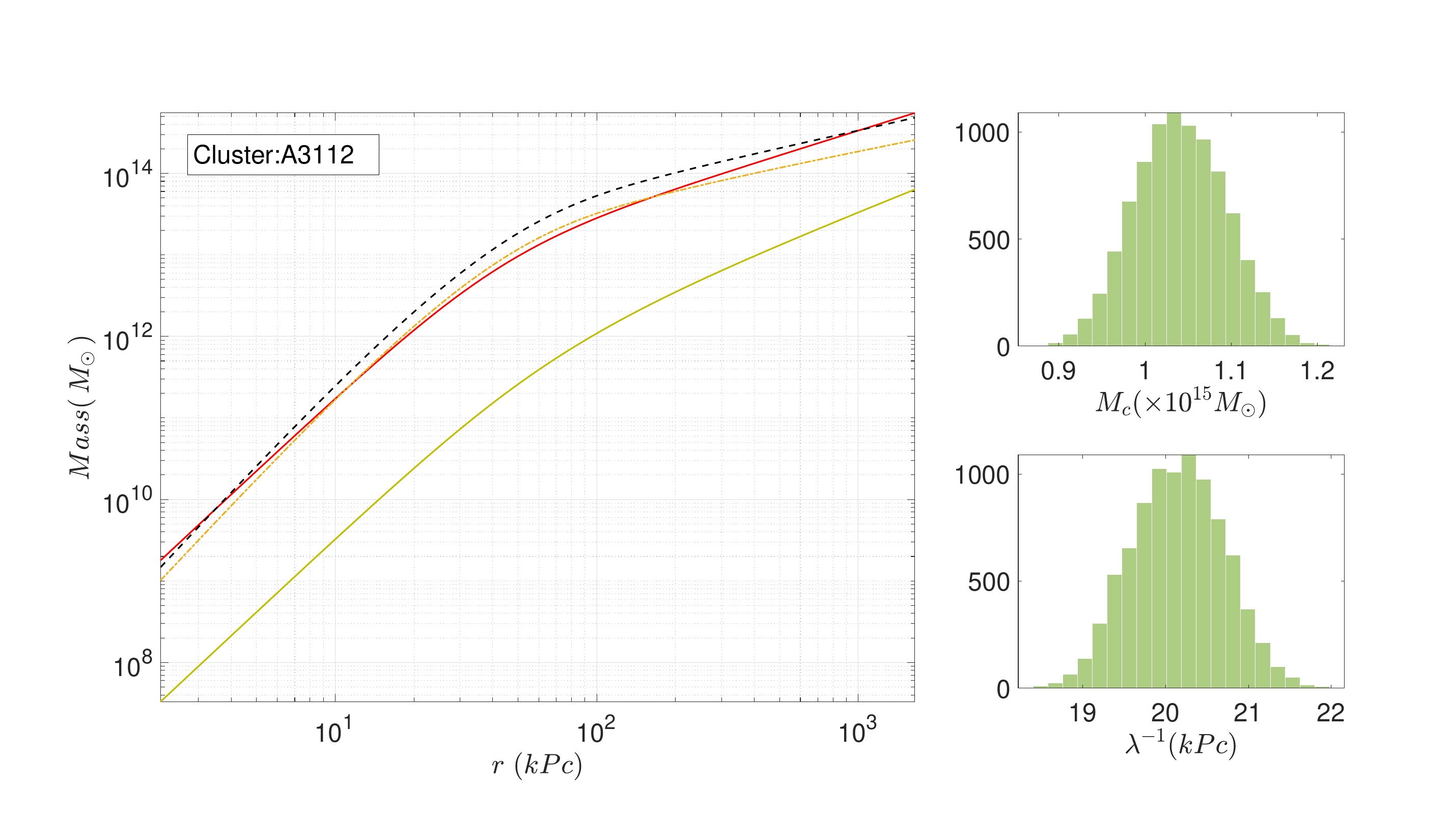}
    \includegraphics[trim=2.0cm 2.0cm 3.0cm 3.0cm, clip=true, width=0.32\columnwidth]{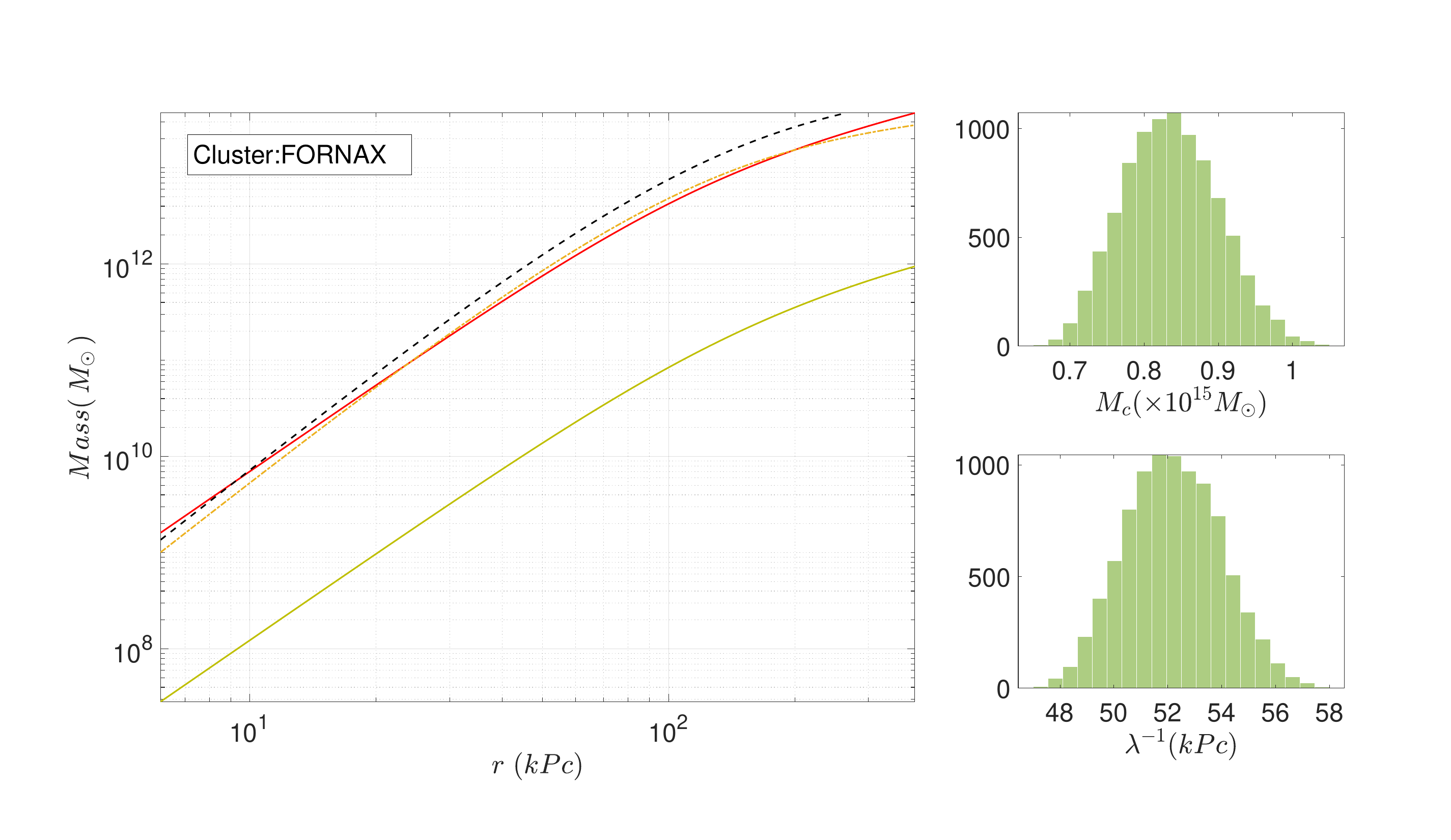}
    \includegraphics[trim=2.0cm 2.0cm 3.0cm 3.0cm, clip=true, width=0.32\columnwidth]{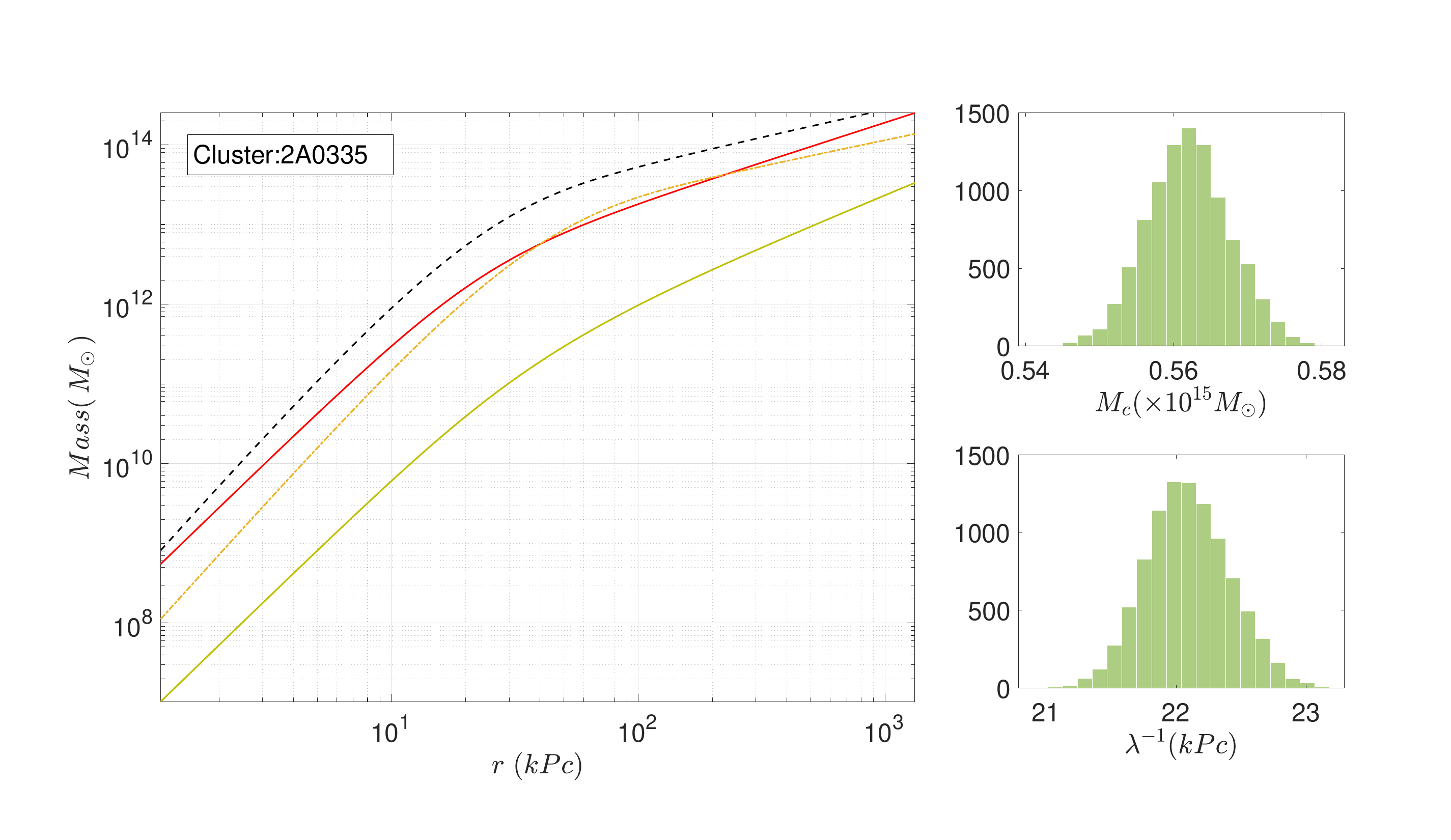}
    \includegraphics[trim=2.0cm 2.0cm 3.0cm 3.0cm, clip=true, width=0.32\columnwidth]{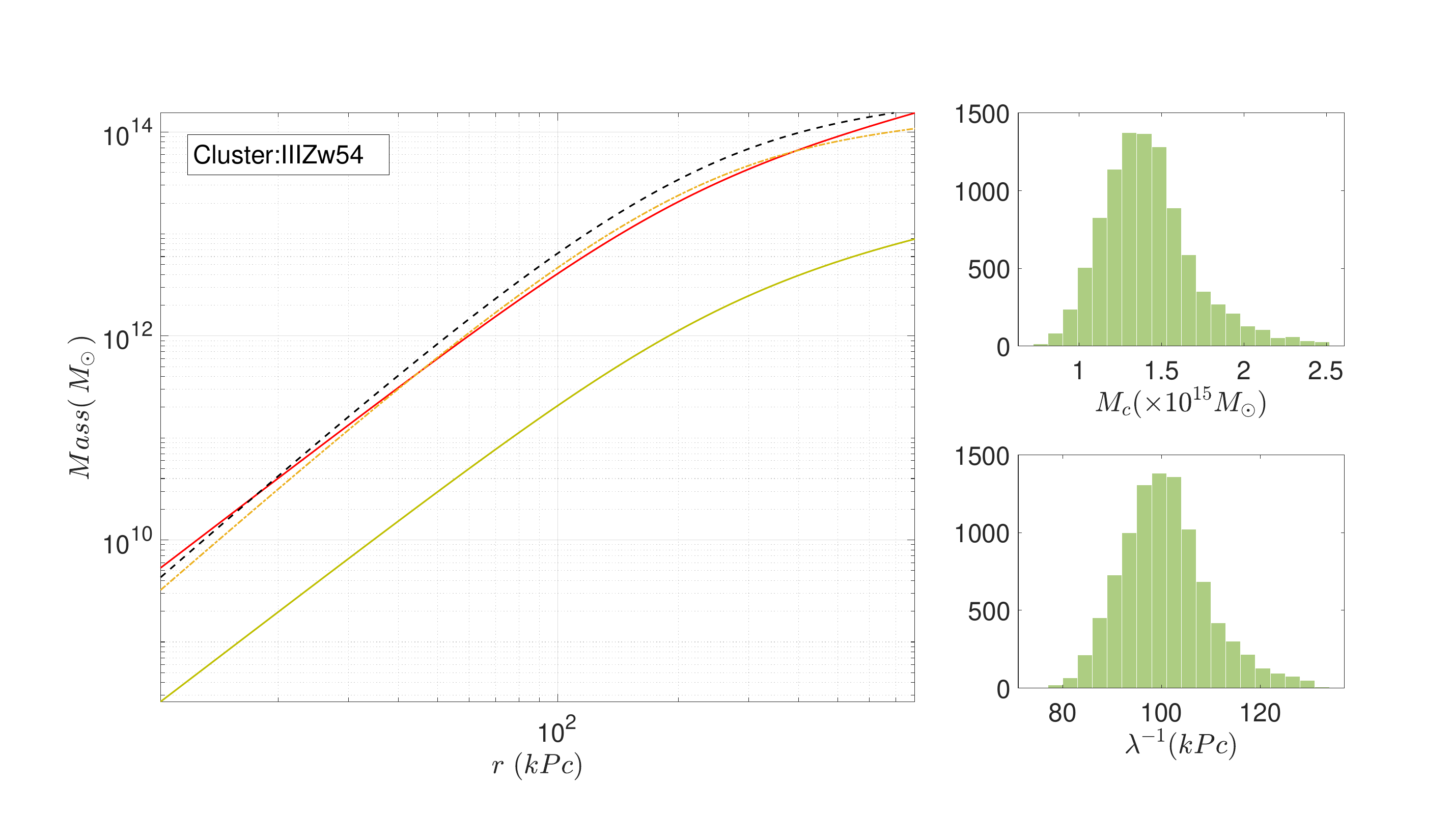}
    \includegraphics[trim=2.0cm 2.0cm 3.0cm 3.0cm, clip=true, width=0.32\columnwidth]{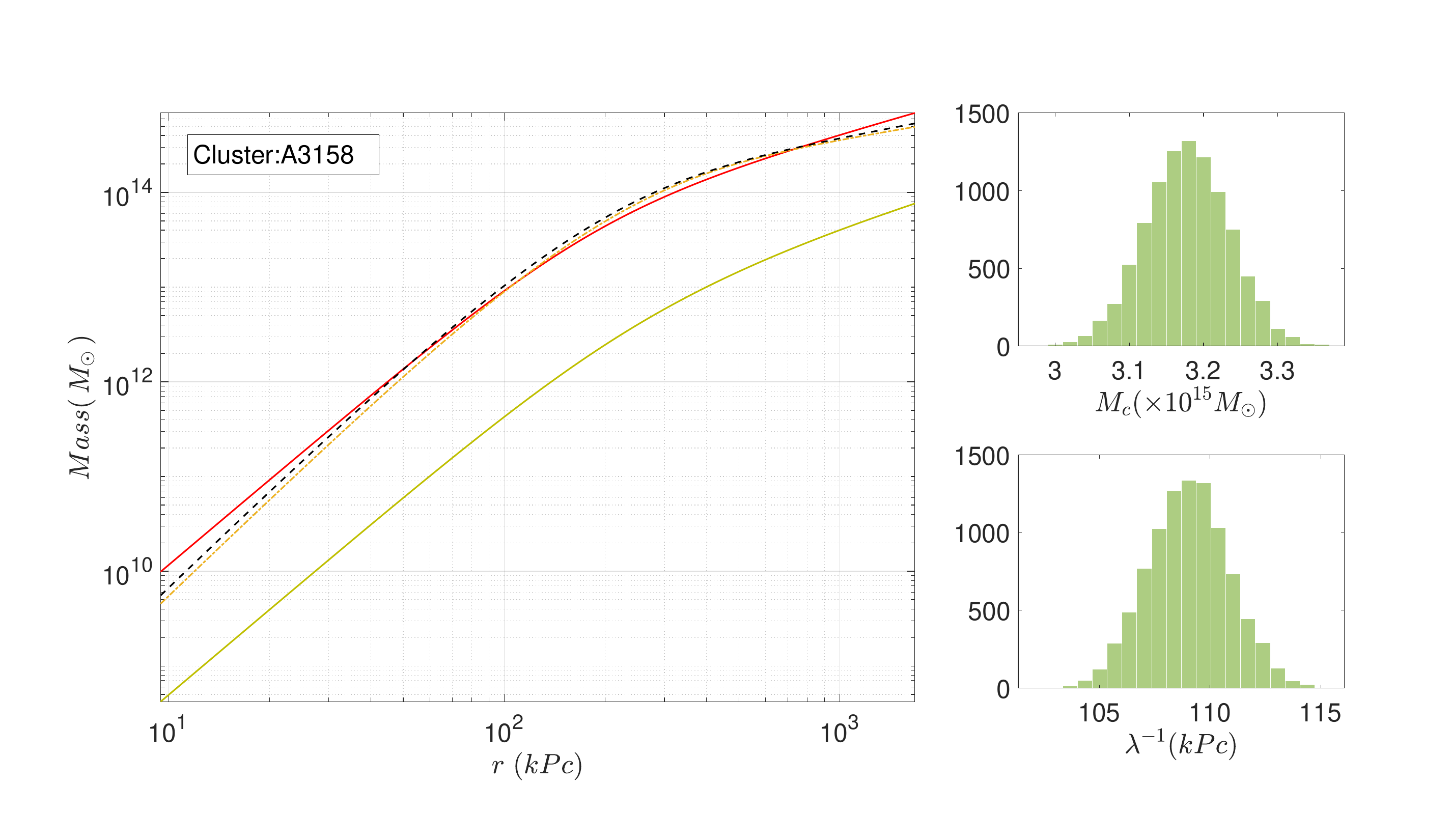}
    \includegraphics[trim=2.0cm 2.0cm 3.0cm 3.0cm, clip=true, width=0.32\columnwidth]{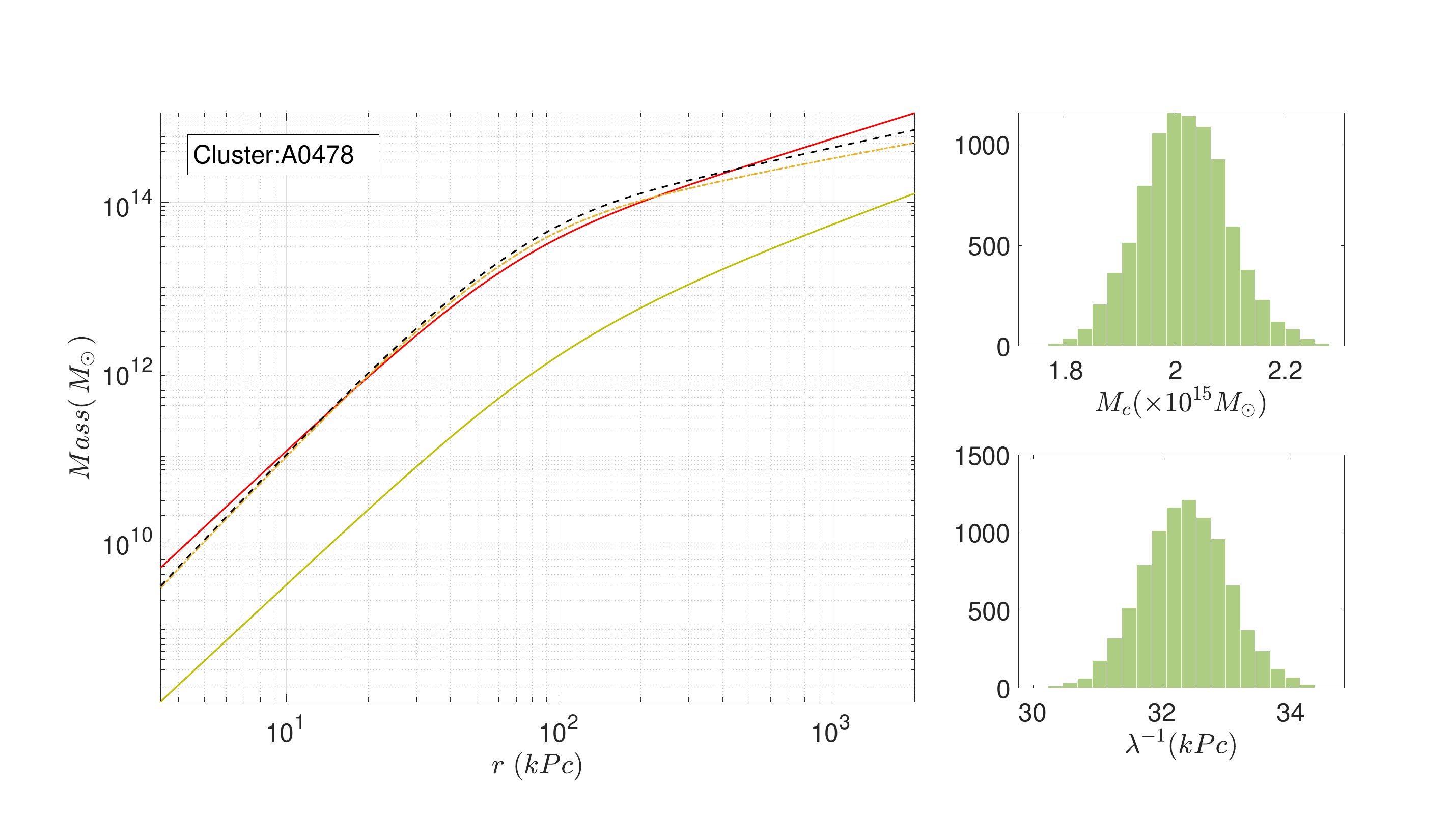}
    \includegraphics[trim=2.0cm 2.0cm 3.0cm 3.0cm, clip=true, width=0.32\columnwidth]{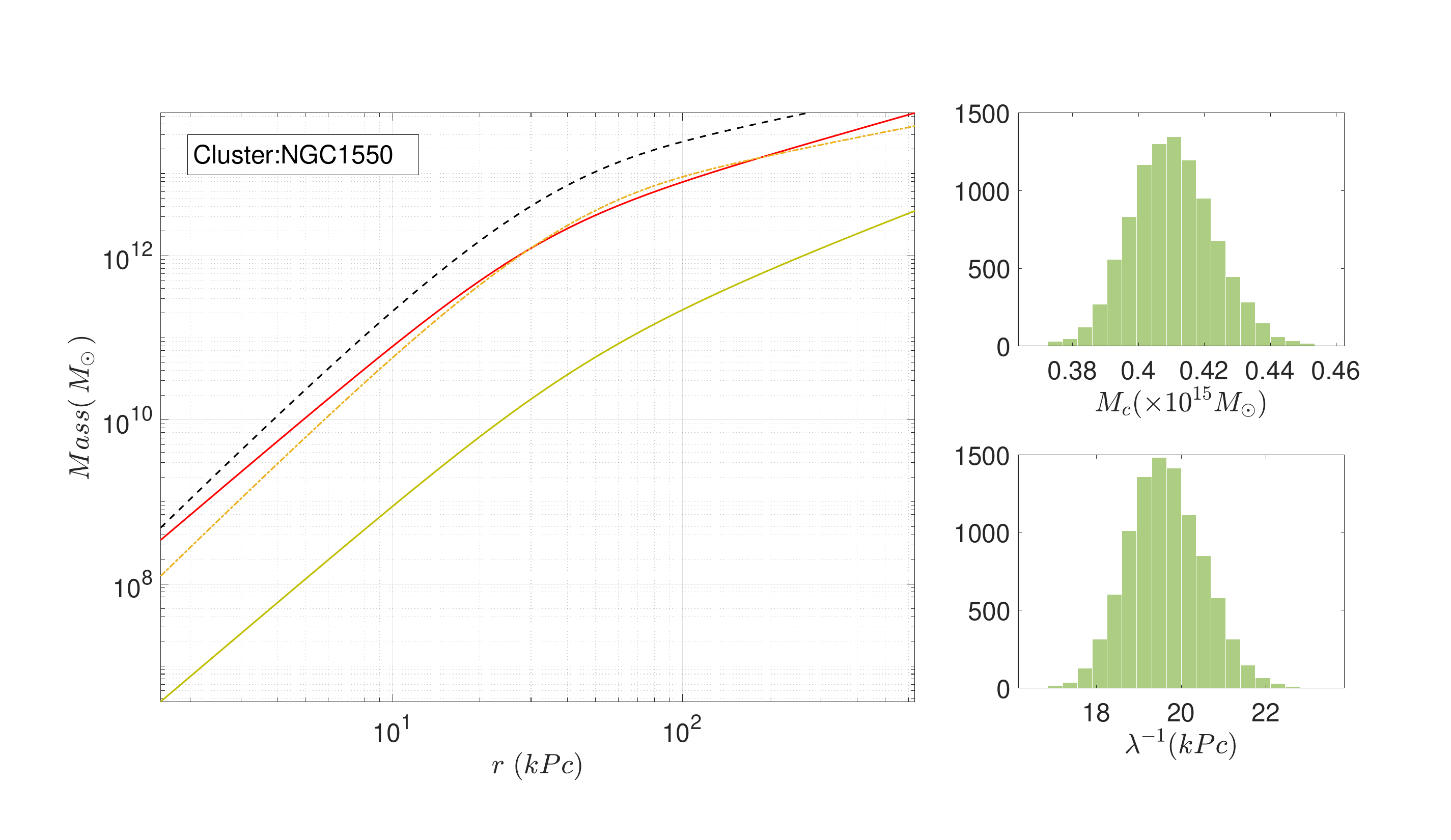}
    \includegraphics[trim=2.0cm 2.0cm 3.0cm 3.0cm, clip=true, width=0.32\columnwidth]{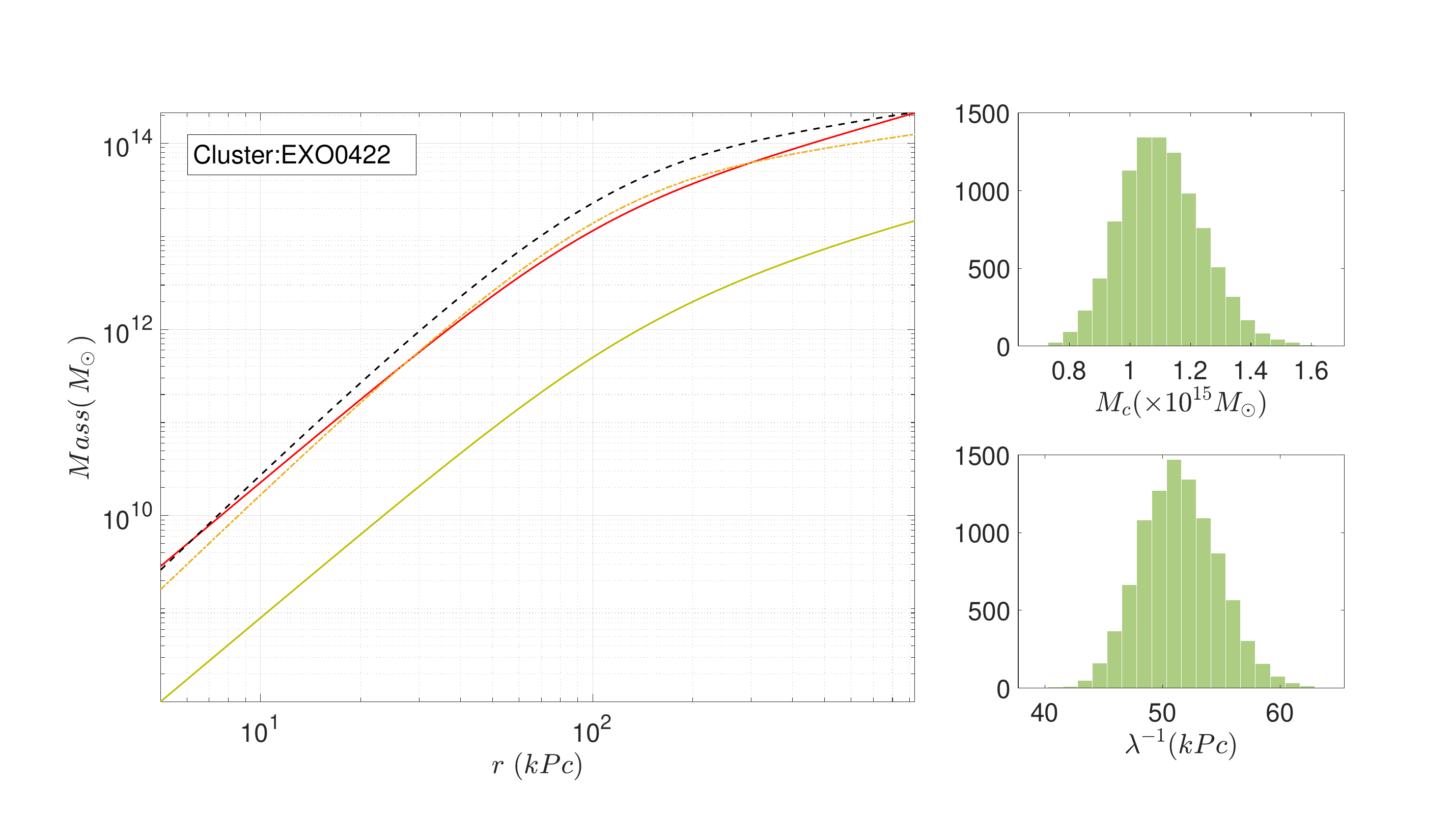}
    \includegraphics[trim=2.0cm 2.0cm 3.0cm 3.0cm, clip=true, width=0.32\columnwidth]{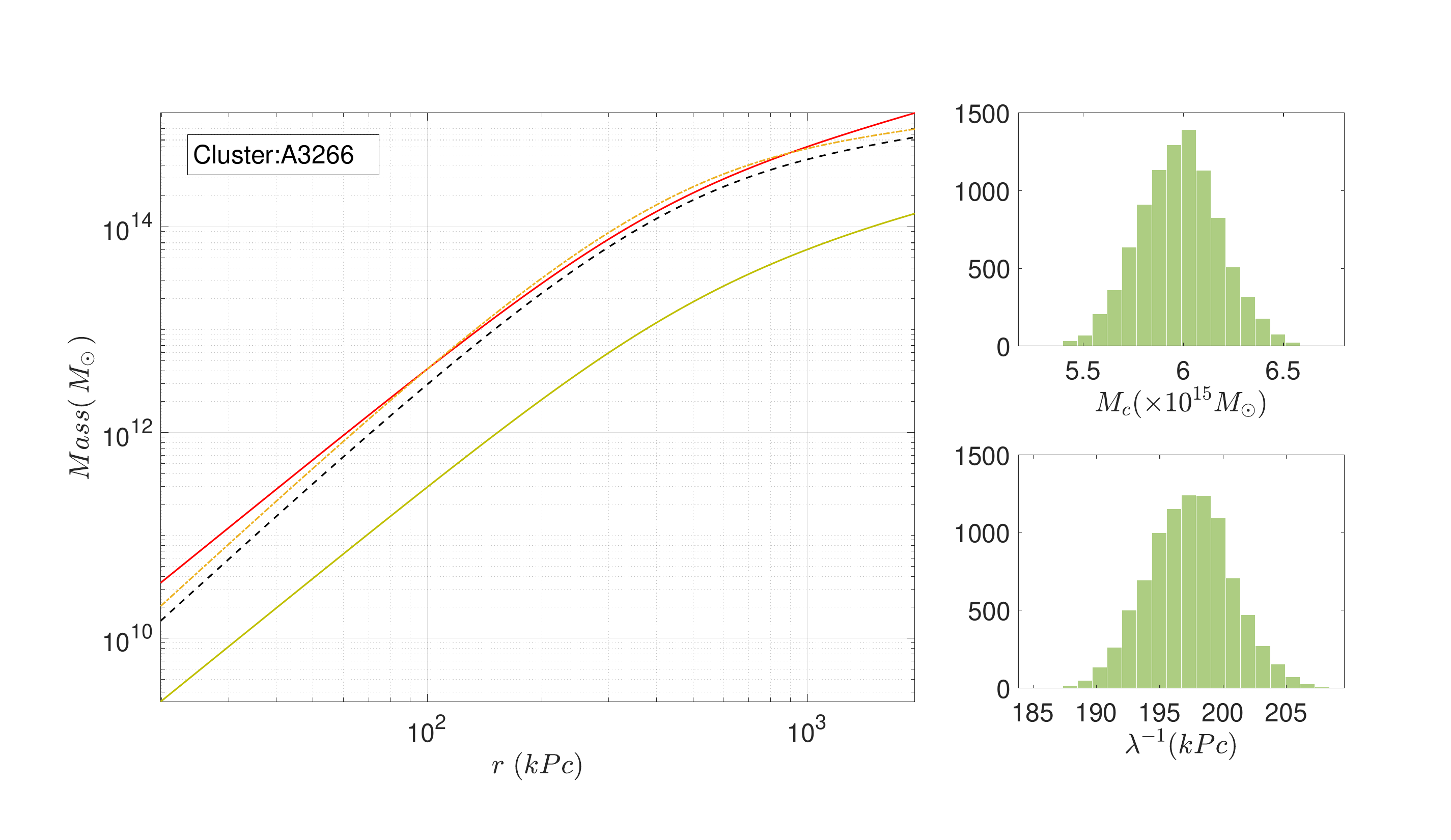}
    \includegraphics[trim=2.0cm 2.0cm 3.0cm 3.0cm, clip=true, width=0.32\columnwidth]{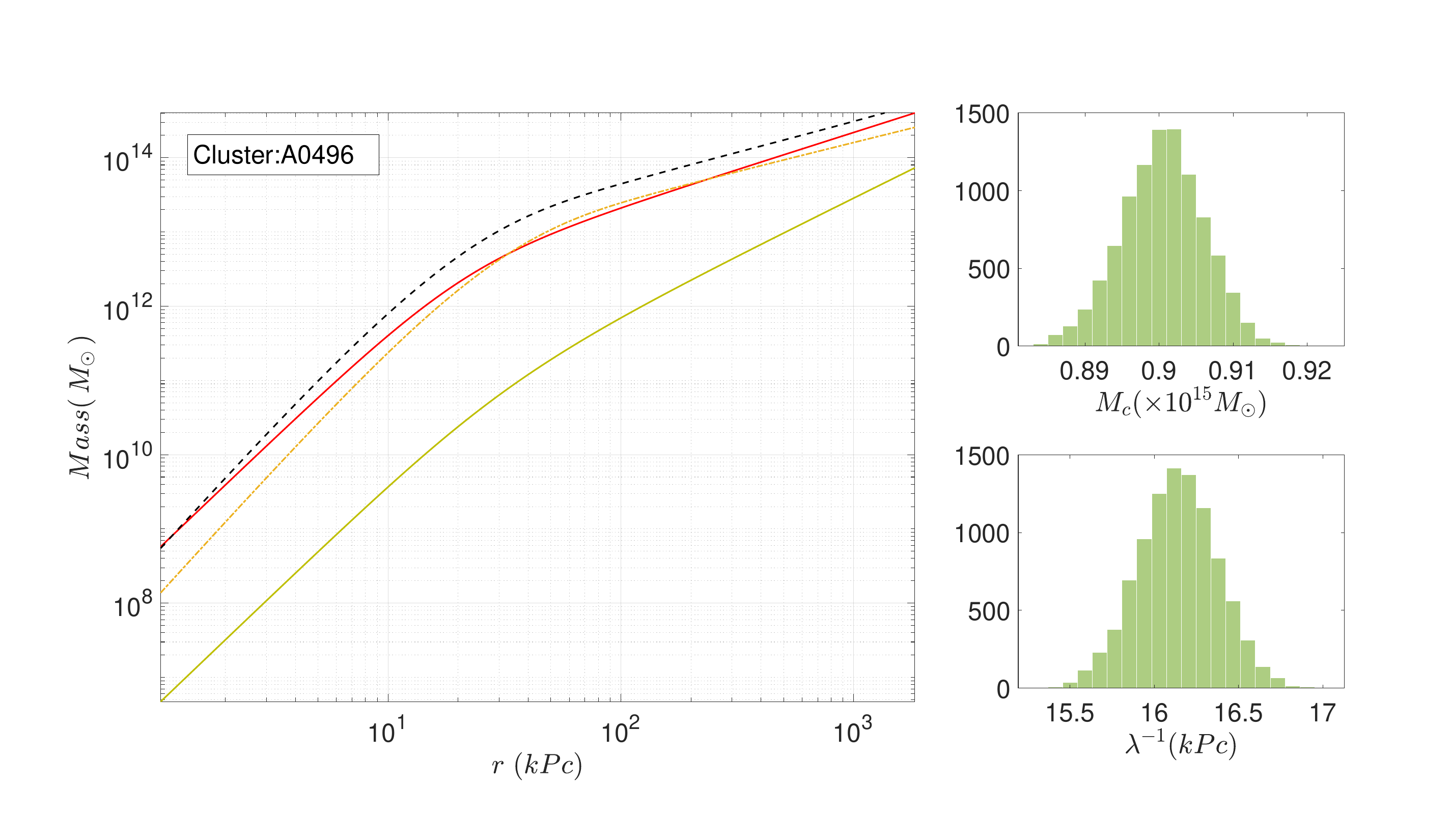}
    \includegraphics[trim=2.0cm 2.0cm 3.0cm 3.0cm, clip=true, width=0.32\columnwidth]{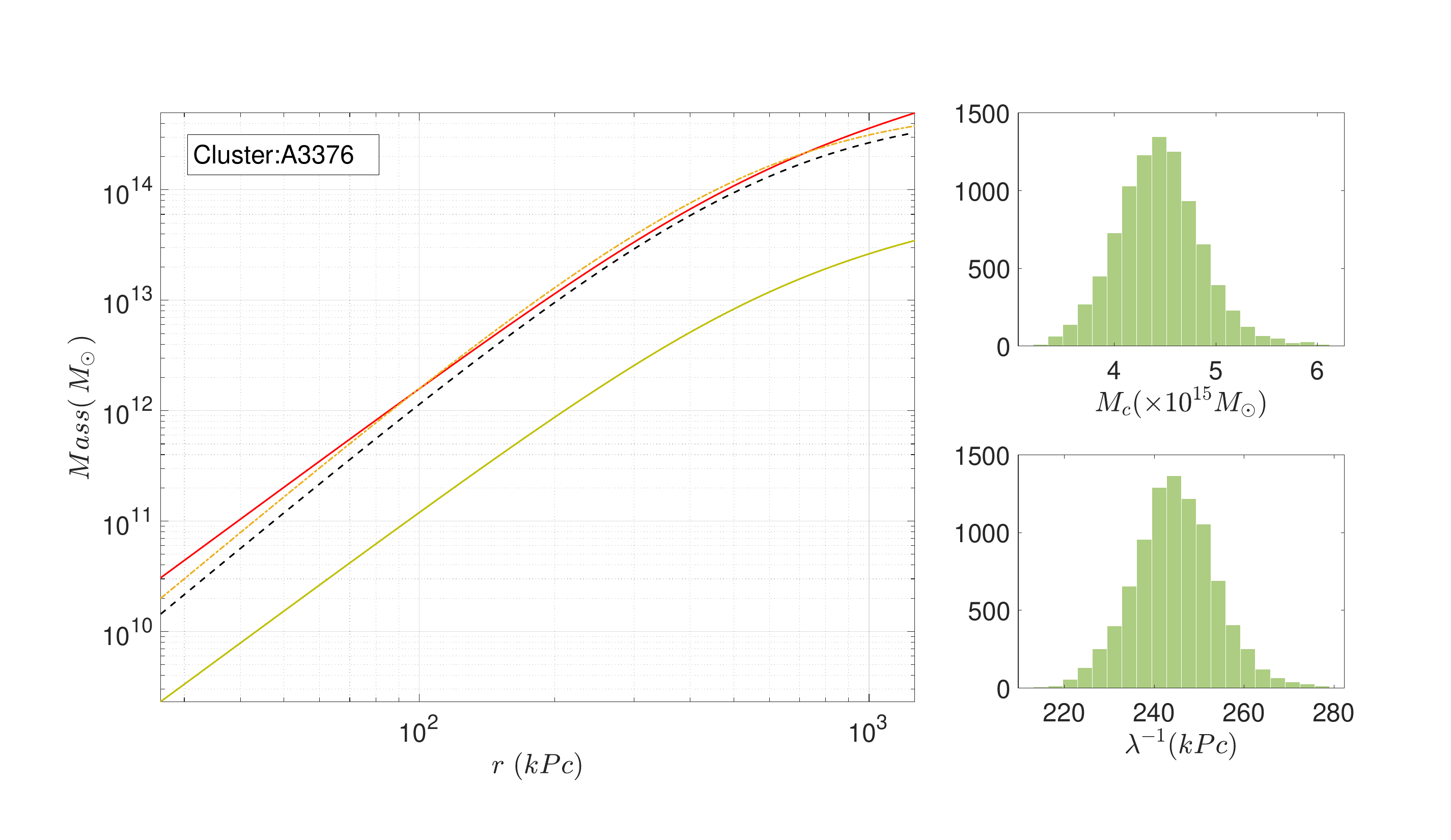}
    \includegraphics[trim=2.0cm 2.0cm 3.0cm 3.0cm, clip=true, width=0.32\columnwidth]{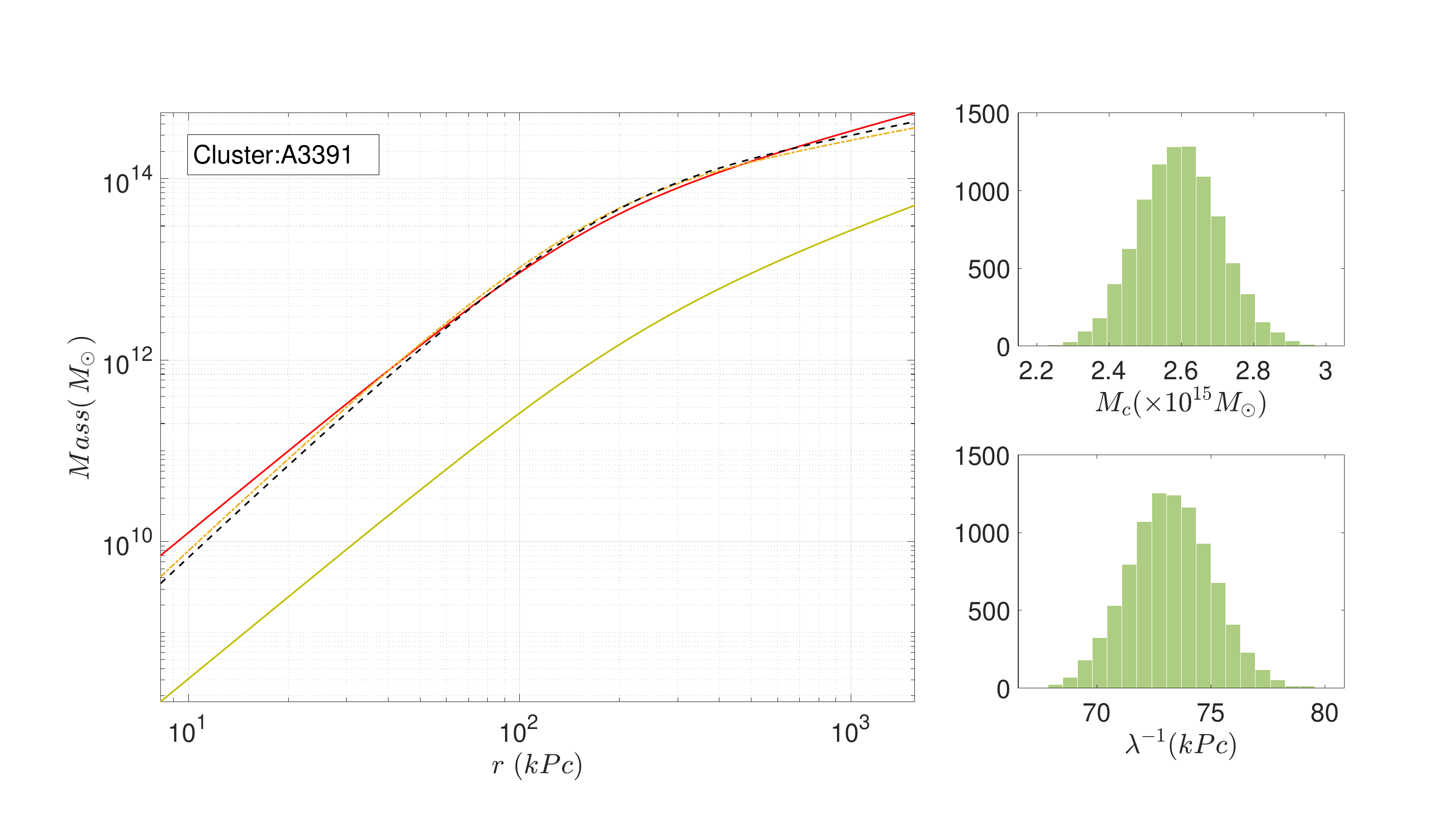}
    \includegraphics[trim=2.0cm 2.0cm 3.0cm 3.0cm, clip=true, width=0.32\columnwidth]{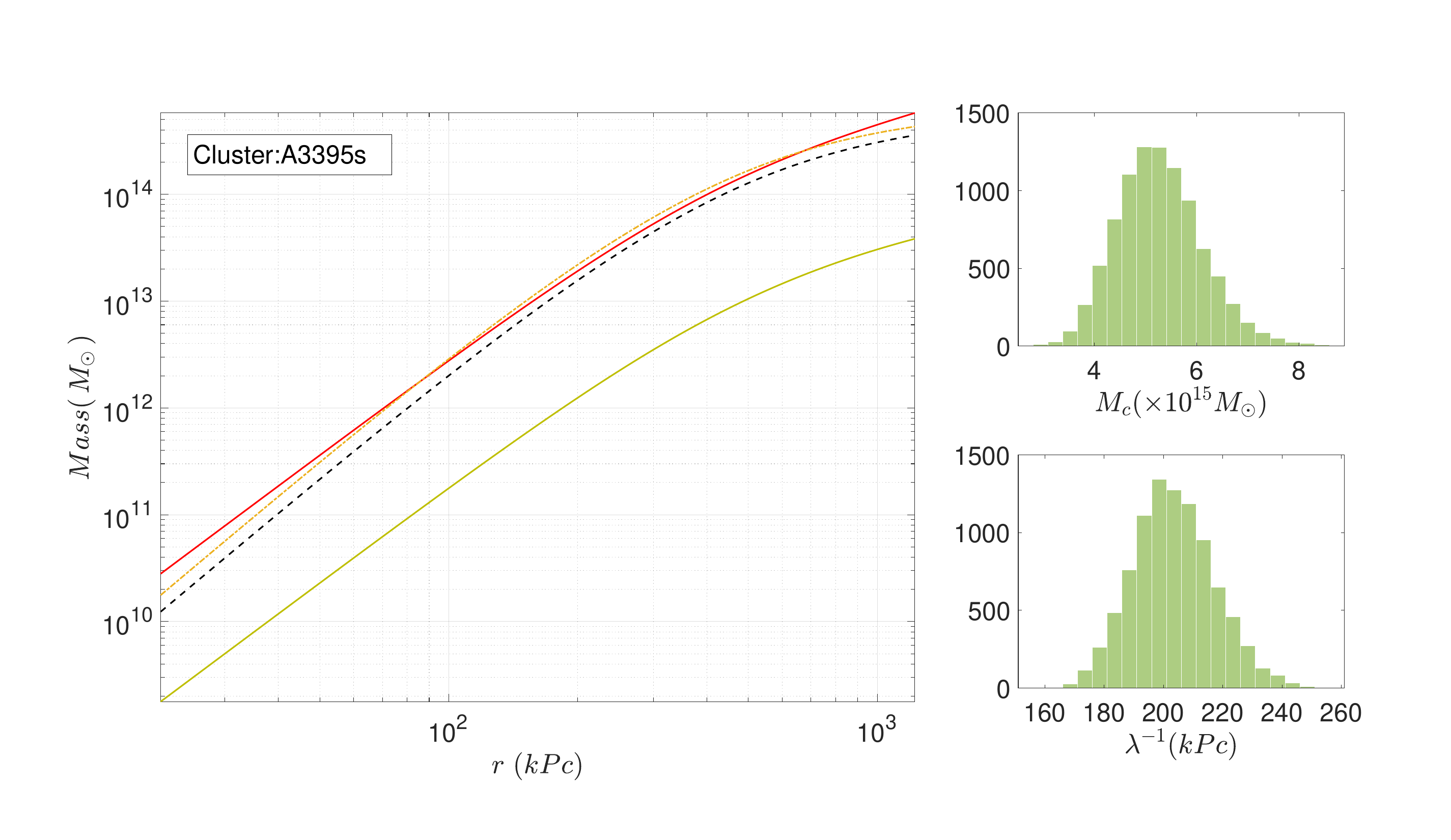}
    \includegraphics[trim=2.0cm 2.0cm 3.0cm 3.0cm, clip=true, width=0.32\columnwidth]{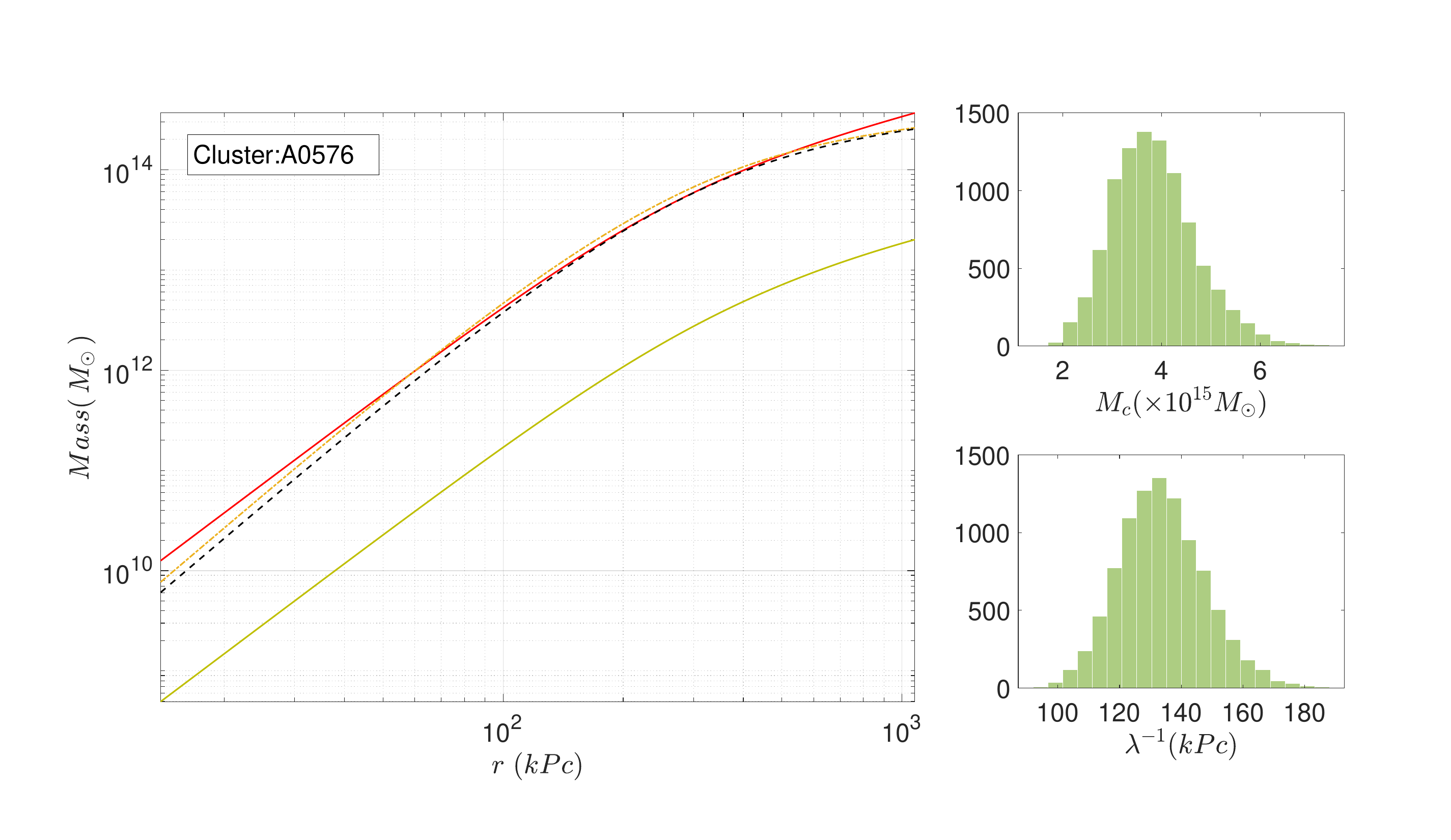}
    \includegraphics[trim=2.0cm 2.0cm 3.0cm 3.0cm, clip=true, width=0.32\columnwidth]{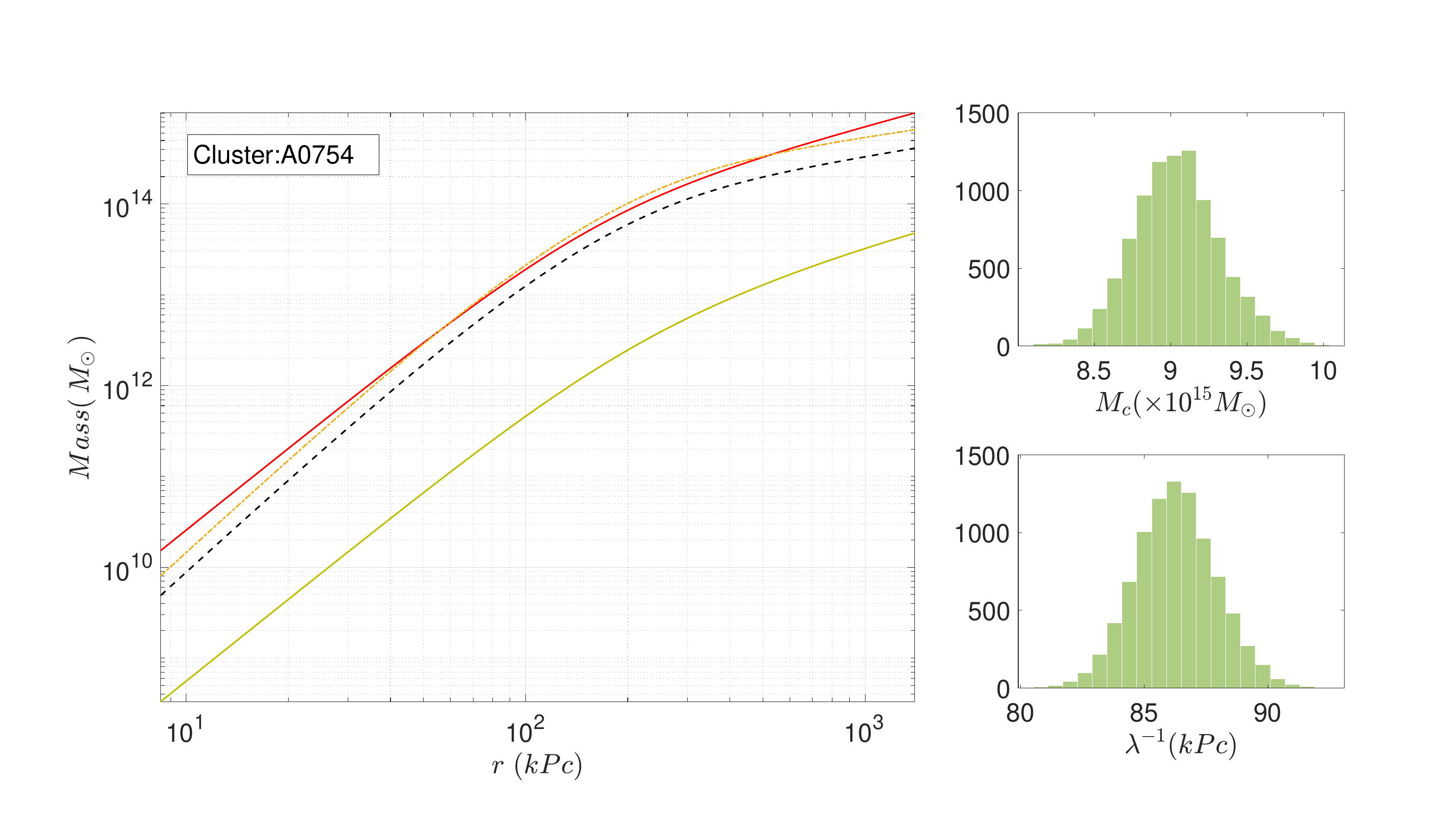}
    \includegraphics[trim=2.0cm 2.0cm 3.0cm 3.0cm, clip=true, width=0.32\columnwidth]{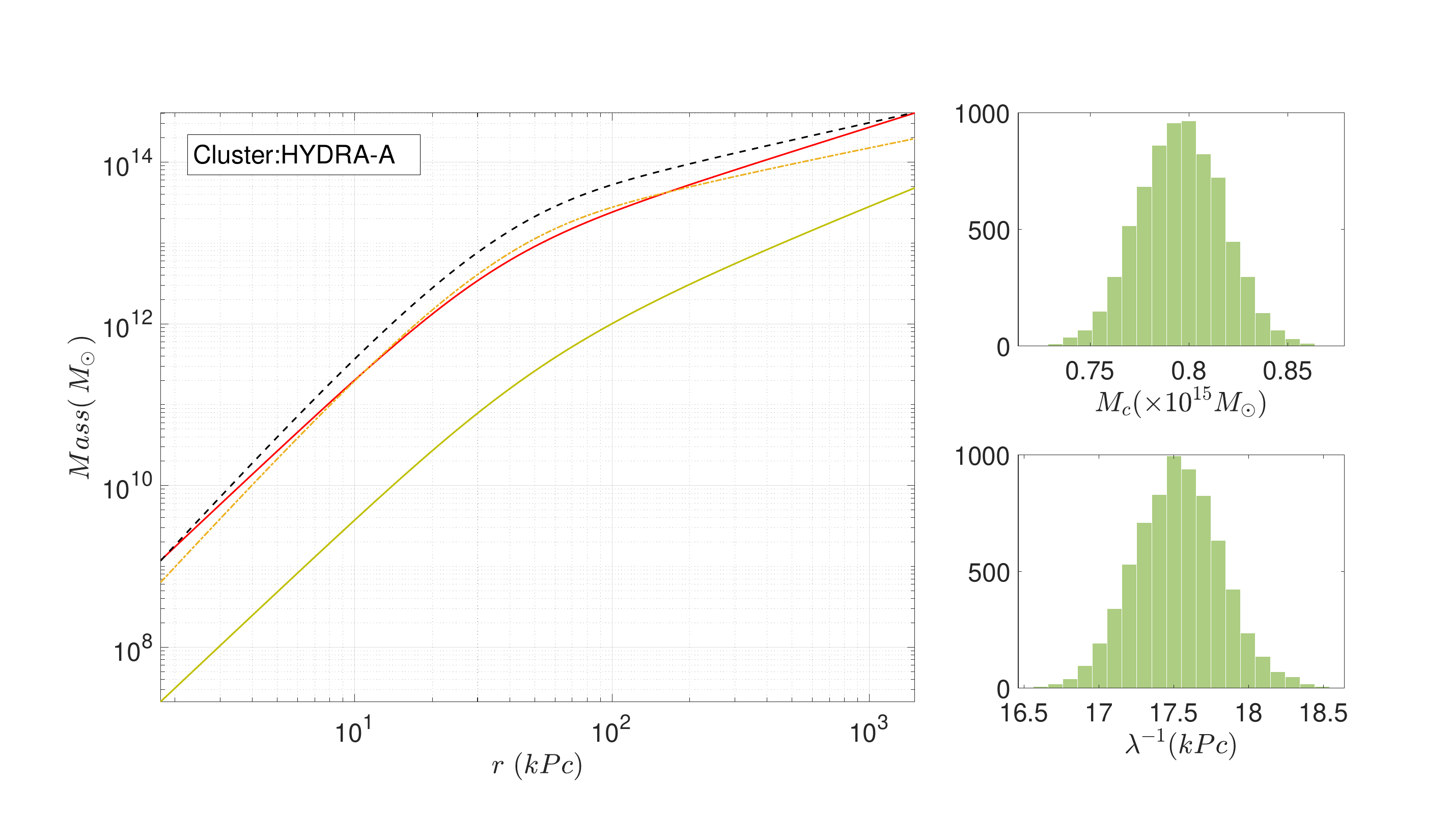}
    \end{figure}

    \begin{figure}
    \centering
    \includegraphics[trim=2.0cm 2.0cm 3.0cm 3.0cm, clip=true, width=0.32\columnwidth]{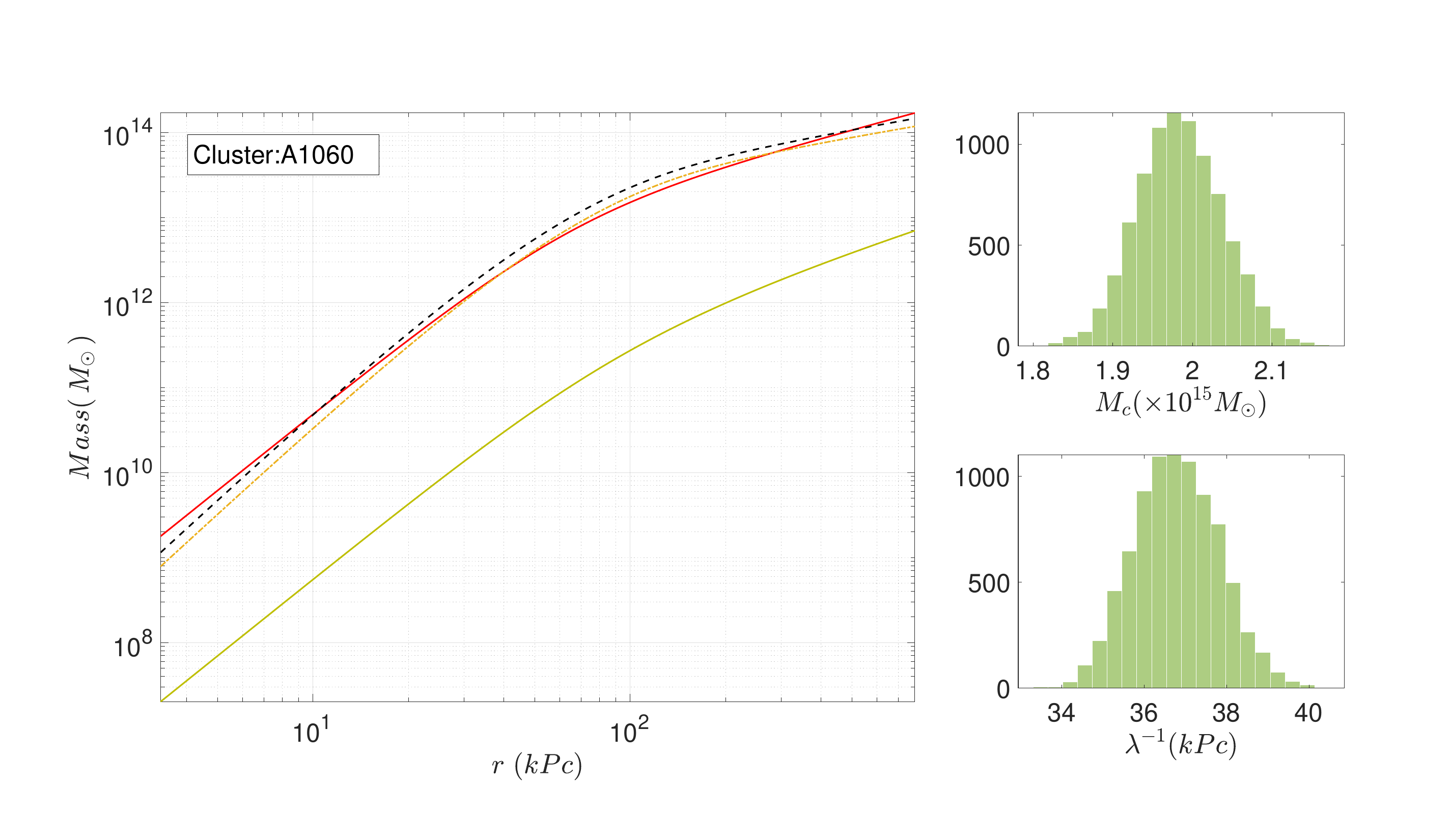}
    \includegraphics[trim=2.0cm 2.0cm 3.0cm 3.0cm, clip=true, width=0.32\columnwidth]{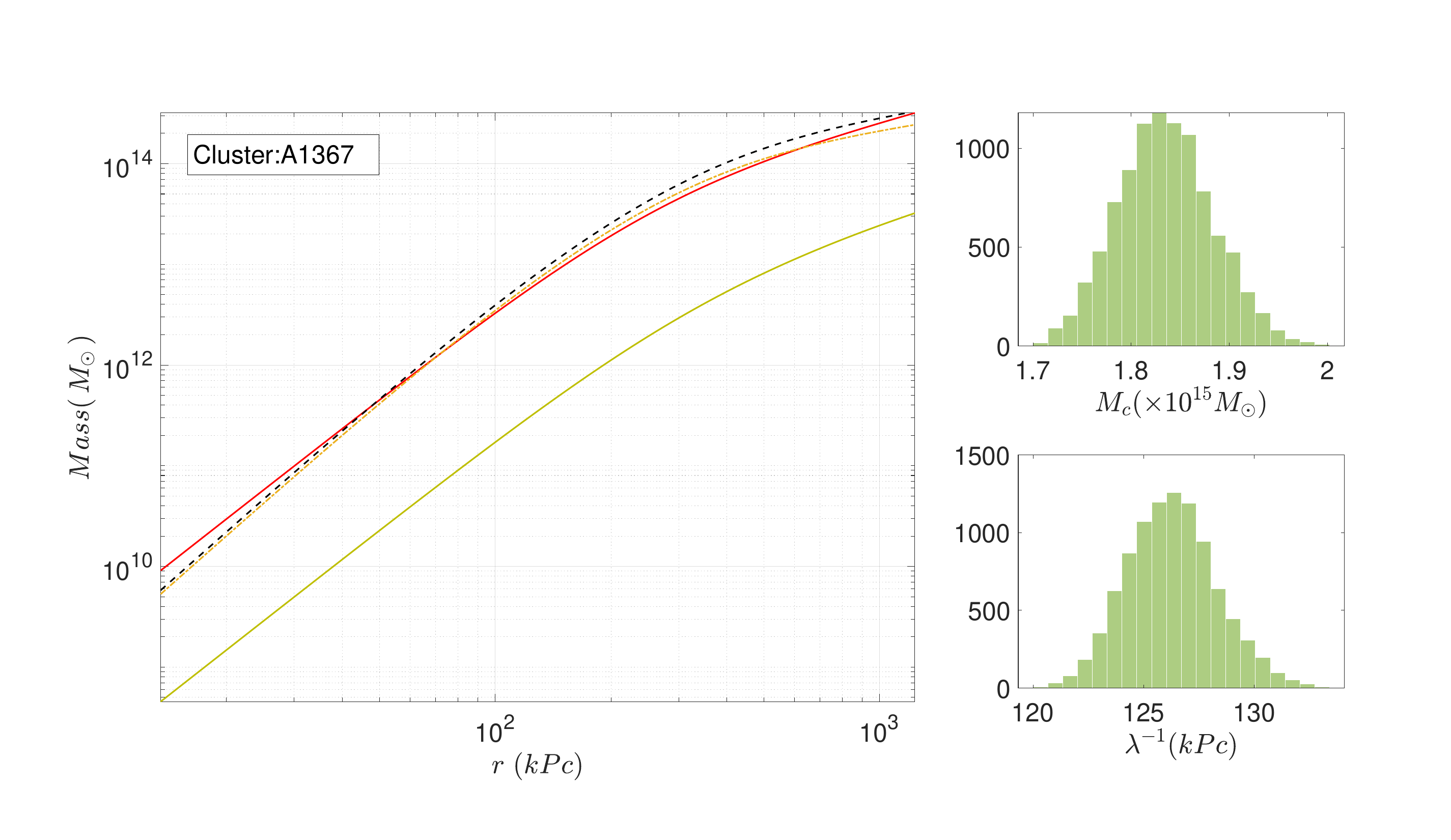}
    \includegraphics[trim=2.0cm 2.0cm 3.0cm 3.0cm, clip=true, width=0.32\columnwidth]{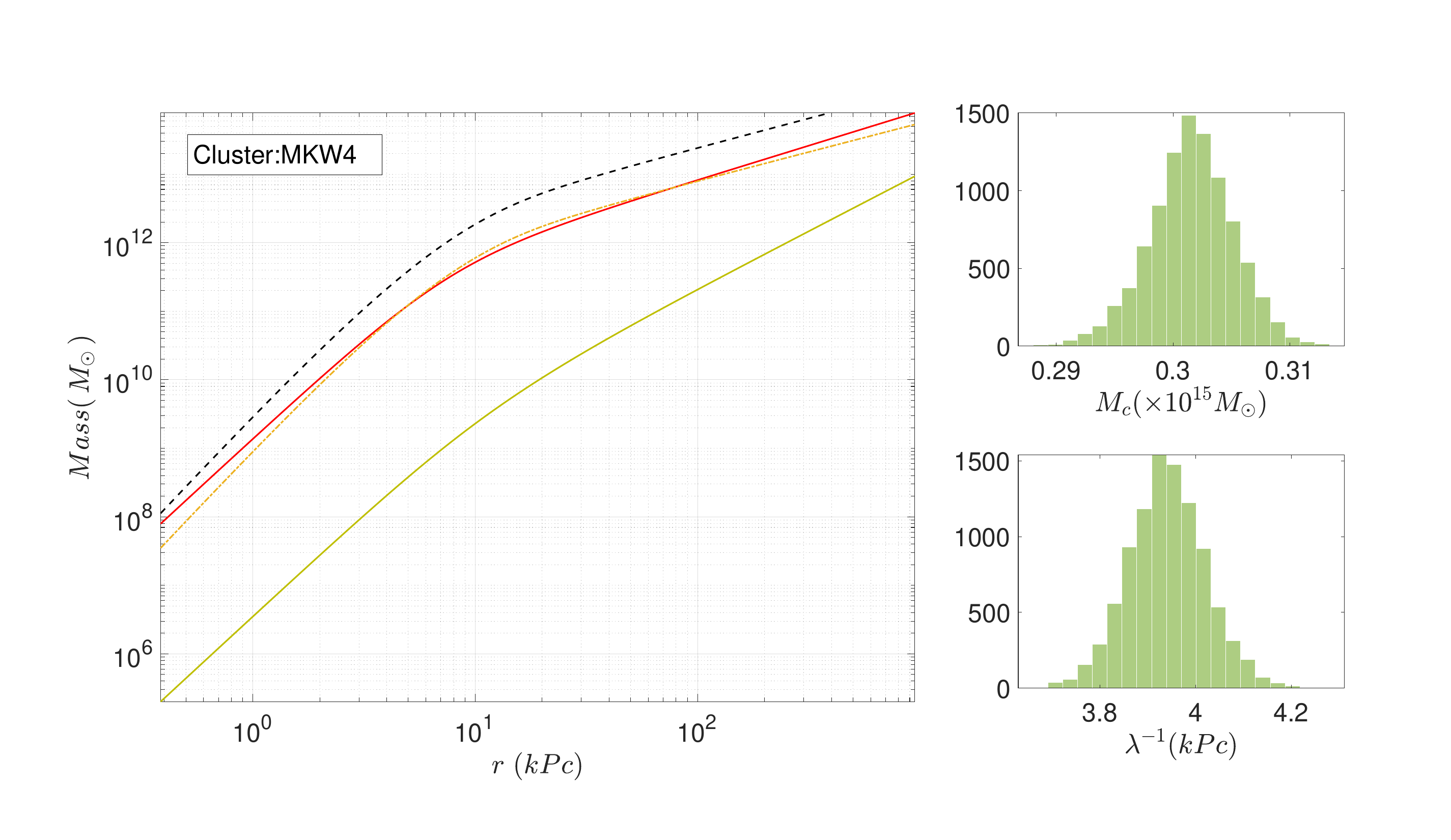}
    \includegraphics[trim=2.0cm 2.0cm 3.0cm 3.0cm, clip=true, width=0.32\columnwidth]{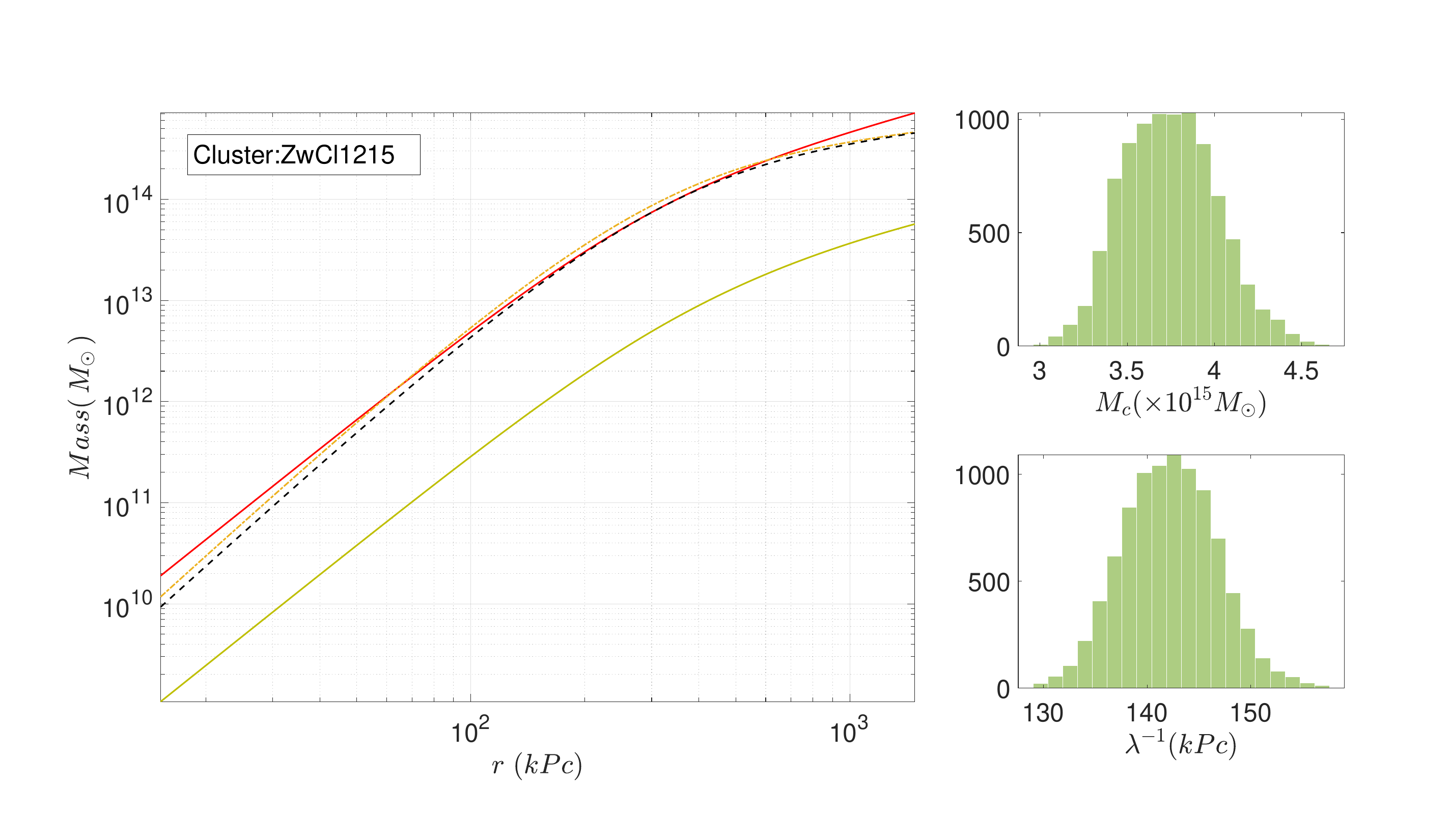}
    \includegraphics[trim=2.0cm 2.0cm 3.0cm 3.0cm, clip=true, width=0.32\columnwidth]{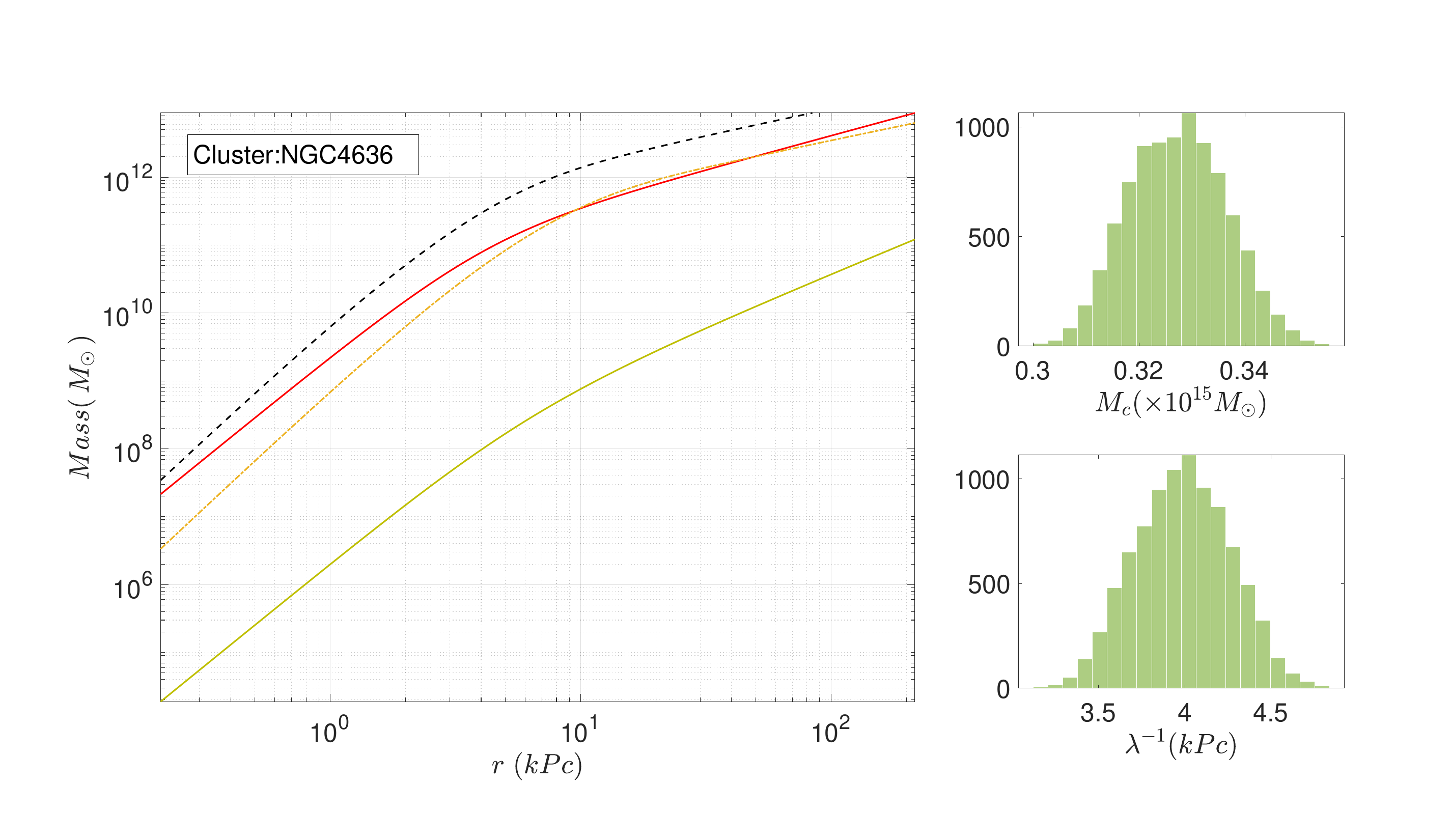}
    \includegraphics[trim=2.0cm 2.0cm 3.0cm 3.0cm, clip=true, width=0.32\columnwidth]{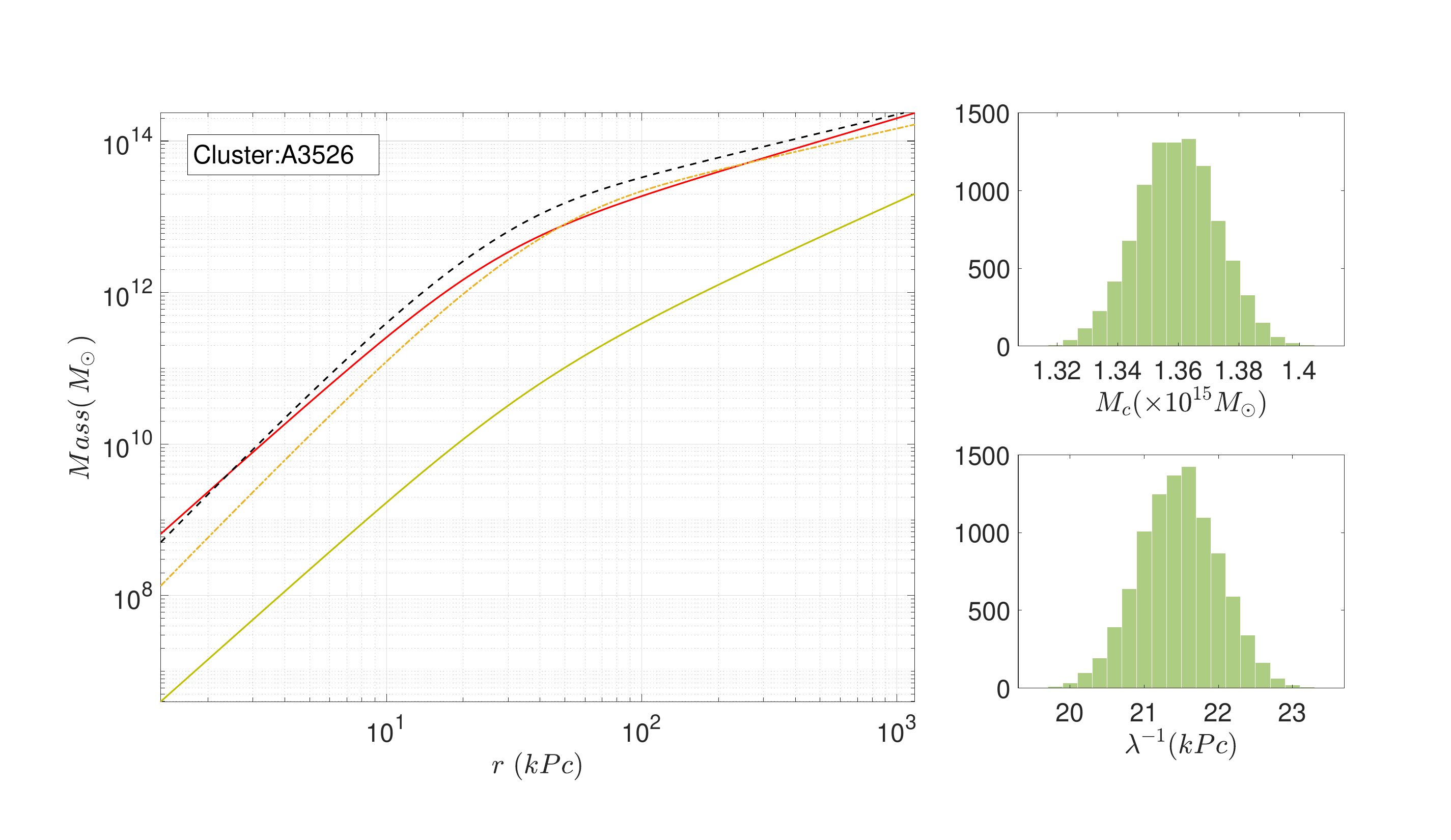}
    \includegraphics[trim=2.0cm 2.0cm 3.0cm 3.0cm, clip=true, width=0.32\columnwidth]{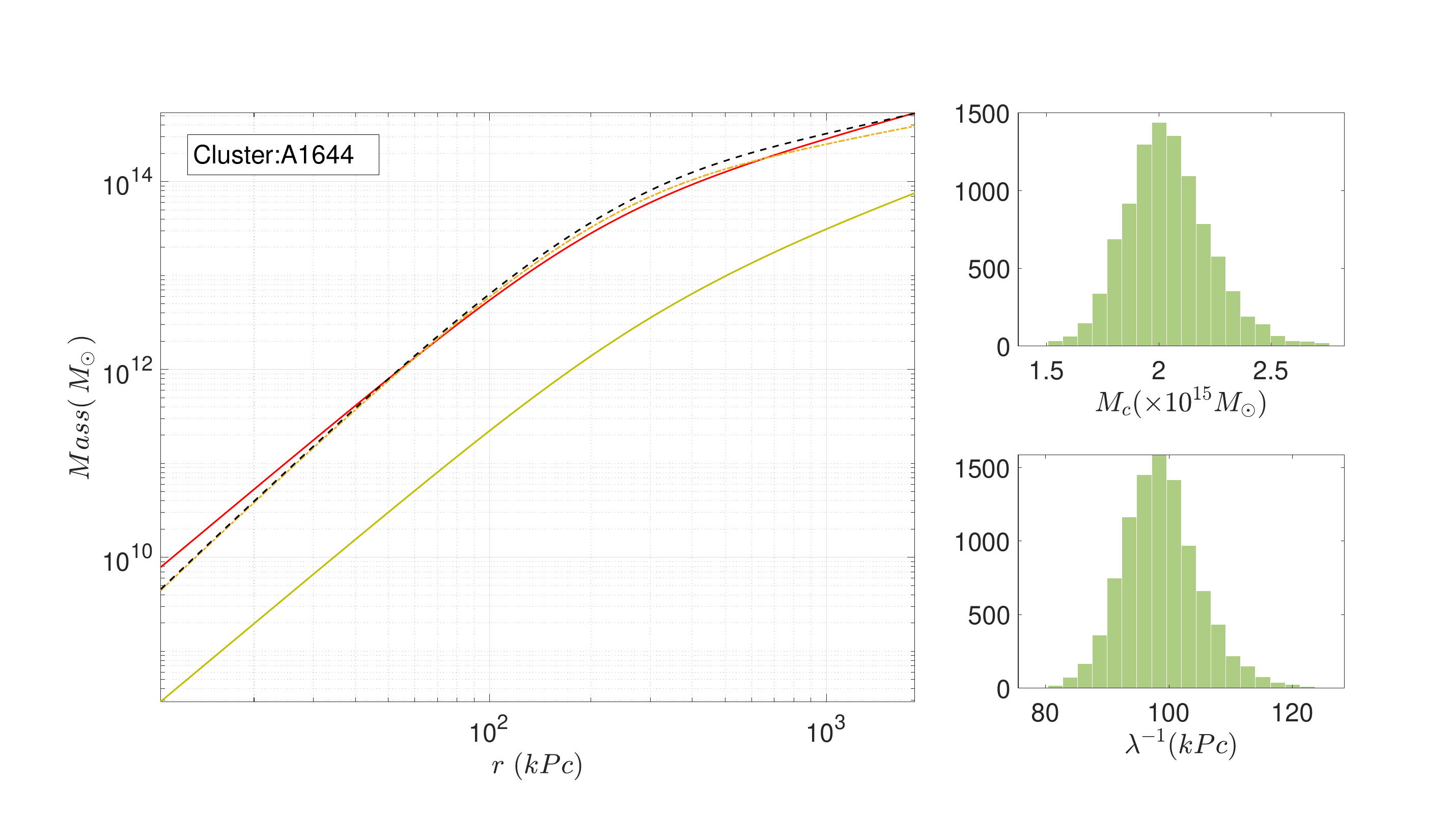}
    \includegraphics[trim=2.0cm 2.0cm 3.0cm 3.0cm, clip=true, width=0.32\columnwidth]{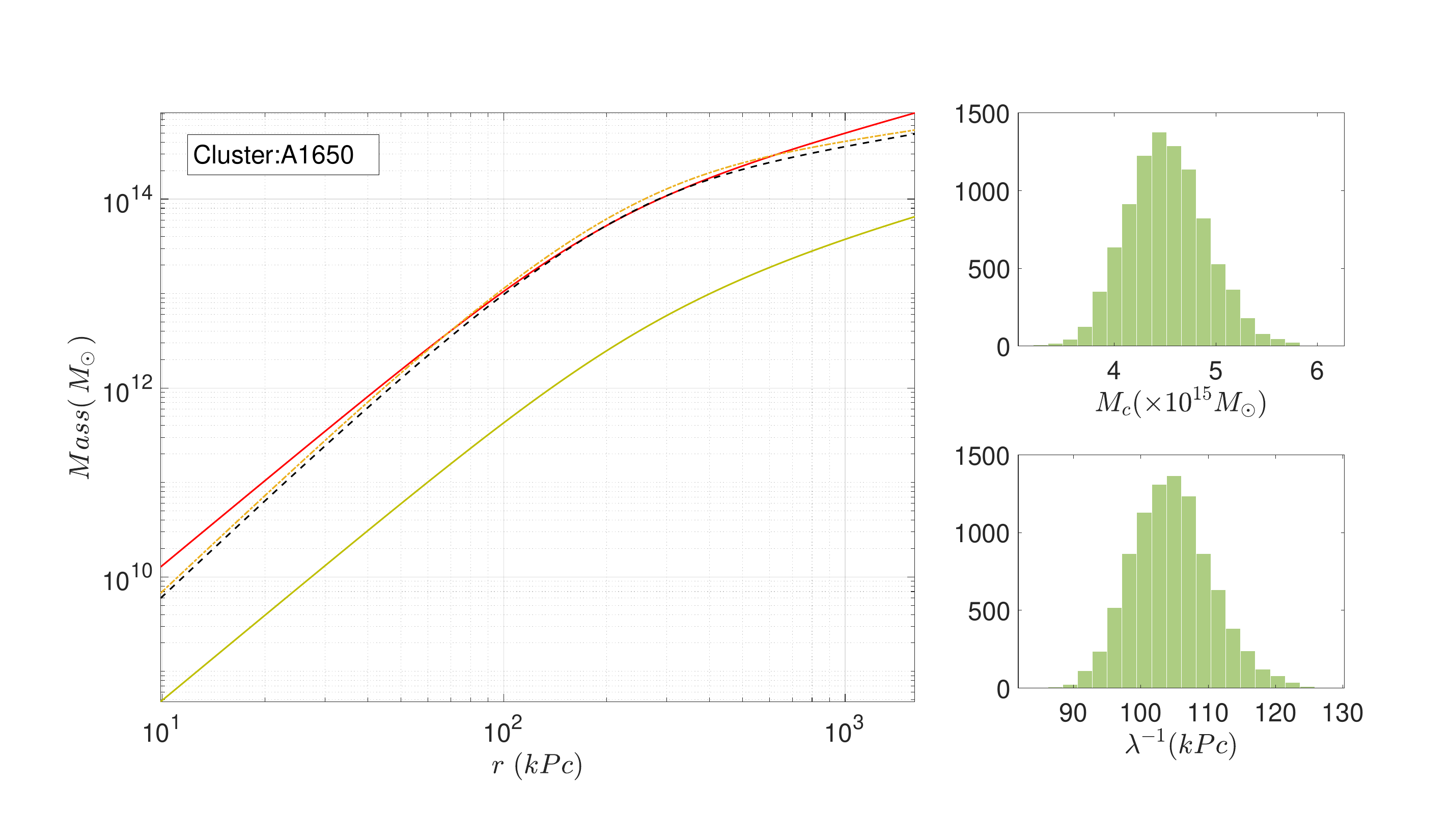}
    \includegraphics[trim=2.0cm 2.0cm 3.0cm 3.0cm, clip=true, width=0.32\columnwidth]{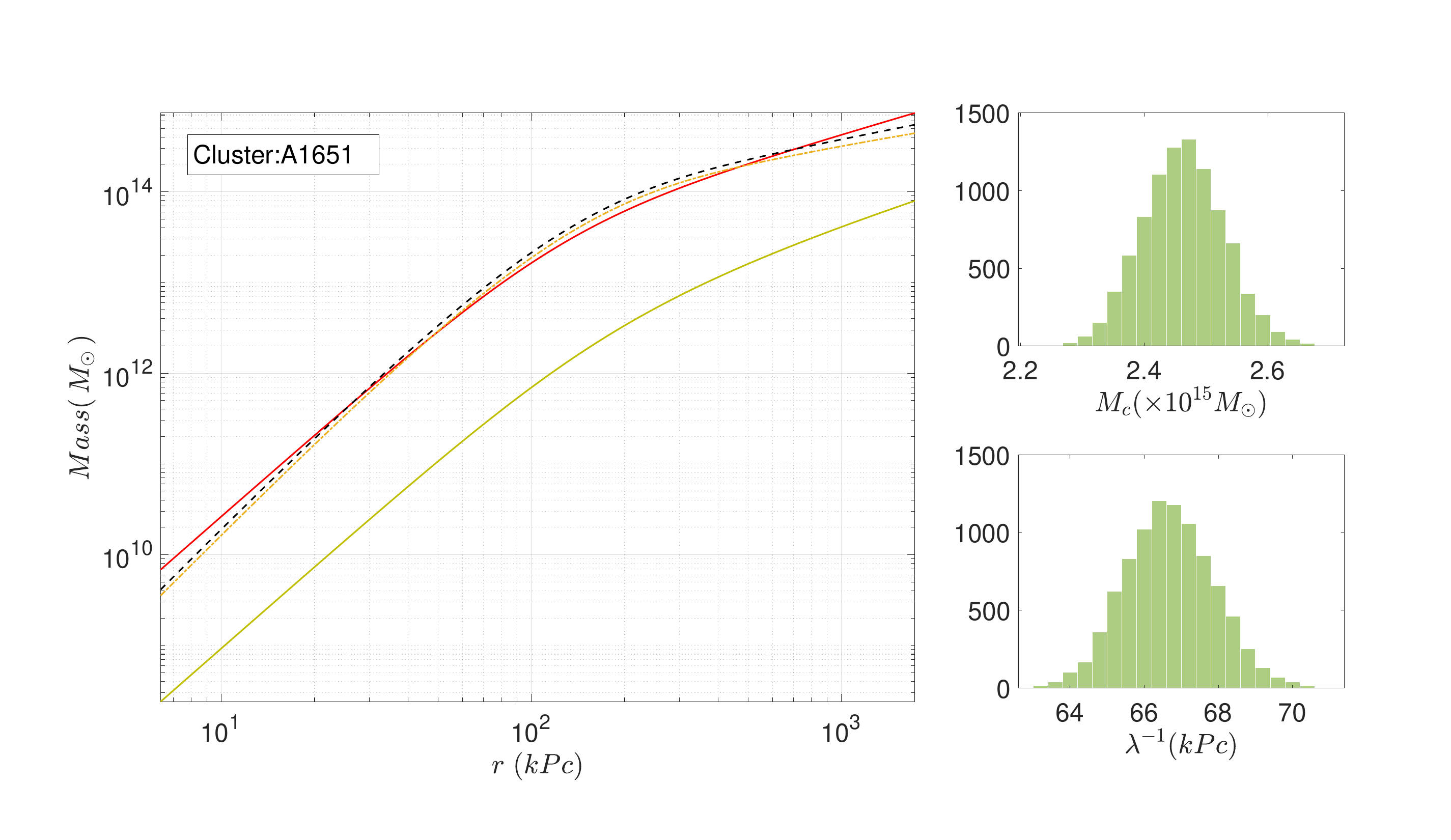}
    \includegraphics[trim=2.0cm 2.0cm 3.0cm 3.0cm, clip=true, width=0.32\columnwidth]{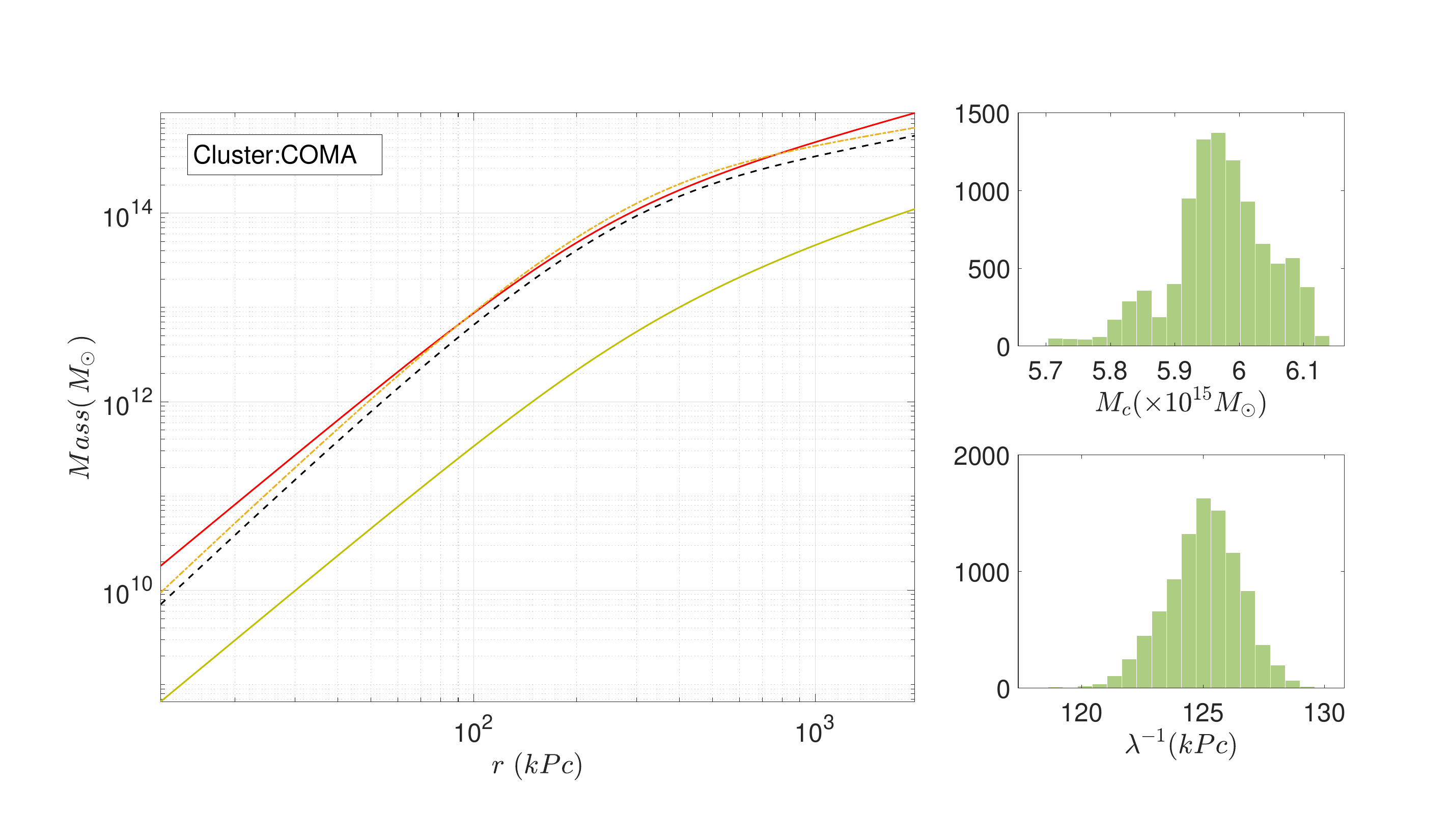}
    \includegraphics[trim=2.0cm 2.0cm 3.0cm 3.0cm, clip=true, width=0.32\columnwidth]{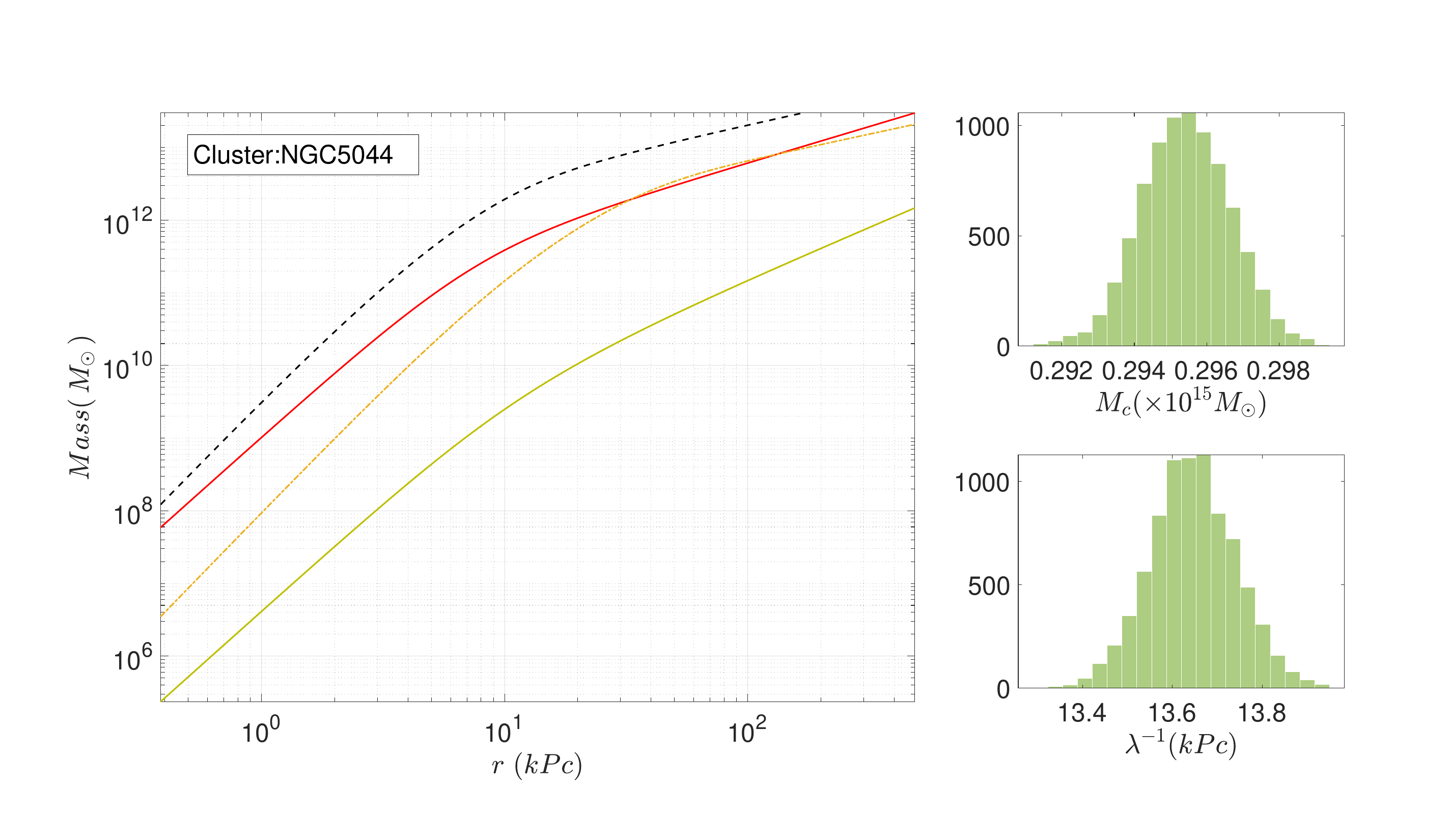}
    \includegraphics[trim=2.0cm 2.0cm 3.0cm 3.0cm, clip=true, width=0.32\columnwidth]{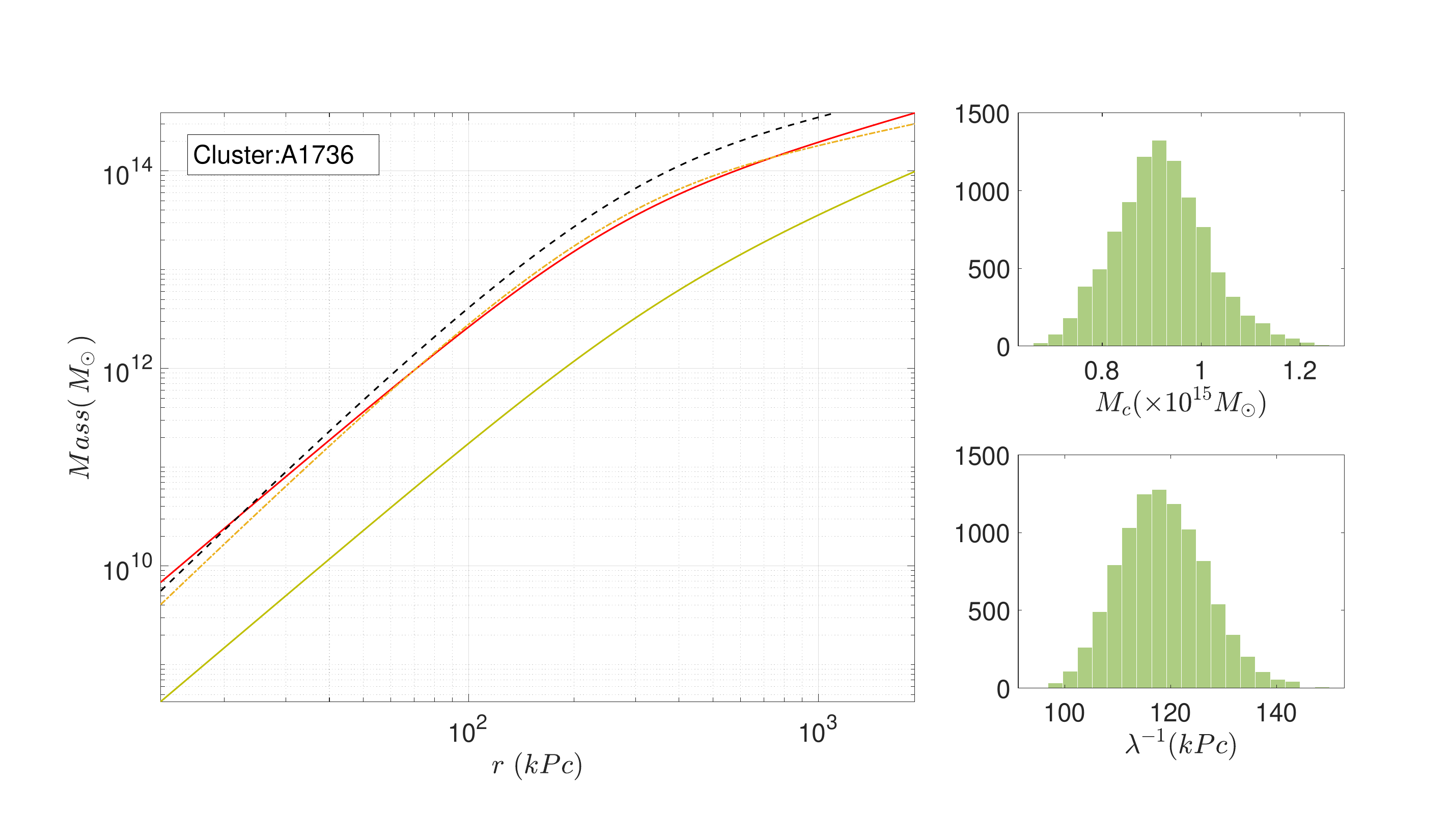}
    \includegraphics[trim=2.0cm 2.0cm 3.0cm 3.0cm, clip=true, width=0.32\columnwidth]{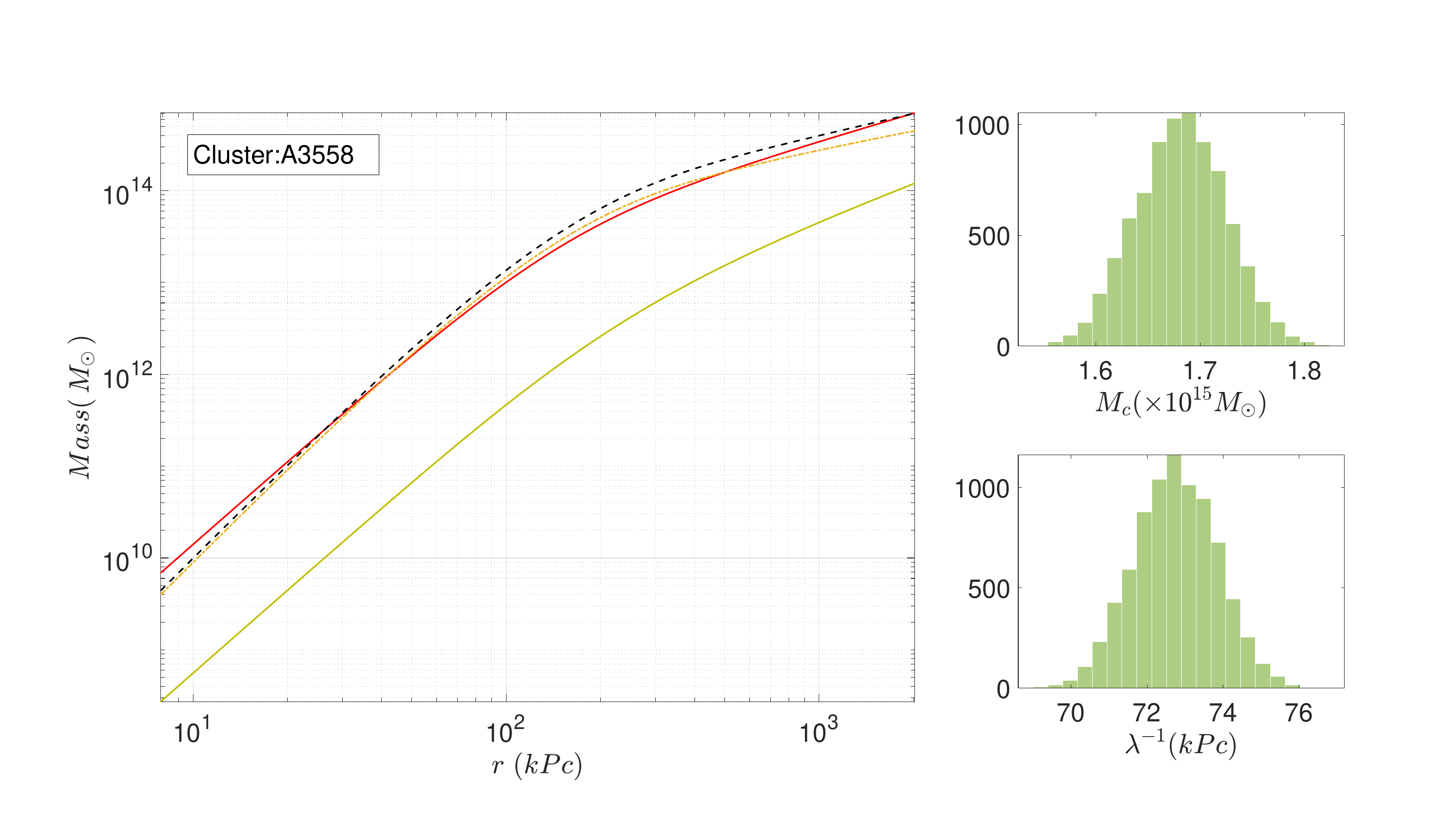}
    \includegraphics[trim=2.0cm 2.0cm 3.0cm 3.0cm, clip=true, width=0.32\columnwidth]{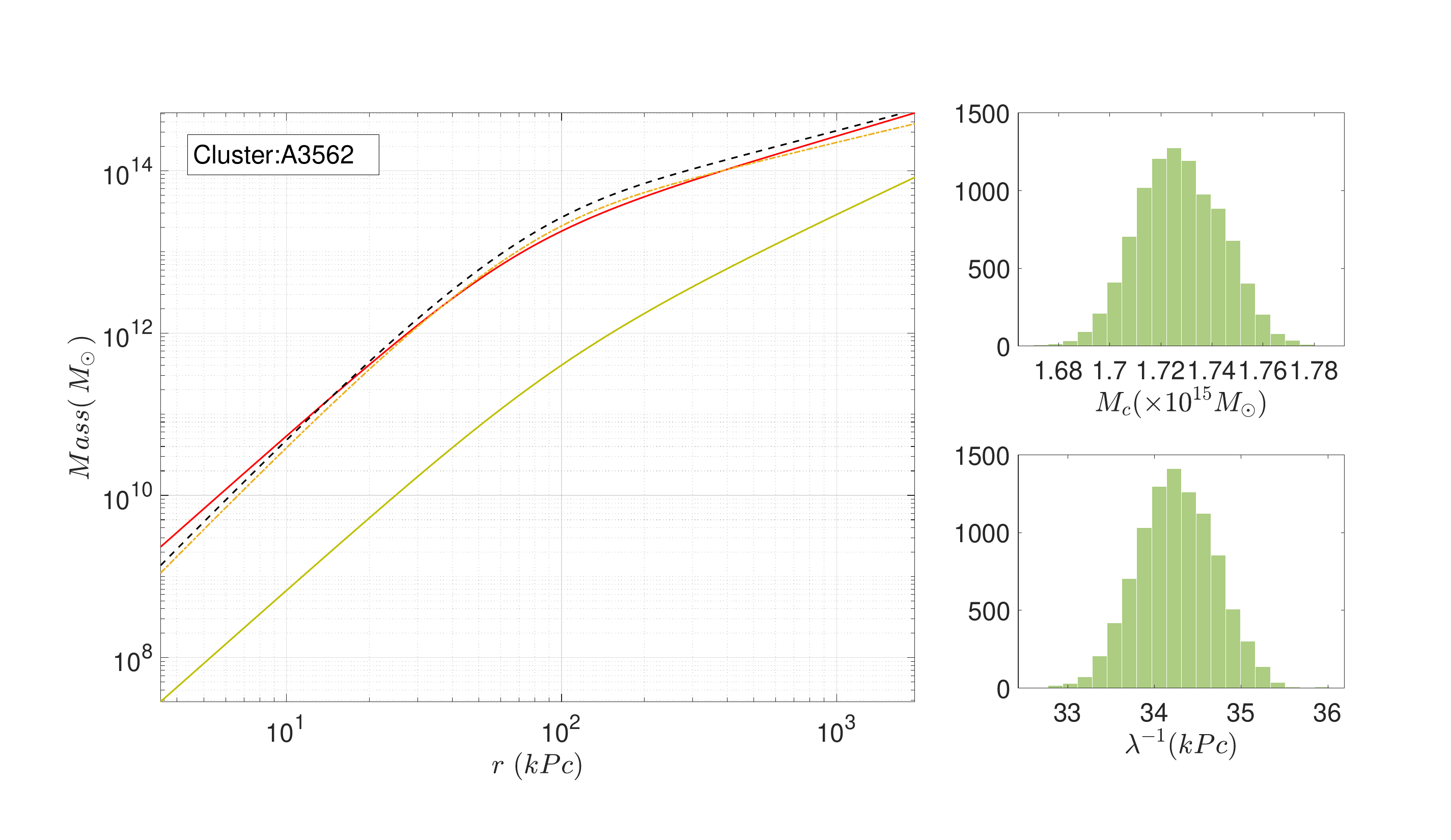}
    \includegraphics[trim=2.0cm 2.0cm 3.0cm 3.0cm, clip=true, width=0.32\columnwidth]{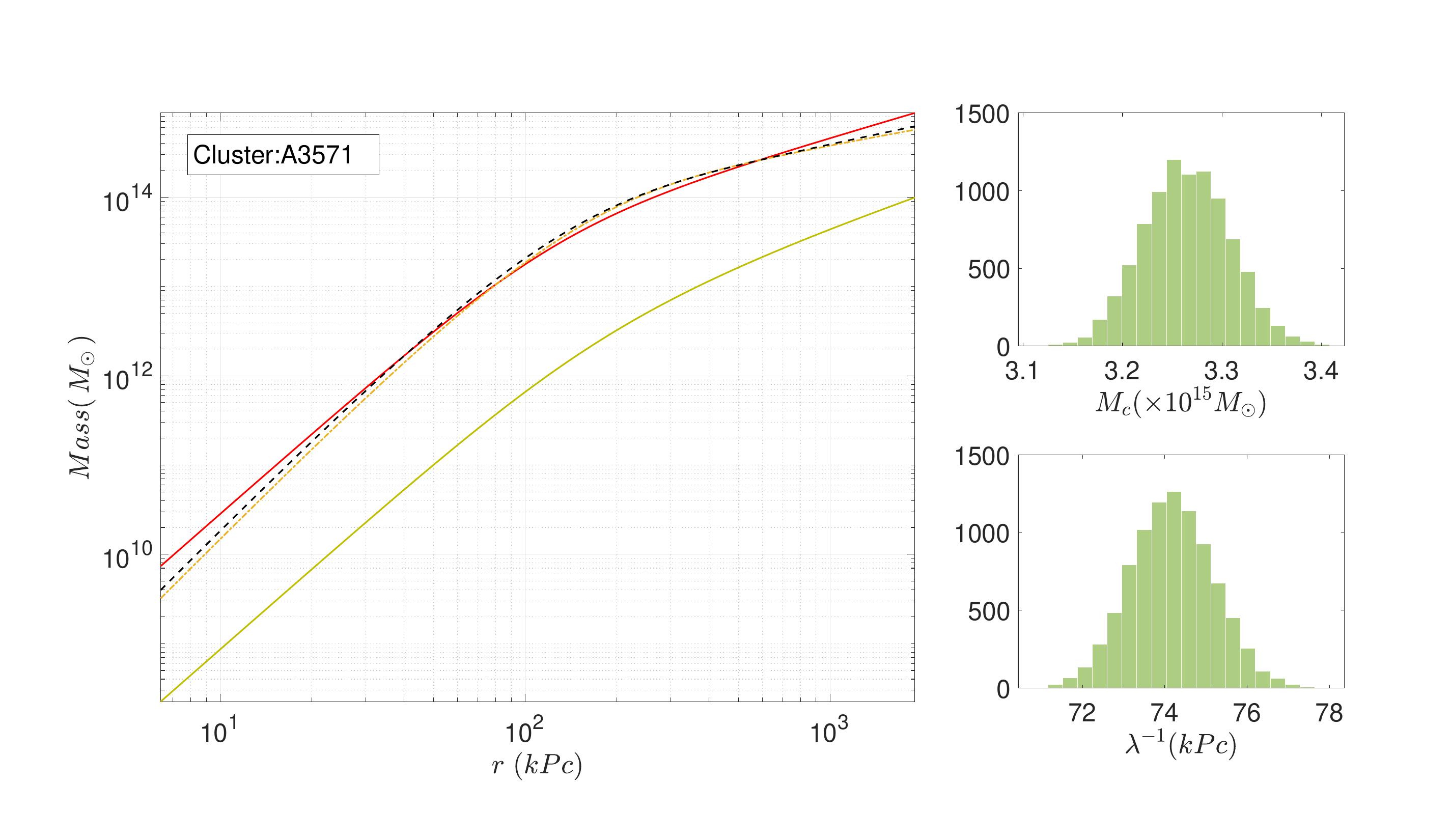}
    \includegraphics[trim=2.0cm 2.0cm 3.0cm 3.0cm, clip=true, width=0.32\columnwidth]{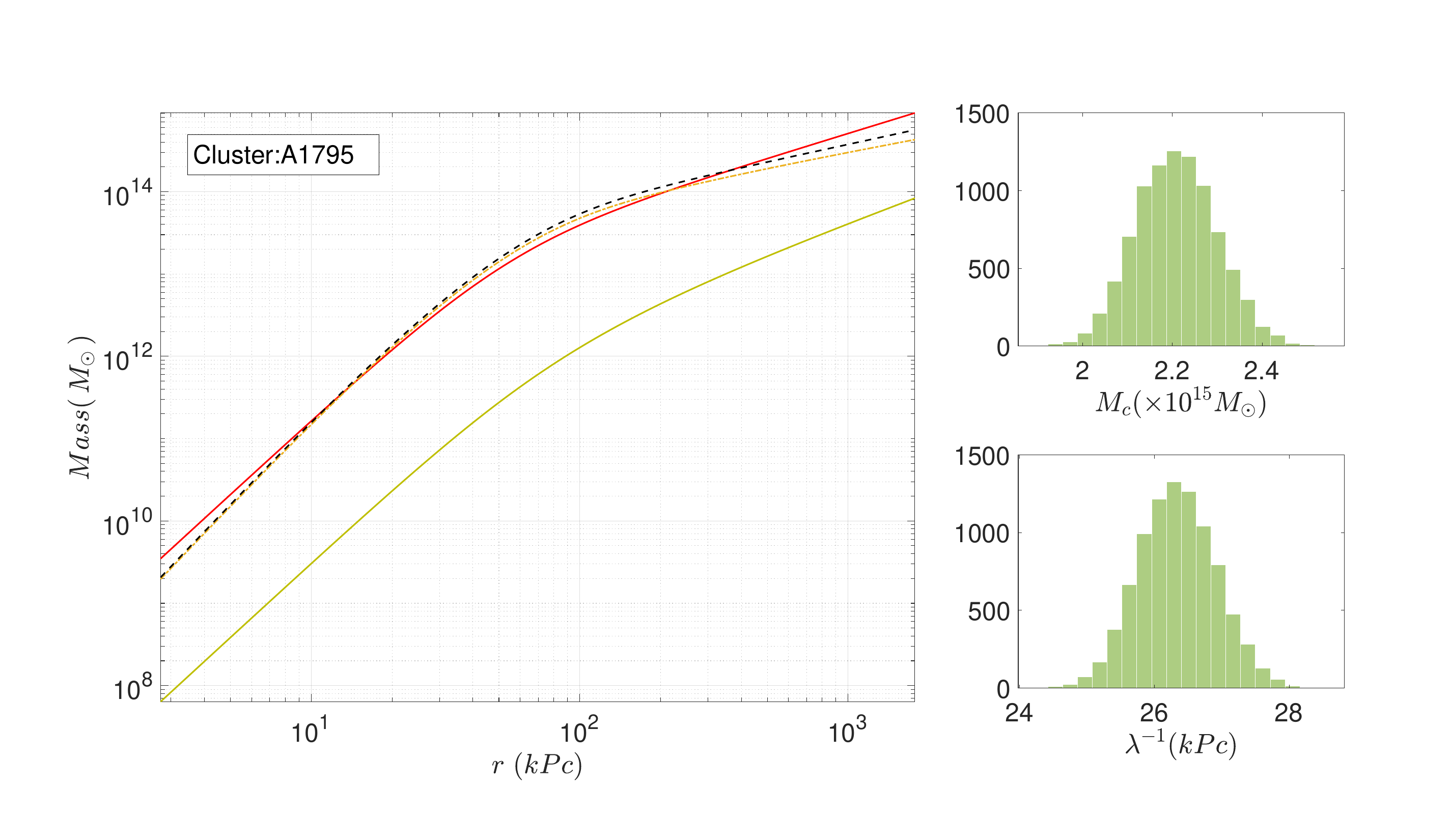}
    \includegraphics[trim=2.0cm 2.0cm 3.0cm 3.0cm, clip=true, width=0.32\columnwidth]{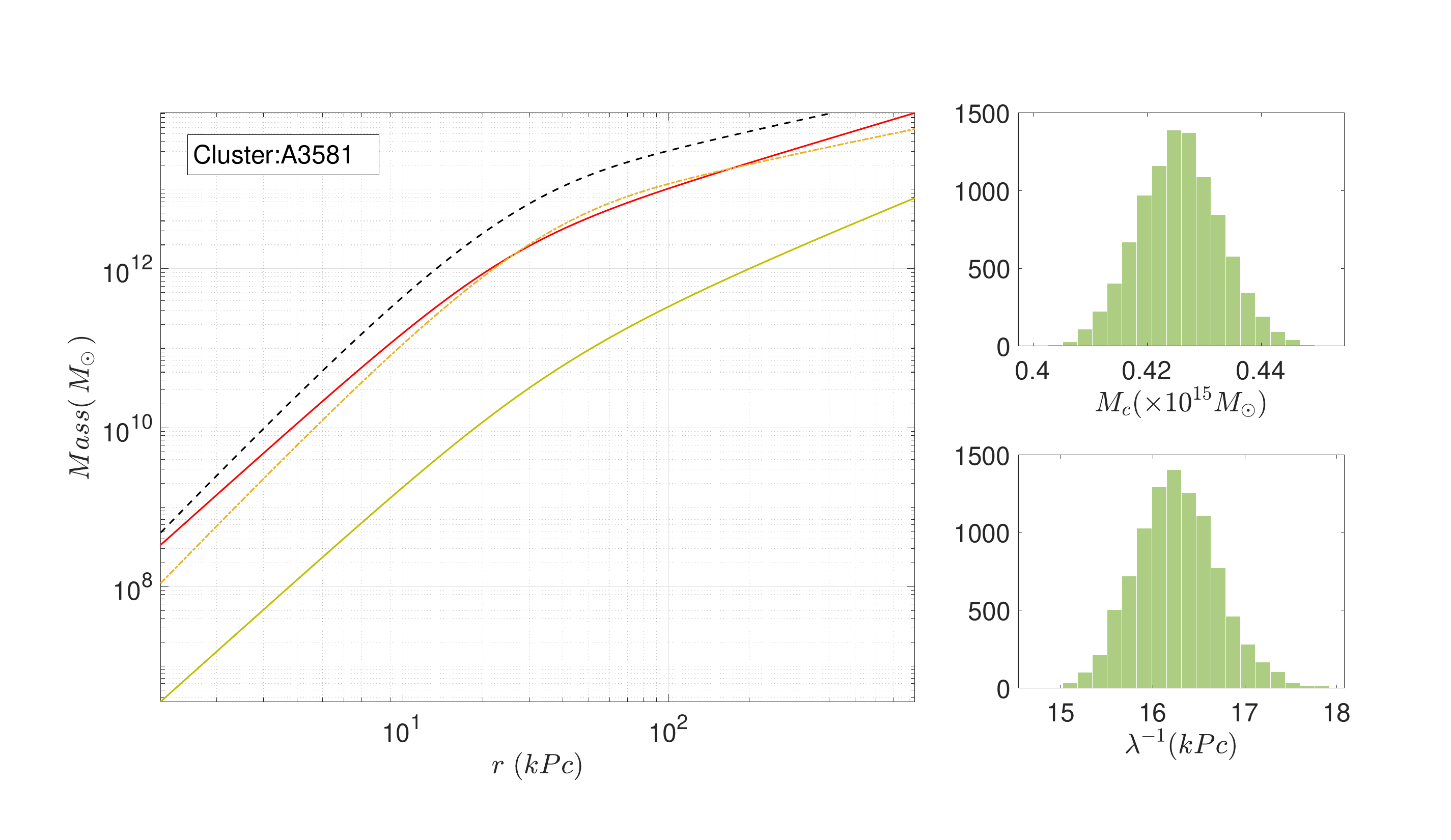}
    \includegraphics[trim=2.0cm 2.0cm 3.0cm 3.0cm, clip=true, width=0.32\columnwidth]{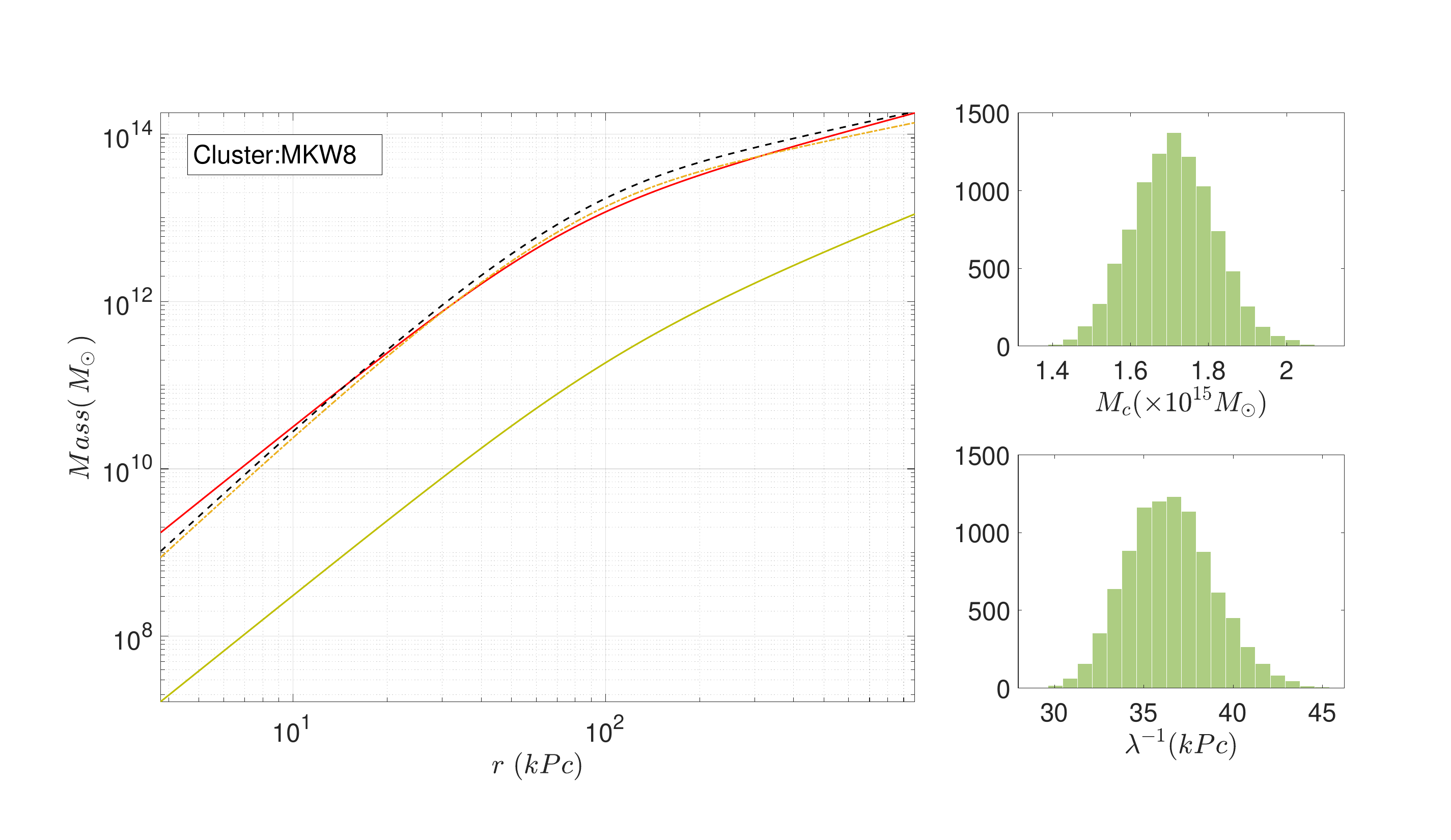}
    \includegraphics[trim=2.0cm 2.0cm 3.0cm 3.0cm, clip=true, width=0.32\columnwidth]{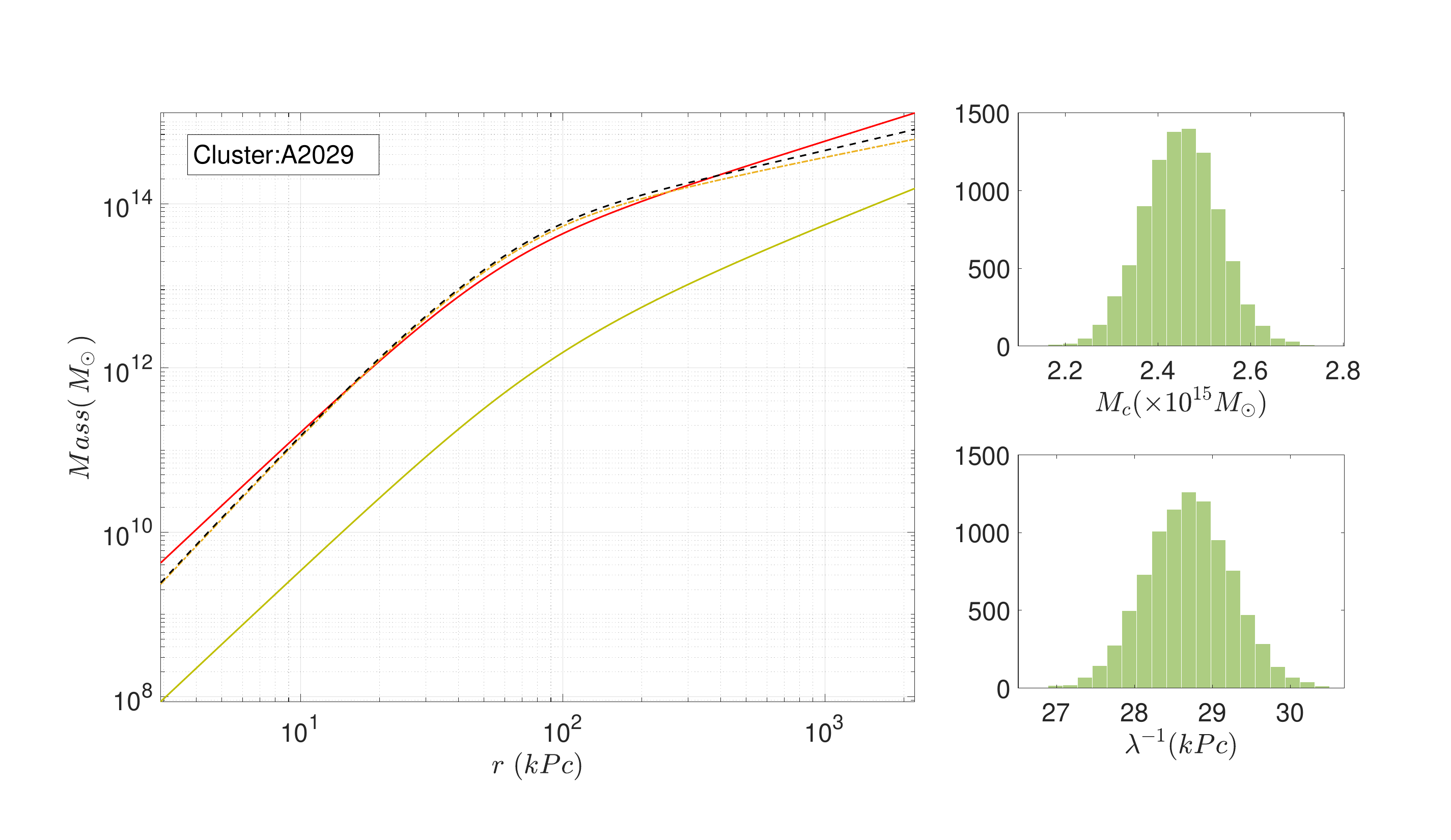}
    \includegraphics[trim=2.0cm 2.0cm 3.0cm 3.0cm, clip=true, width=0.32\columnwidth]{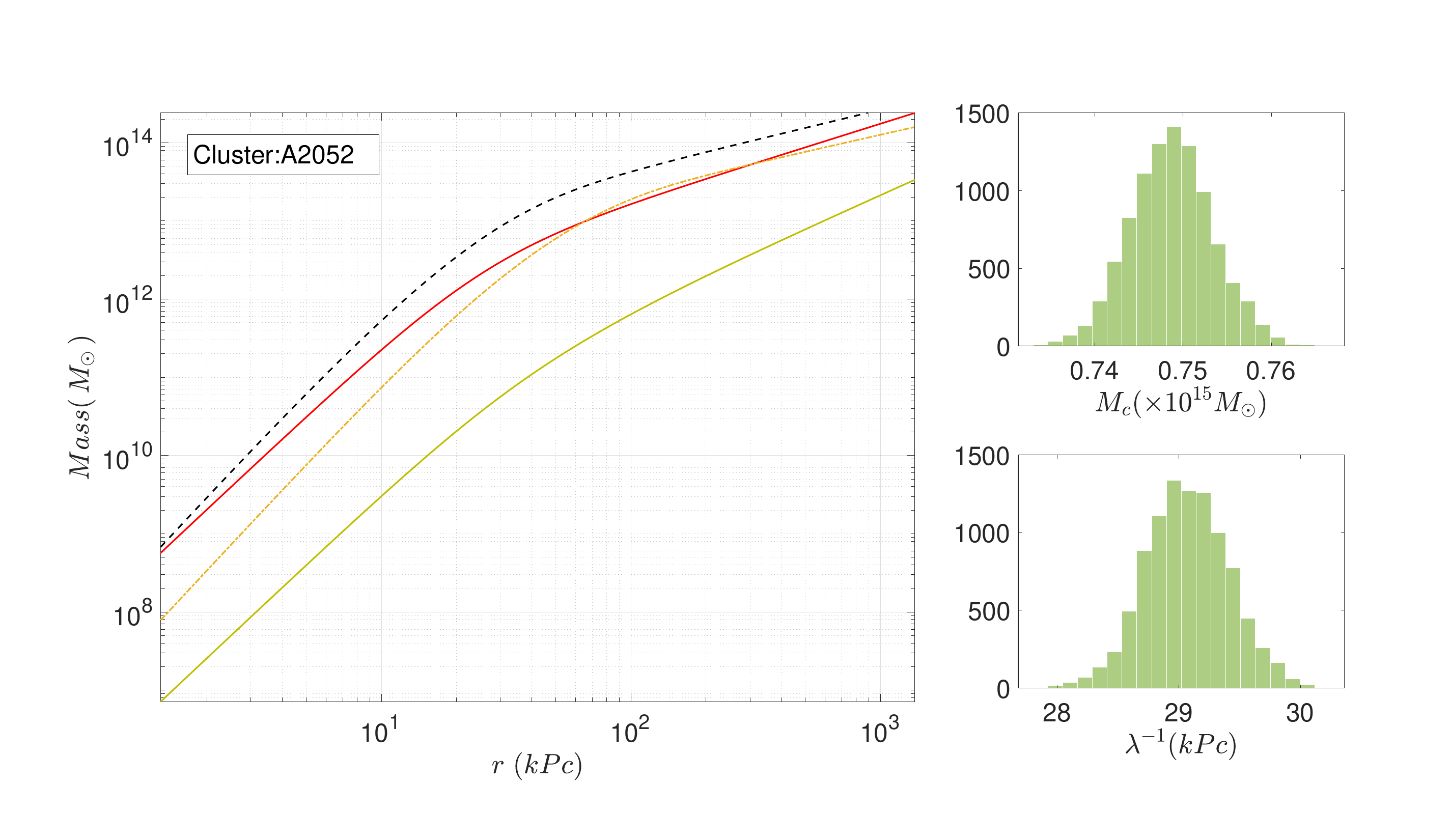}
    \includegraphics[trim=2.0cm 2.0cm 3.0cm 3.0cm, clip=true, width=0.32\columnwidth]{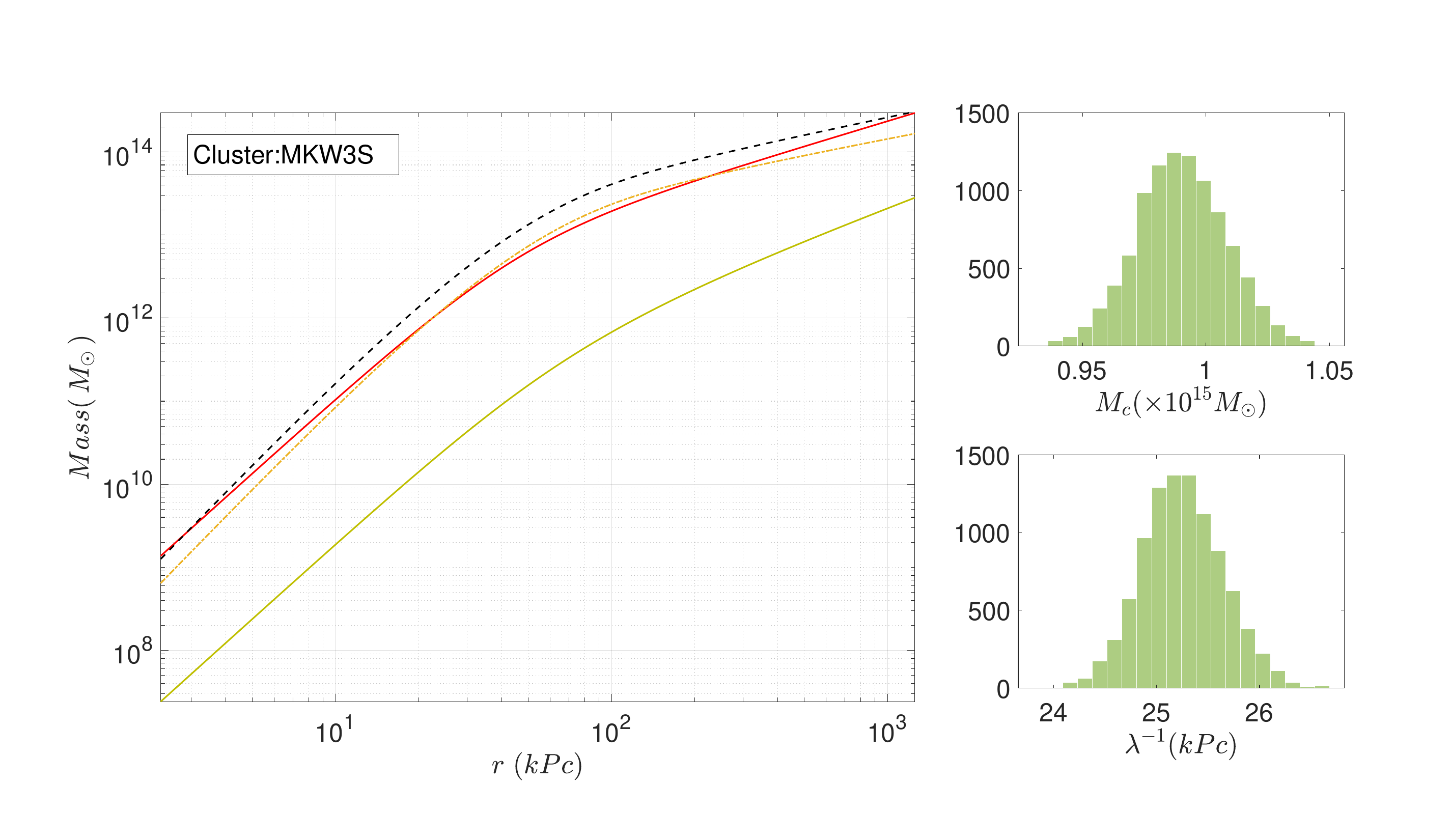}
    \includegraphics[trim=2.0cm 2.0cm 3.0cm 3.0cm, clip=true, width=0.32\columnwidth]{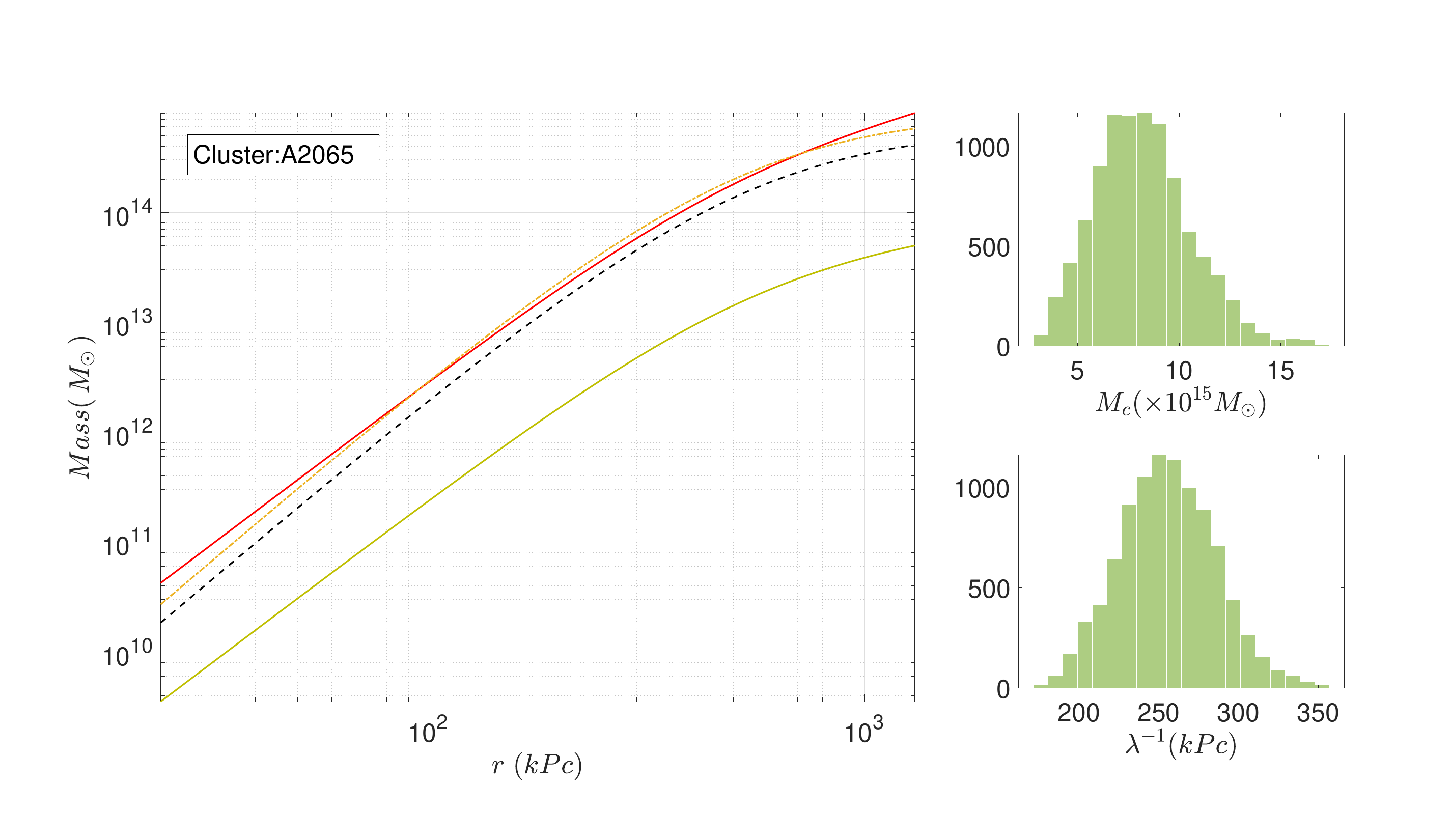}
    \includegraphics[trim=2.0cm 2.0cm 3.0cm 3.0cm, clip=true, width=0.32\columnwidth]{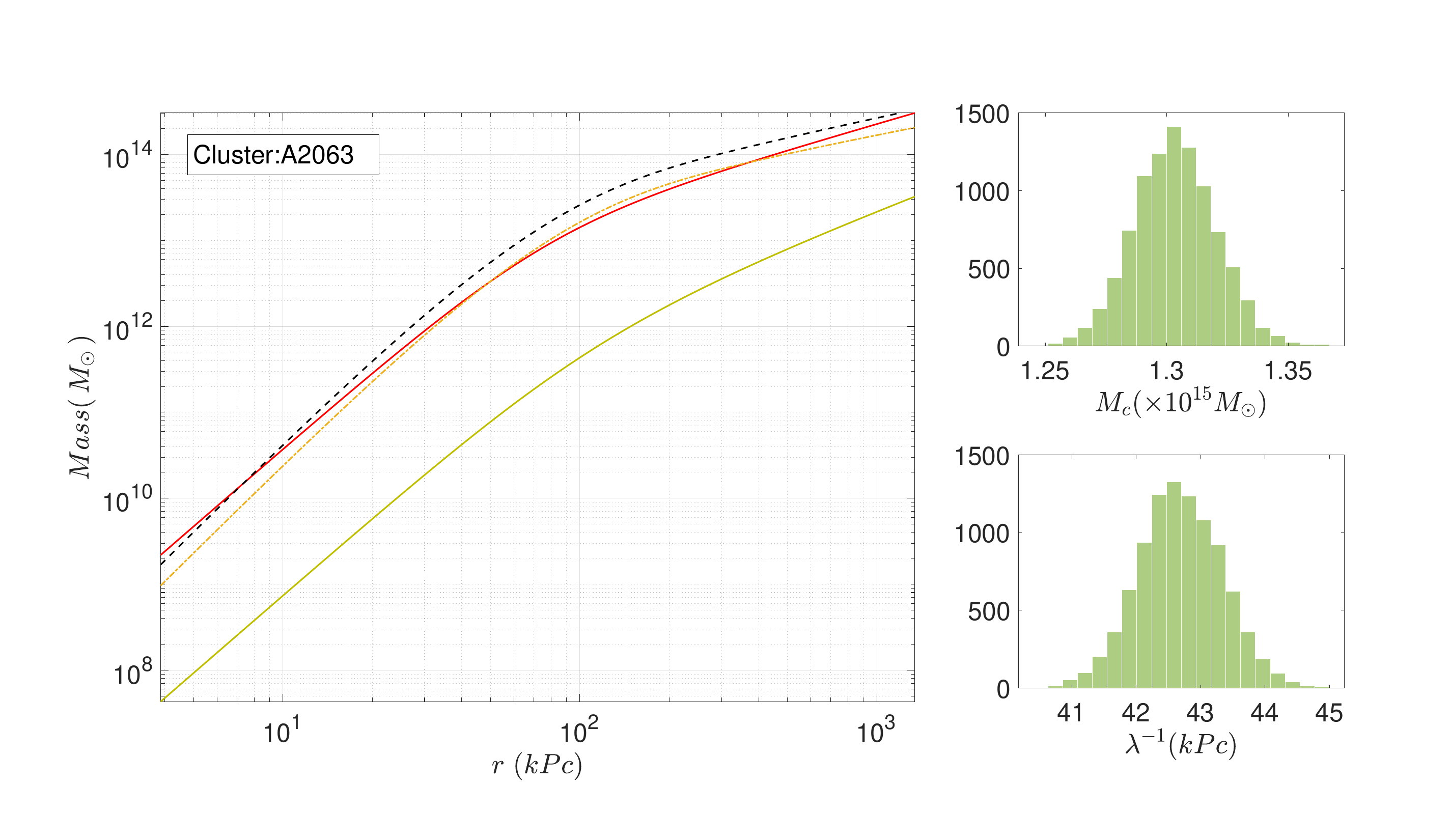}
    \includegraphics[trim=2.0cm 2.0cm 3.0cm 3.0cm, clip=true, width=0.32\columnwidth]{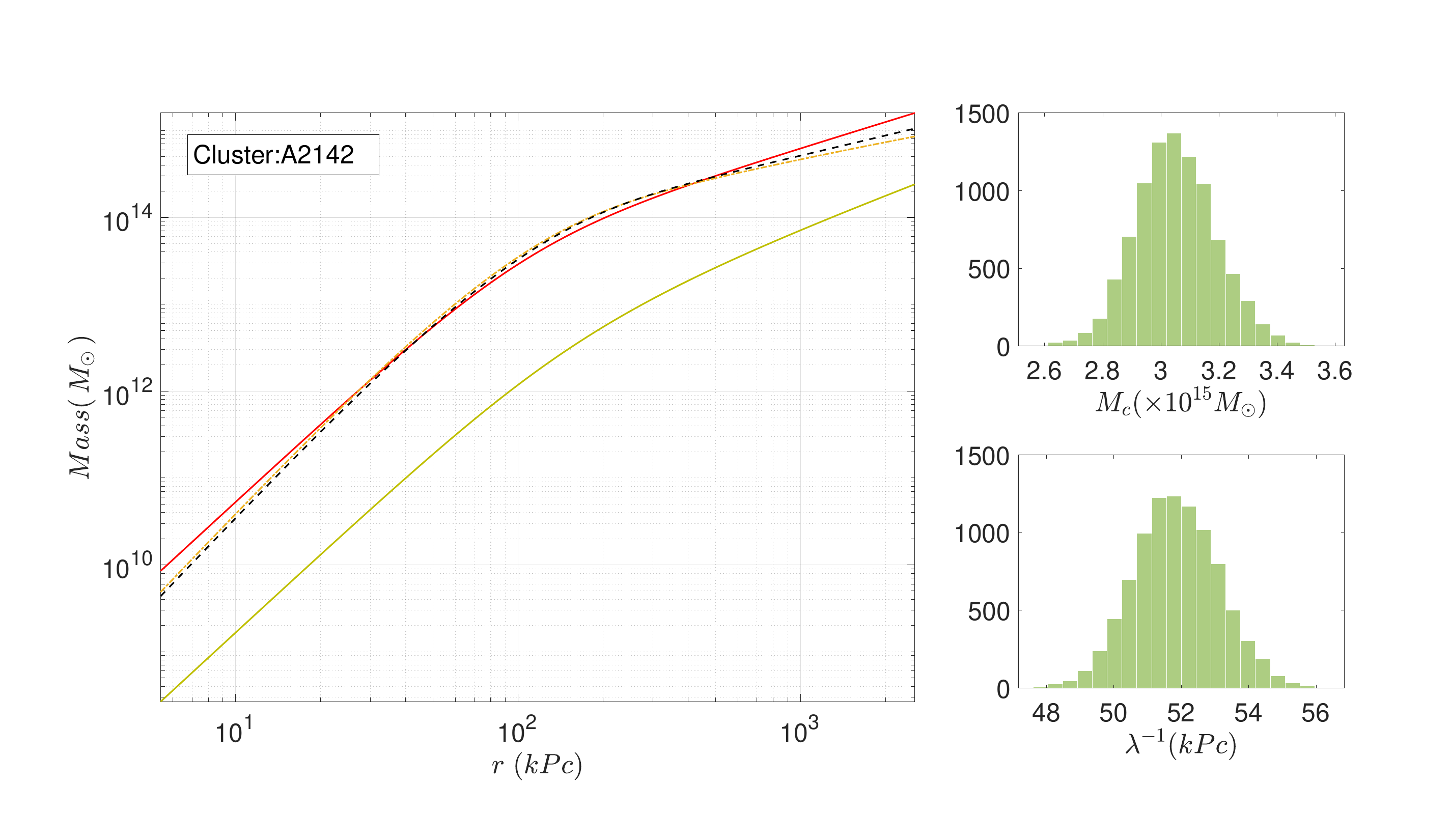}
    \end{figure}
    \begin{figure}
    \centering
    \includegraphics[trim=2.0cm 2.0cm 3.0cm 3.0cm, clip=true, width=0.32\columnwidth]{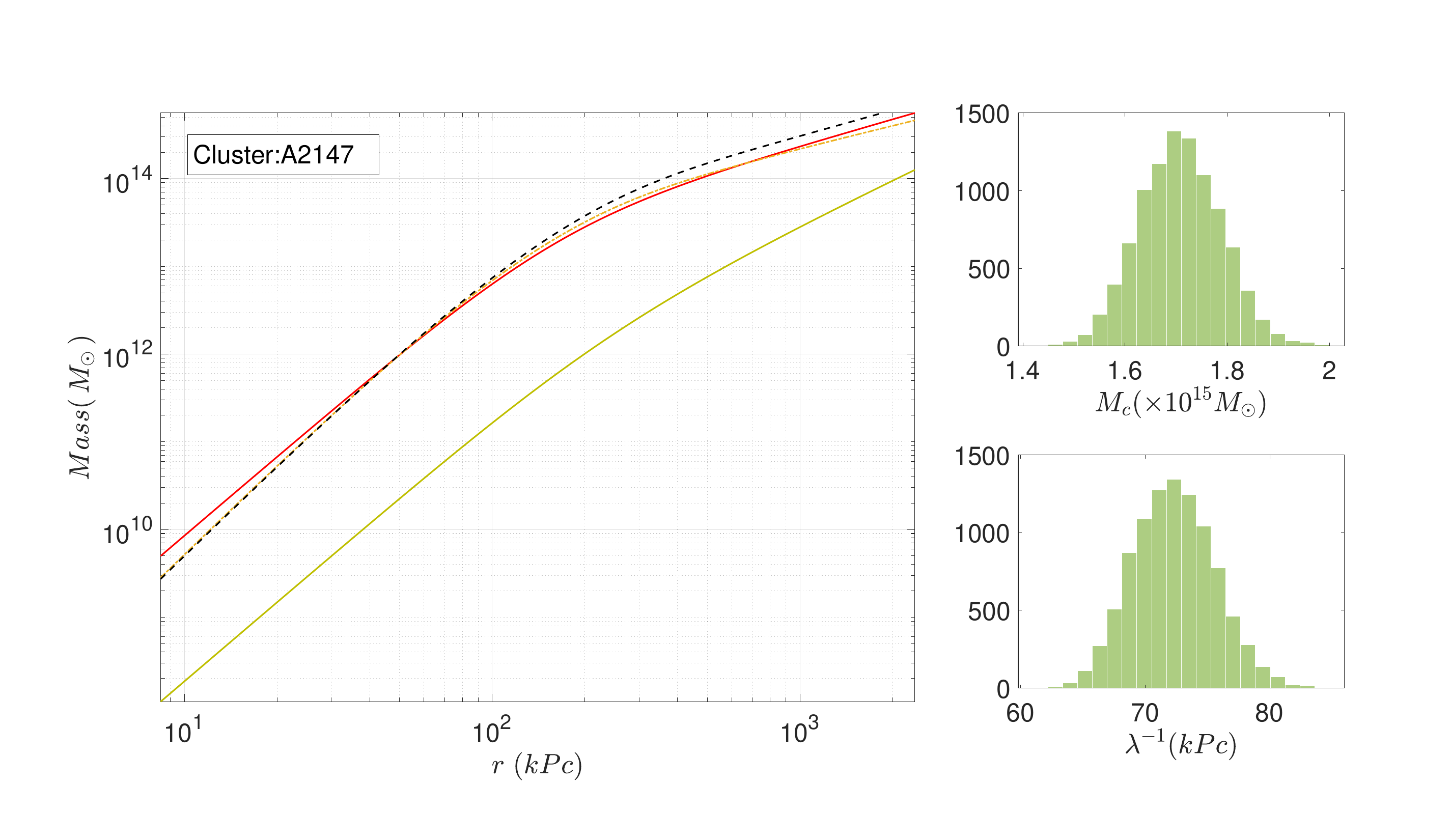}
    \includegraphics[trim=2.0cm 2.0cm 3.0cm 3.0cm, clip=true, width=0.32\columnwidth]{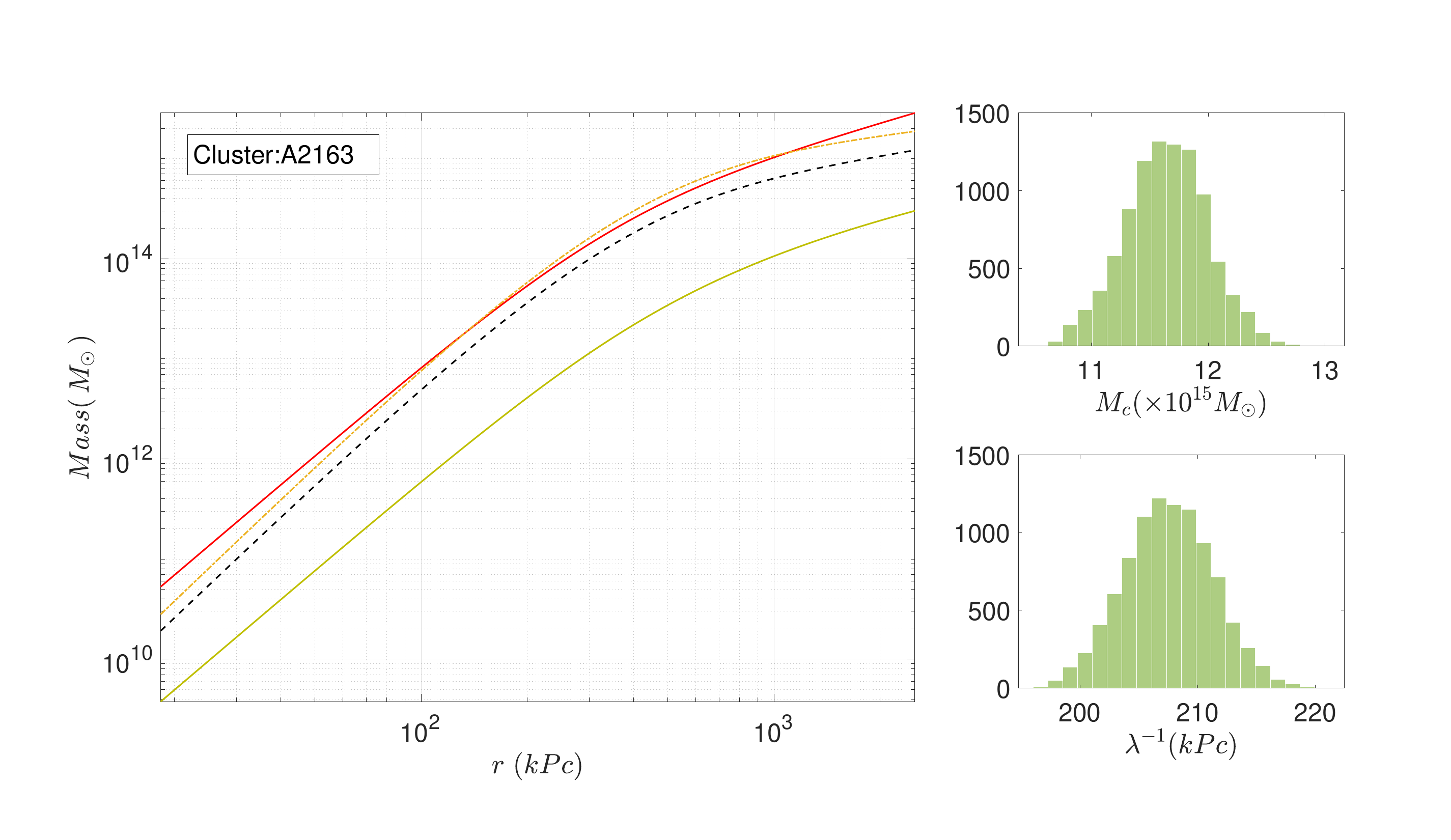}
    \includegraphics[trim=2.0cm 2.0cm 3.0cm 3.0cm, clip=true, width=0.32\columnwidth]{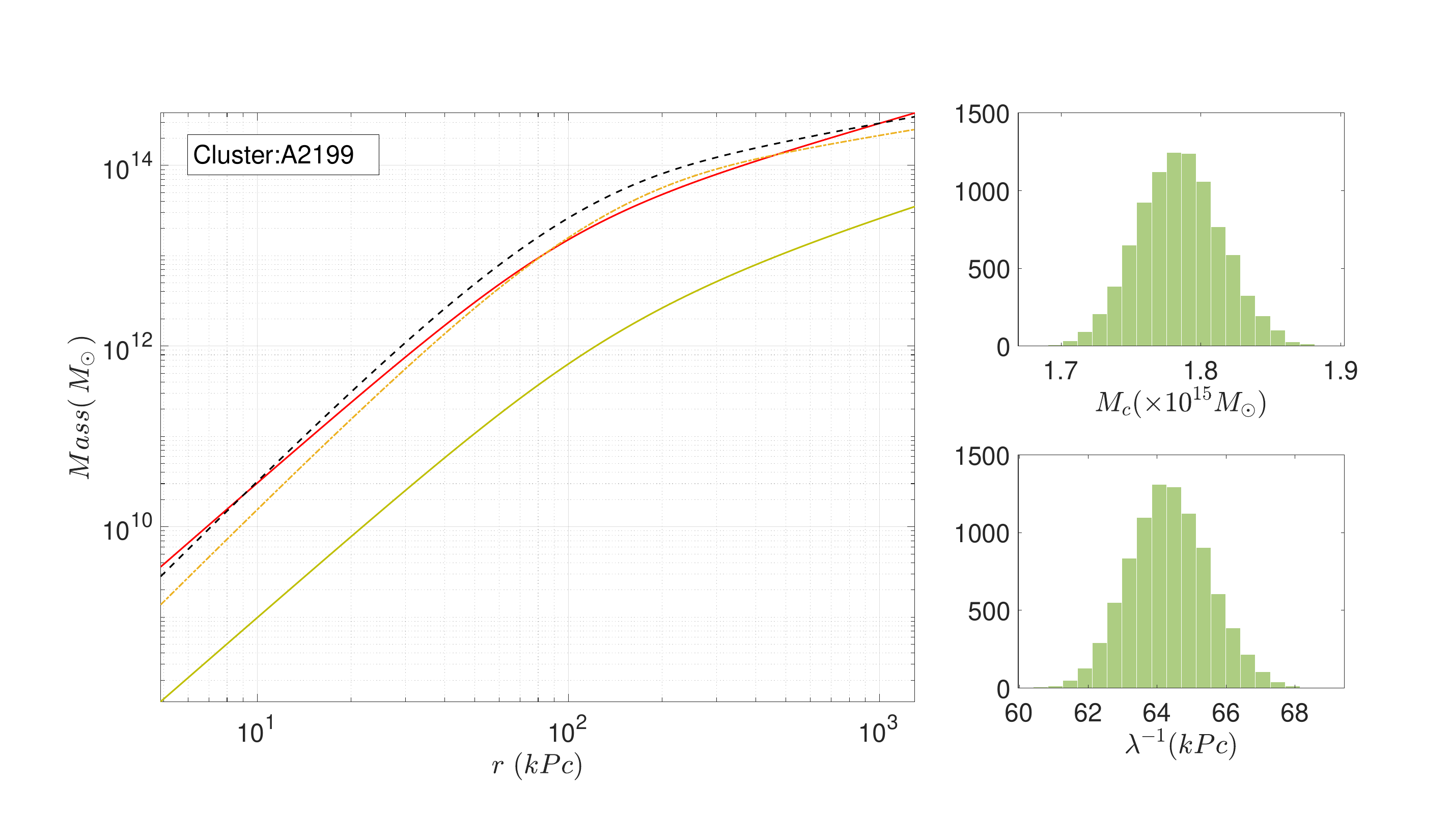}
    \includegraphics[trim=2.0cm 2.0cm 3.0cm 3.0cm, clip=true, width=0.32\columnwidth]{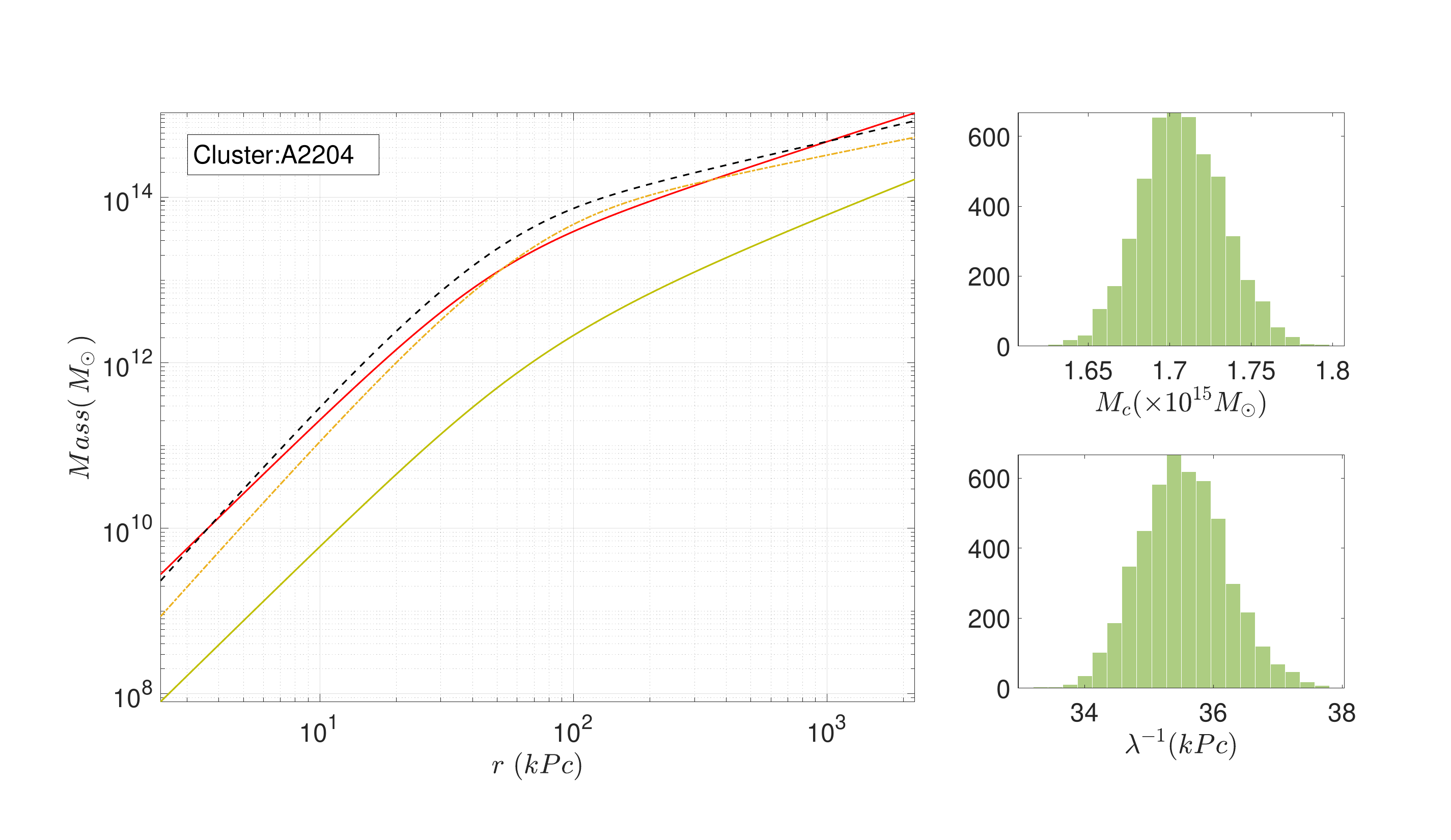}
    \includegraphics[trim=2.0cm 2.0cm 3.0cm 3.0cm, clip=true, width=0.32\columnwidth]{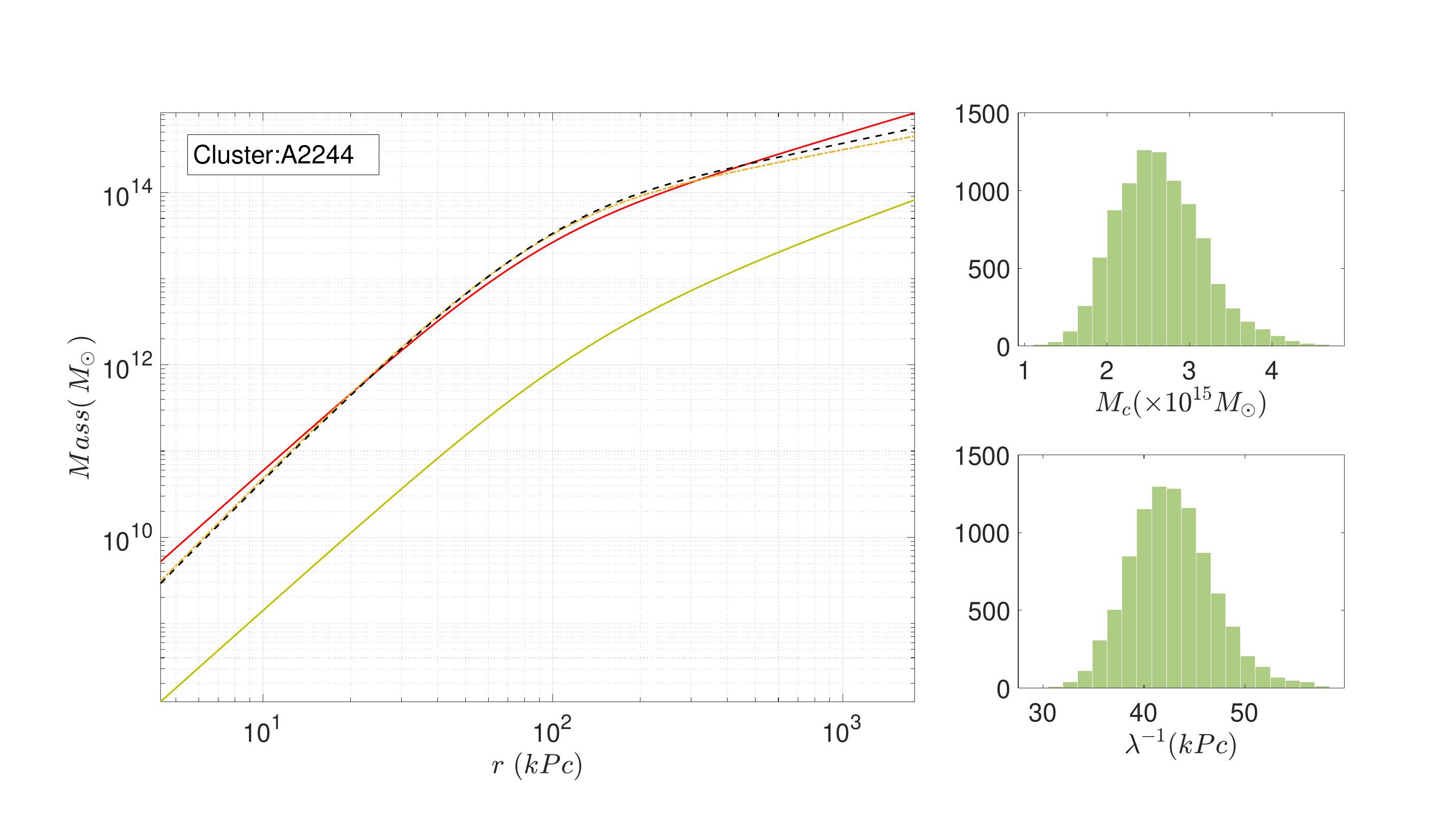}
    \includegraphics[trim=2.0cm 2.0cm 3.0cm 3.0cm, clip=true, width=0.32\columnwidth]{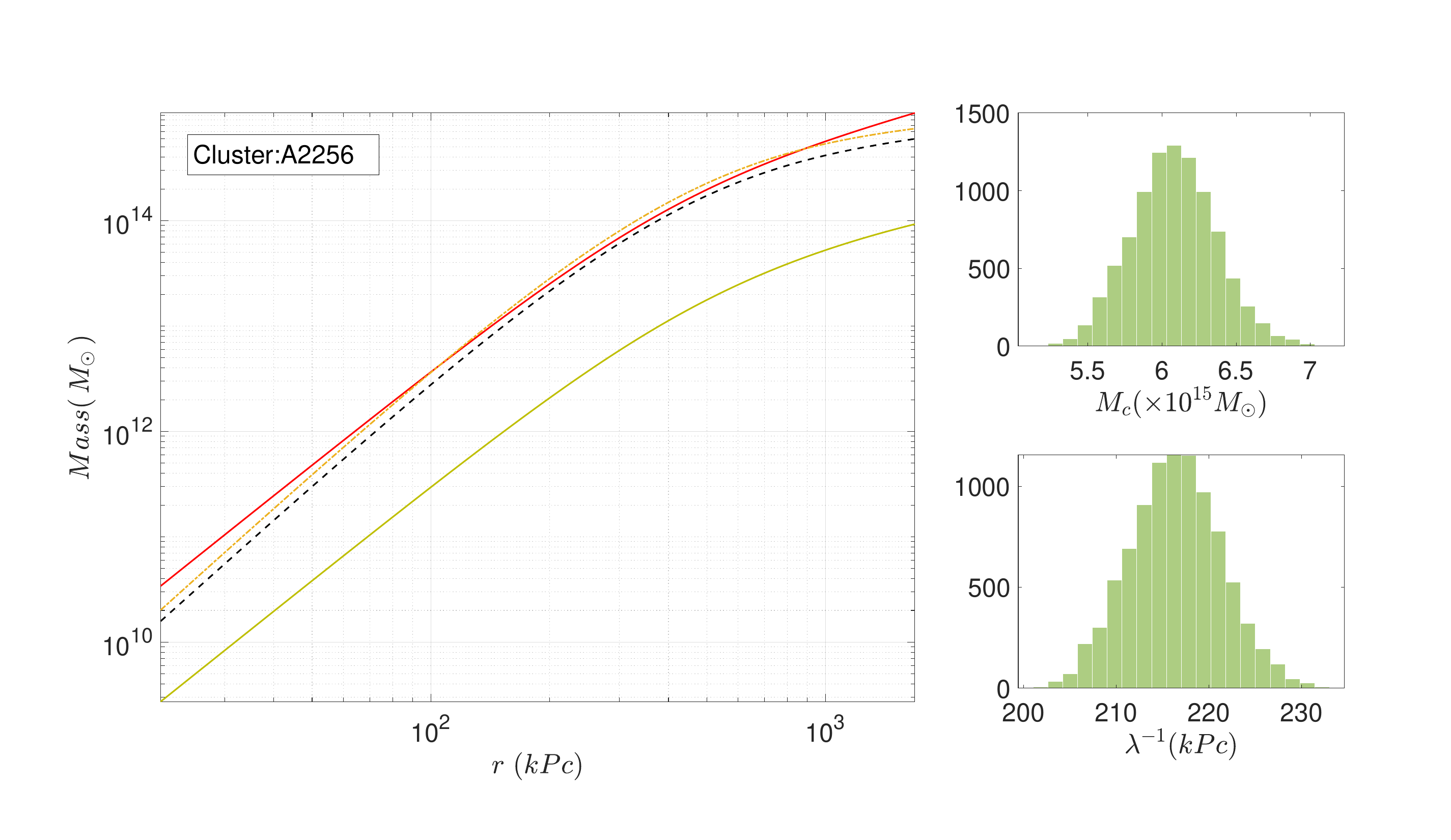}
    \includegraphics[trim=2.0cm 2.0cm 3.0cm 3.0cm, clip=true, width=0.32\columnwidth]{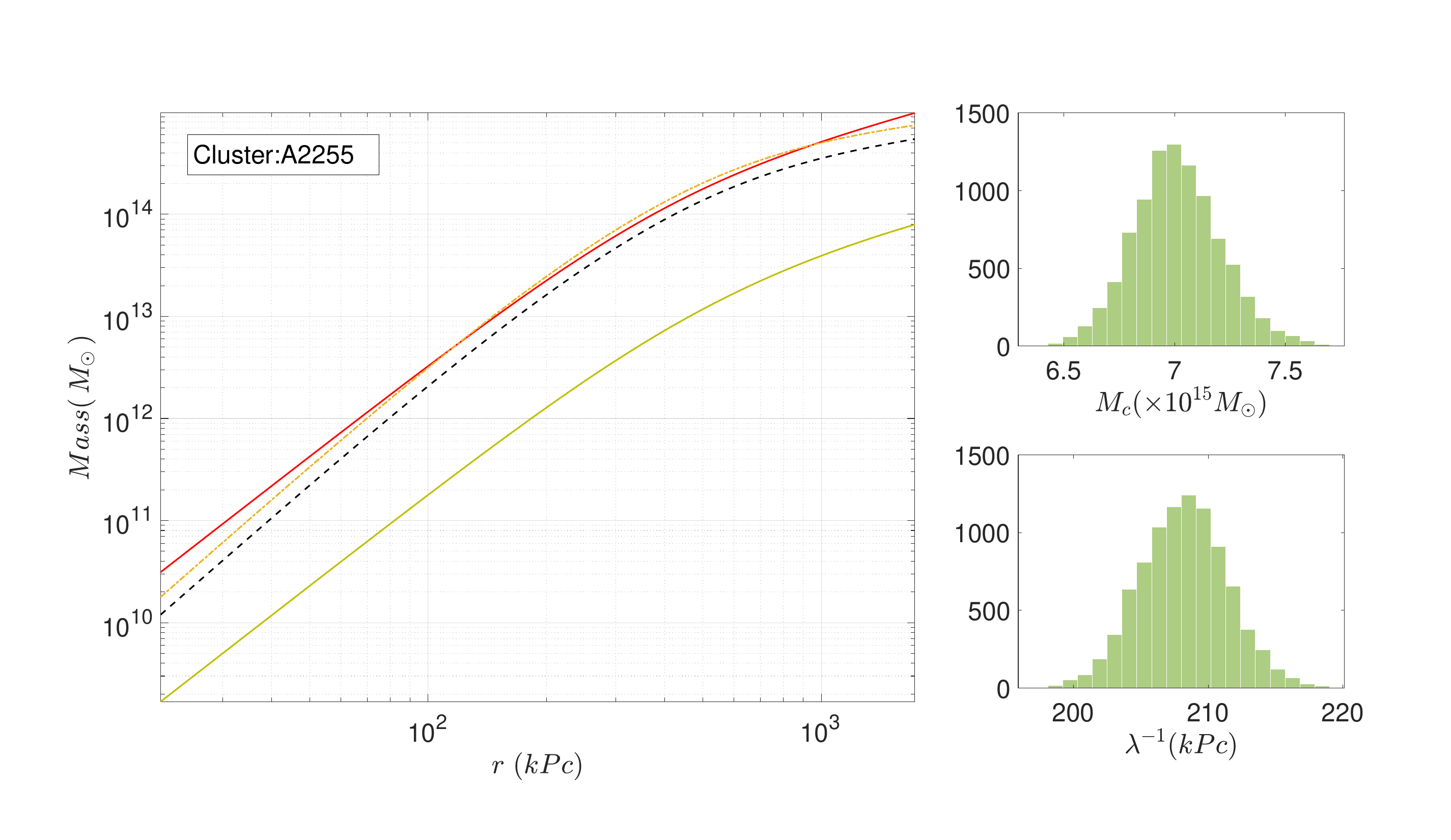}
    \includegraphics[trim=2.0cm 2.0cm 3.0cm 3.0cm, clip=true, width=0.32\columnwidth]{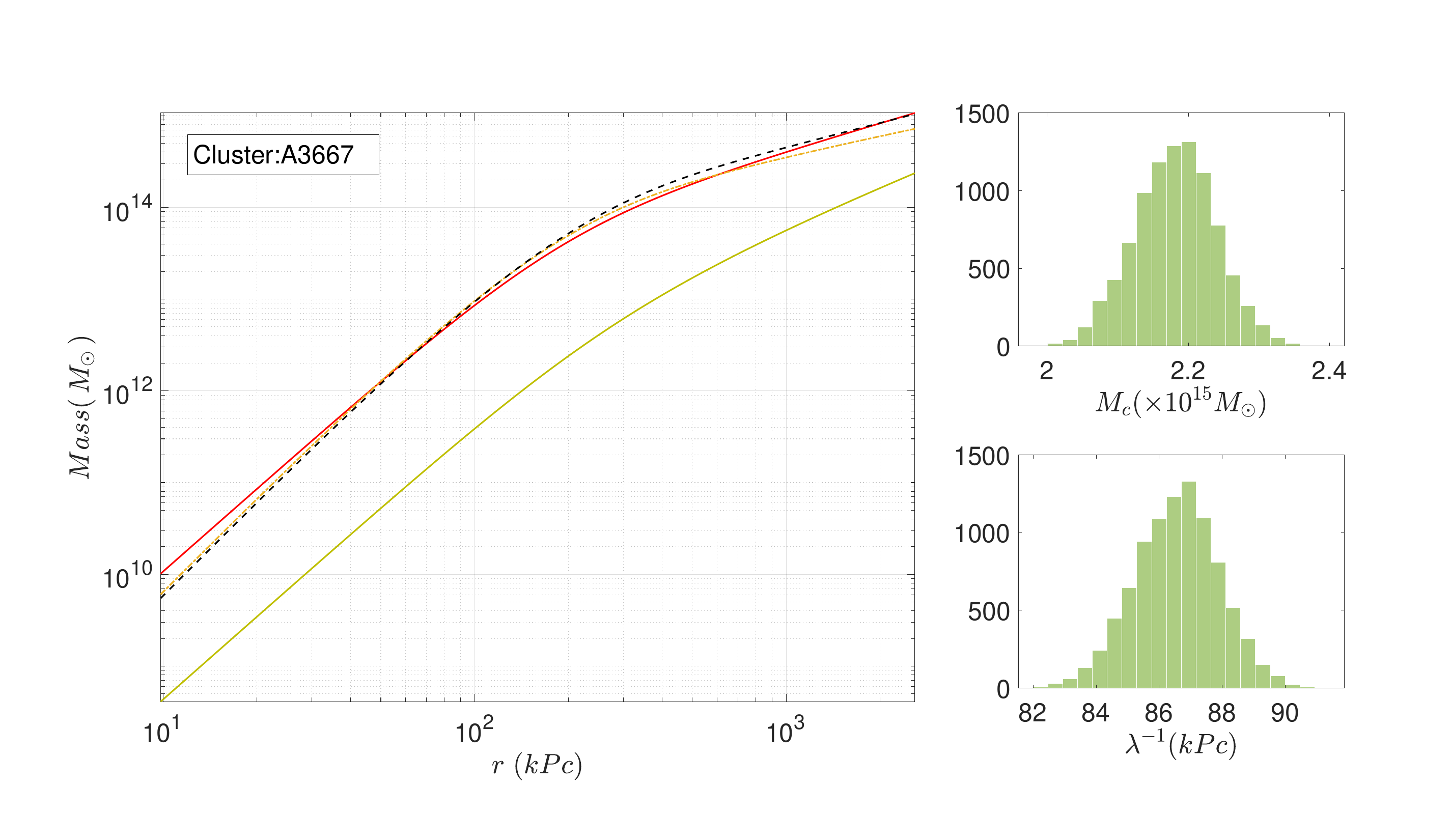}
    \includegraphics[trim=2.0cm 2.0cm 3.0cm 3.0cm, clip=true, width=0.32\columnwidth]{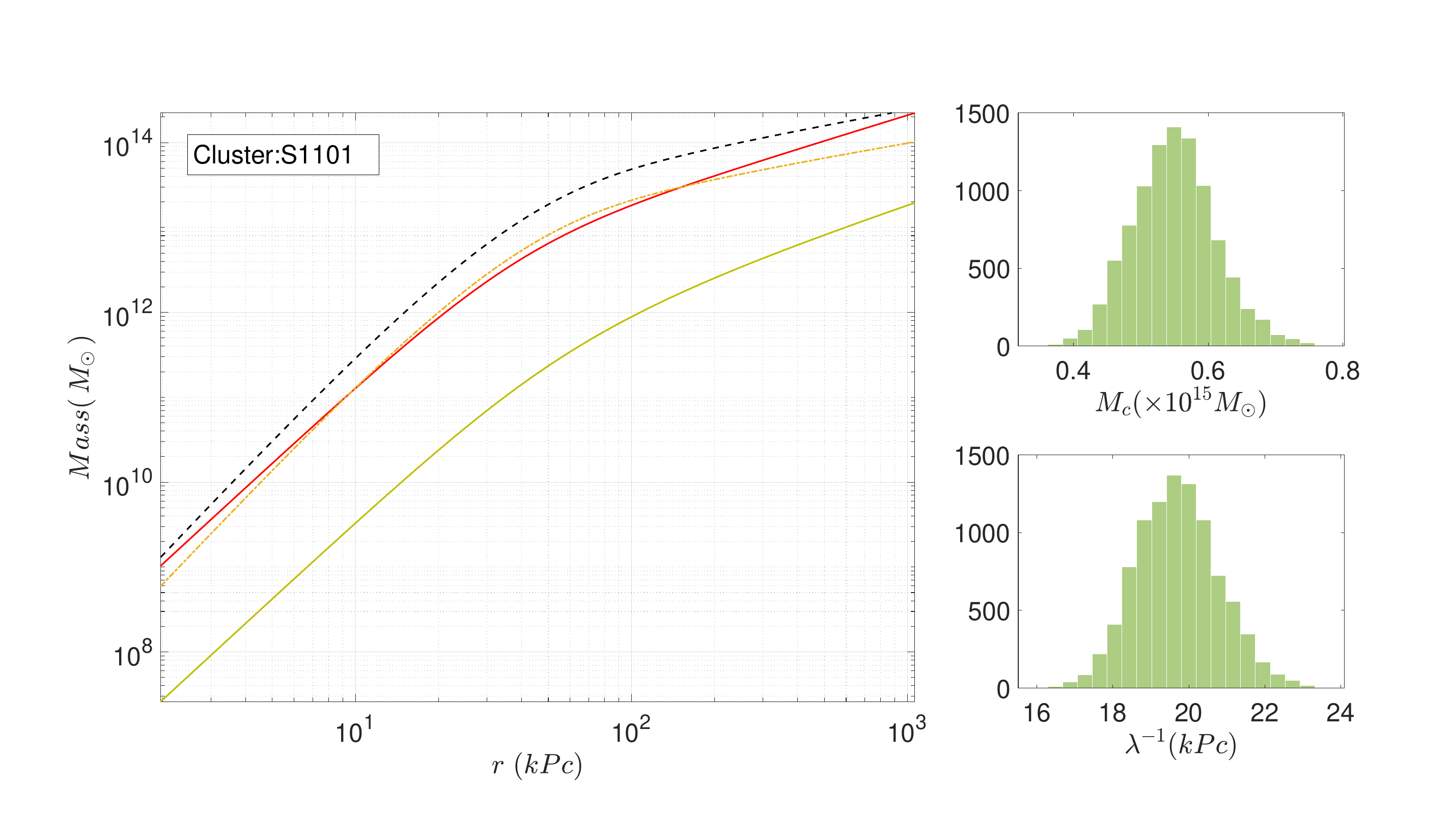}
    \includegraphics[trim=2.0cm 2.0cm 3.0cm 3.0cm, clip=true, width=0.32\columnwidth]{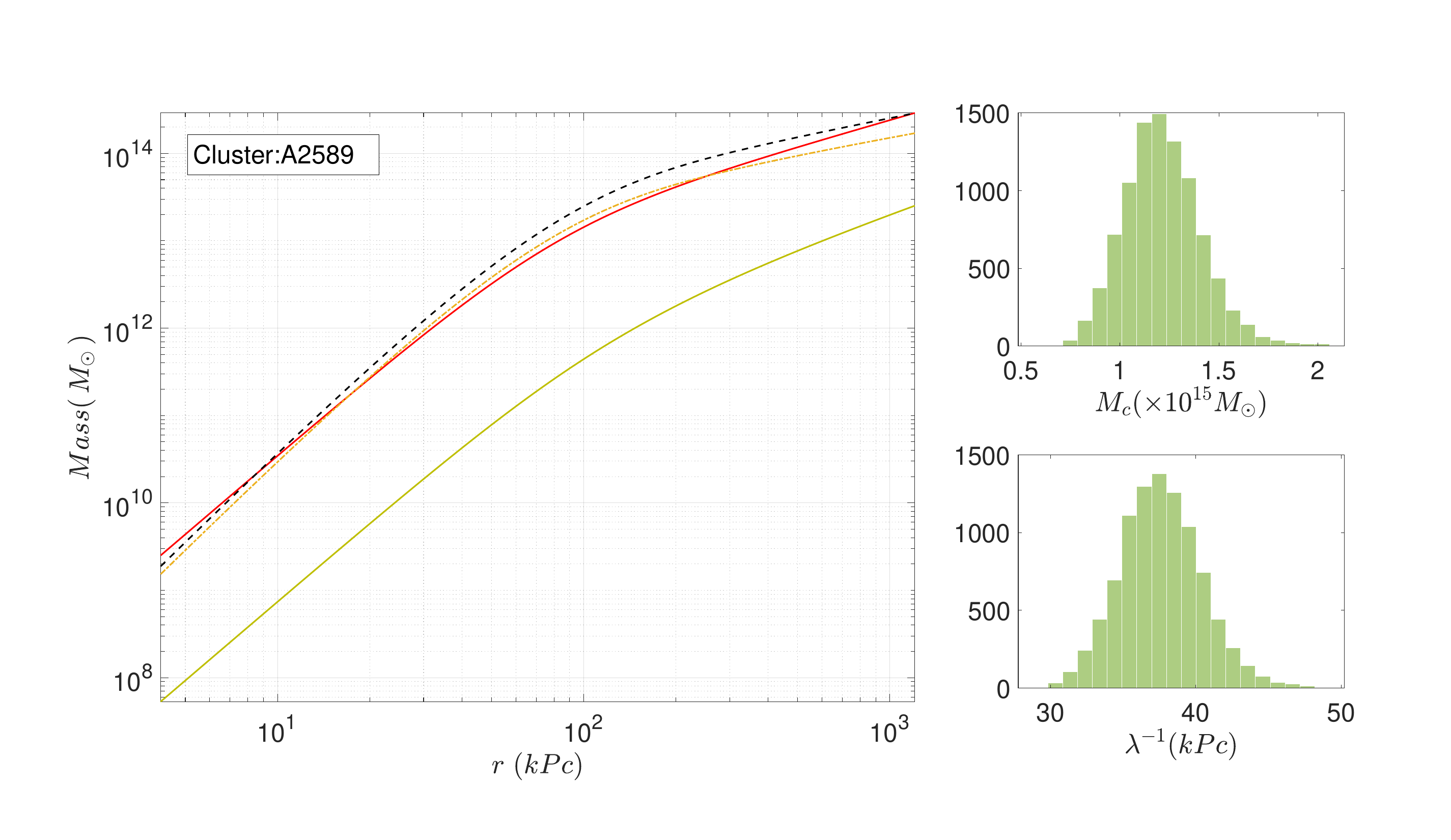}
    \includegraphics[trim=2.0cm 2.0cm 3.0cm 3.0cm, clip=true, width=0.32\columnwidth]{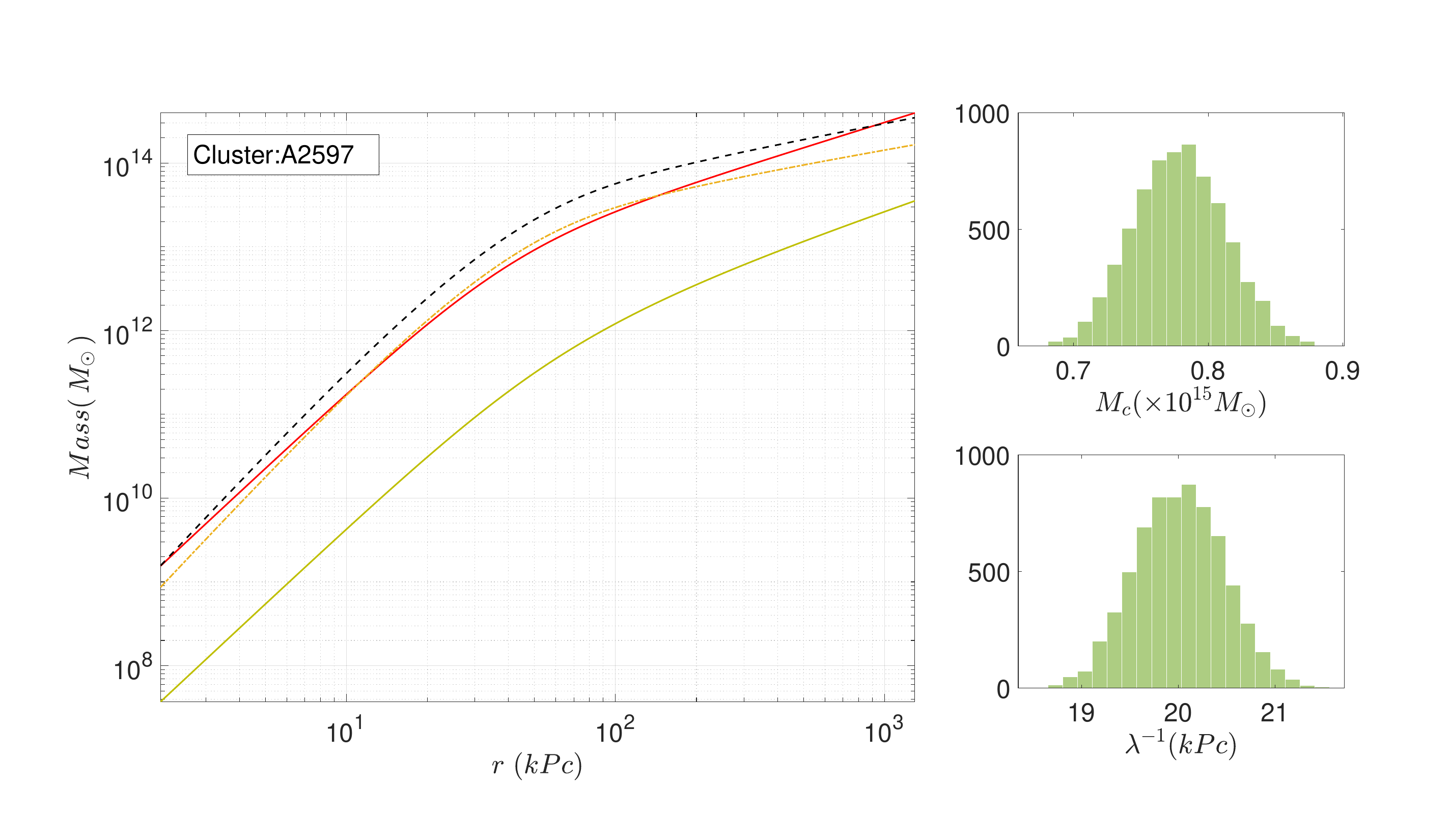}
    \includegraphics[trim=2.0cm 2.0cm 3.0cm 3.0cm, clip=true, width=0.32\columnwidth]{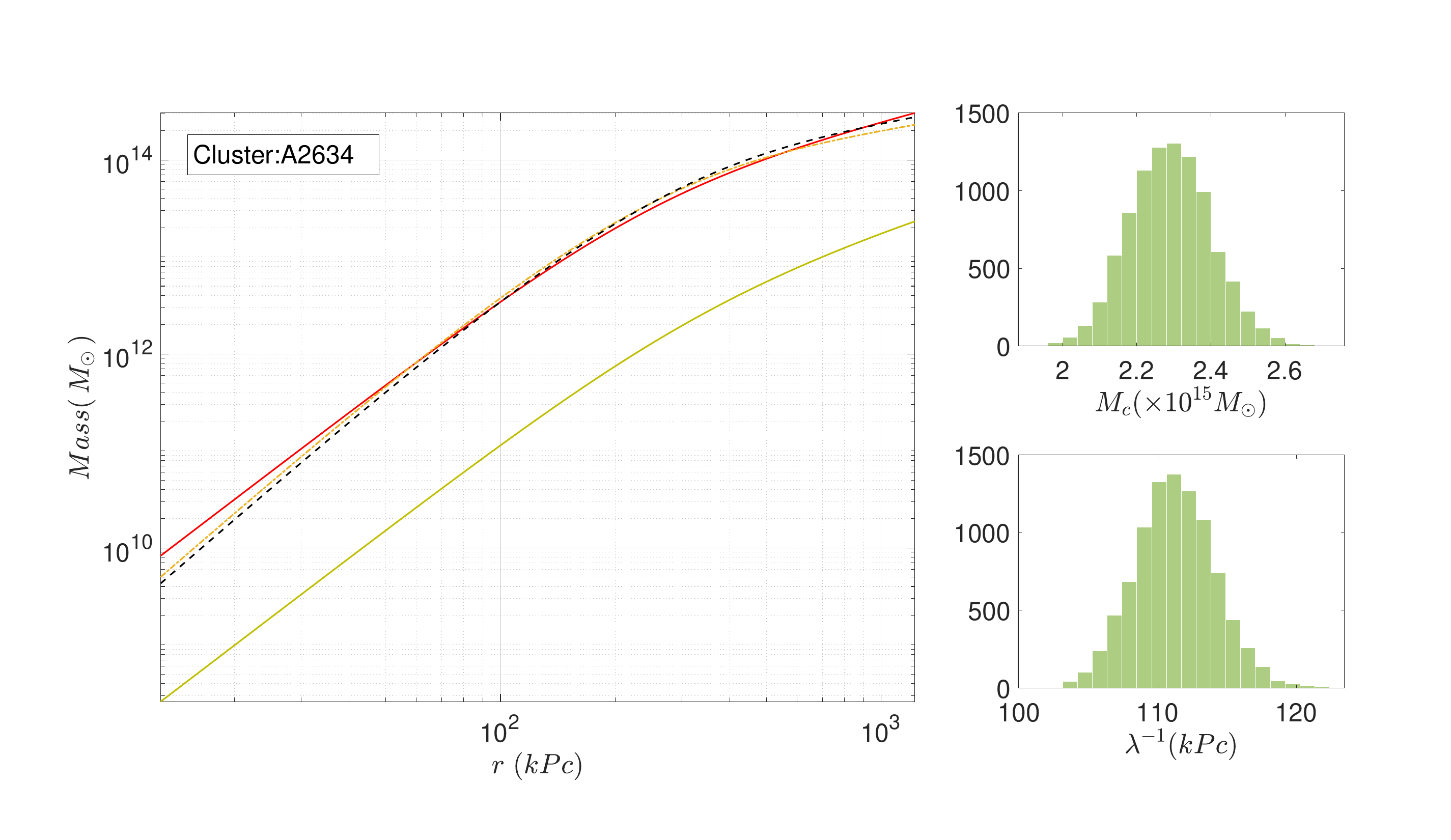}
    \includegraphics[trim=2.0cm 2.0cm 3.0cm 3.0cm, clip=true, width=0.32\columnwidth]{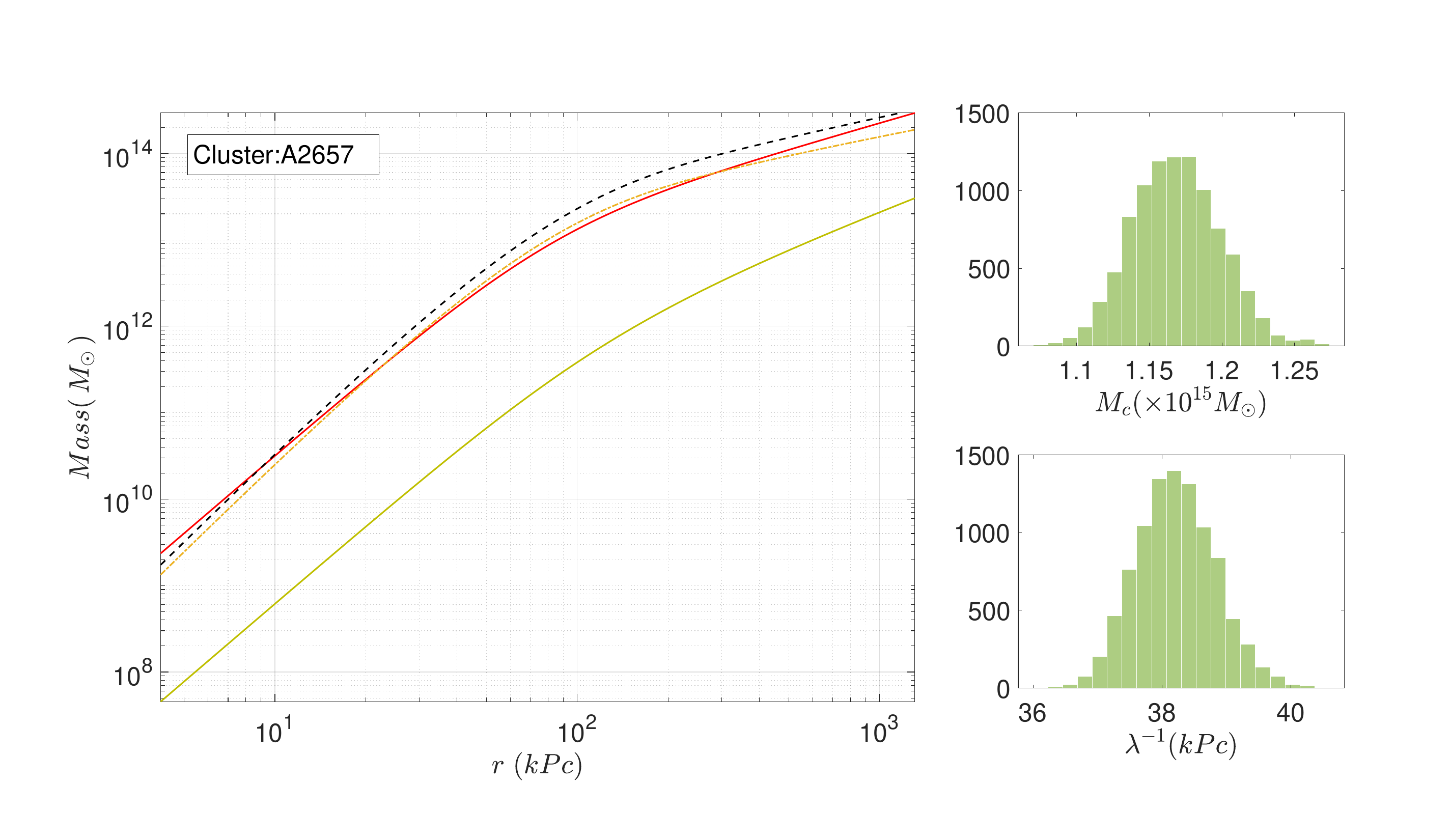}
    \includegraphics[trim=2.0cm 2.0cm 3.0cm 3.0cm, clip=true, width=0.32\columnwidth]{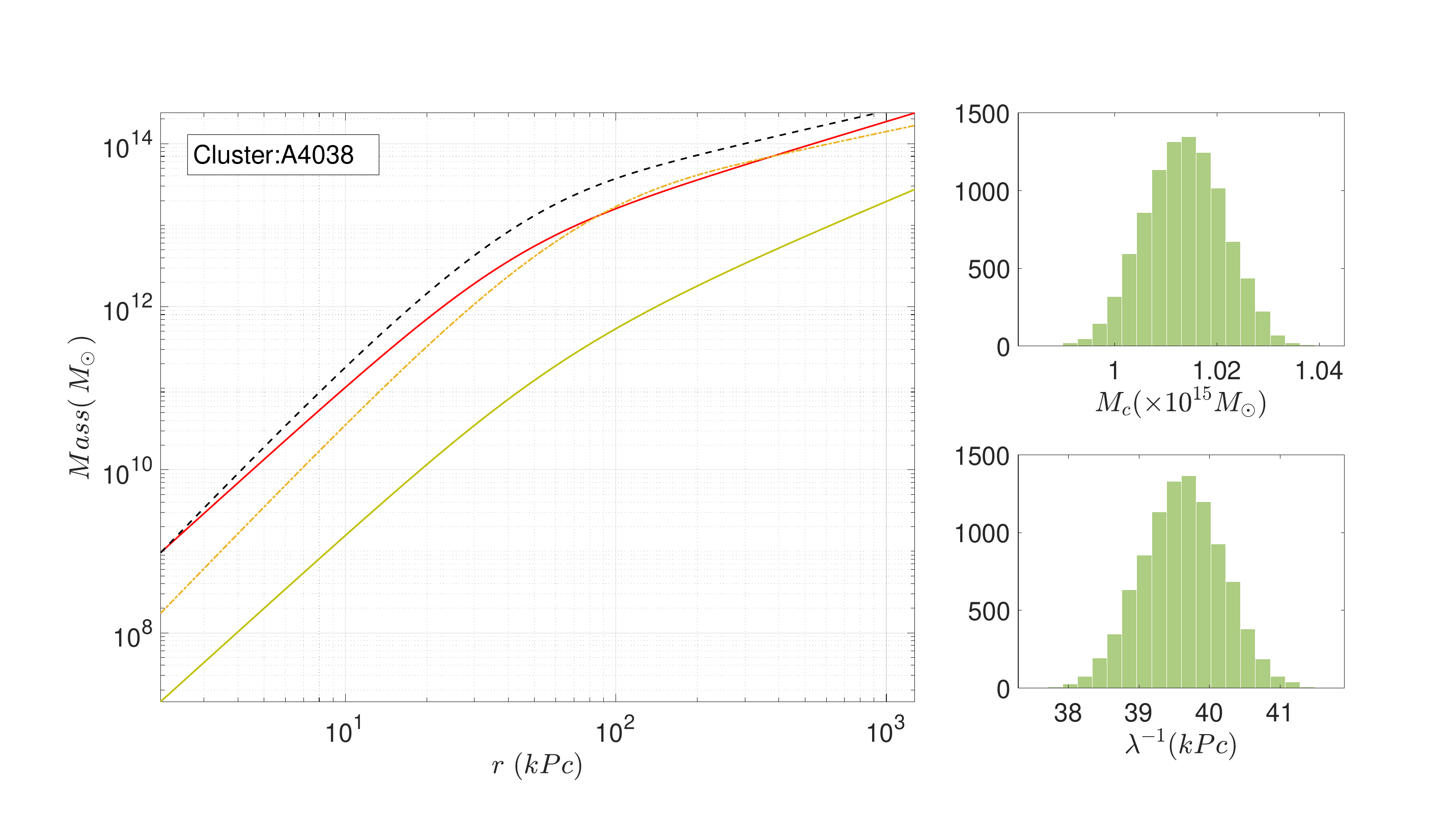}
    \includegraphics[trim=2.0cm 2.0cm 3.0cm 3.0cm, clip=true, width=0.32\columnwidth]{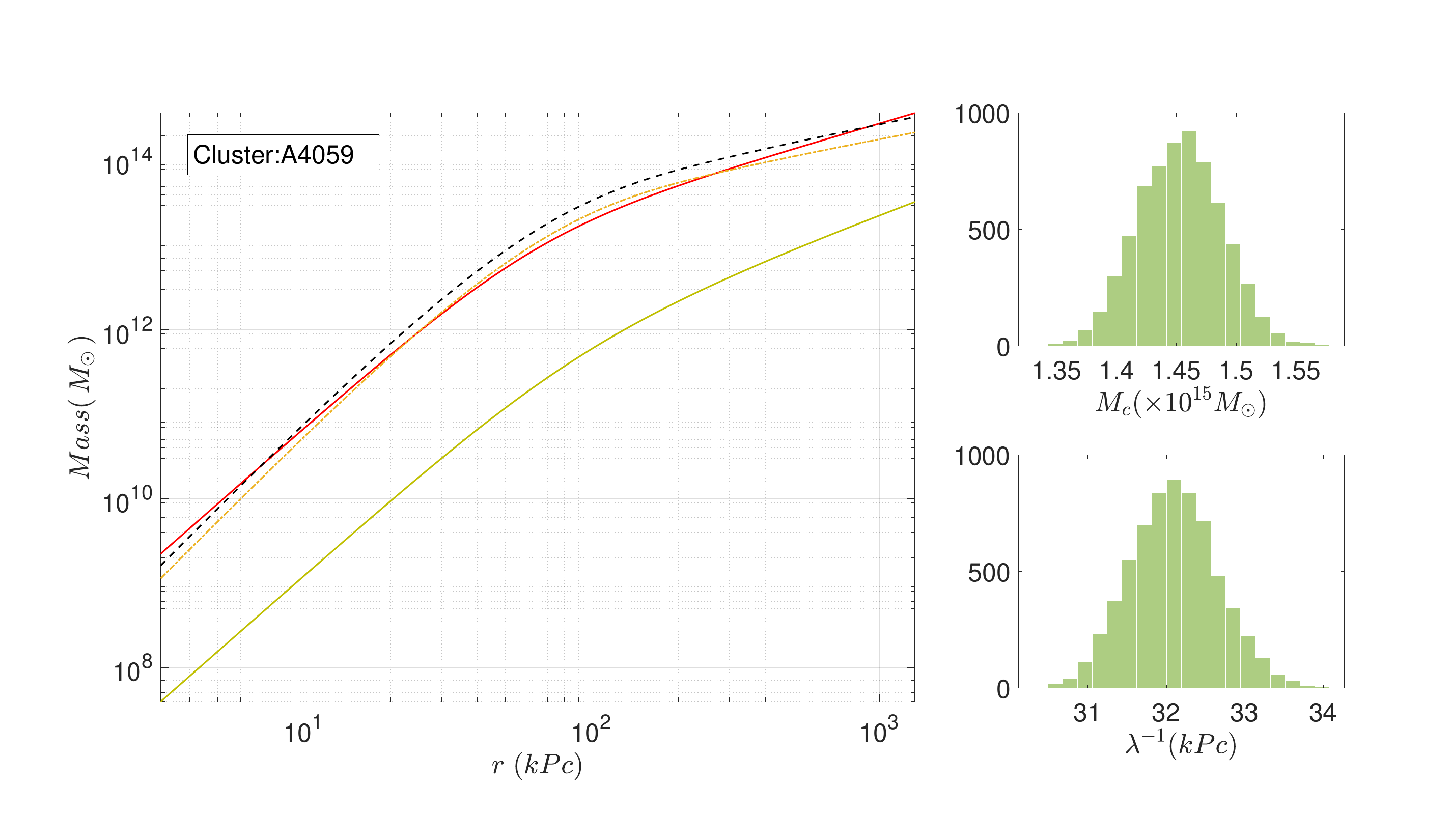}
    \includegraphics[trim=2.0cm 2.0cm 3.0cm 3.0cm, clip=true, width=0.32\columnwidth]{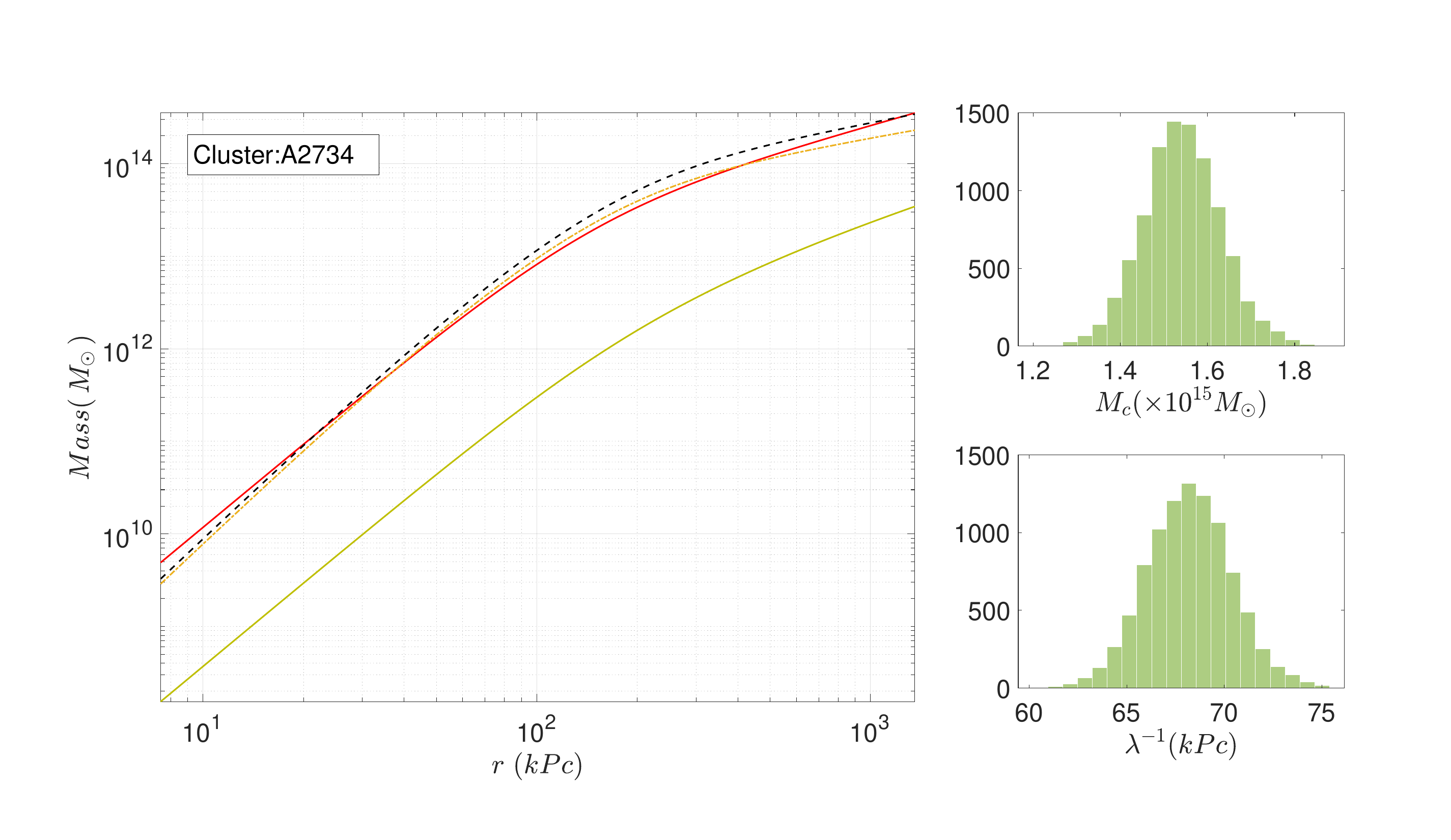}
    \includegraphics[trim=2.0cm 2.0cm 3.0cm 3.0cm, clip=true, width=0.32\columnwidth]{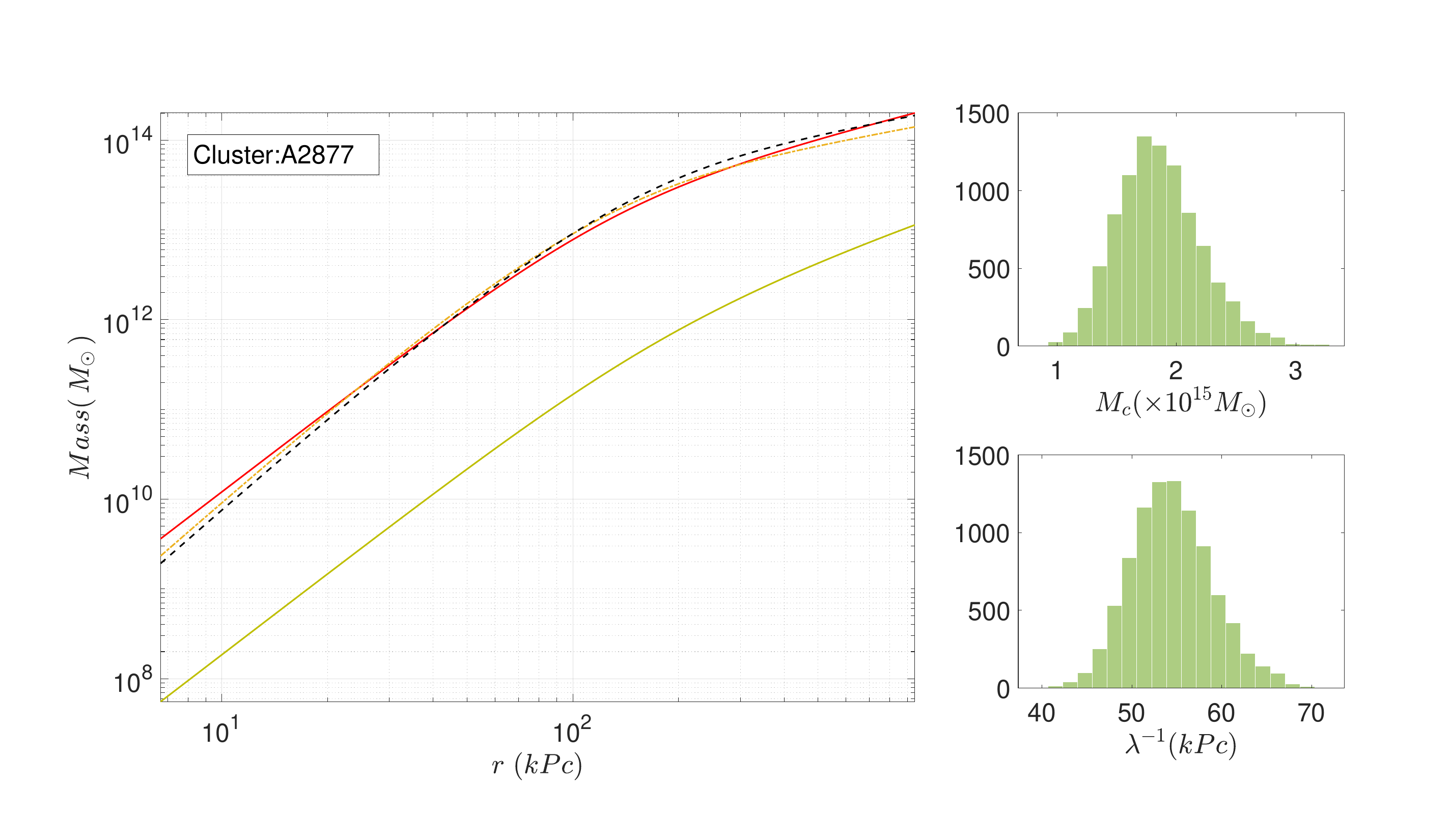}
    \includegraphics[trim=2.0cm 2.0cm 3.0cm 3.0cm, clip=true, width=0.32\columnwidth]{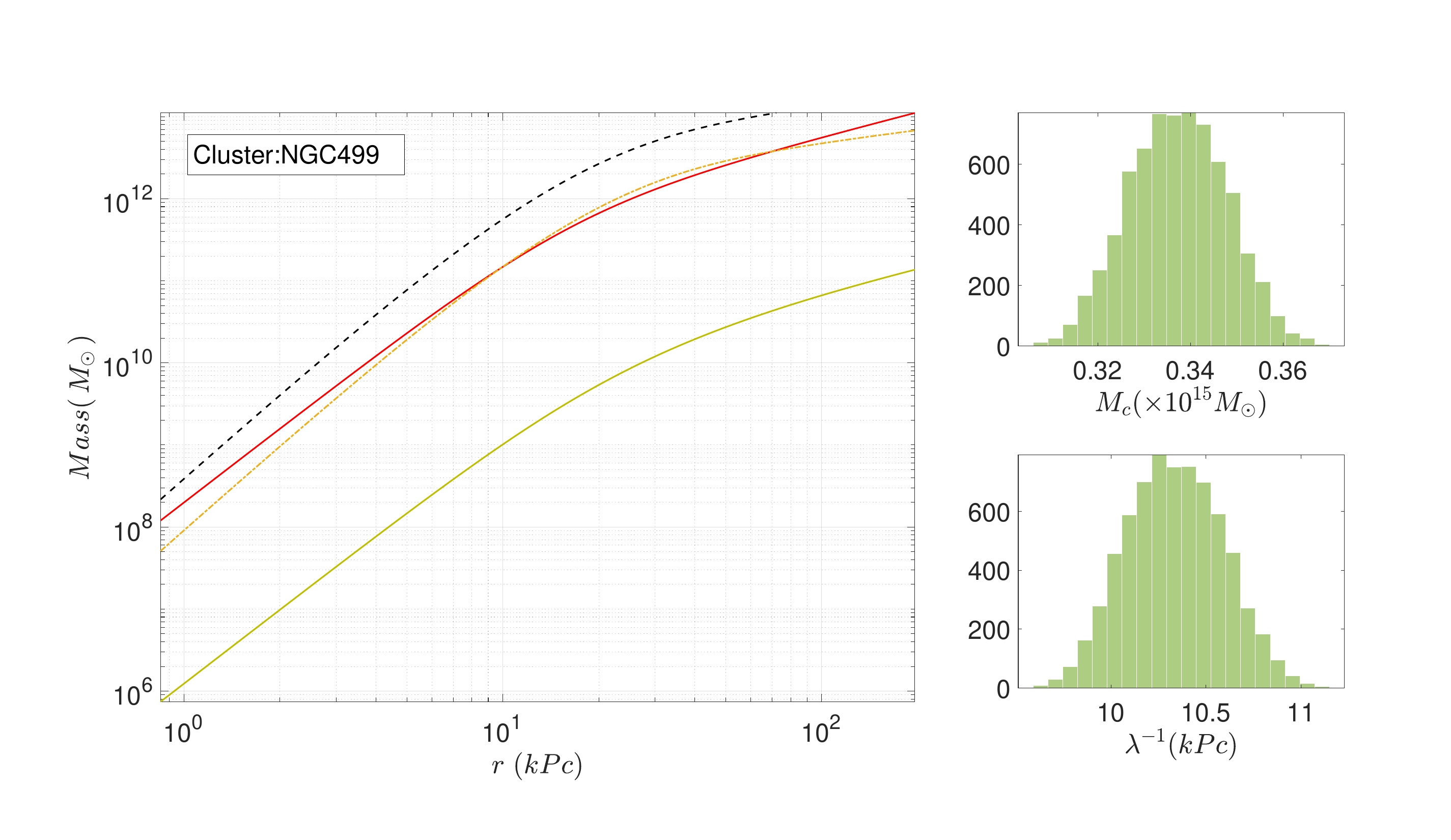}
    \includegraphics[trim=2.0cm 2.0cm 3.0cm 3.0cm, clip=true, width=0.32\columnwidth]{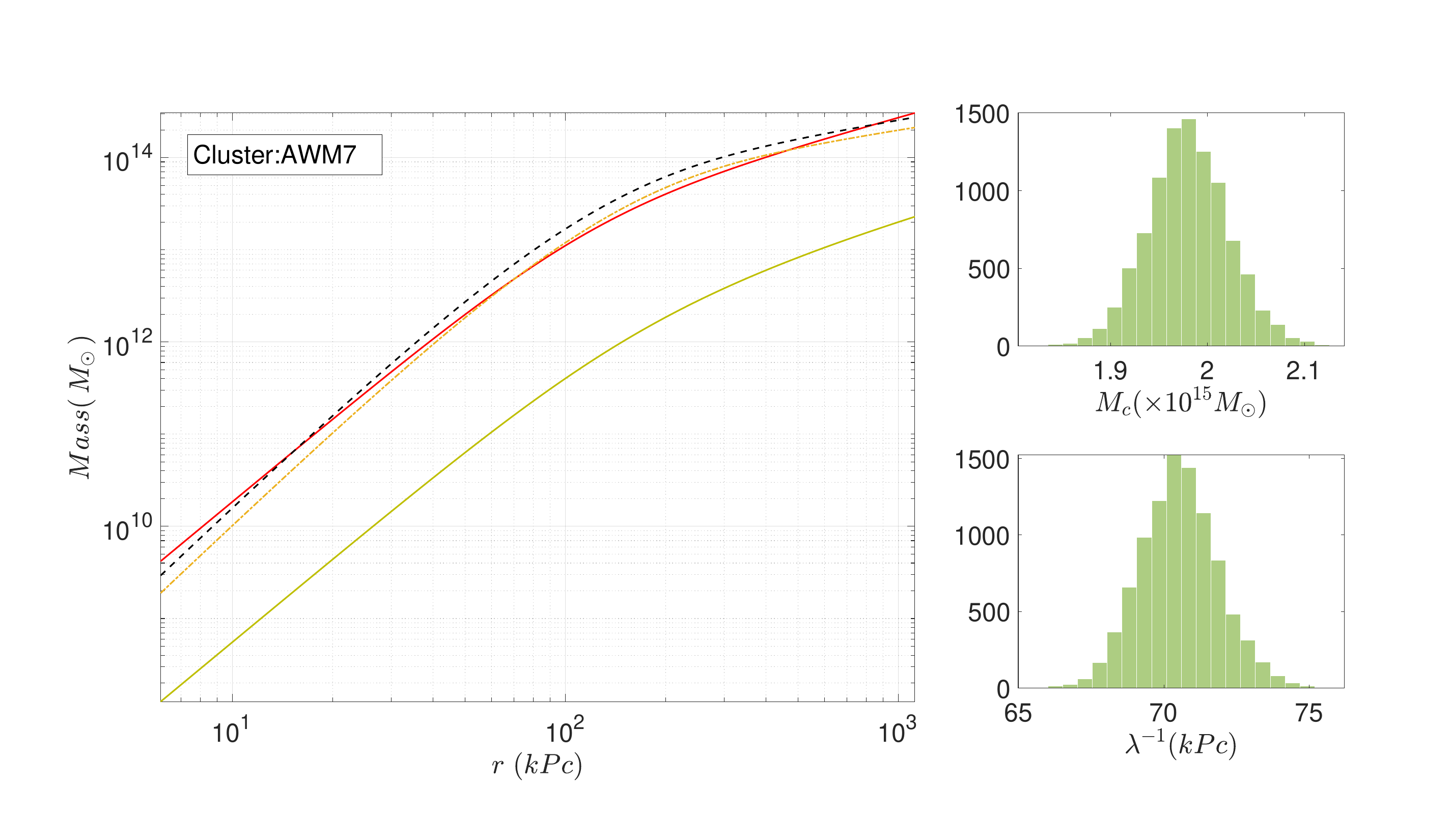}
    \includegraphics[trim=2.0cm 2.0cm 3.0cm 3.0cm, clip=true, width=0.32\columnwidth]{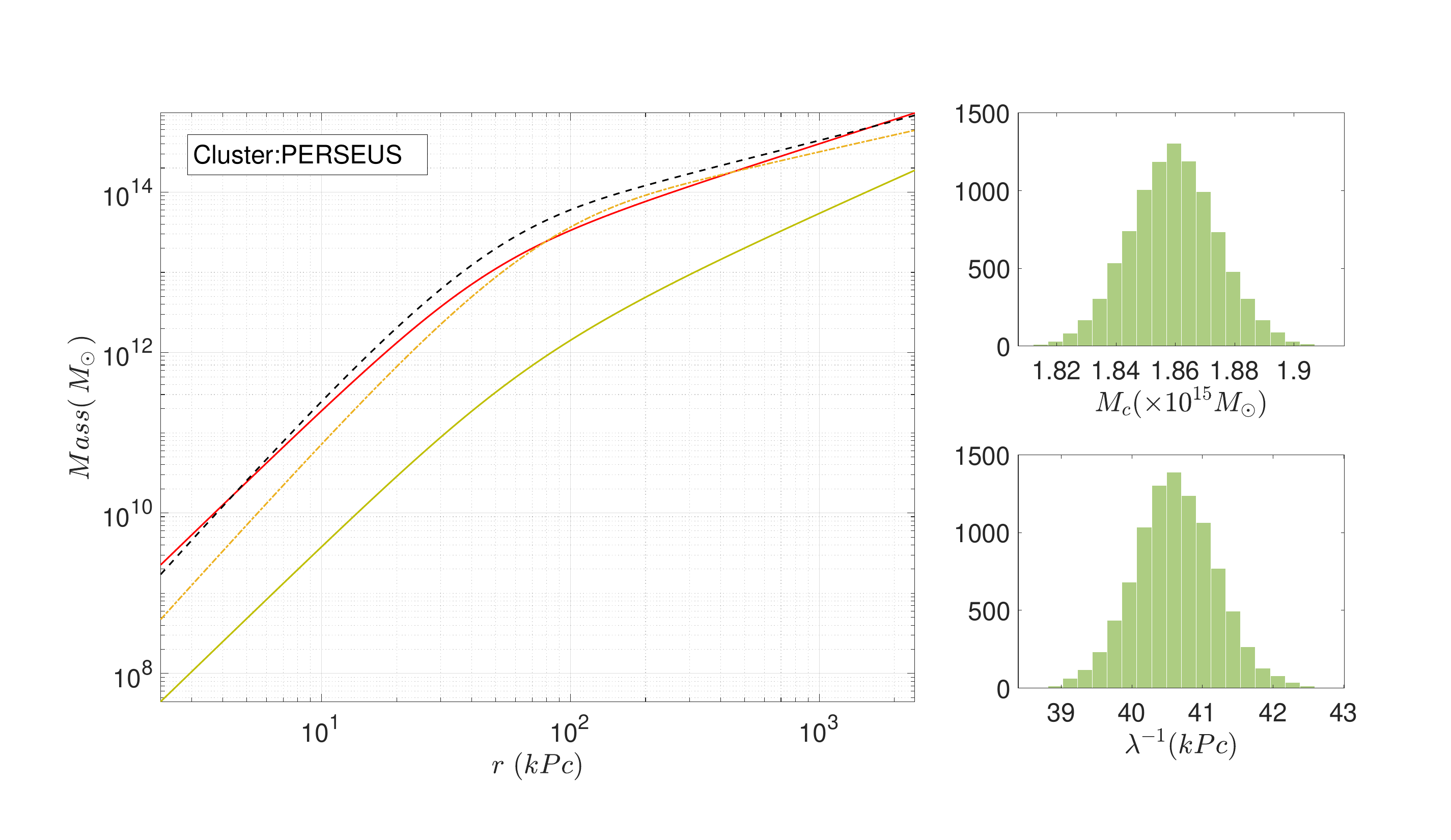}
    \includegraphics[trim=2.0cm 2.0cm 3.0cm 3.0cm, clip=true, width=0.32\columnwidth]{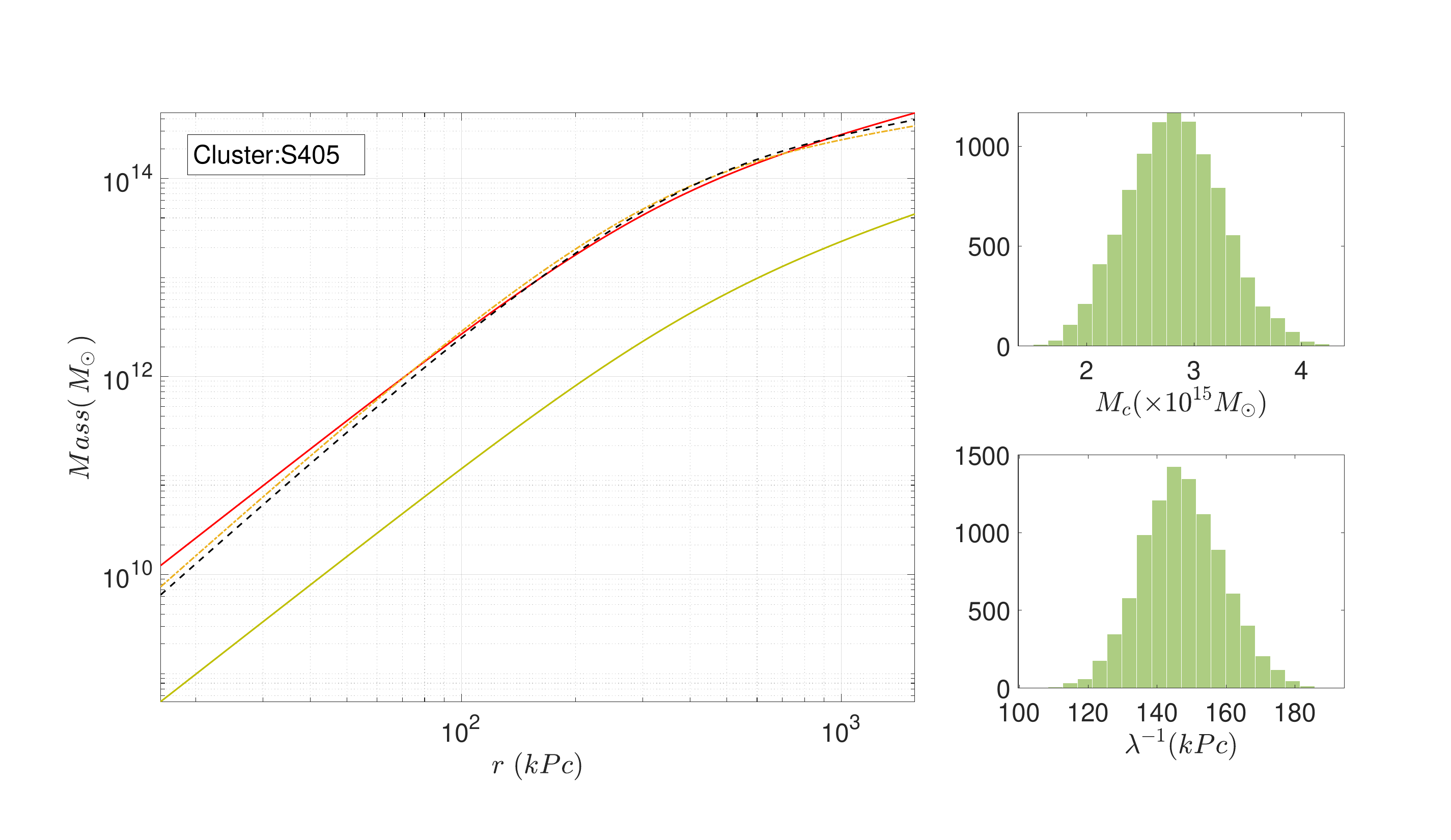}
    \includegraphics[trim=2.0cm 2.0cm 3.0cm 3.0cm, clip=true, width=0.32\columnwidth]{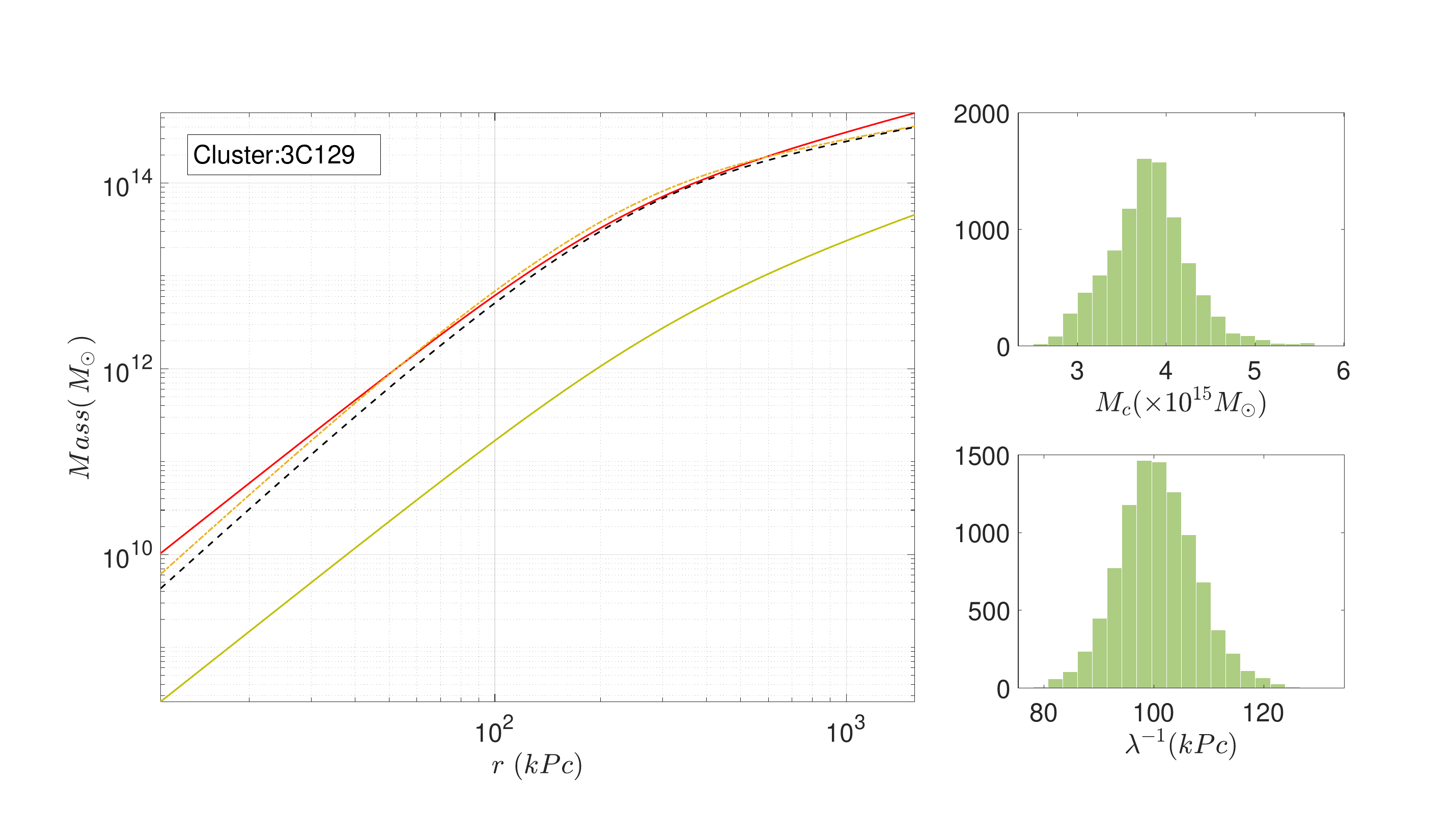}
    \includegraphics[trim=2.0cm 2.0cm 3.0cm 3.0cm, clip=true, width=0.32\columnwidth]{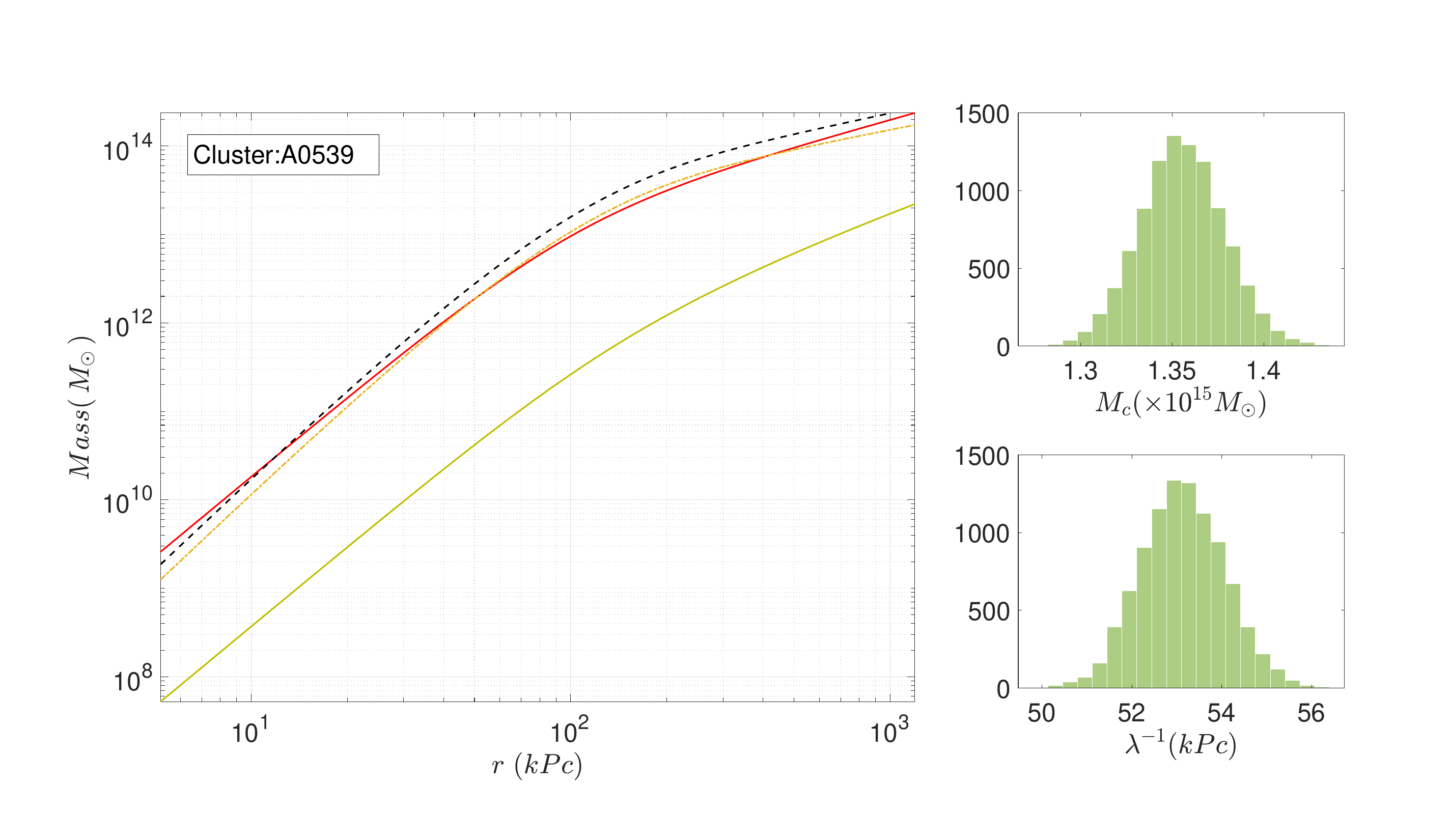}
    \includegraphics[trim=2.0cm 2.0cm 3.0cm 3.0cm, clip=true, width=0.32\columnwidth]{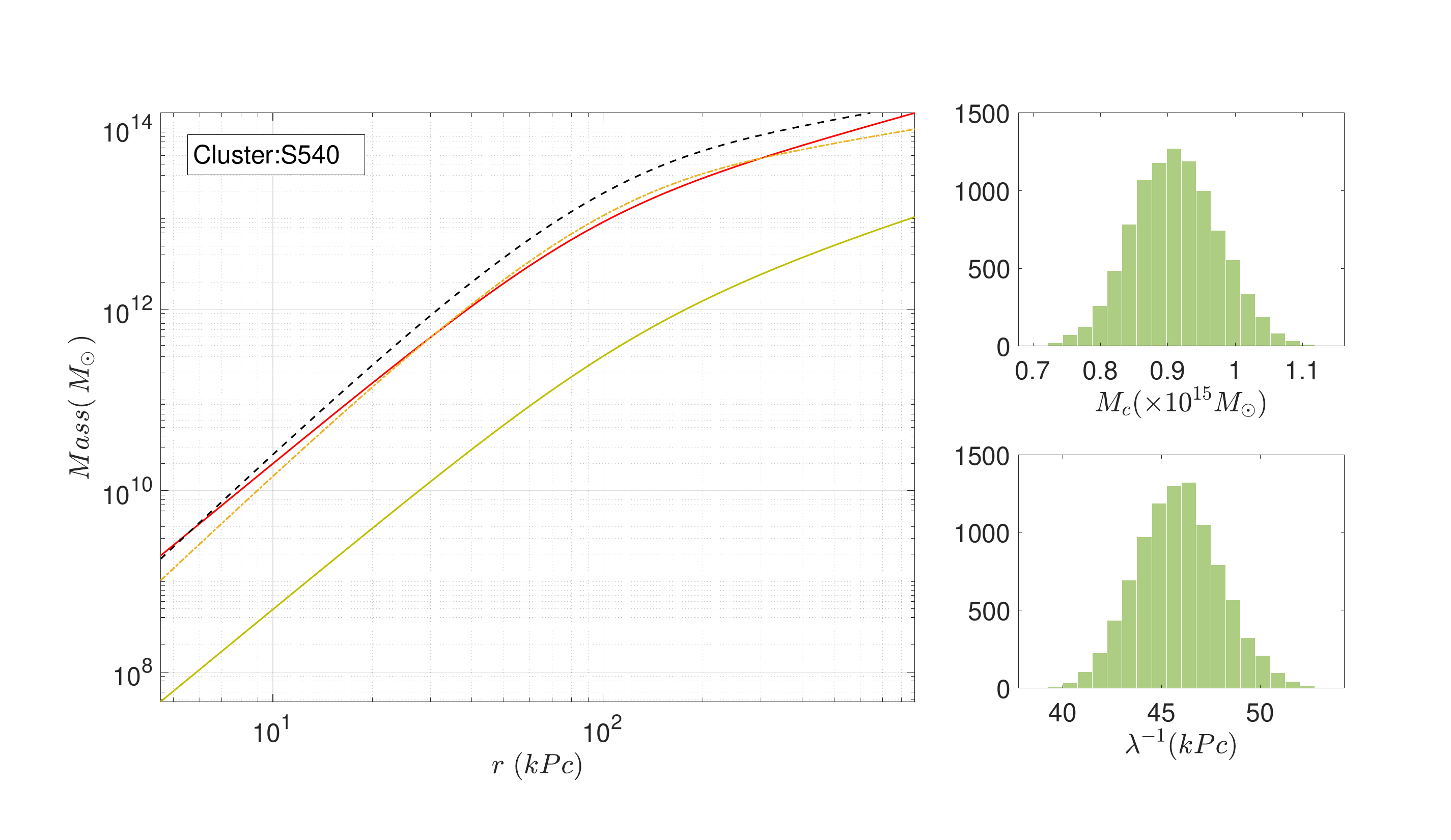}
    \end{figure}
    \begin{figure}
    \centering
    \includegraphics[trim=2.0cm 2.0cm 3.0cm 3.0cm, clip=true, width=0.32\columnwidth]{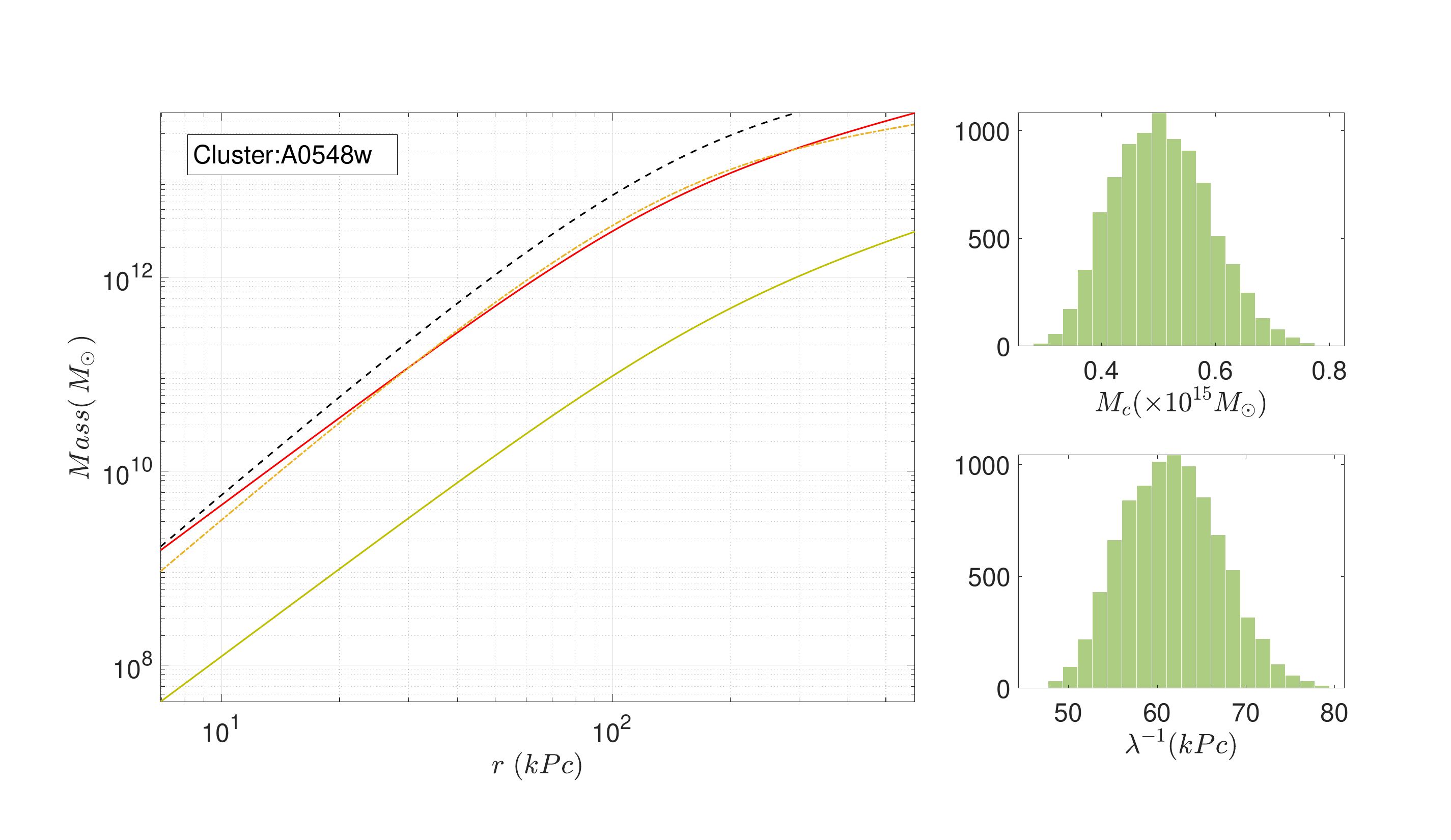}
    \includegraphics[trim=2.0cm 2.0cm 3.0cm 3.0cm, clip=true, width=0.32\columnwidth]{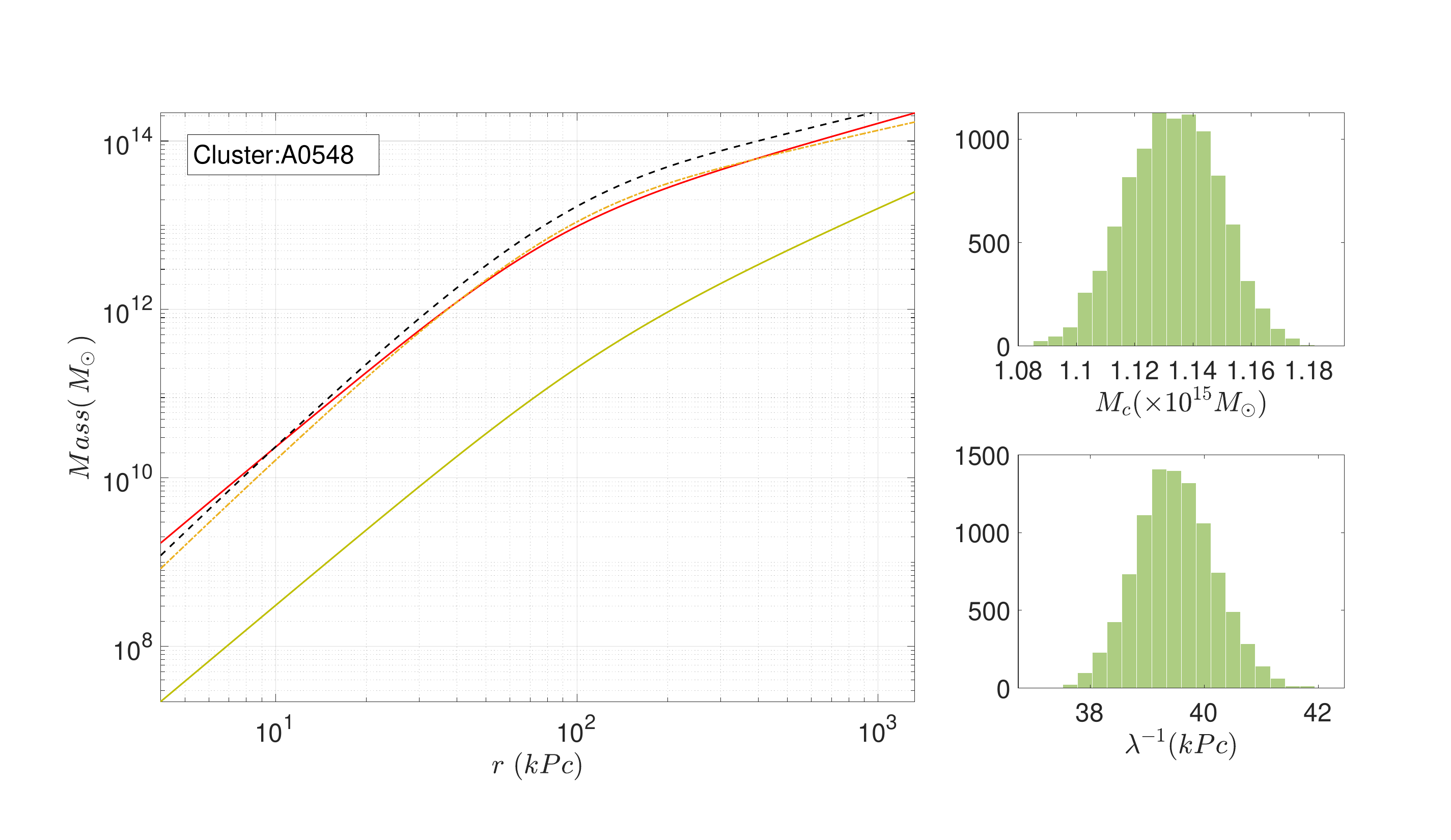}
    \includegraphics[trim=2.0cm 2.0cm 3.0cm 3.0cm, clip=true, width=0.32\columnwidth]{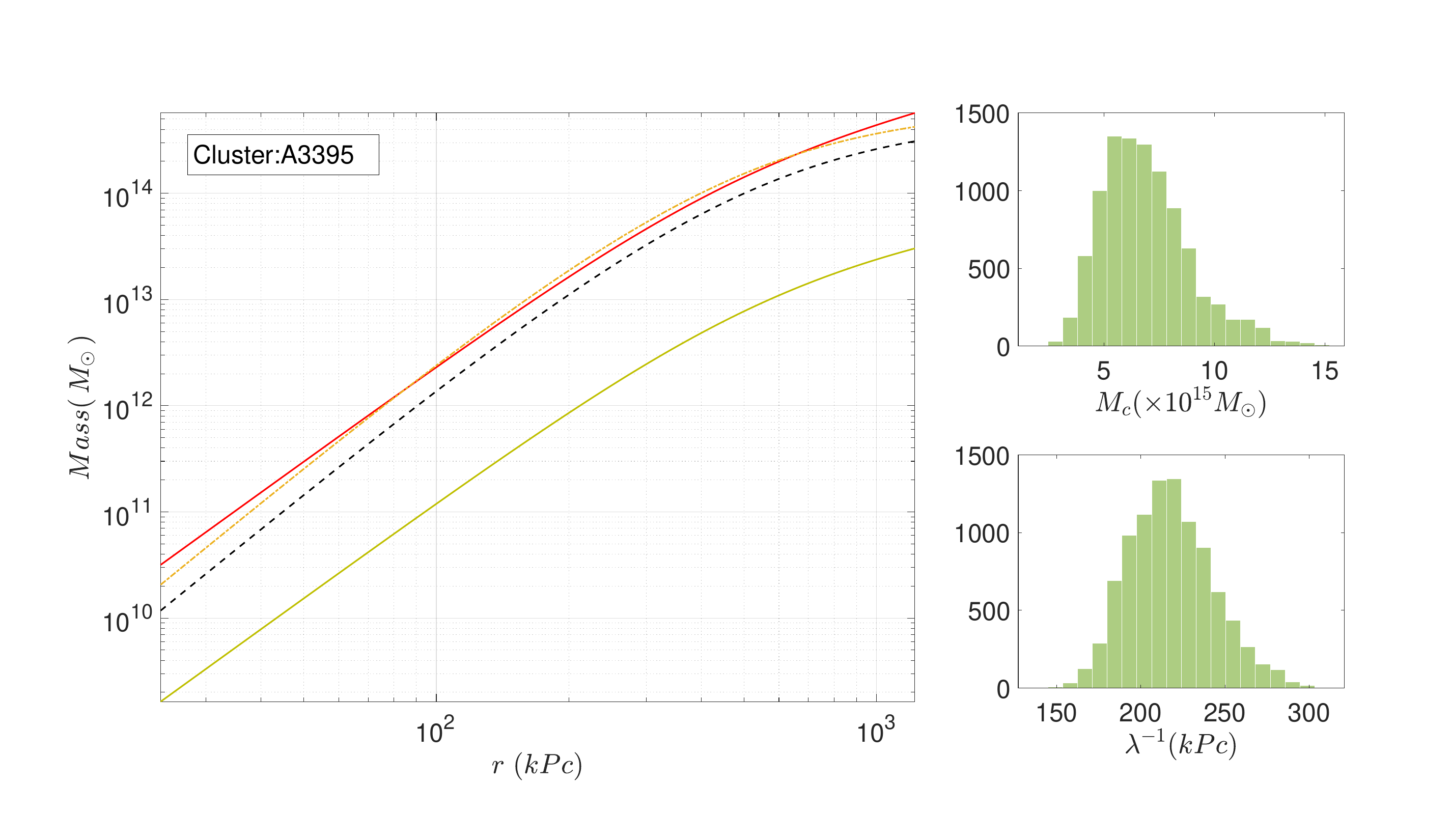}
    \includegraphics[trim=2.0cm 2.0cm 3.0cm 3.0cm, clip=true, width=0.32\columnwidth]{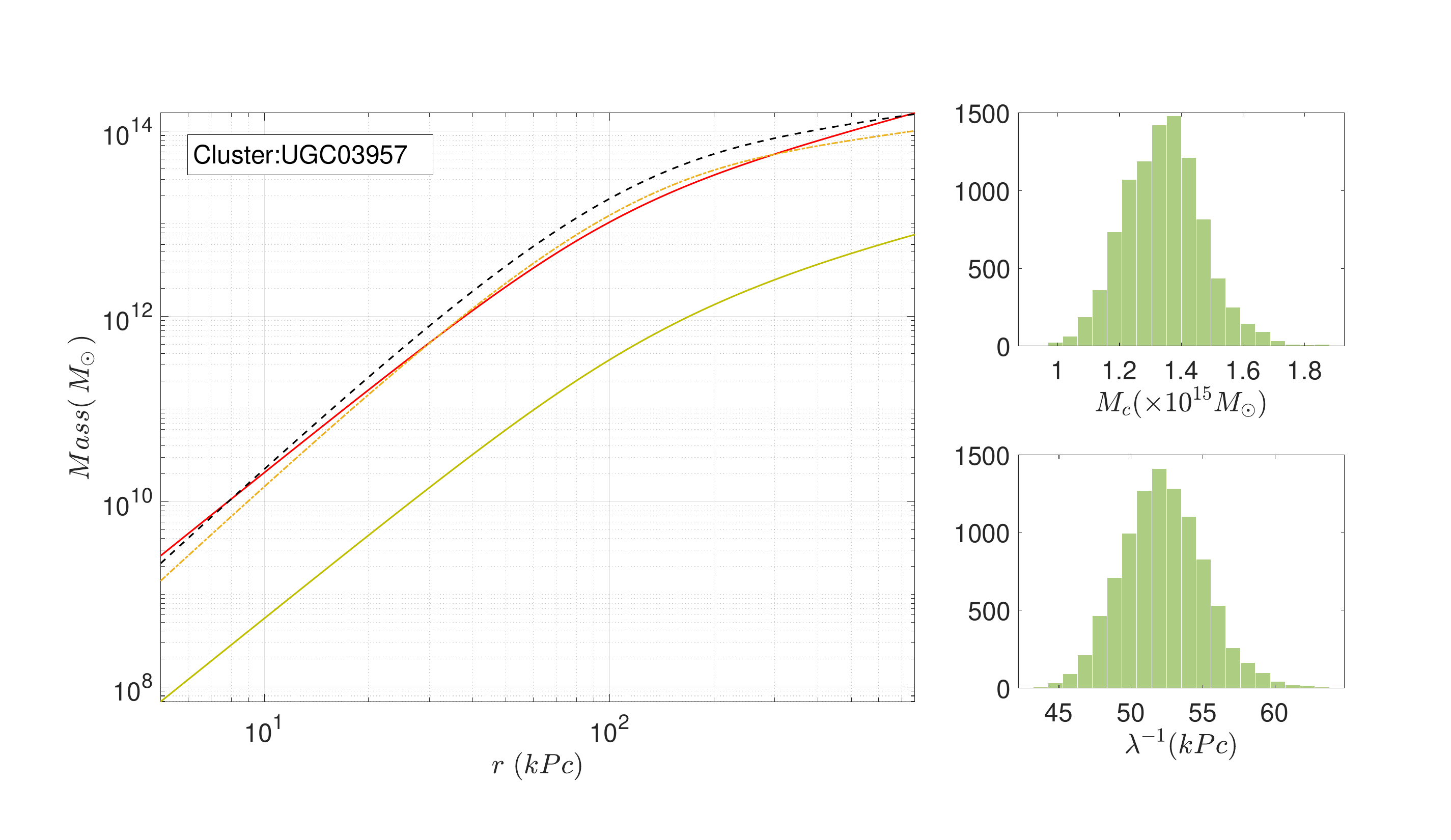}
    \includegraphics[trim=2.0cm 2.0cm 3.0cm 3.0cm, clip=true, width=0.32\columnwidth]{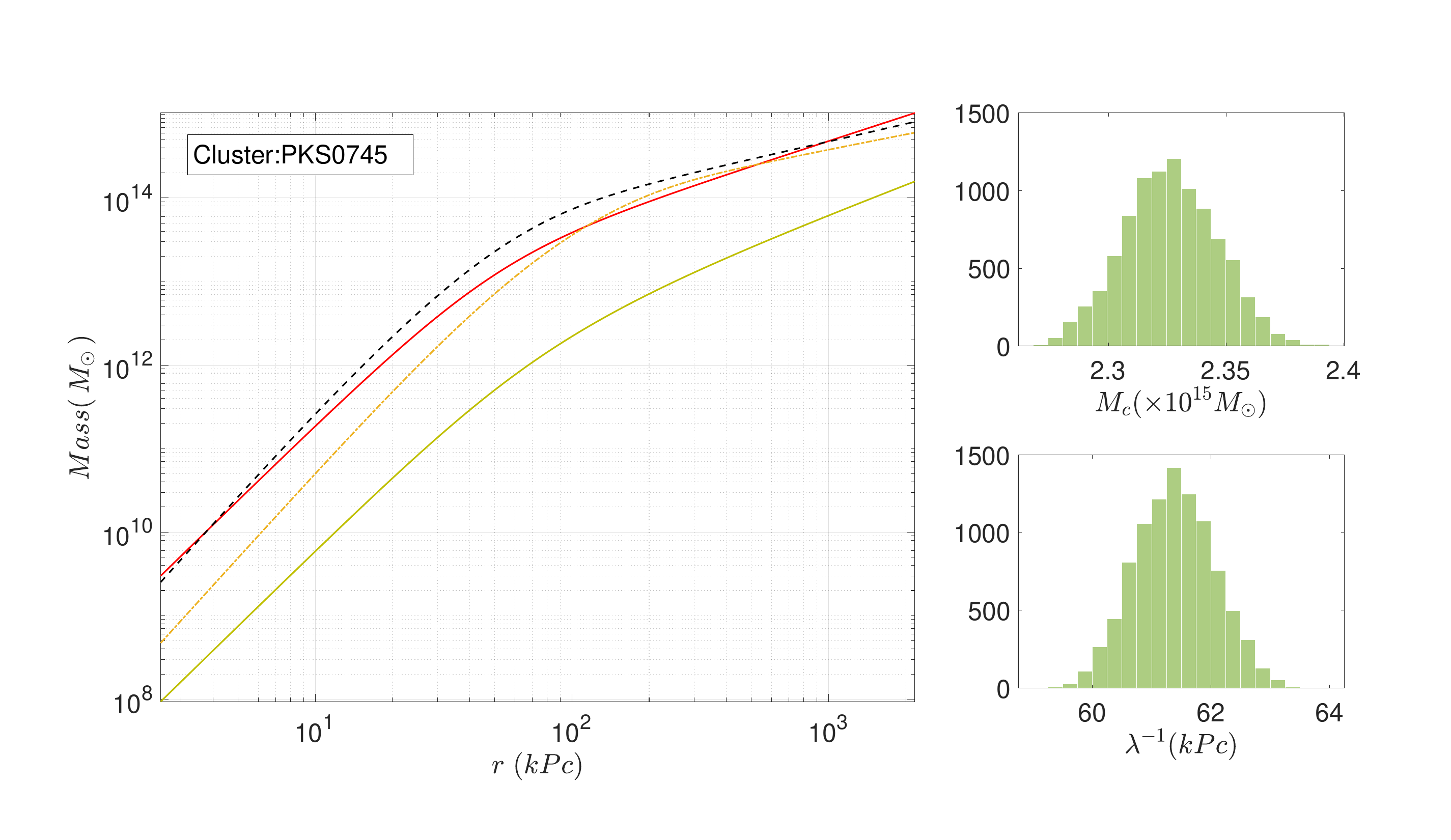}
    \includegraphics[trim=2.0cm 2.0cm 3.0cm 3.0cm, clip=true, width=0.32\columnwidth]{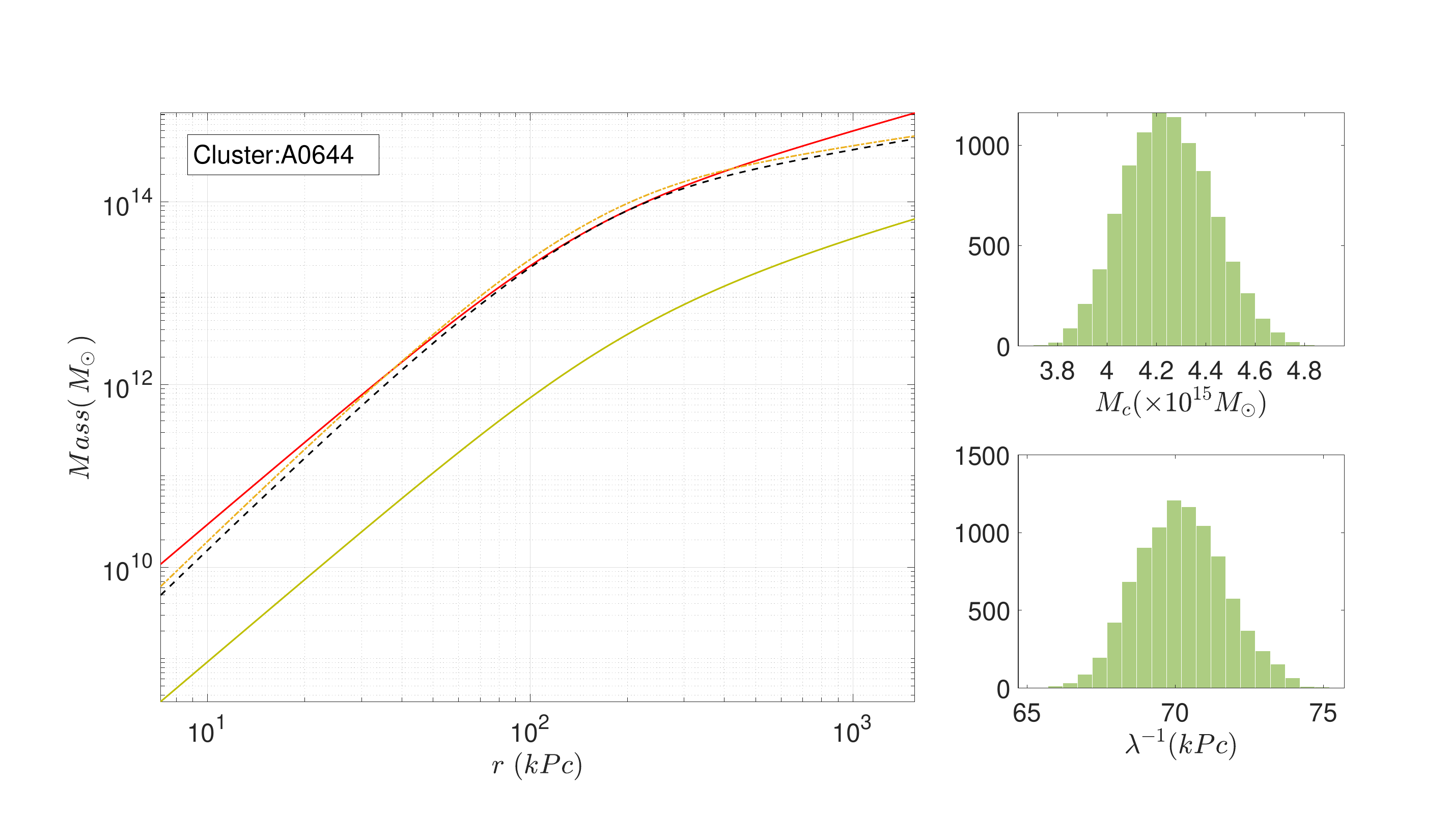}
    \includegraphics[trim=2.0cm 2.0cm 3.0cm 3.0cm, clip=true, width=0.32\columnwidth]{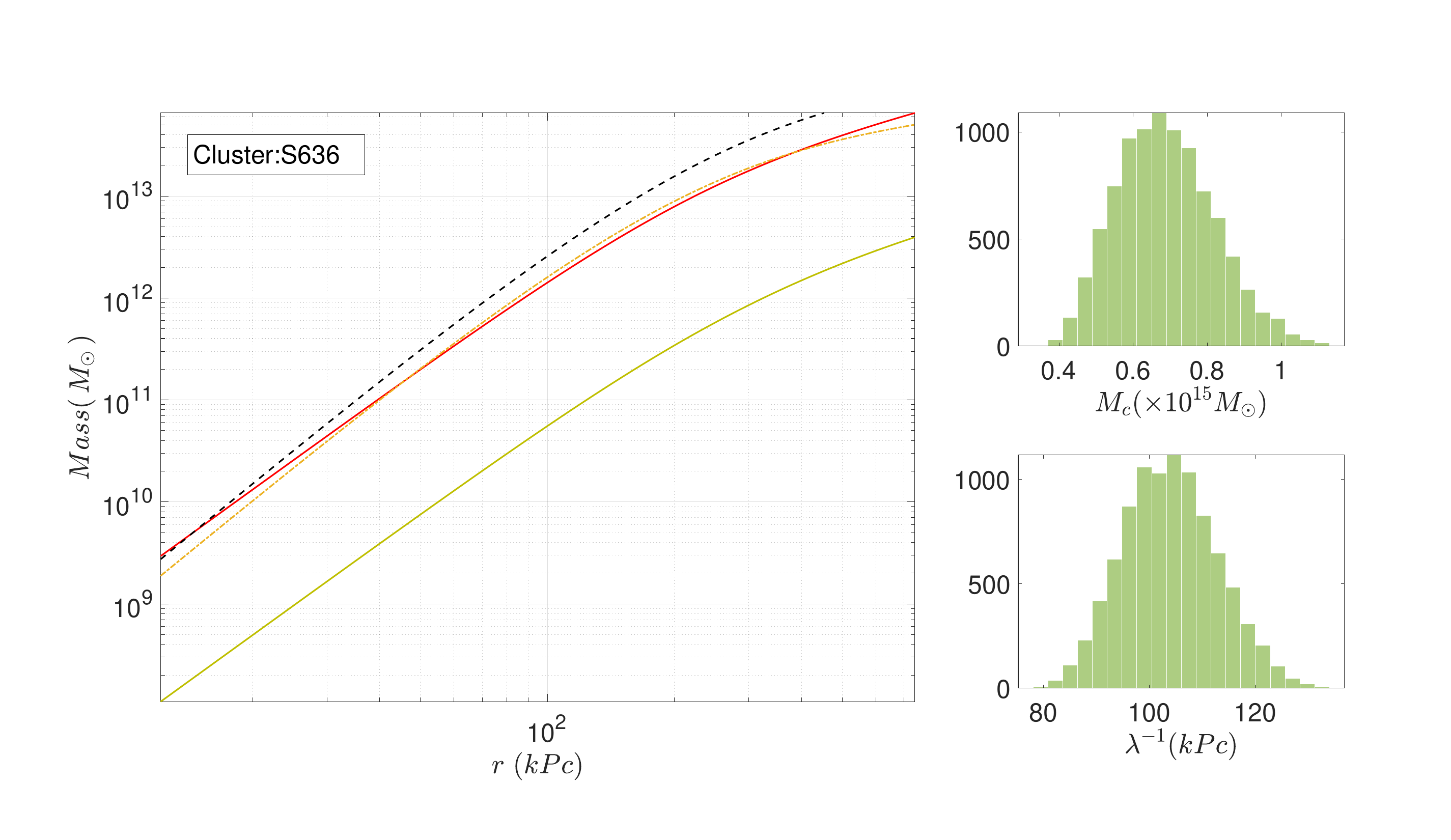}
    \includegraphics[trim=2.0cm 2.0cm 3.0cm 3.0cm, clip=true, width=0.32\columnwidth]{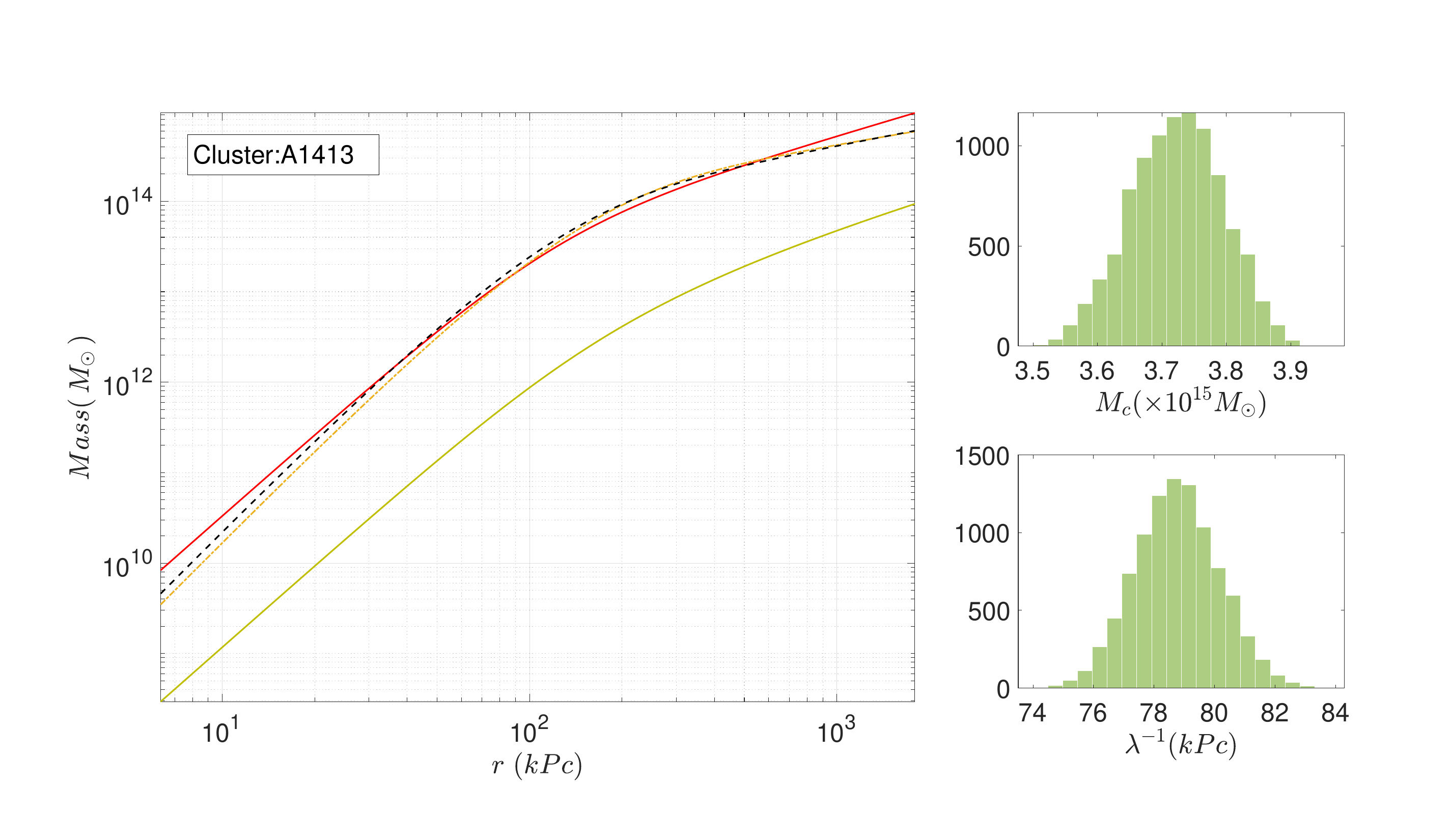}
    \includegraphics[trim=2.0cm 2.0cm 3.0cm 3.0cm, clip=true, width=0.32\columnwidth]{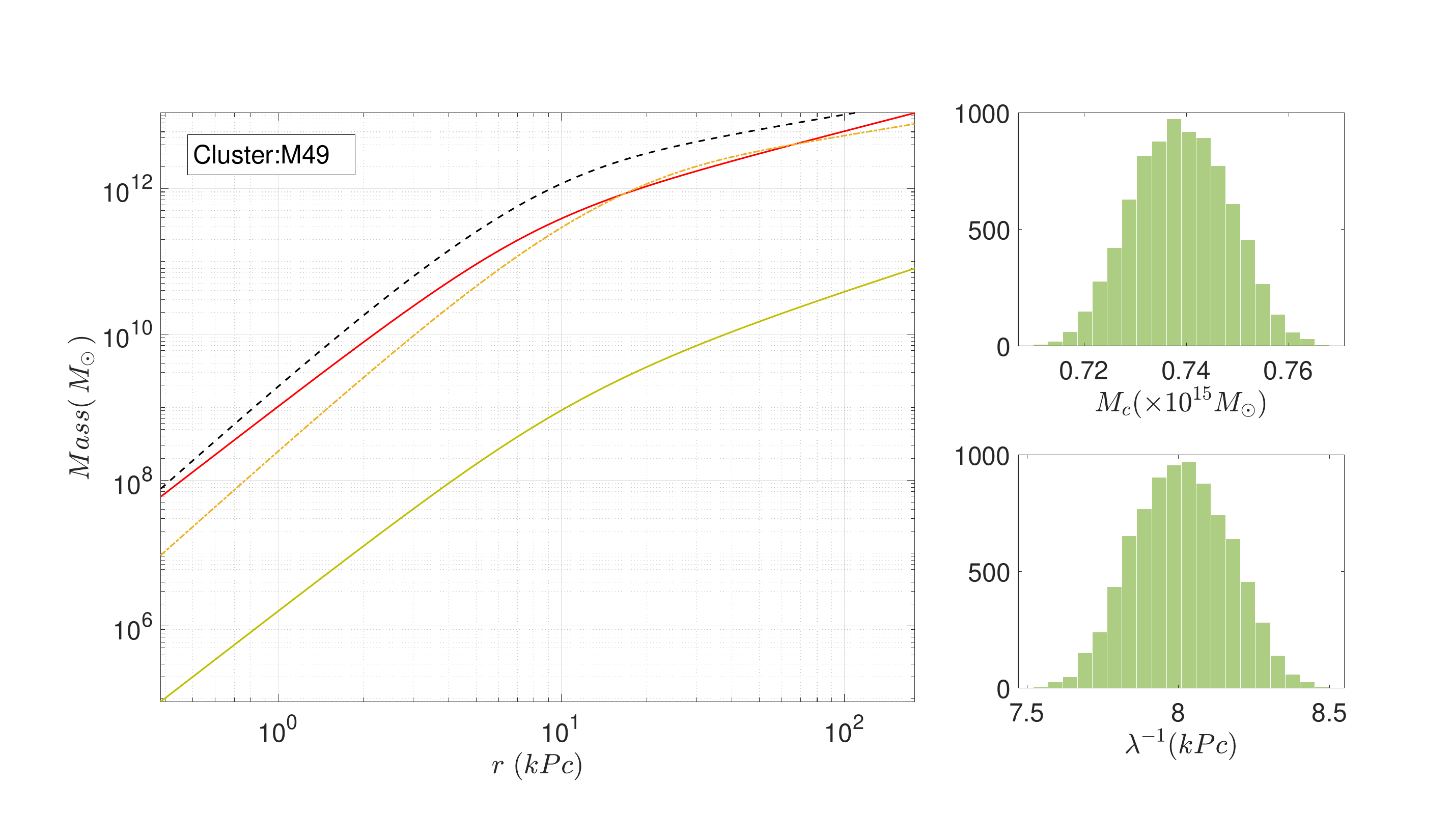}
    \includegraphics[trim=2.0cm 2.0cm 3.0cm 3.0cm, clip=true, width=0.32\columnwidth]{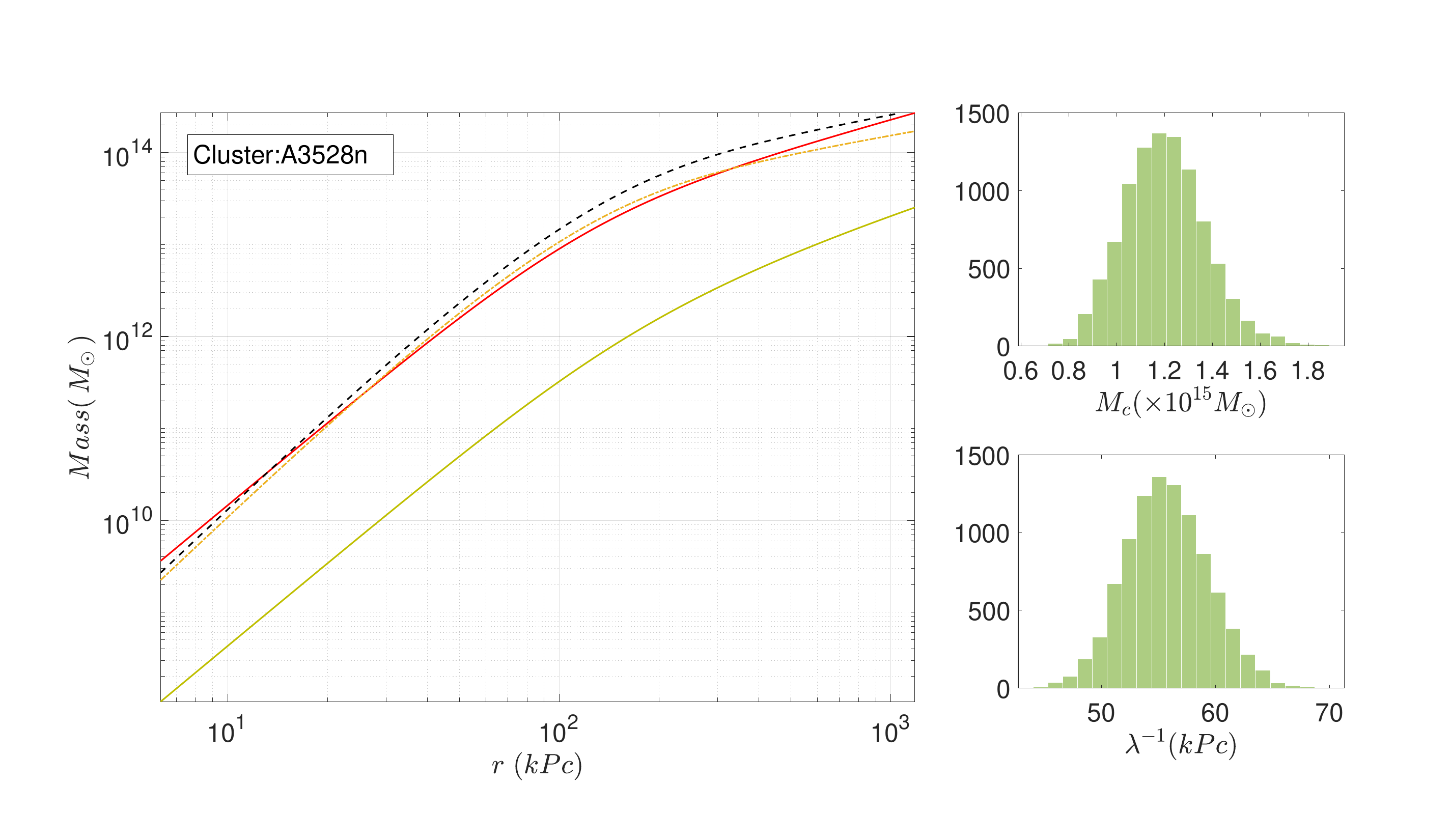}
    \includegraphics[trim=2.0cm 2.0cm 3.0cm 3.0cm, clip=true, width=0.32\columnwidth]{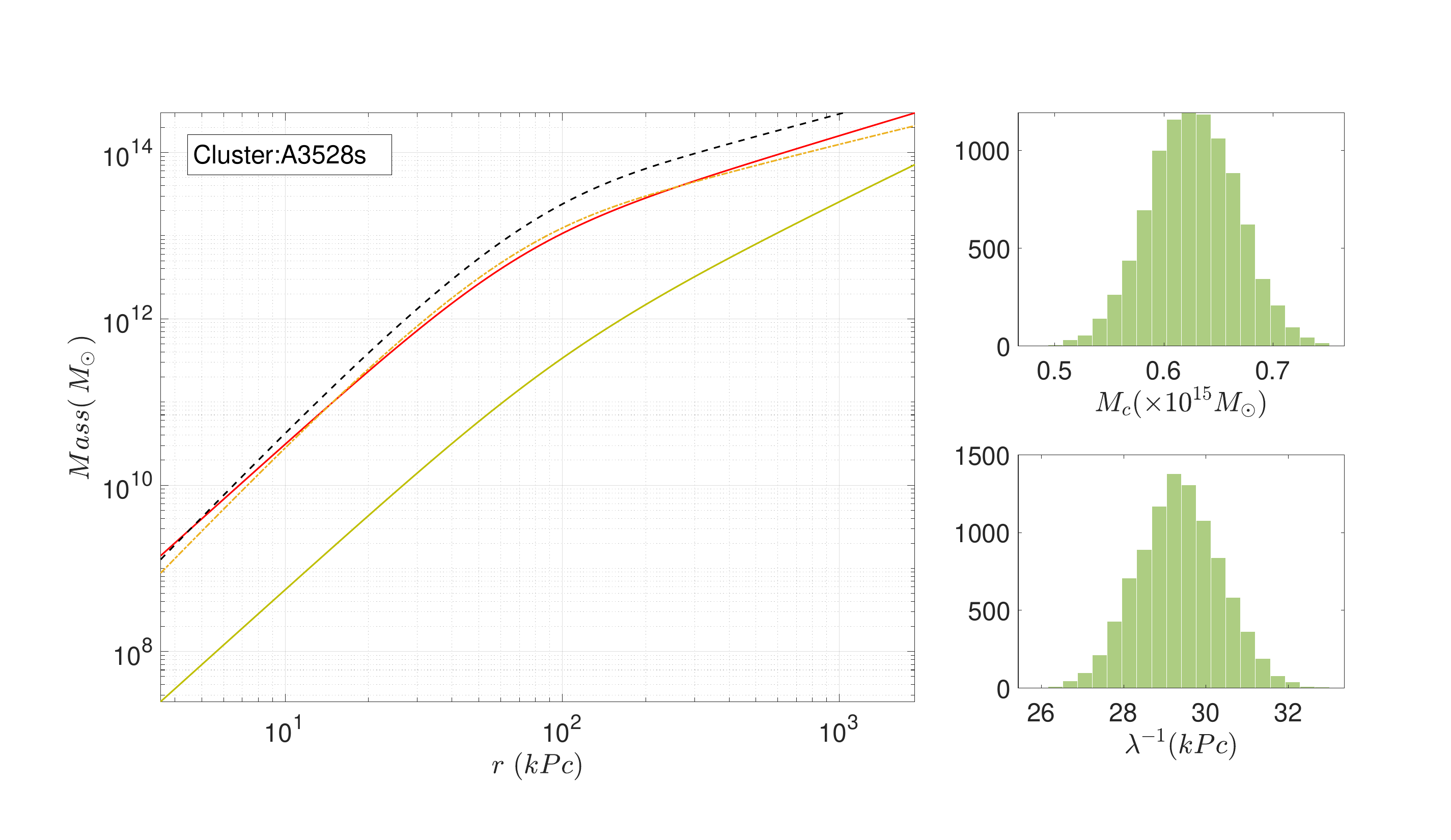}
    \includegraphics[trim=2.0cm 2.0cm 3.0cm 3.0cm, clip=true, width=0.32\columnwidth]{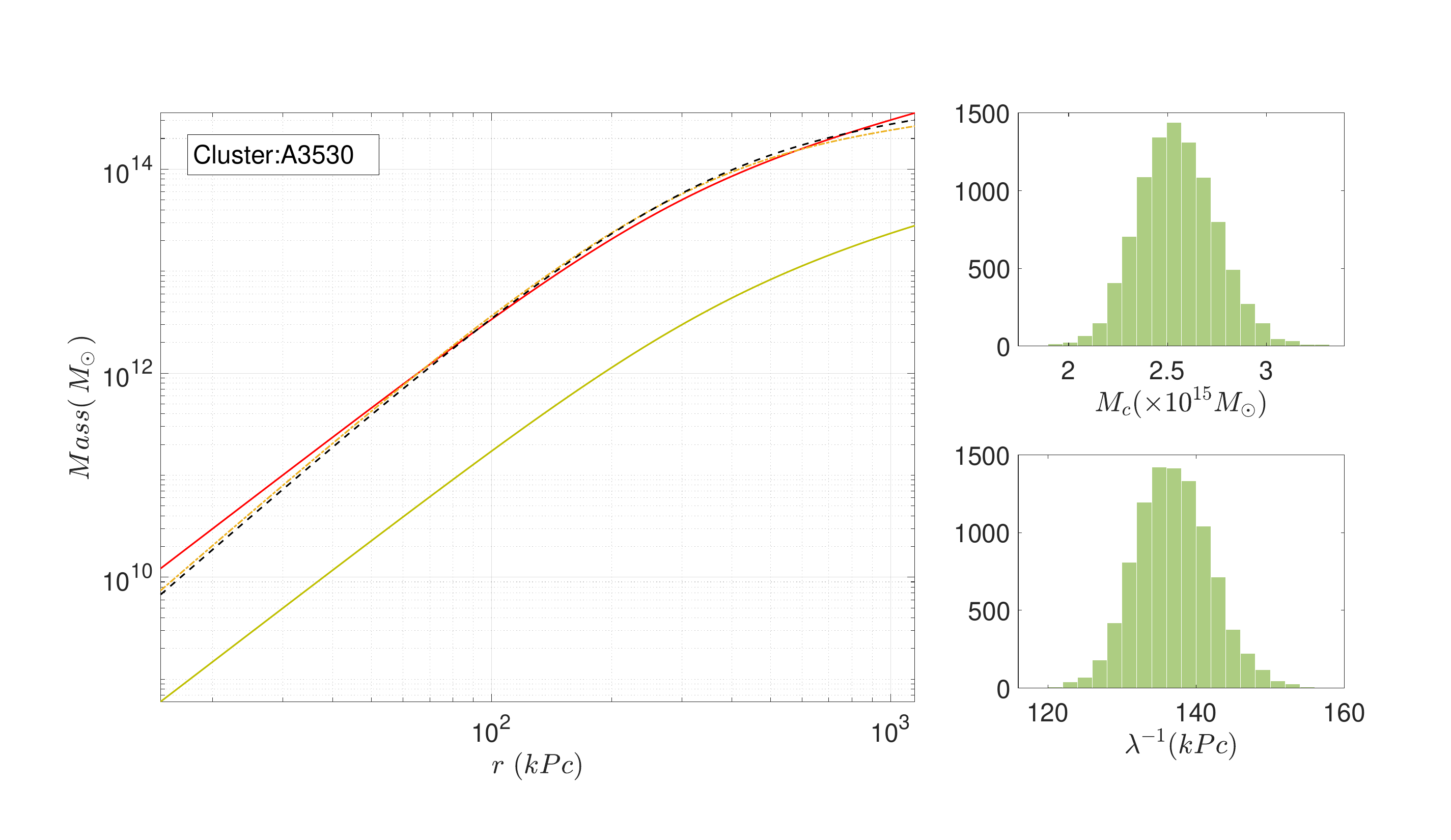}
    \includegraphics[trim=2.0cm 2.0cm 3.0cm 3.0cm, clip=true, width=0.32\columnwidth]{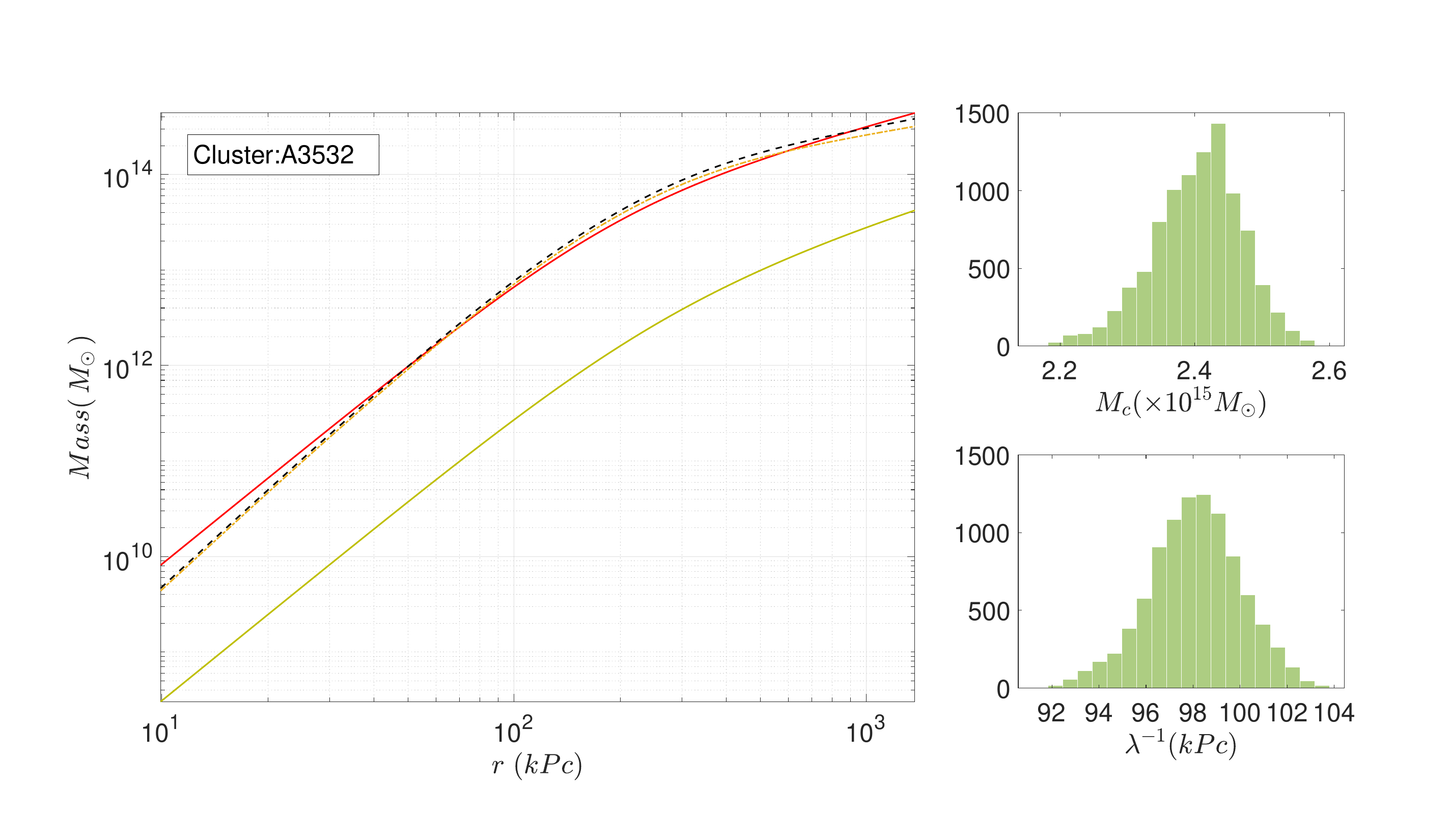}
    \includegraphics[trim=2.0cm 2.0cm 3.0cm 3.0cm, clip=true, width=0.32\columnwidth]{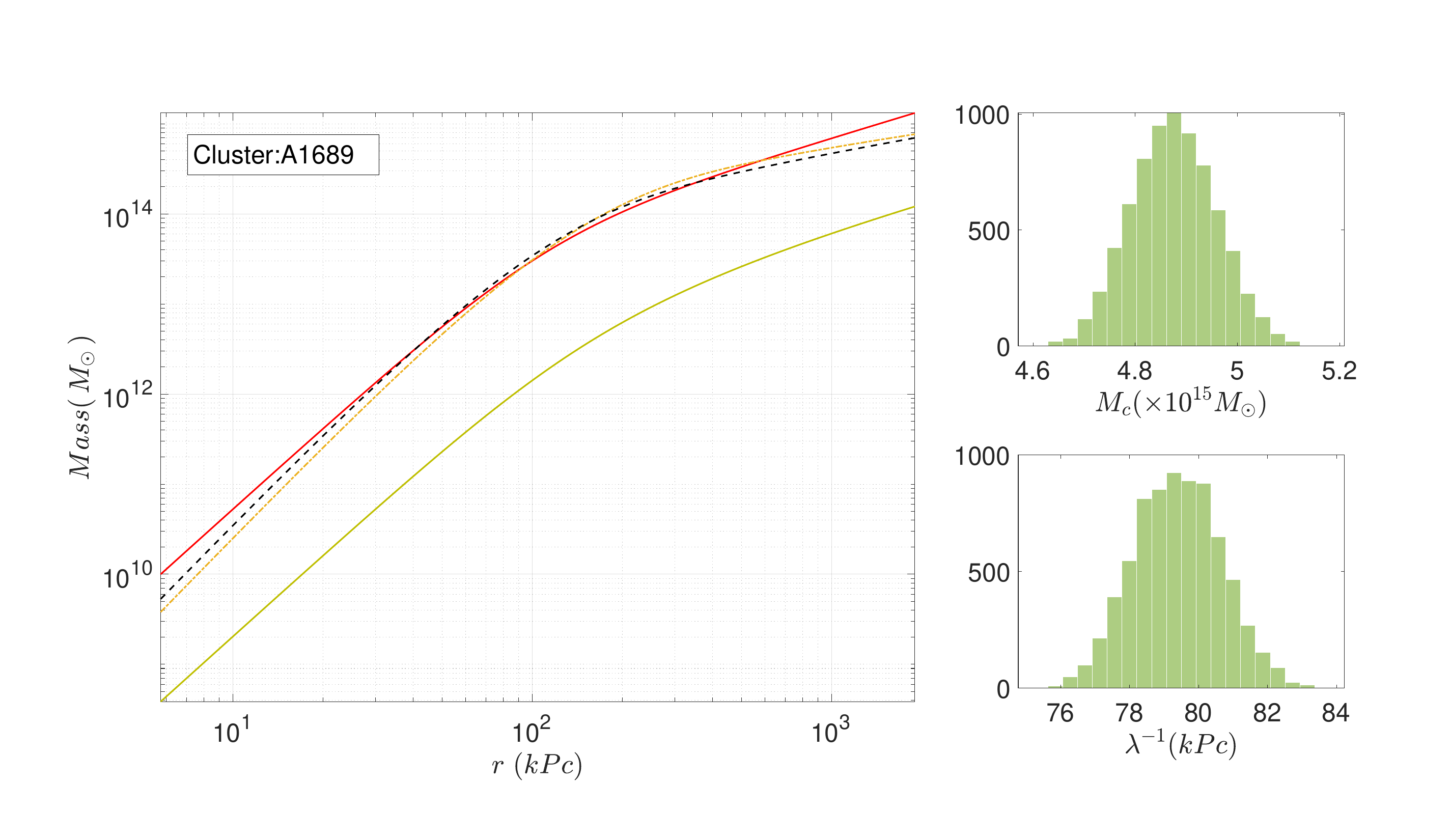}
    \includegraphics[trim=2.0cm 2.0cm 3.0cm 3.0cm, clip=true, width=0.32\columnwidth]{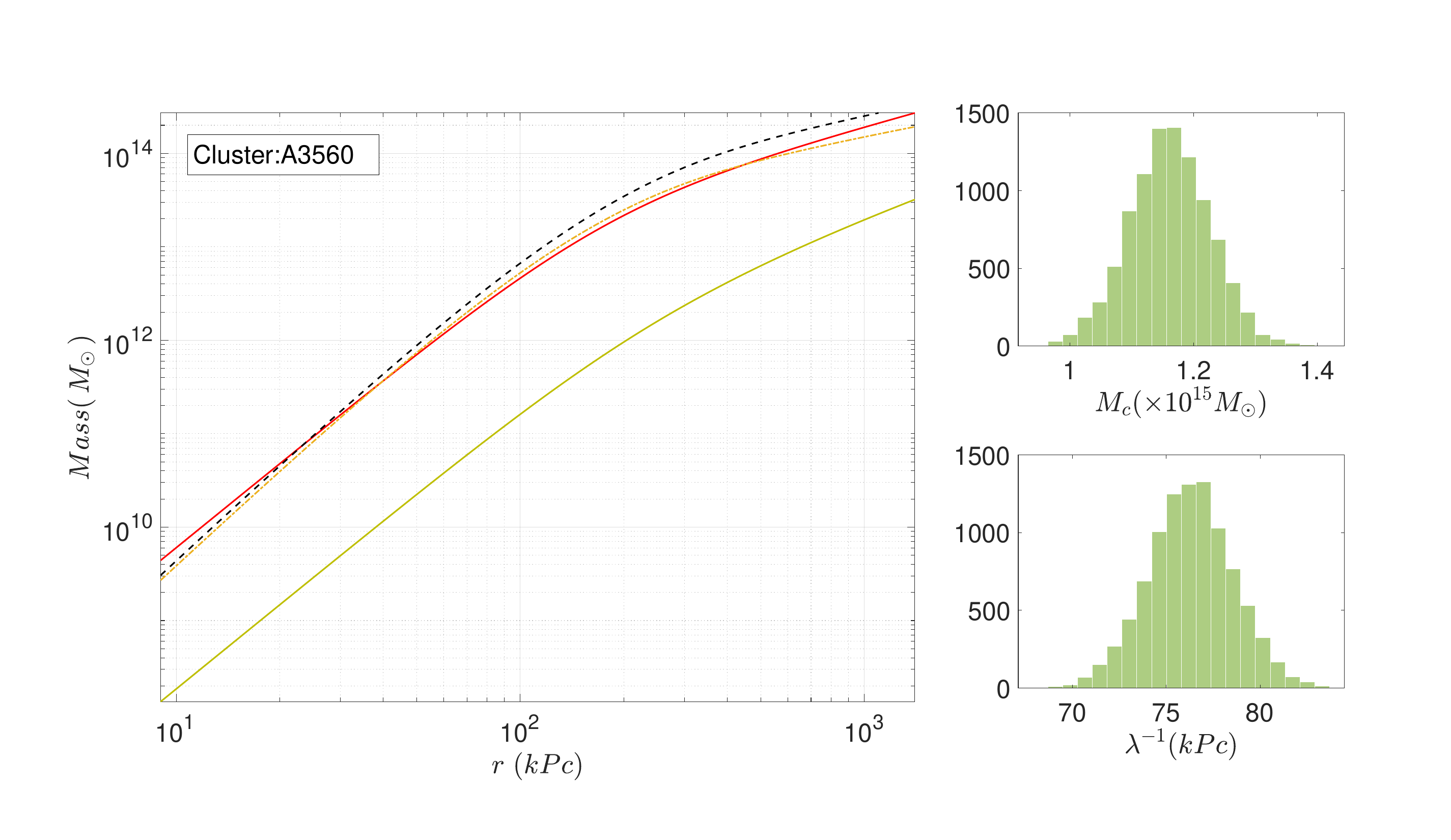}
    \includegraphics[trim=2.0cm 2.0cm 3.0cm 3.0cm, clip=true, width=0.32\columnwidth]{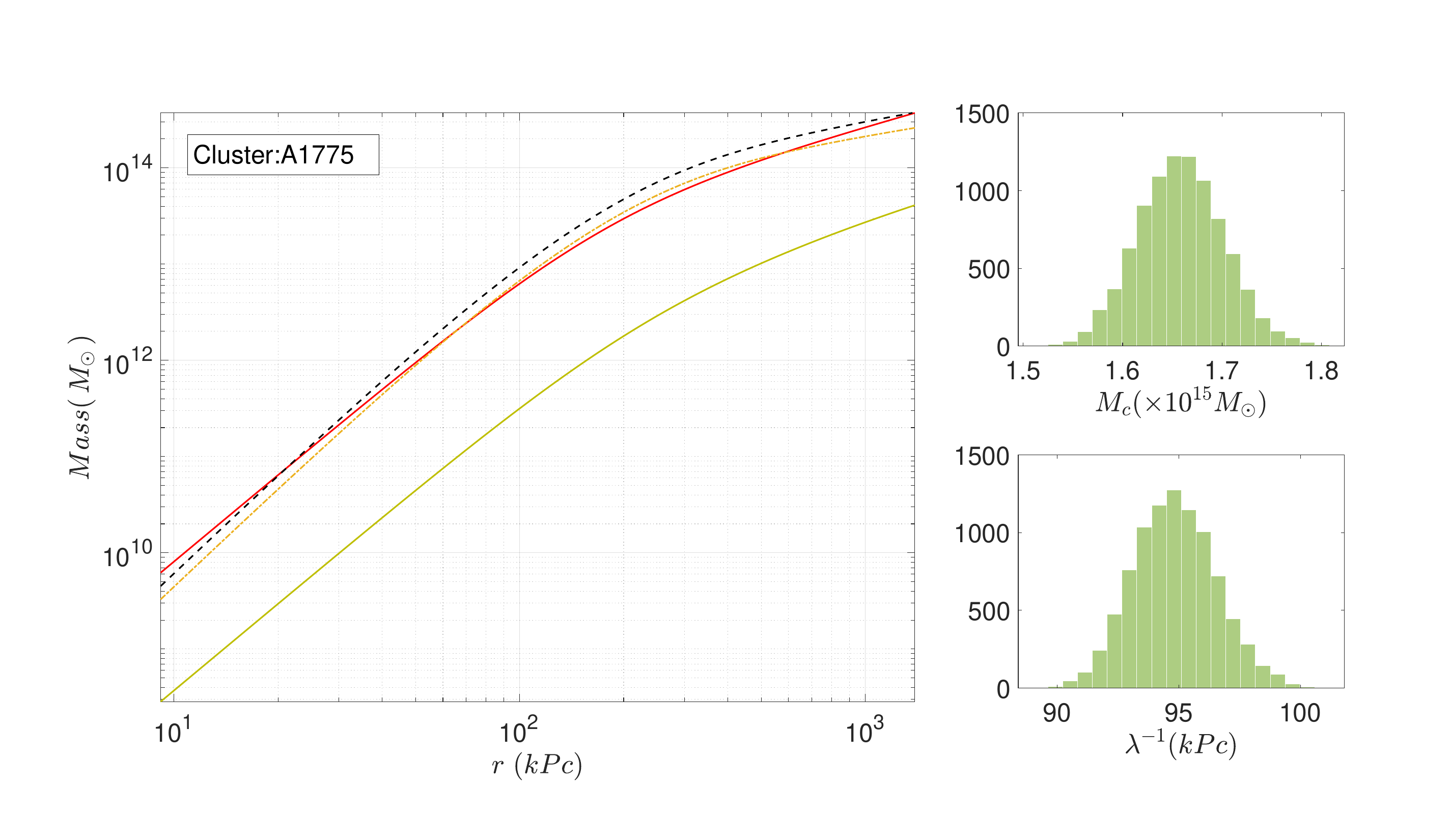}
    \includegraphics[trim=2.0cm 2.0cm 3.0cm 3.0cm, clip=true, width=0.32\columnwidth]{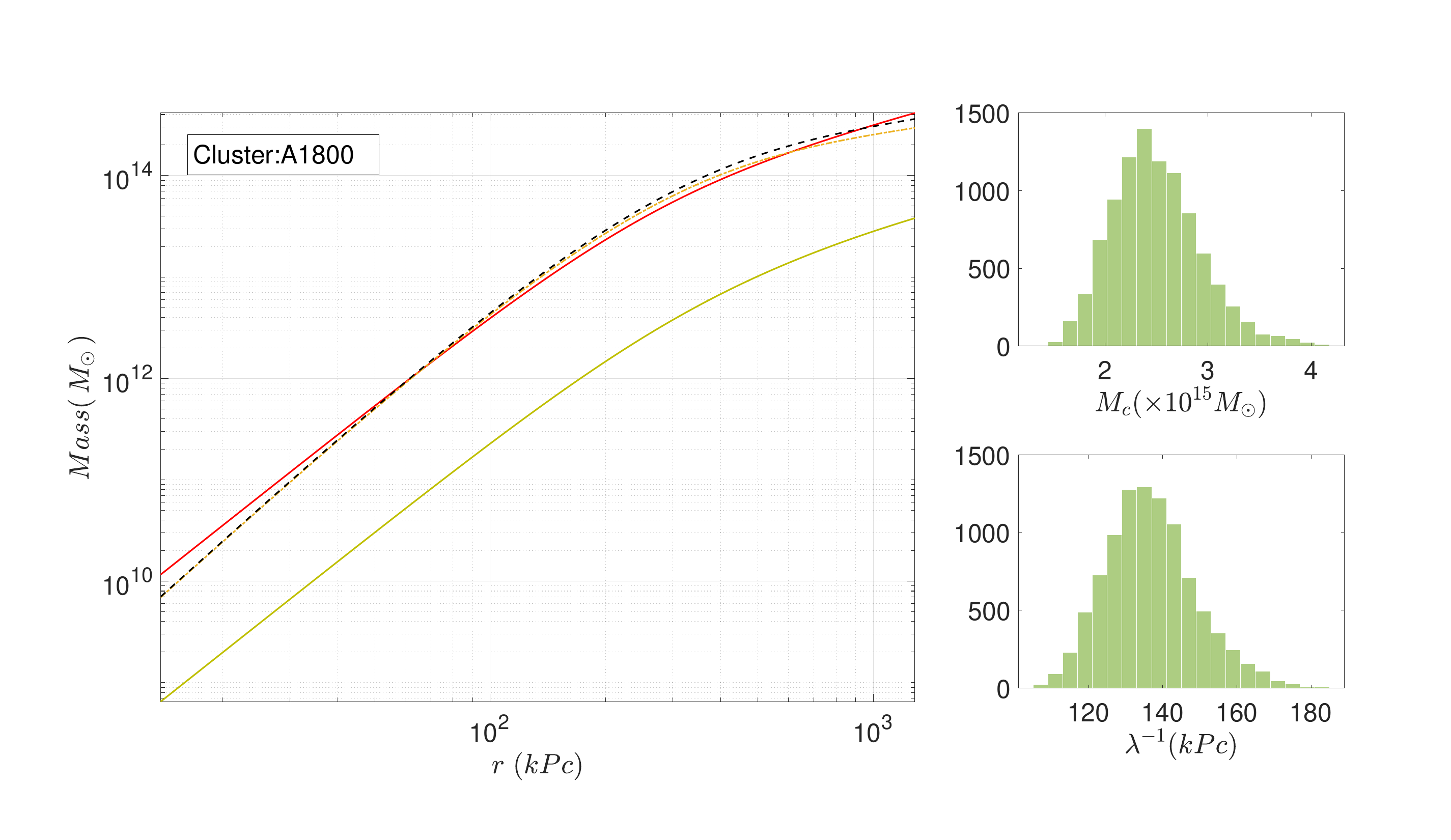}
    \includegraphics[trim=2.0cm 2.0cm 3.0cm 3.0cm, clip=true, width=0.32\columnwidth]{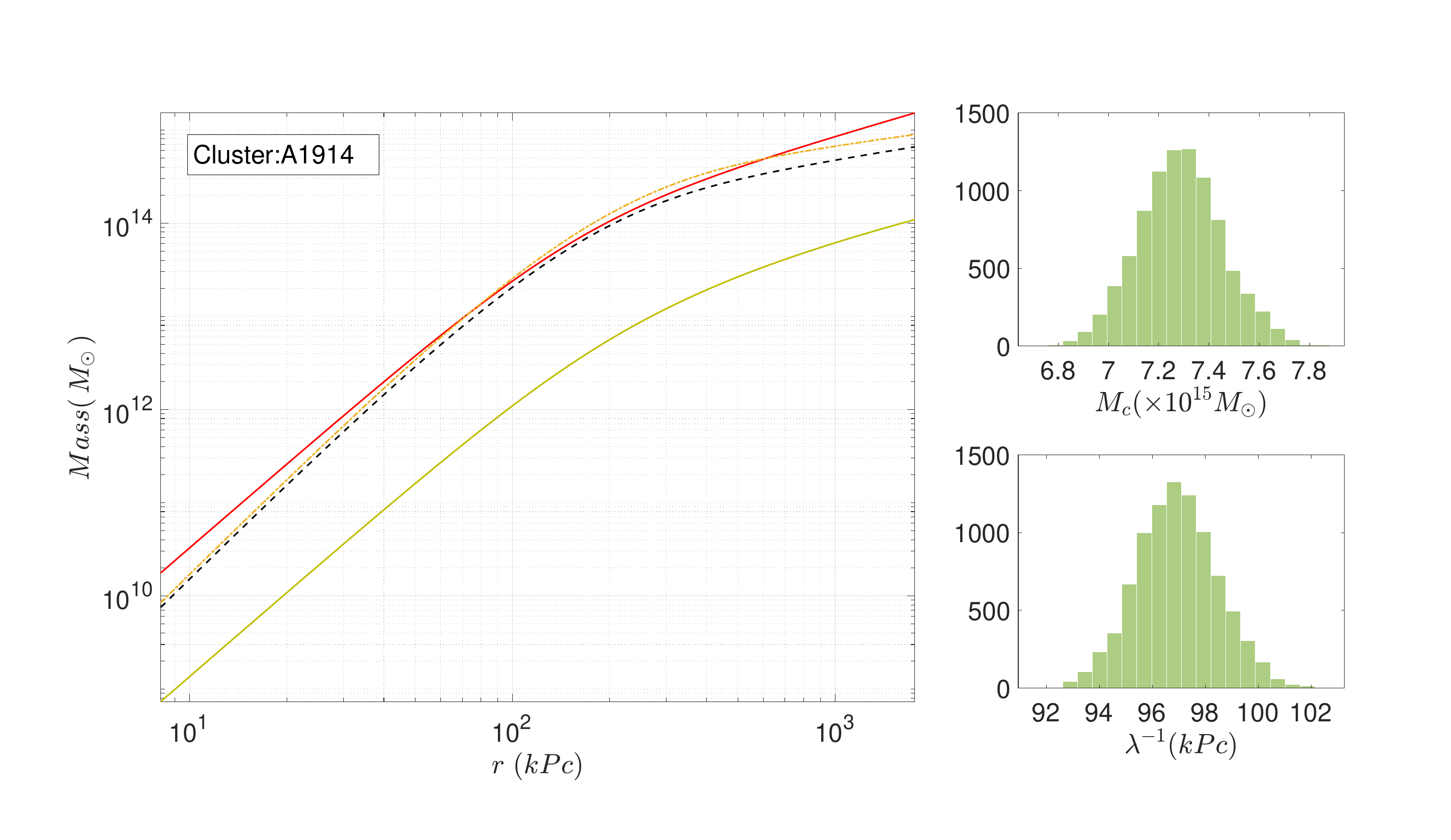}
    \includegraphics[trim=2.0cm 2.0cm 3.0cm 3.0cm, clip=true, width=0.32\columnwidth]{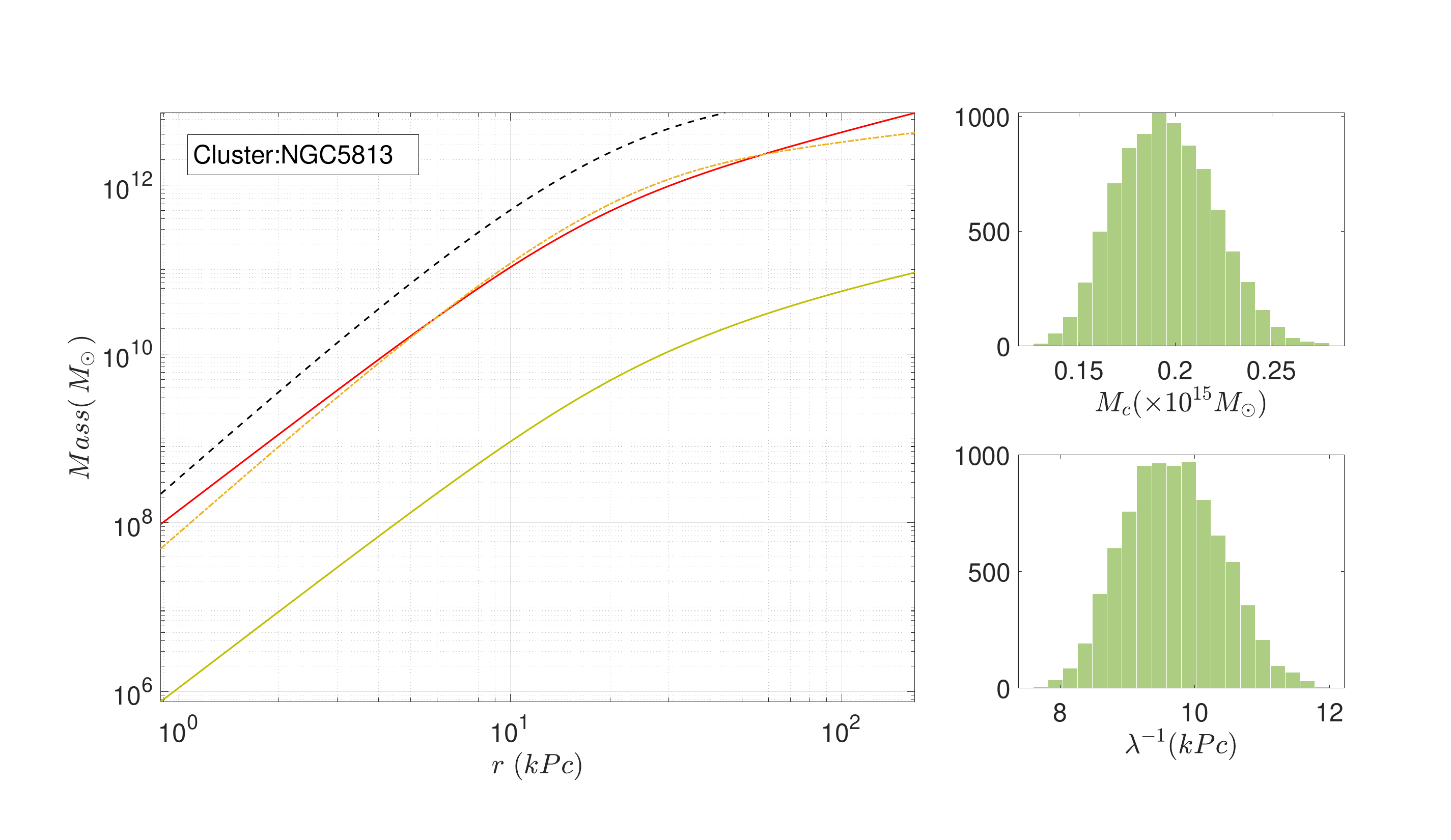}
    \includegraphics[trim=2.0cm 2.0cm 3.0cm 3.0cm, clip=true, width=0.32\columnwidth]{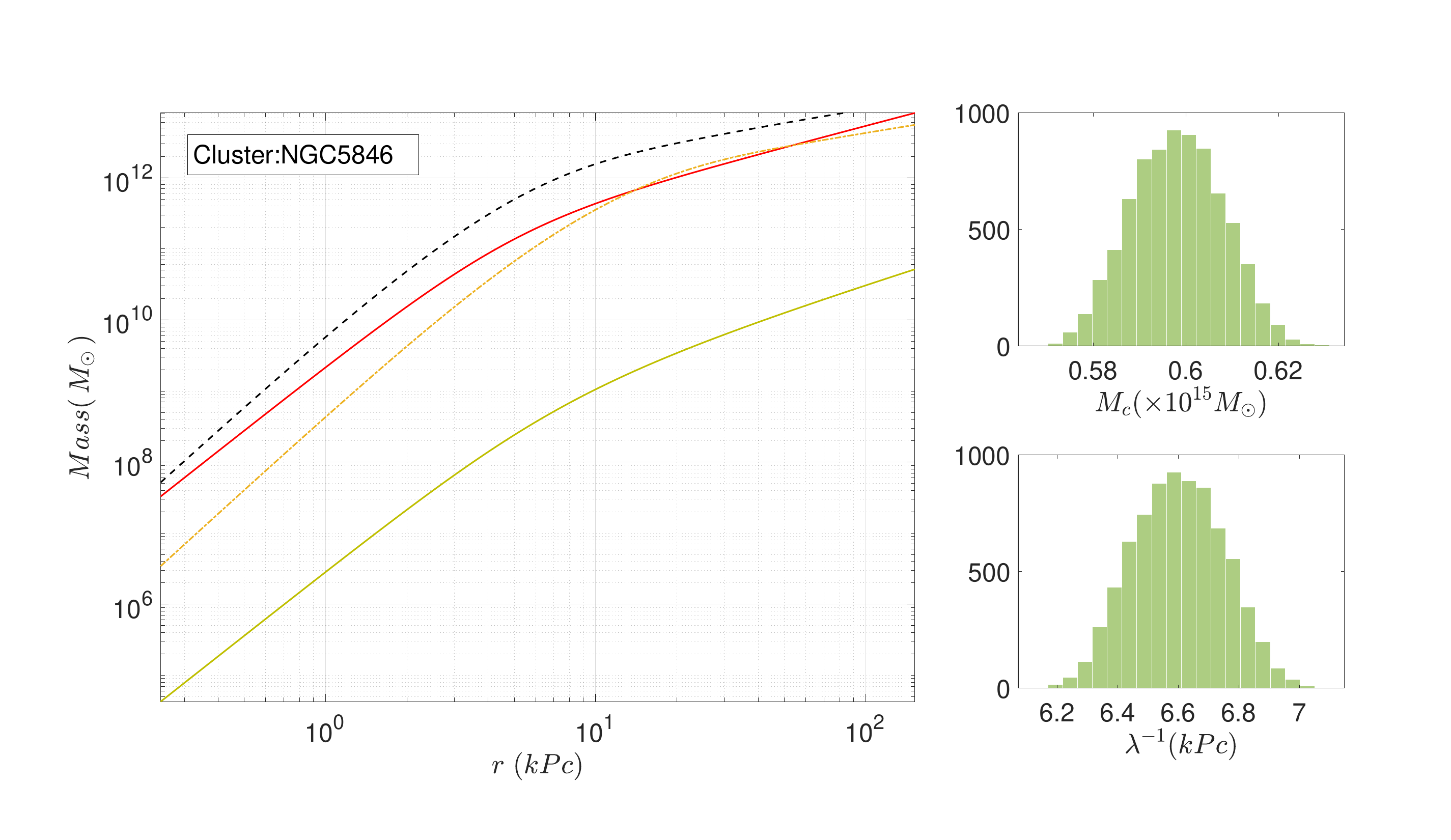}
    \includegraphics[trim=2.0cm 2.0cm 3.0cm 3.0cm, clip=true, width=0.32\columnwidth]{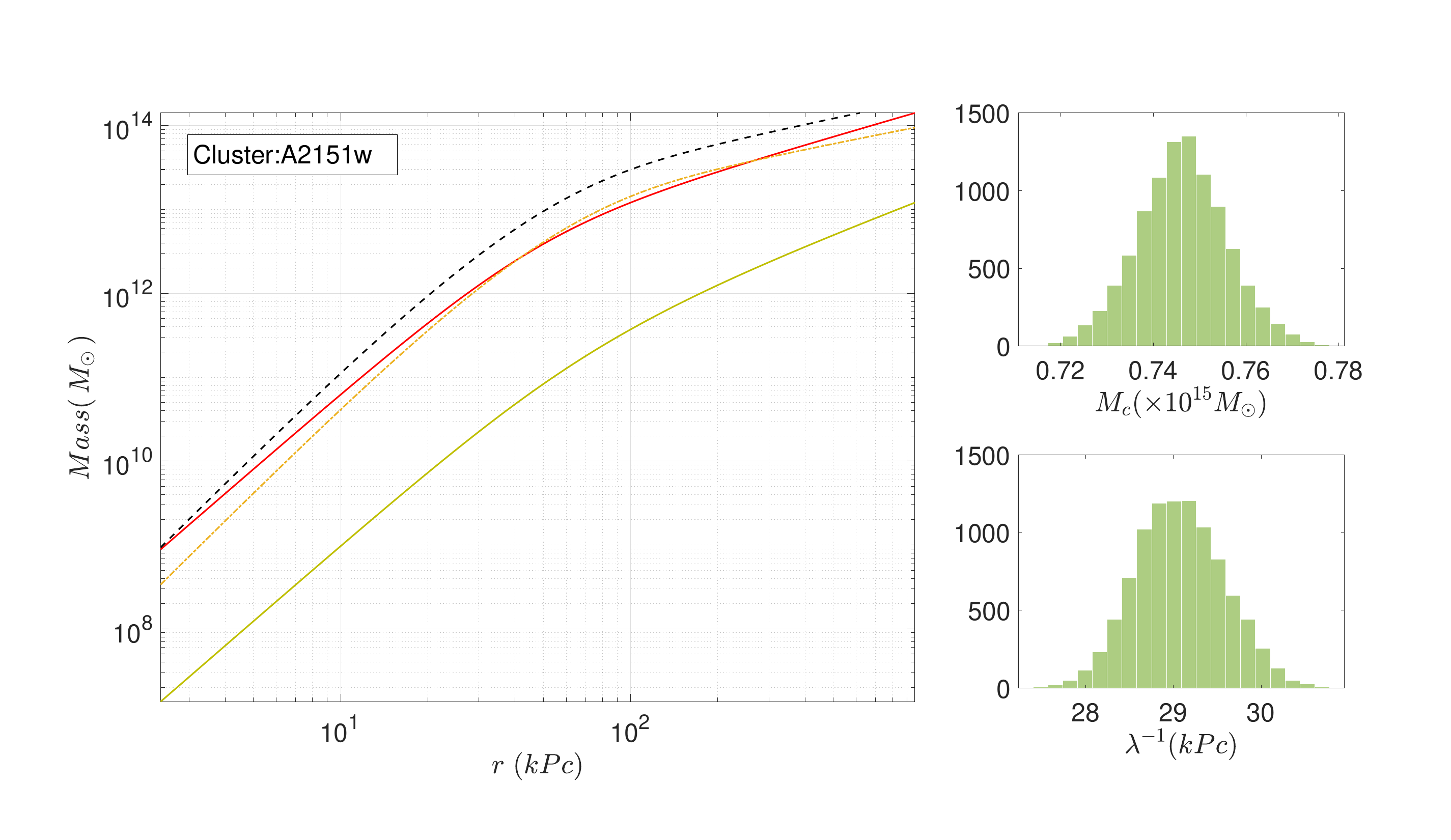}
    \includegraphics[trim=2.0cm 2.0cm 3.0cm 3.0cm, clip=true, width=0.32\columnwidth]{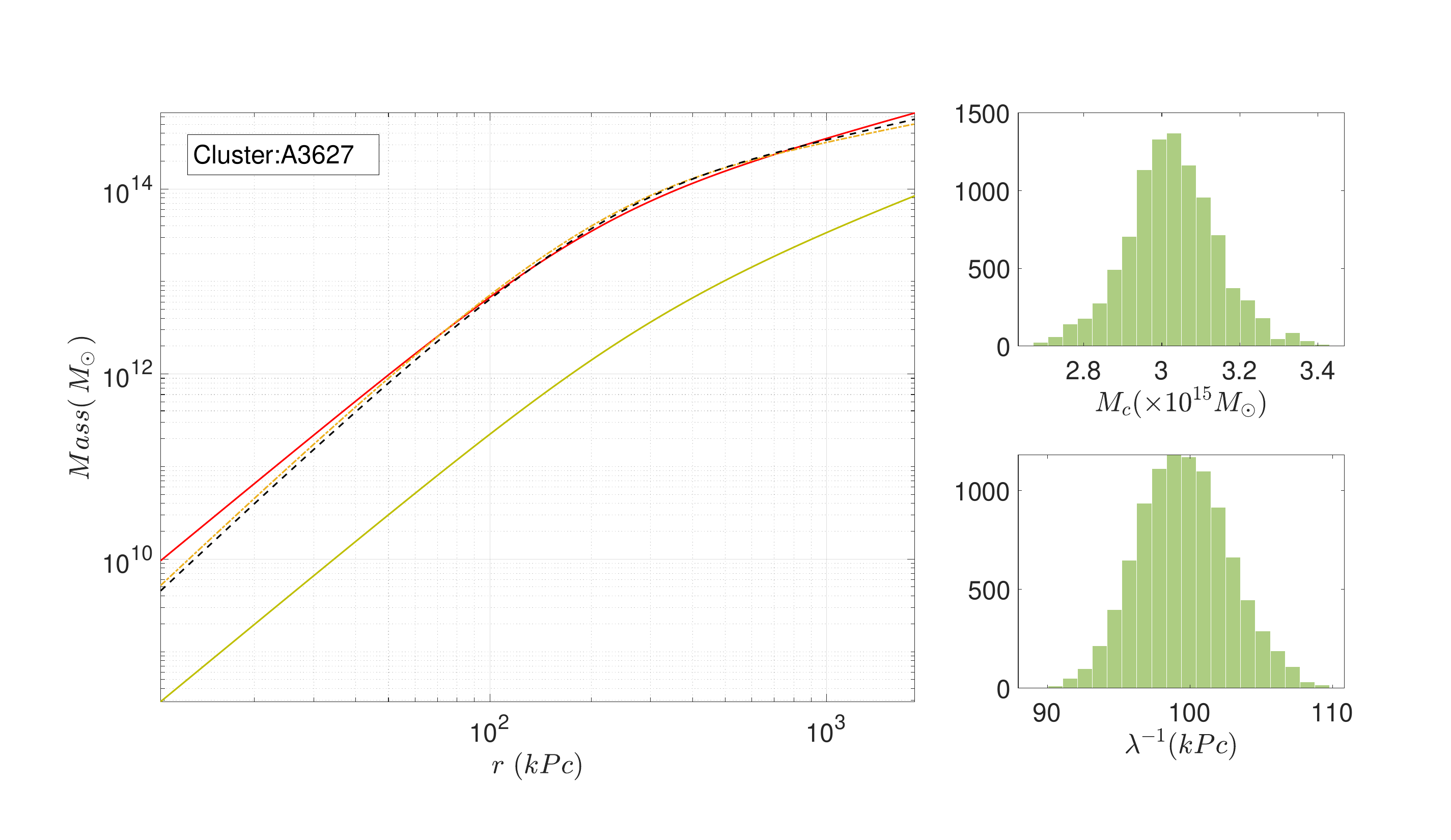}
    \includegraphics[trim=2.0cm 2.0cm 3.0cm 3.0cm, clip=true, width=0.32\columnwidth]{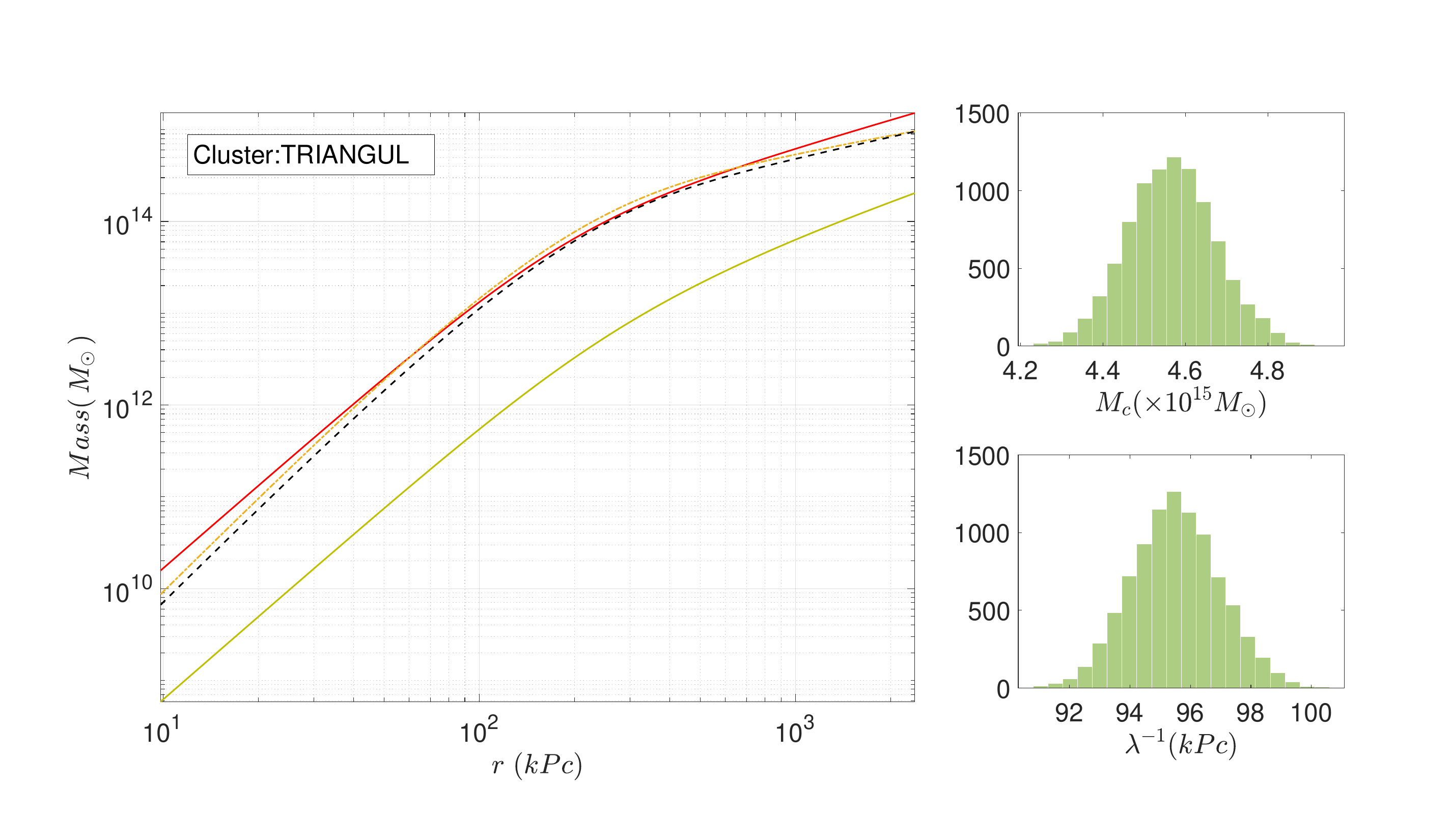}
    \includegraphics[trim=2.0cm 2.0cm 3.0cm 3.0cm, clip=true, width=0.32\columnwidth]{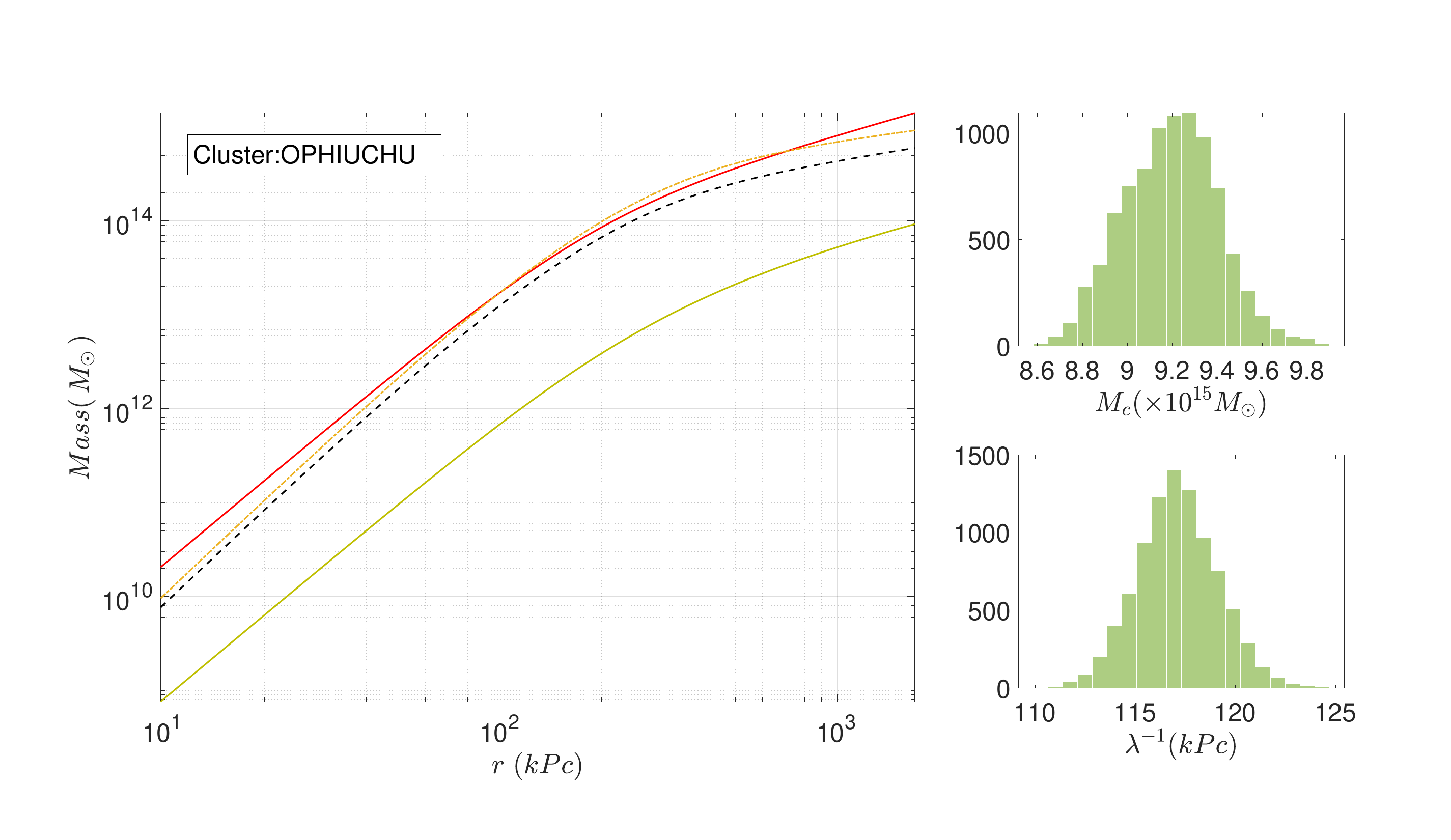}
    \end{figure}
    \begin{figure}
    \centering
    \includegraphics[trim=2.0cm 2.0cm 3.0cm 3.0cm, clip=true, width=0.32\columnwidth]{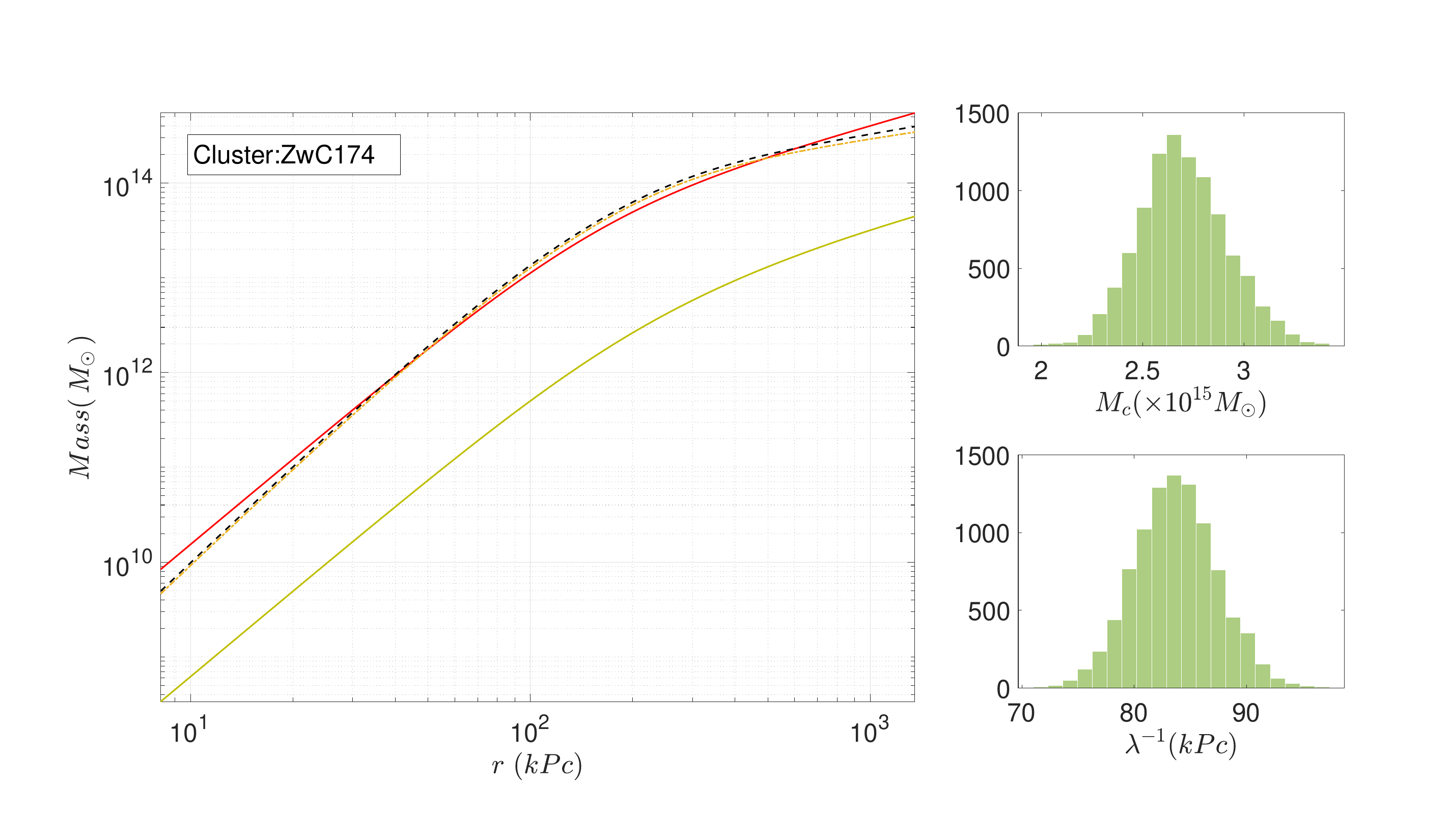}
    \includegraphics[trim=2.0cm 2.0cm 3.0cm 3.0cm, clip=true, width=0.32\columnwidth]{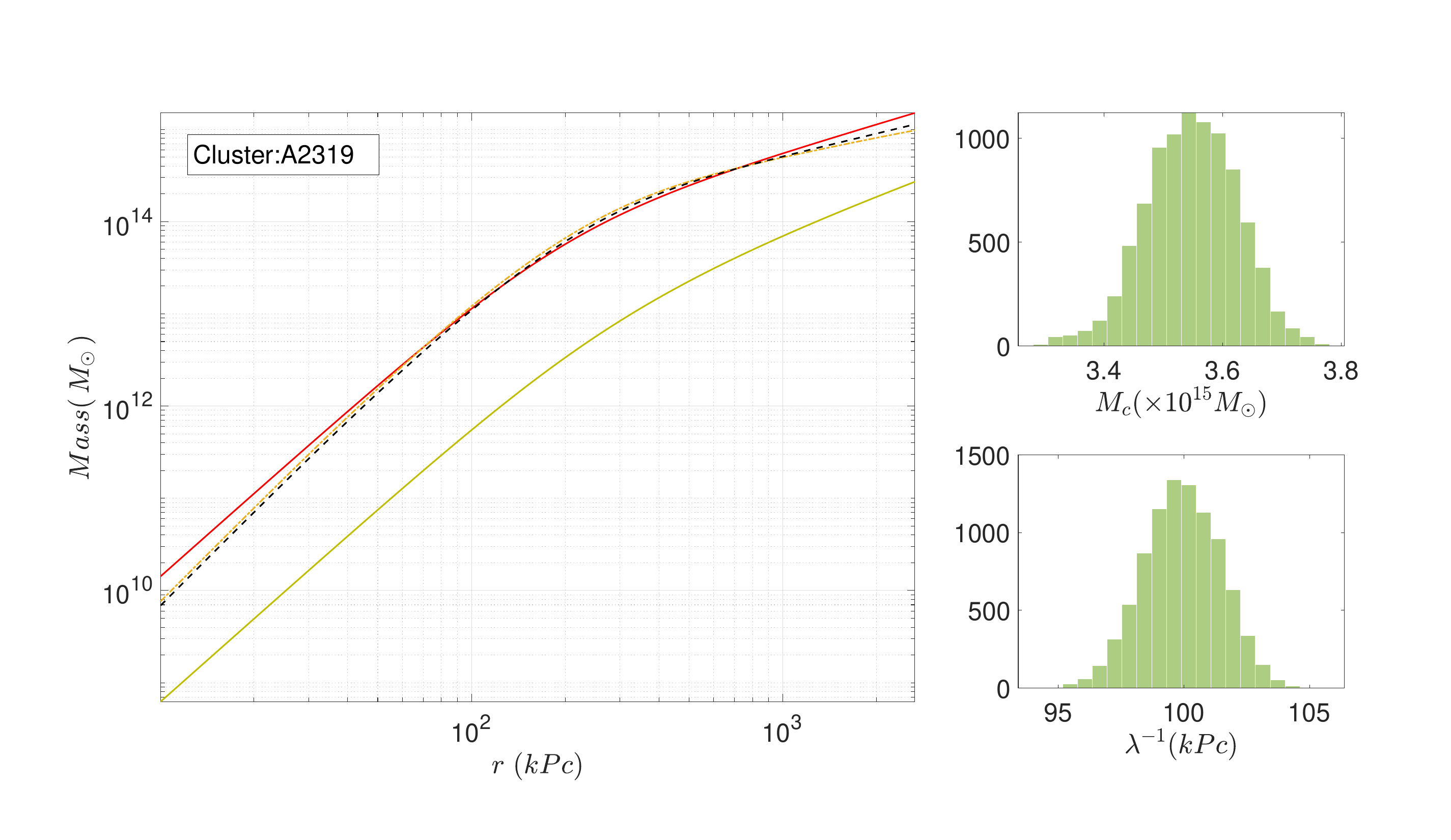}
    \includegraphics[trim=2.0cm 2.0cm 3.0cm 3.0cm, clip=true, width=0.32\columnwidth]{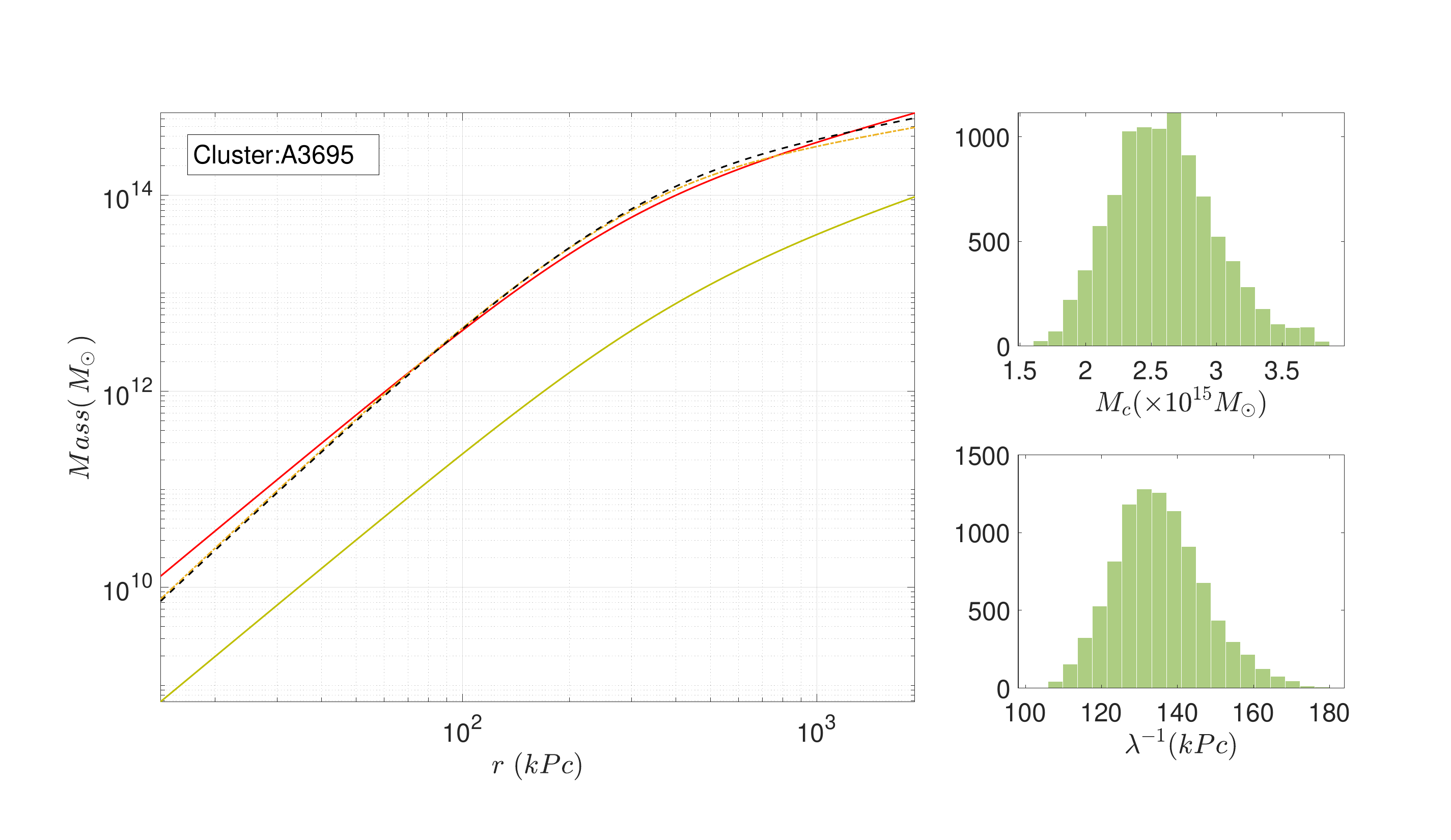}
    \includegraphics[trim=2.0cm 2.0cm 3.0cm 3.0cm, clip=true, width=0.32\columnwidth]{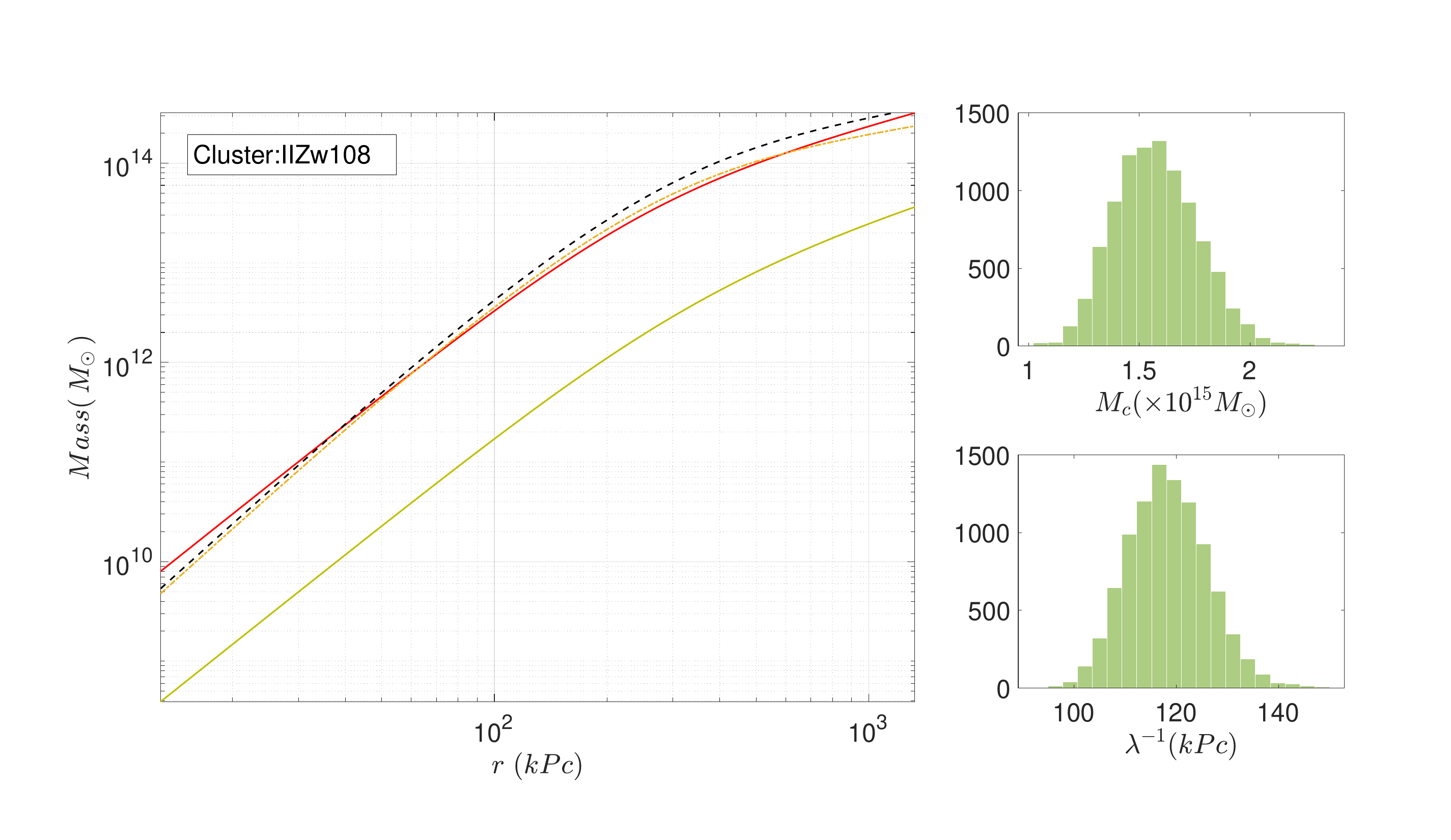}
    \includegraphics[trim=2.0cm 2.0cm 3.0cm 3.0cm, clip=true, width=0.32\columnwidth]{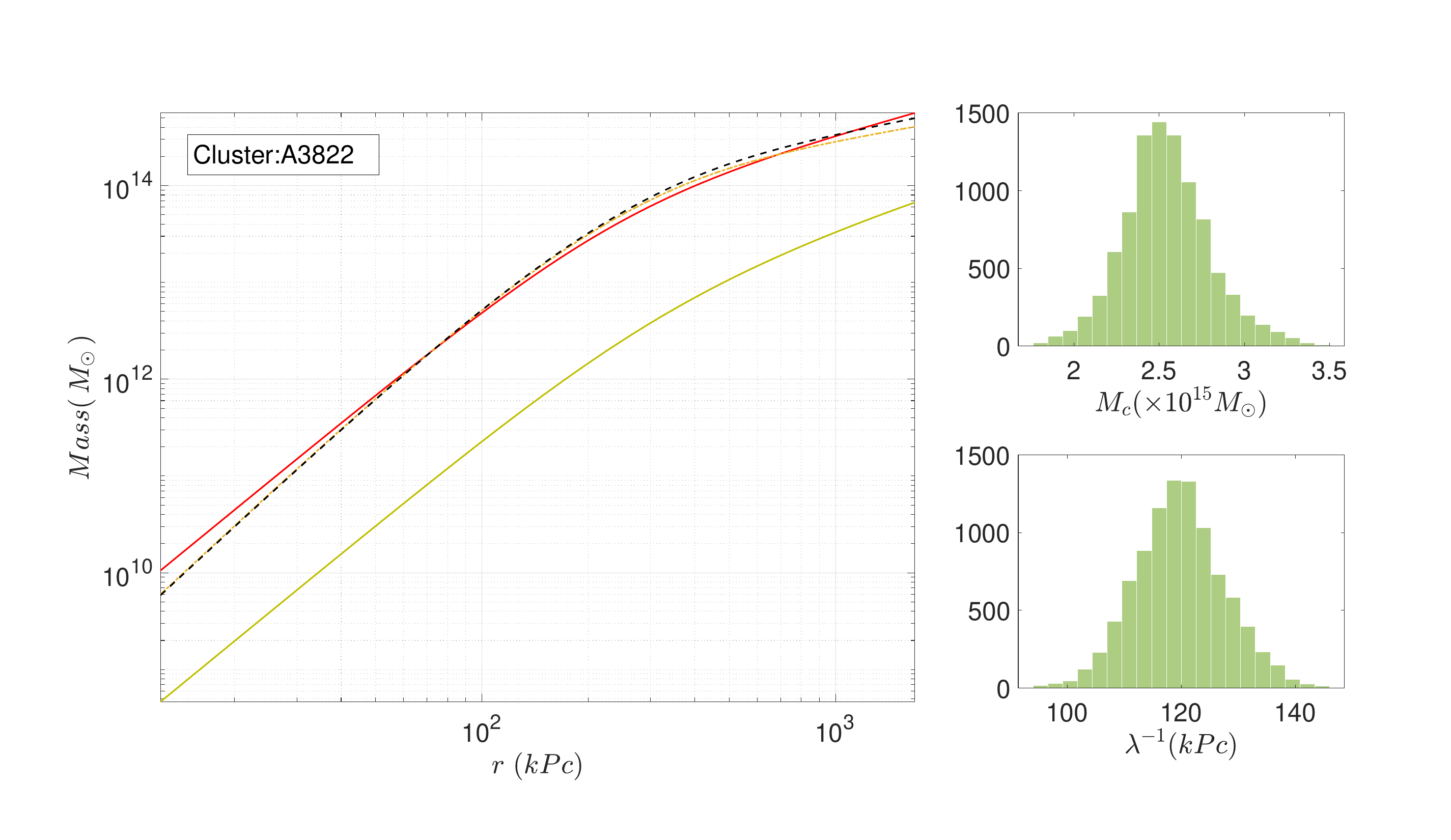}
    \includegraphics[trim=2.0cm 2.0cm 3.0cm 3.0cm, clip=true, width=0.32\columnwidth]{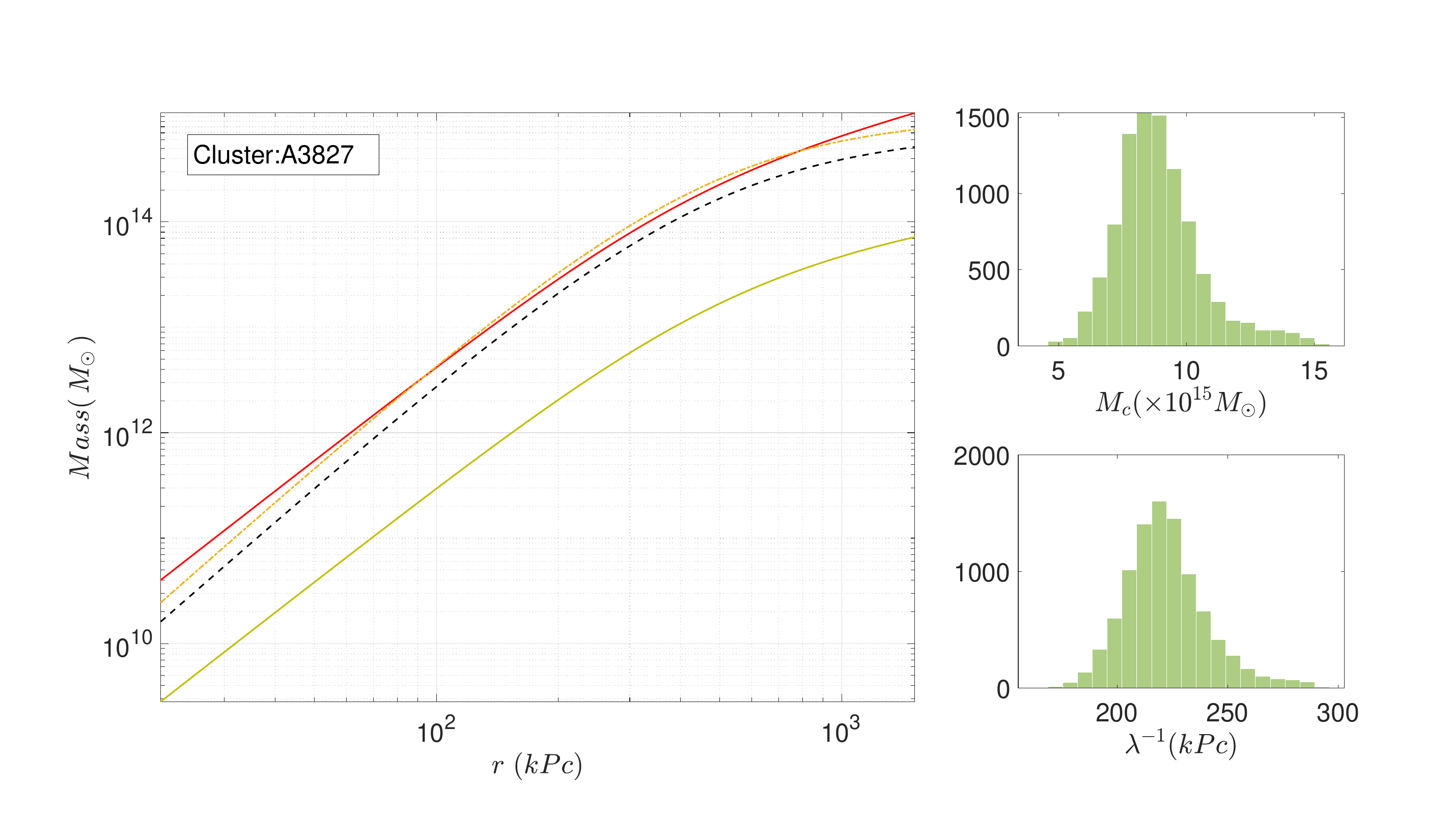}
    \includegraphics[trim=2.0cm 2.0cm 3.0cm 3.0cm, clip=true, width=0.32\columnwidth]{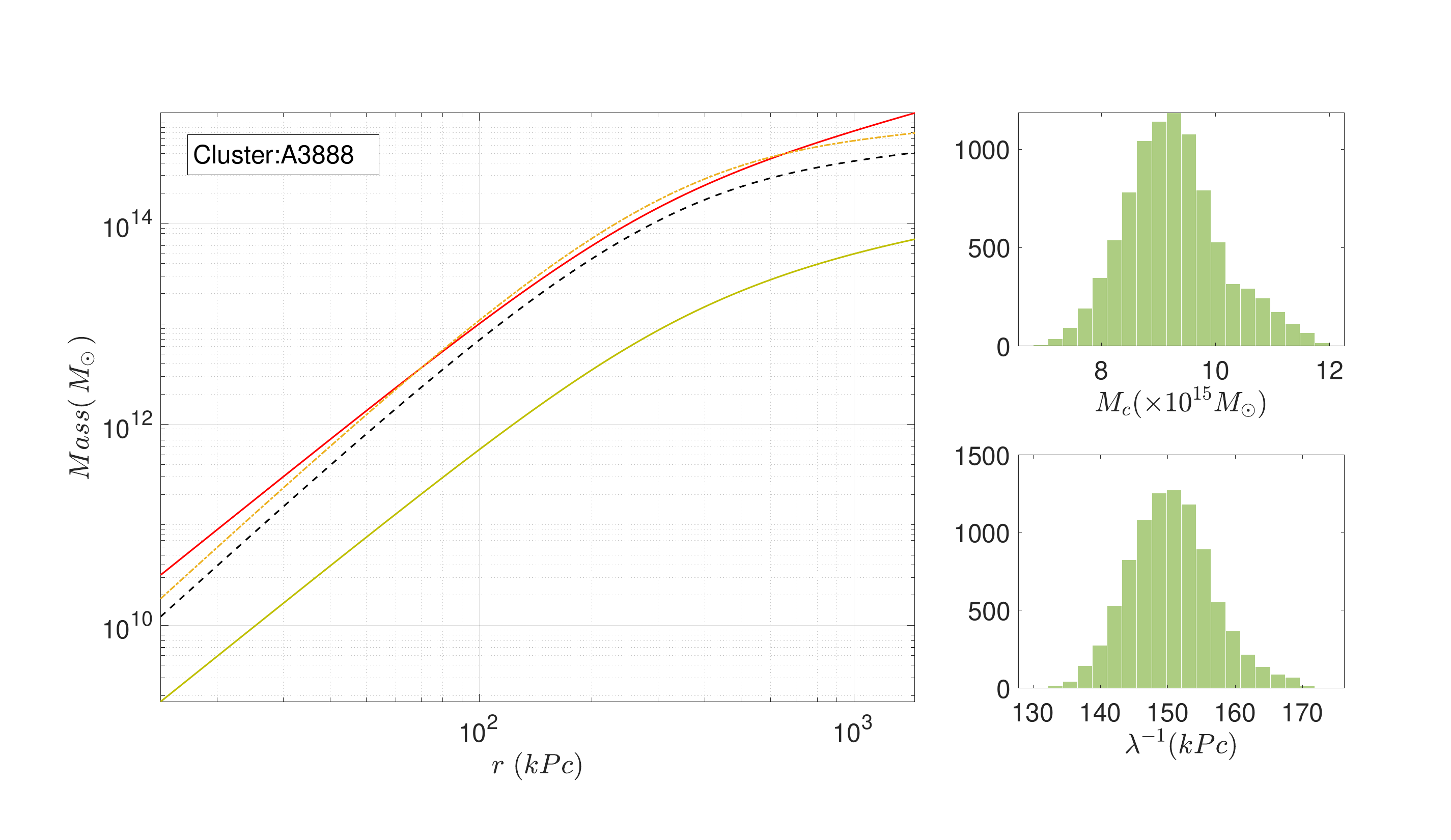}
    \includegraphics[trim=2.0cm 2.0cm 3.0cm 3.0cm, clip=true, width=0.32\columnwidth]{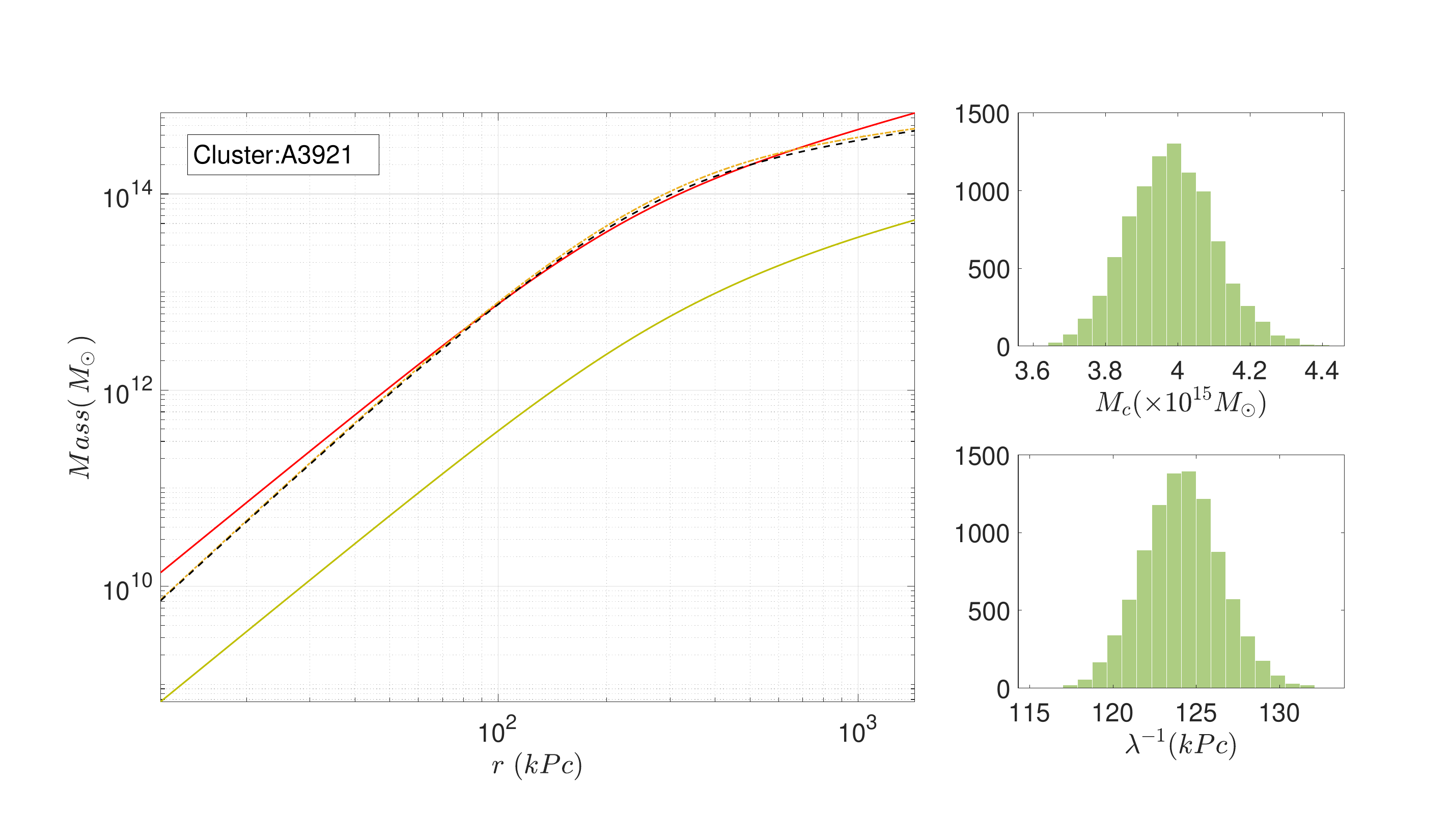}
    \includegraphics[trim=2.0cm 2.0cm 3.0cm 3.0cm, clip=true, width=0.32\columnwidth]{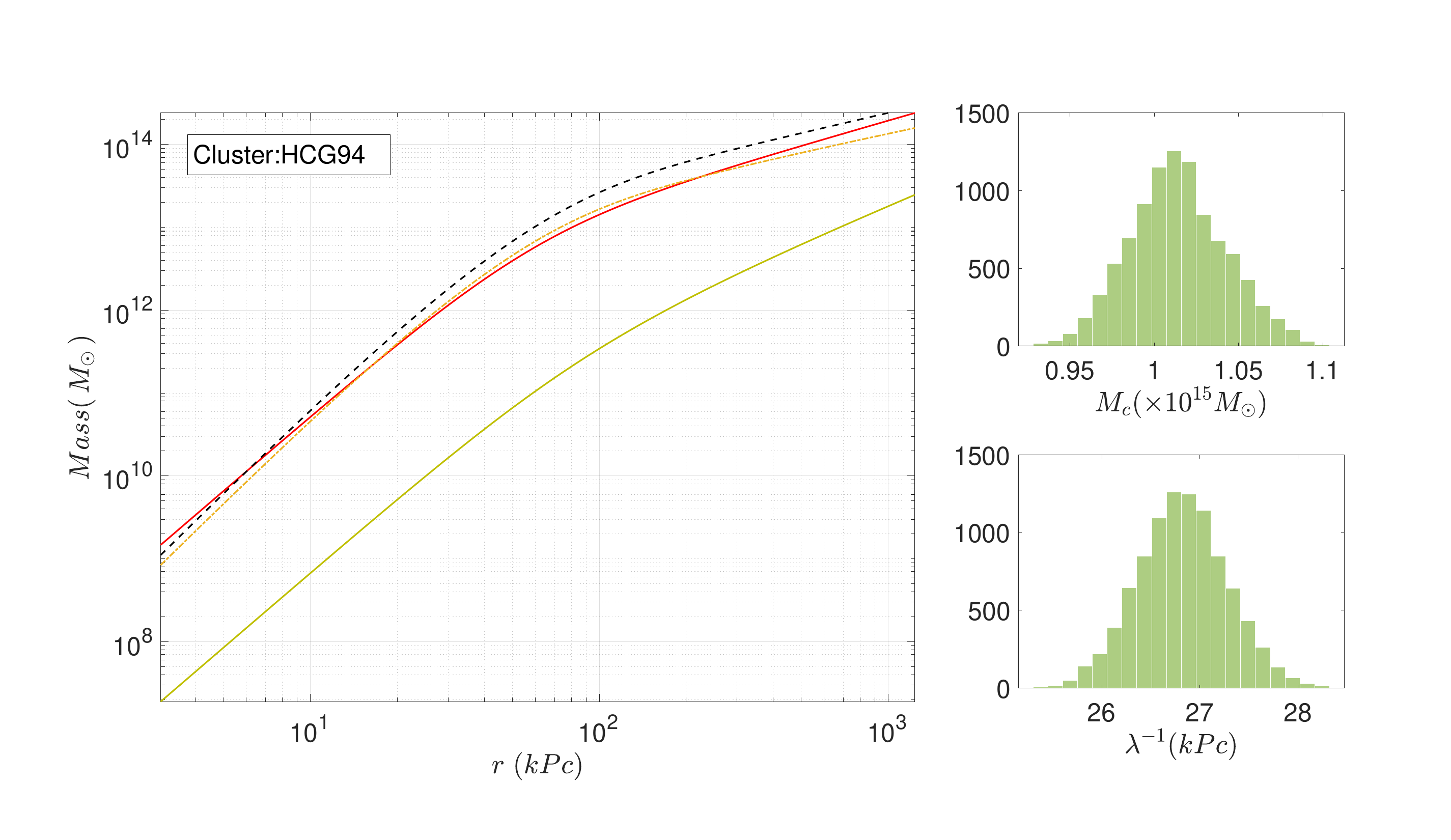}
    \includegraphics[trim=2.0cm 2.0cm 3.0cm 3.0cm, clip=true, width=0.32\columnwidth]{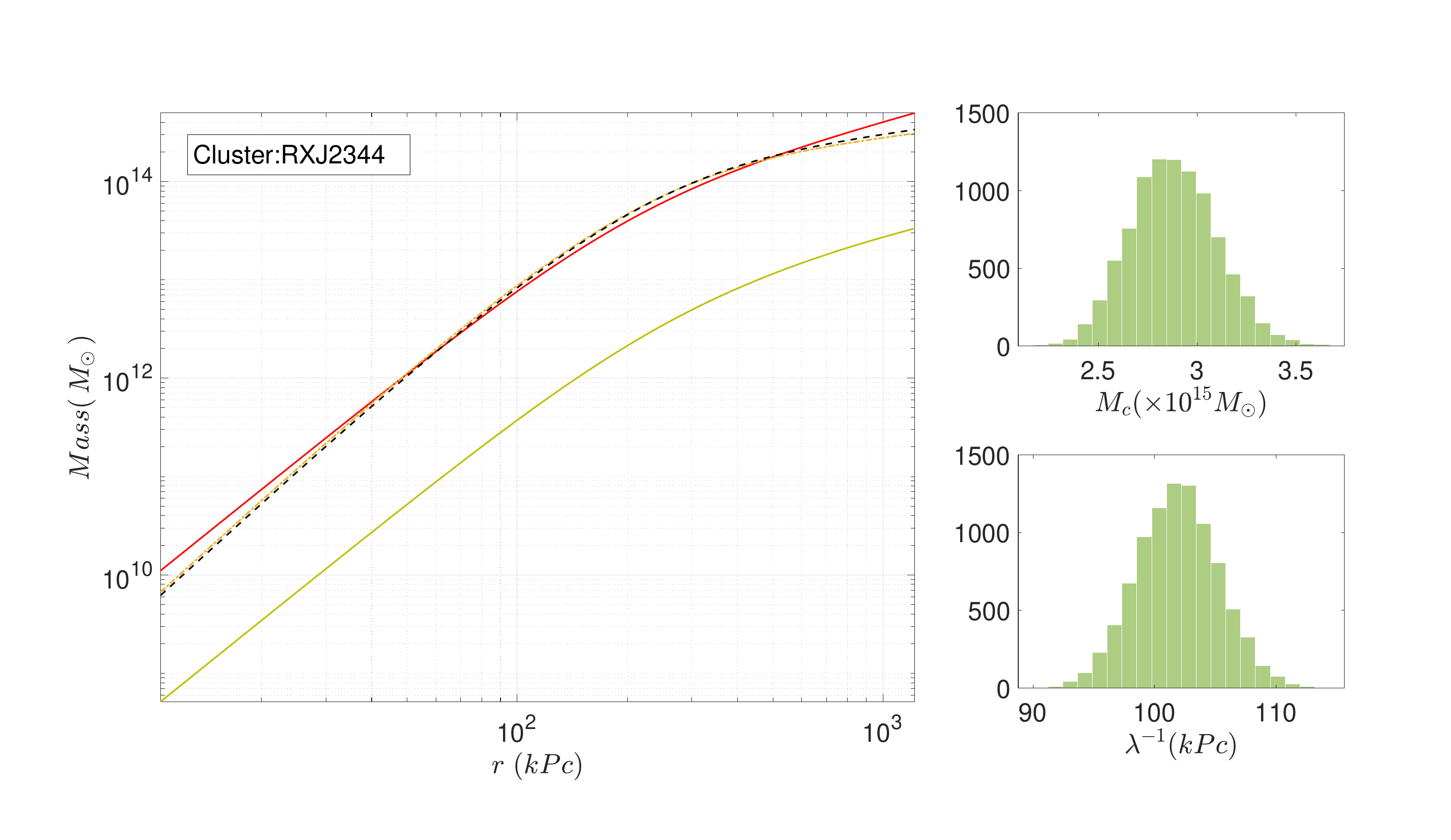}
    \caption{\label{fig:GalaxyCusterMass}We present here the radial mass profiles for 106 galaxy clusters, taken from~\cite{reiprich2003cosmological}. The greenish-yellow curve shows the baryonic mass of each cluster, calculated using Eq.~\ref{BaryonMass1}. The solid red curve depicts the Newtonian dynamical mass derived from Eq.~\ref{eq:M_N}. The dotted yellow-ochre curve corresponds to the left-hand side of Eq.~\ref{eq:Machiangrav}, representing the dynamical mass predicted by the Machian Gravity framework. The comparison demonstrates that these different mass estimators are broadly consistent with one another. We also plot, as a black dotted line, a fitted profile obtained by fixing $M_c = 2.759\times 10^{15} M_{\odot}$ and $\lambda^{-1} = 0.4949 r_c$.}
    \end{figure}
    \newpage

\setlength{\leftmargin}{0.18cm}

\begin{figure}
\includegraphics[trim=0.0cm 0.0cm 1.0cm 0.0cm, clip=true, width=0.33\columnwidth]{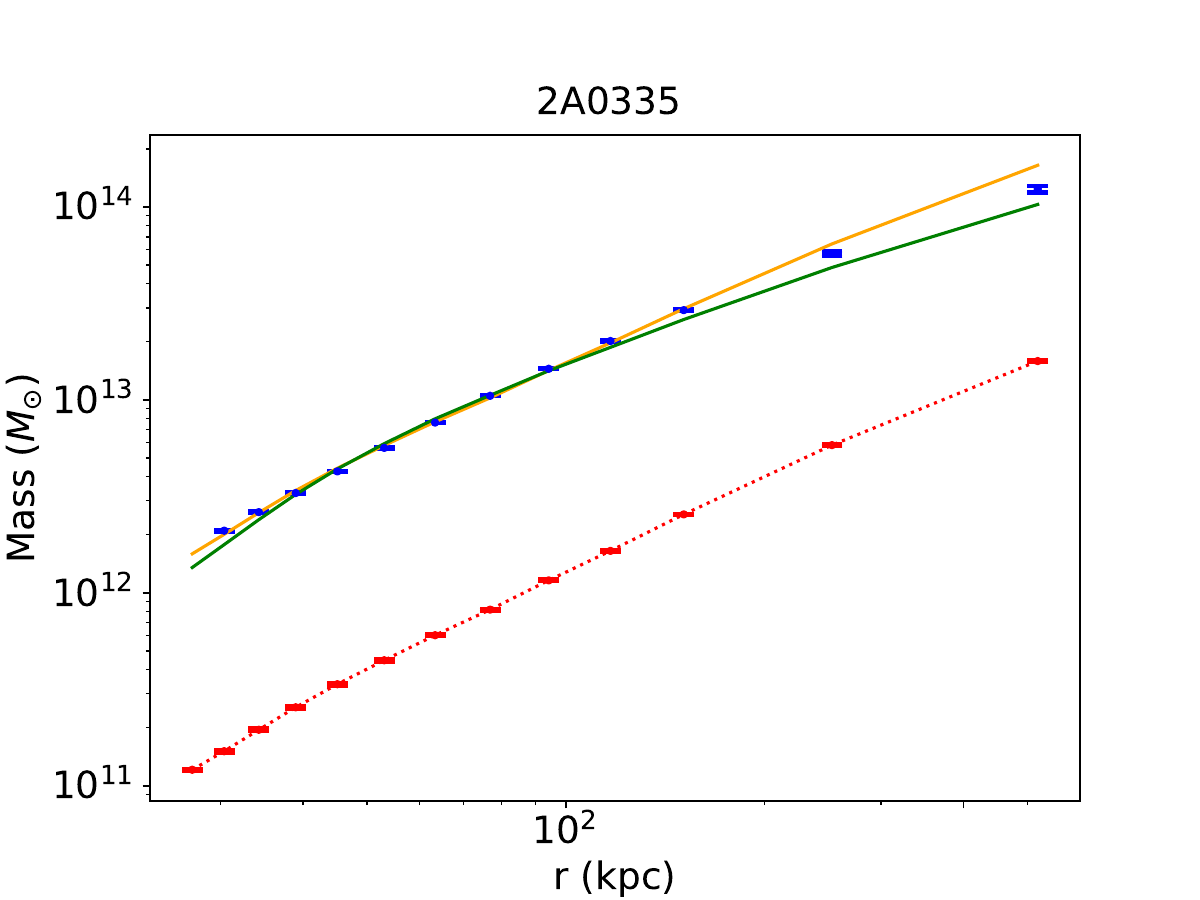}
\includegraphics[trim=0.0cm 0.0cm 1.0cm 0.0cm, clip=true, width=0.33\columnwidth]{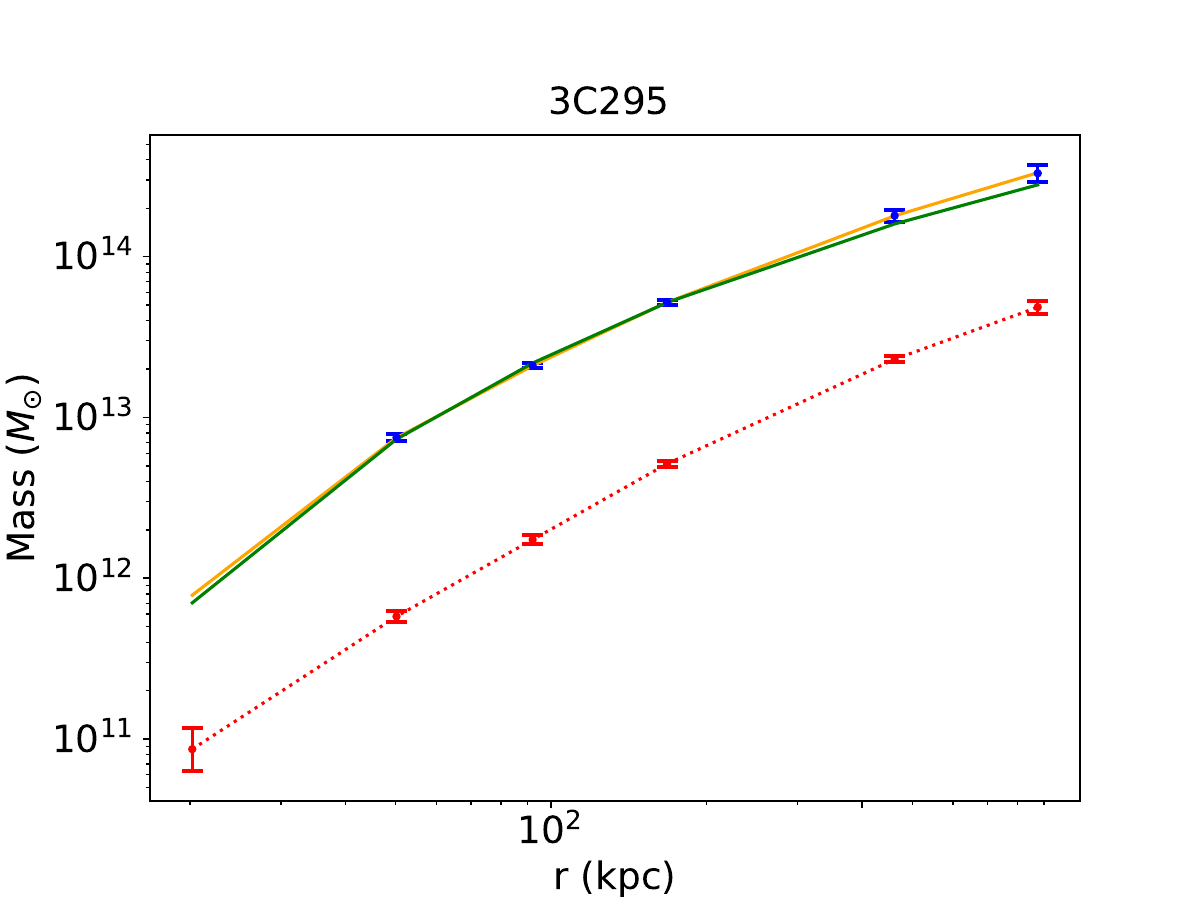}
\includegraphics[trim=0.0cm 0.0cm 1.0cm 0.0cm, clip=true, width=0.33\columnwidth]{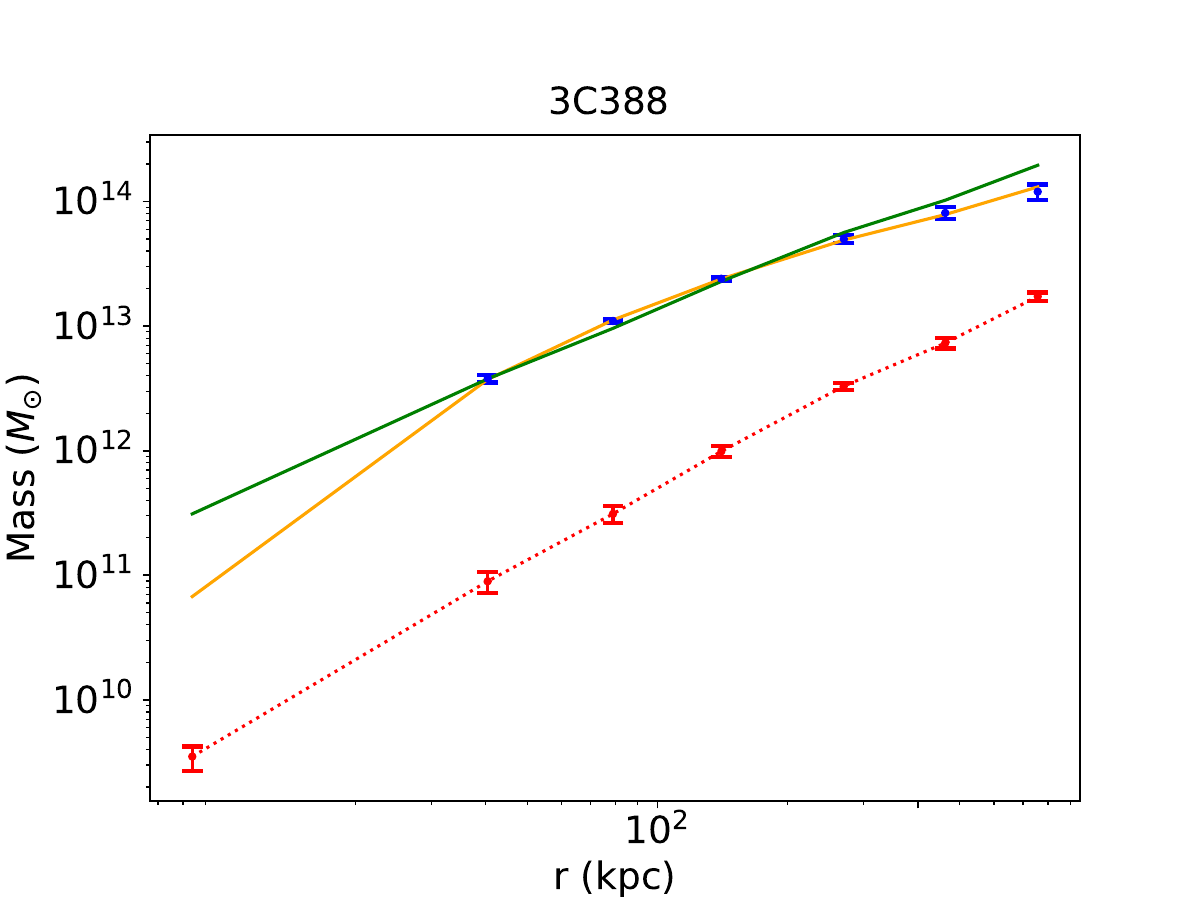}
\includegraphics[trim=0.0cm 0.0cm 1.0cm 0.0cm, clip=true, width=0.33\columnwidth]{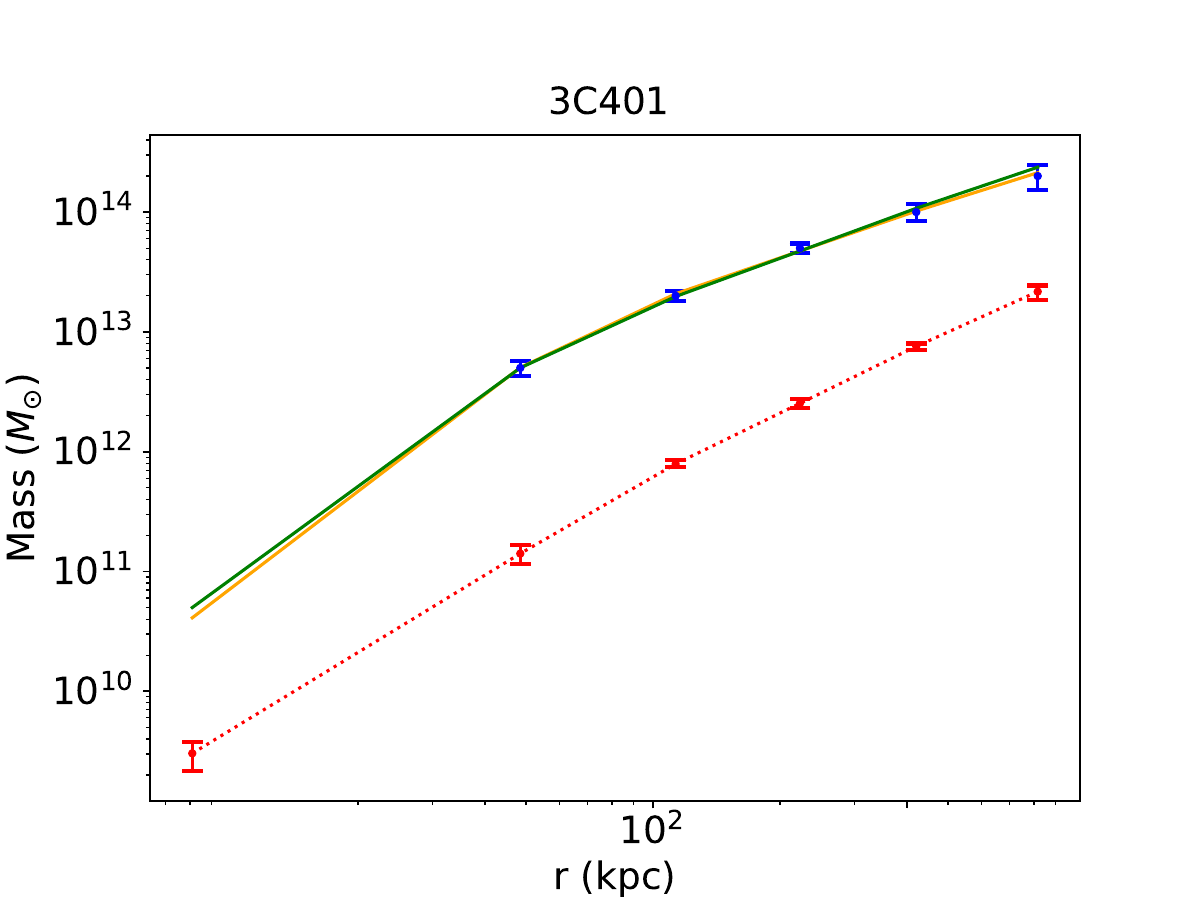}
\includegraphics[trim=0.0cm 0.0cm 1.0cm 0.0cm, clip=true, width=0.33\columnwidth]{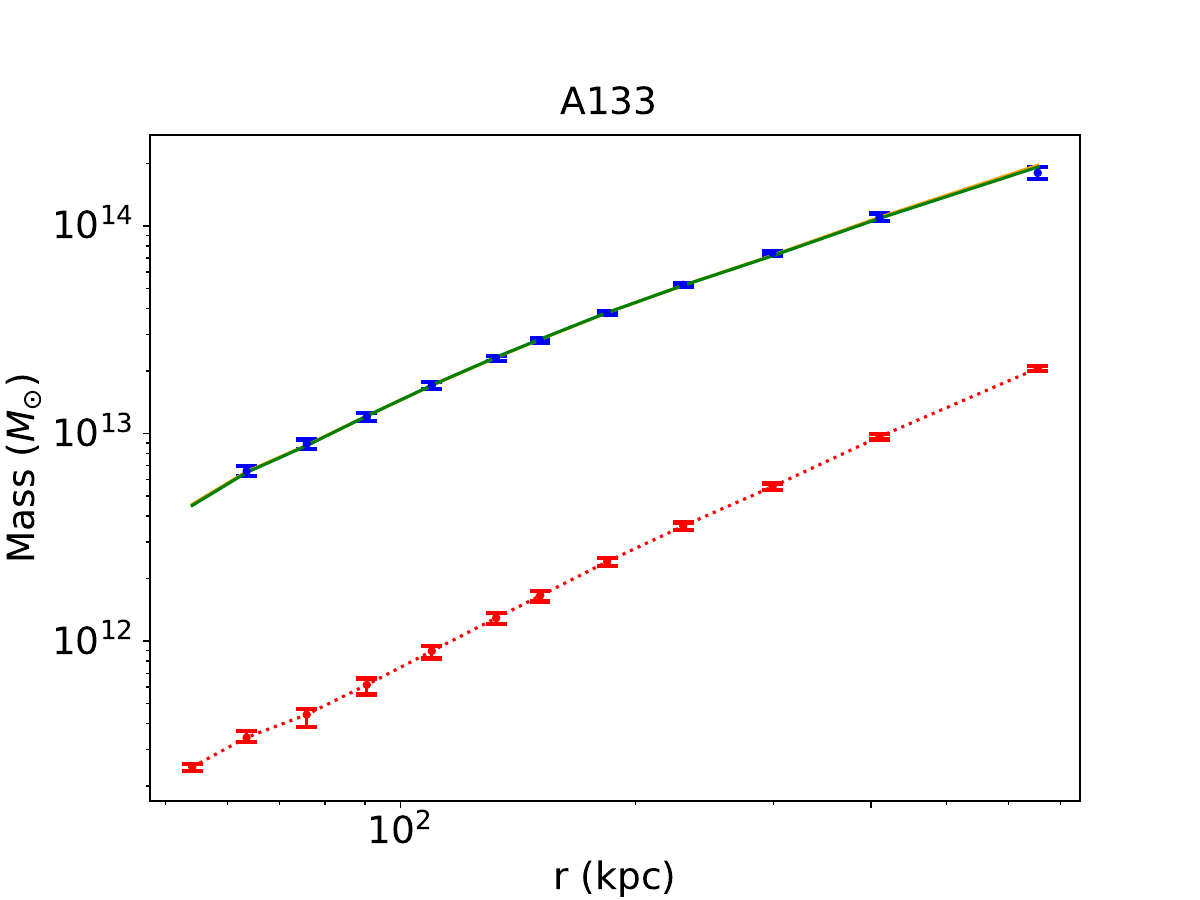}
\includegraphics[trim=0.0cm 0.0cm 1.0cm 0.0cm, clip=true, width=0.33\columnwidth]{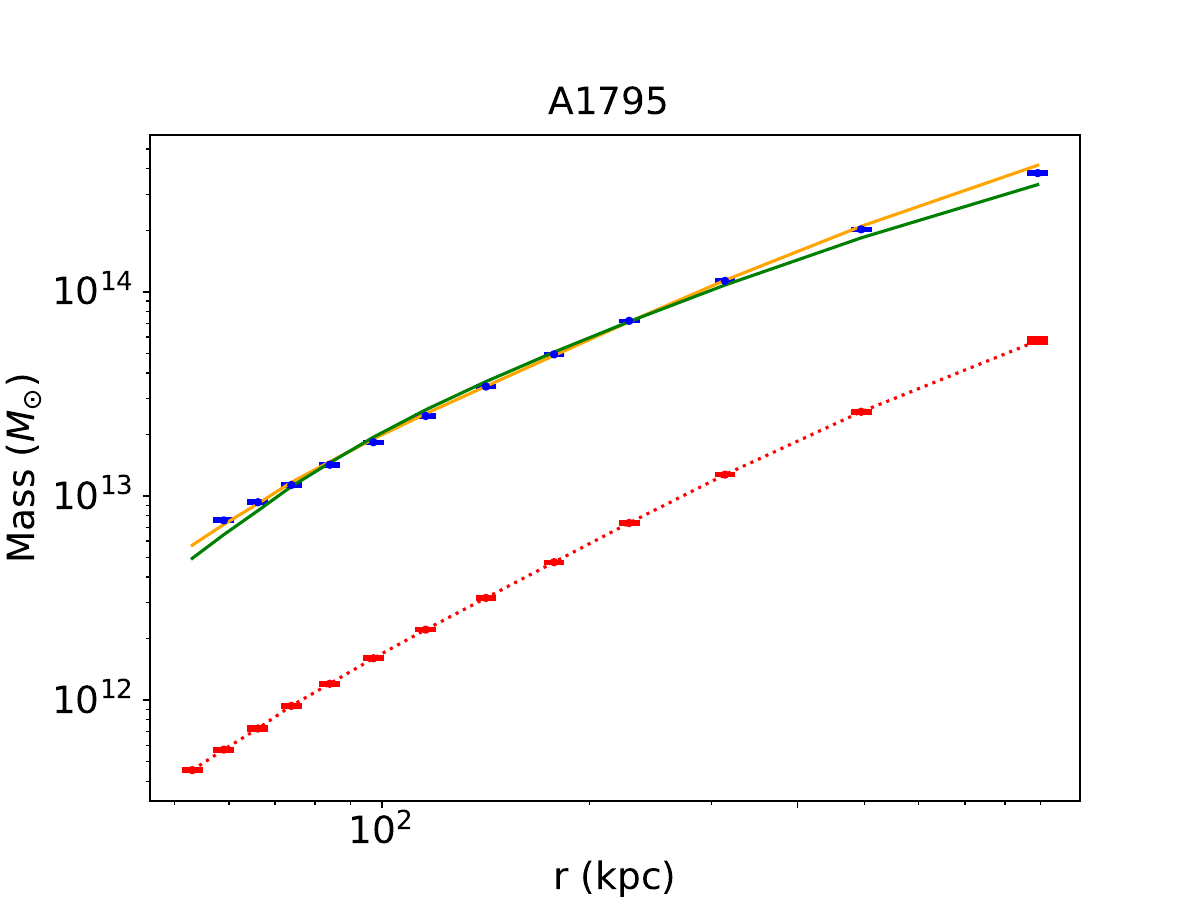}
\includegraphics[trim=0.0cm 0.0cm 1.0cm 0.0cm, clip=true, width=0.33\columnwidth]{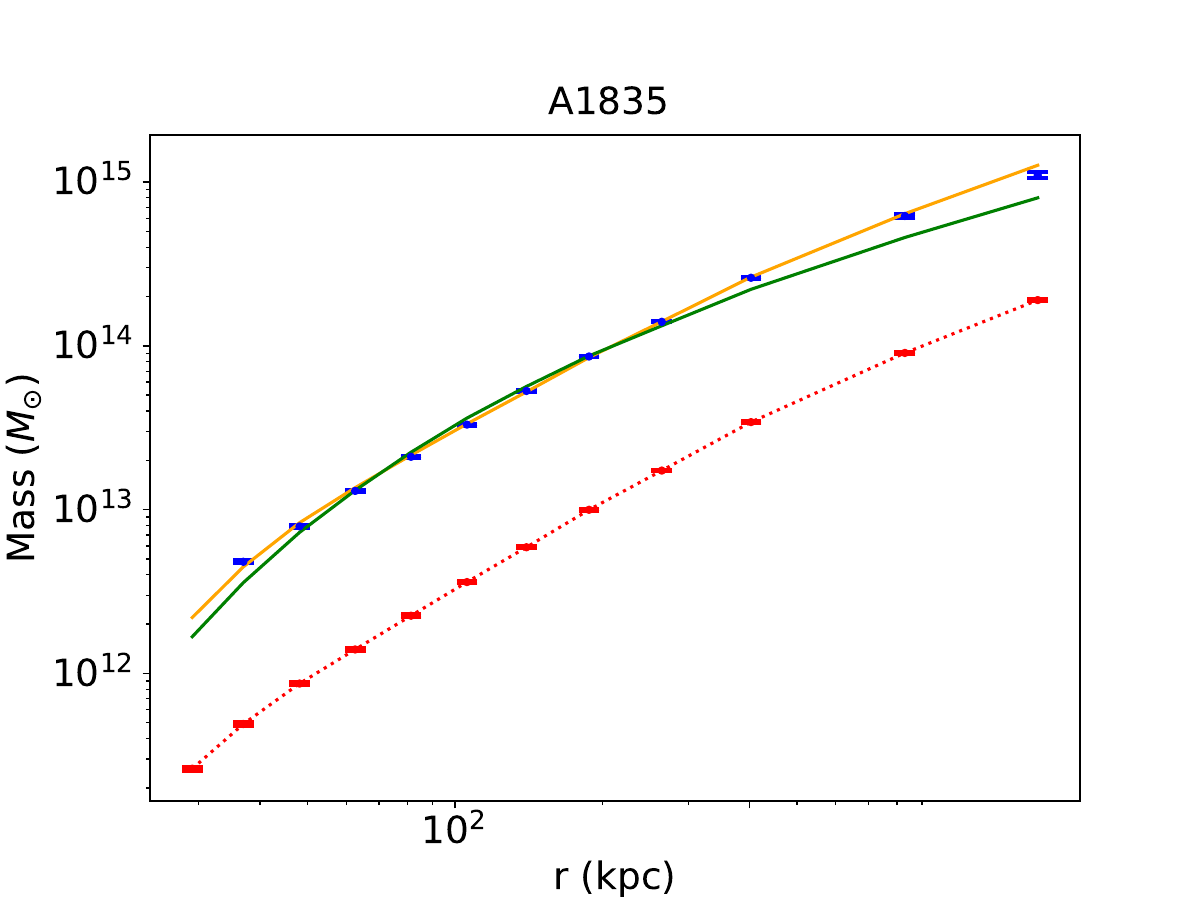}
\includegraphics[trim=0.0cm 0.0cm 1.0cm 0.0cm, clip=true, width=0.33\columnwidth]{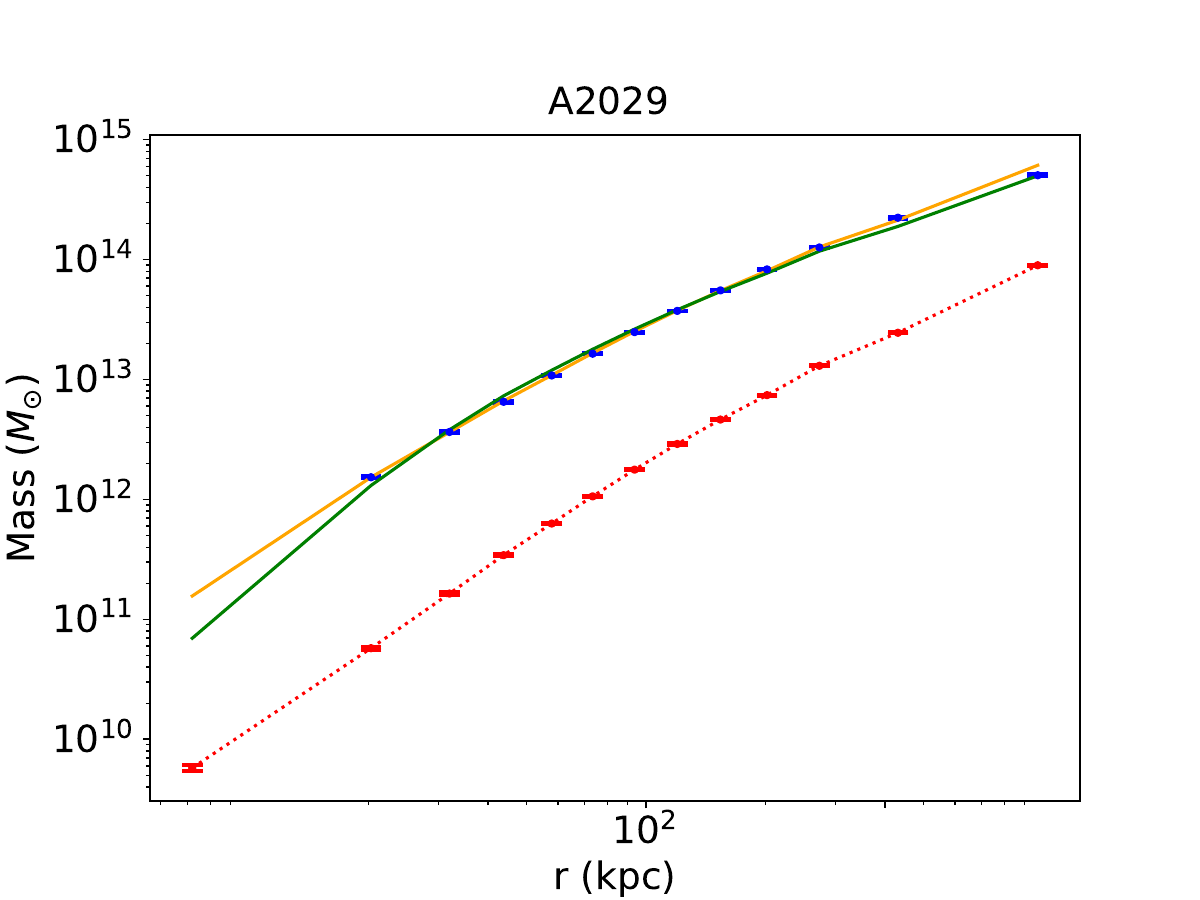}
\includegraphics[trim=0.0cm 0.0cm 1.0cm 0.0cm, clip=true, width=0.33\columnwidth]{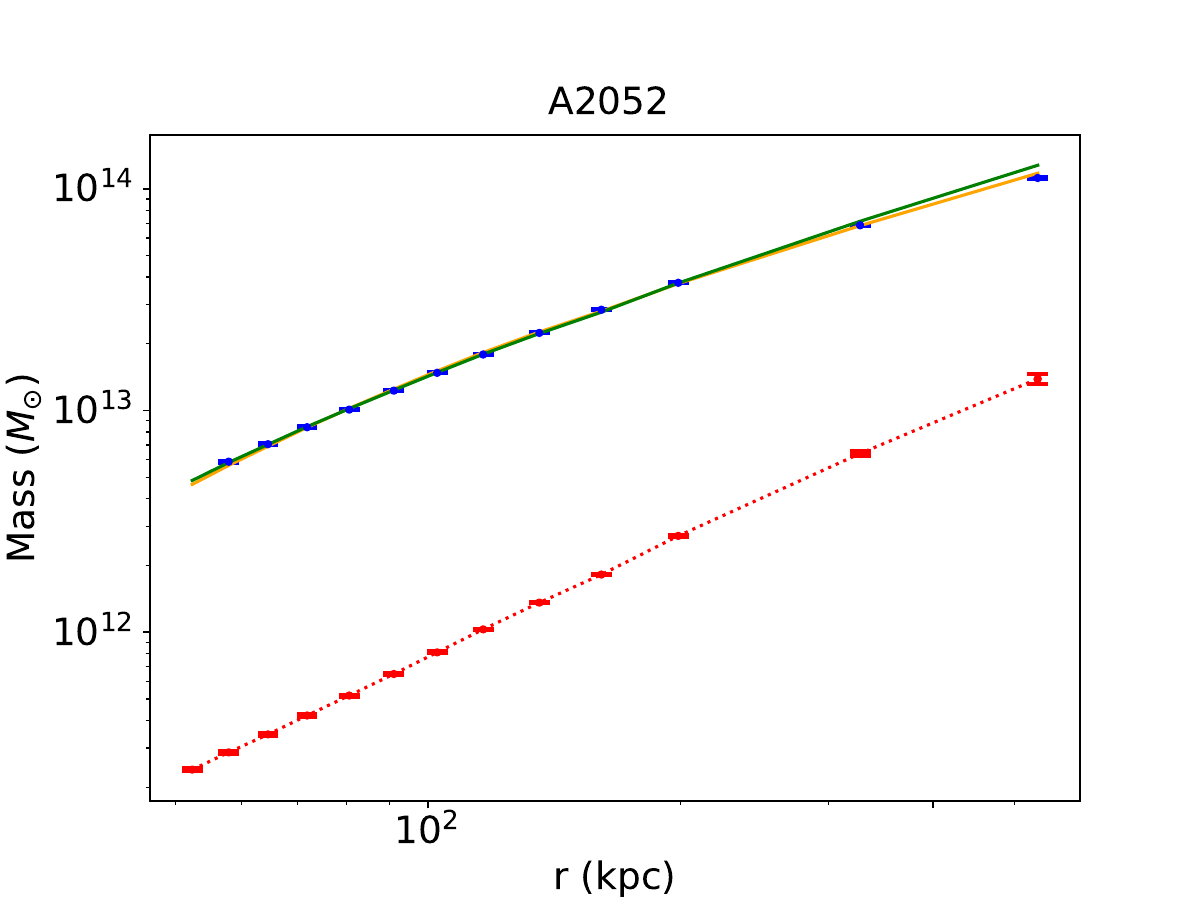}
\includegraphics[trim=0.0cm 0.0cm 1.0cm 0.0cm, clip=true, width=0.33\columnwidth]{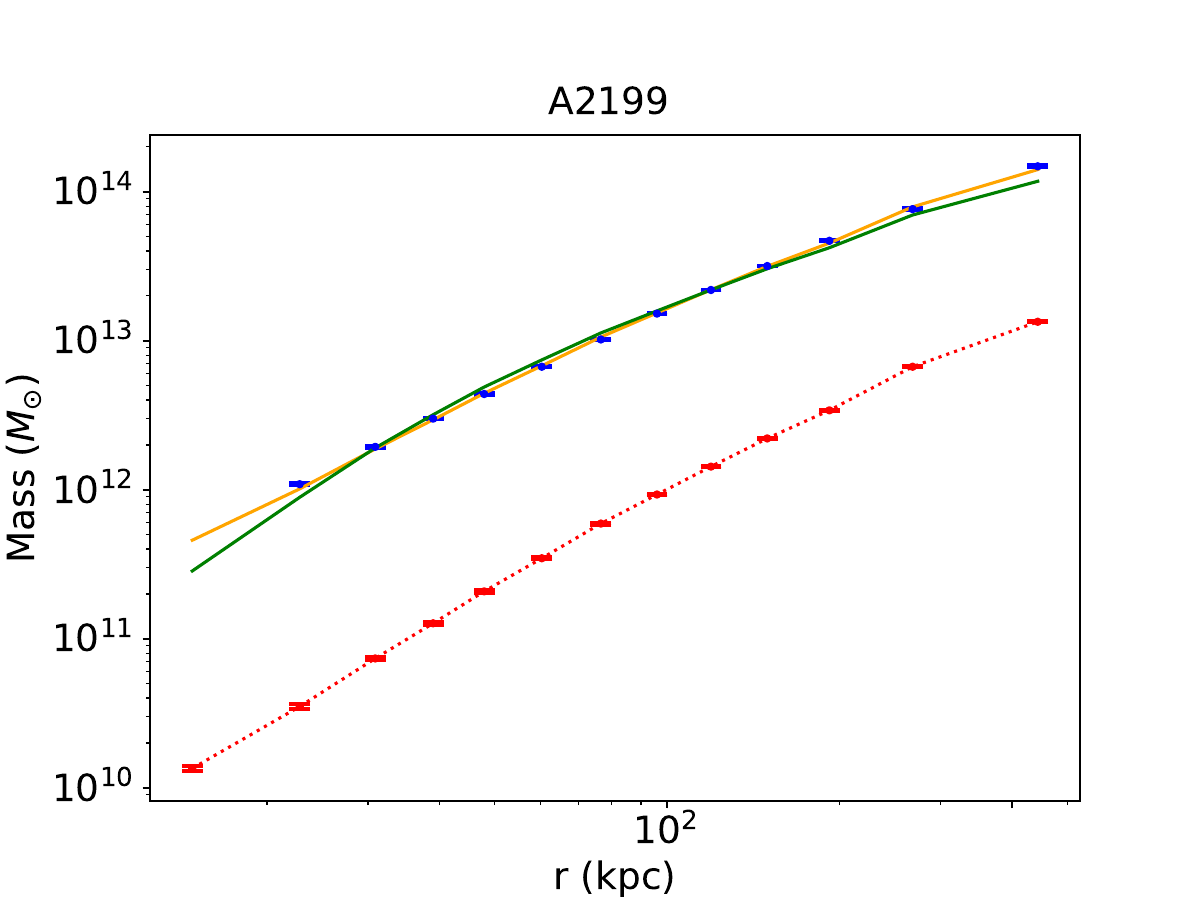}
\includegraphics[trim=0.0cm 0.0cm 1.0cm 0.0cm, clip=true, width=0.33\columnwidth]{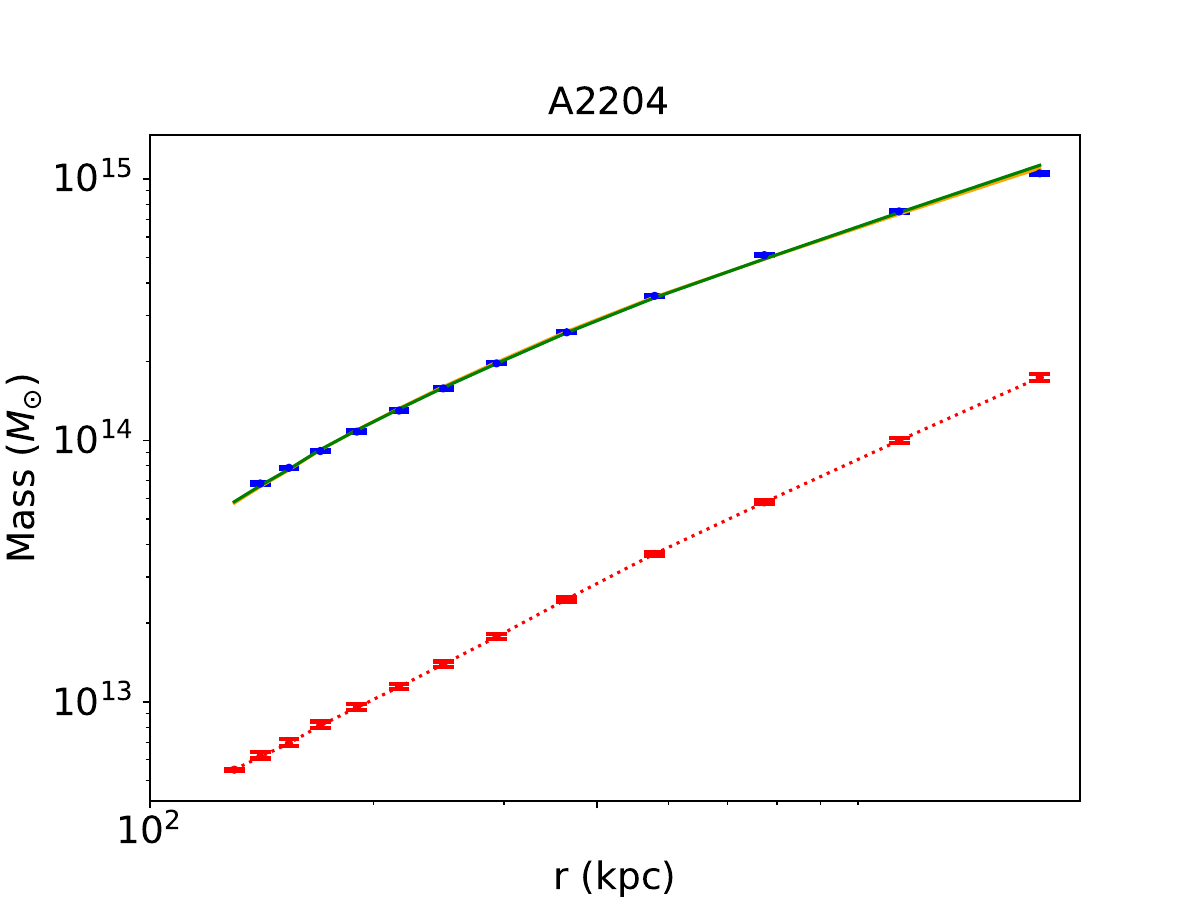}
\includegraphics[trim=0.0cm 0.0cm 1.0cm 0.0cm, clip=true, width=0.33\columnwidth]{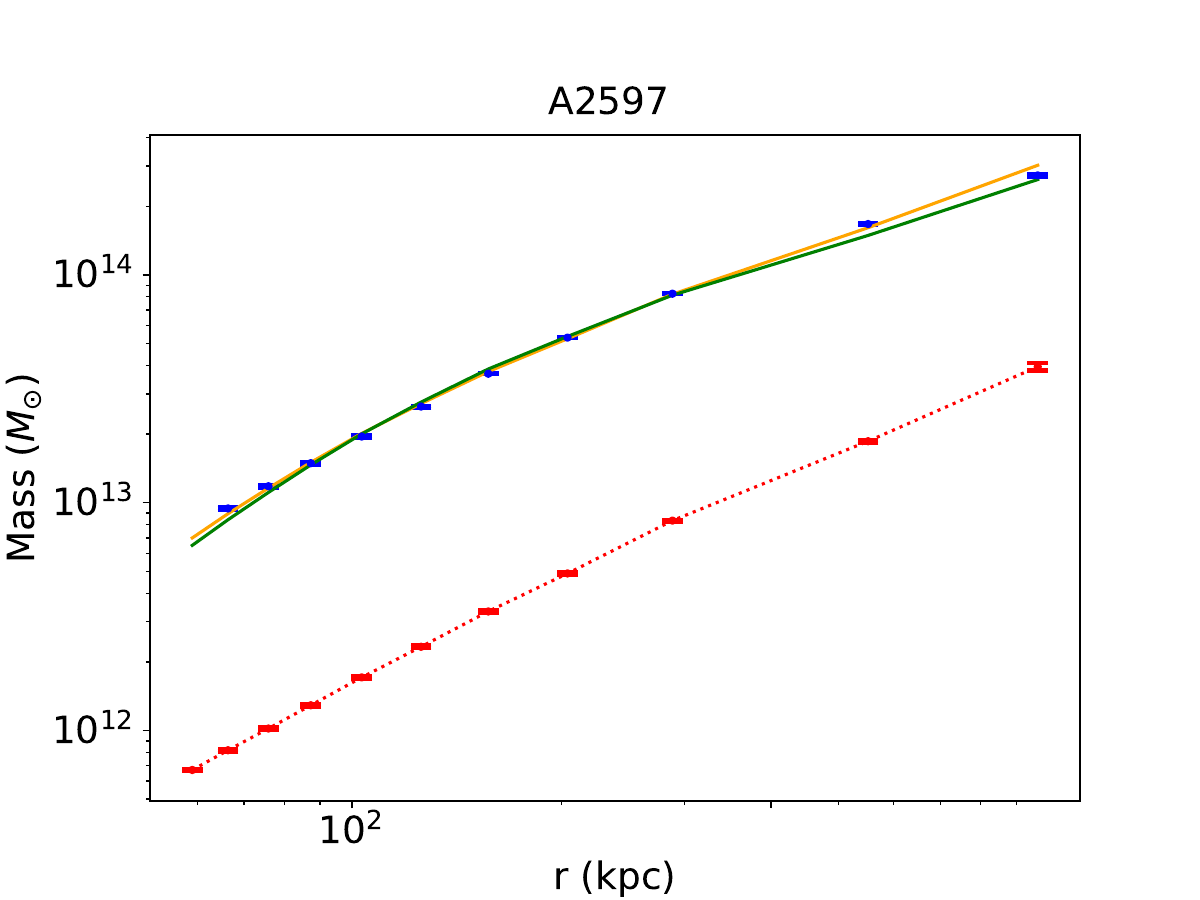}
\includegraphics[trim=0.0cm 0.0cm 1.0cm 0.0cm, clip=true, width=0.33\columnwidth]{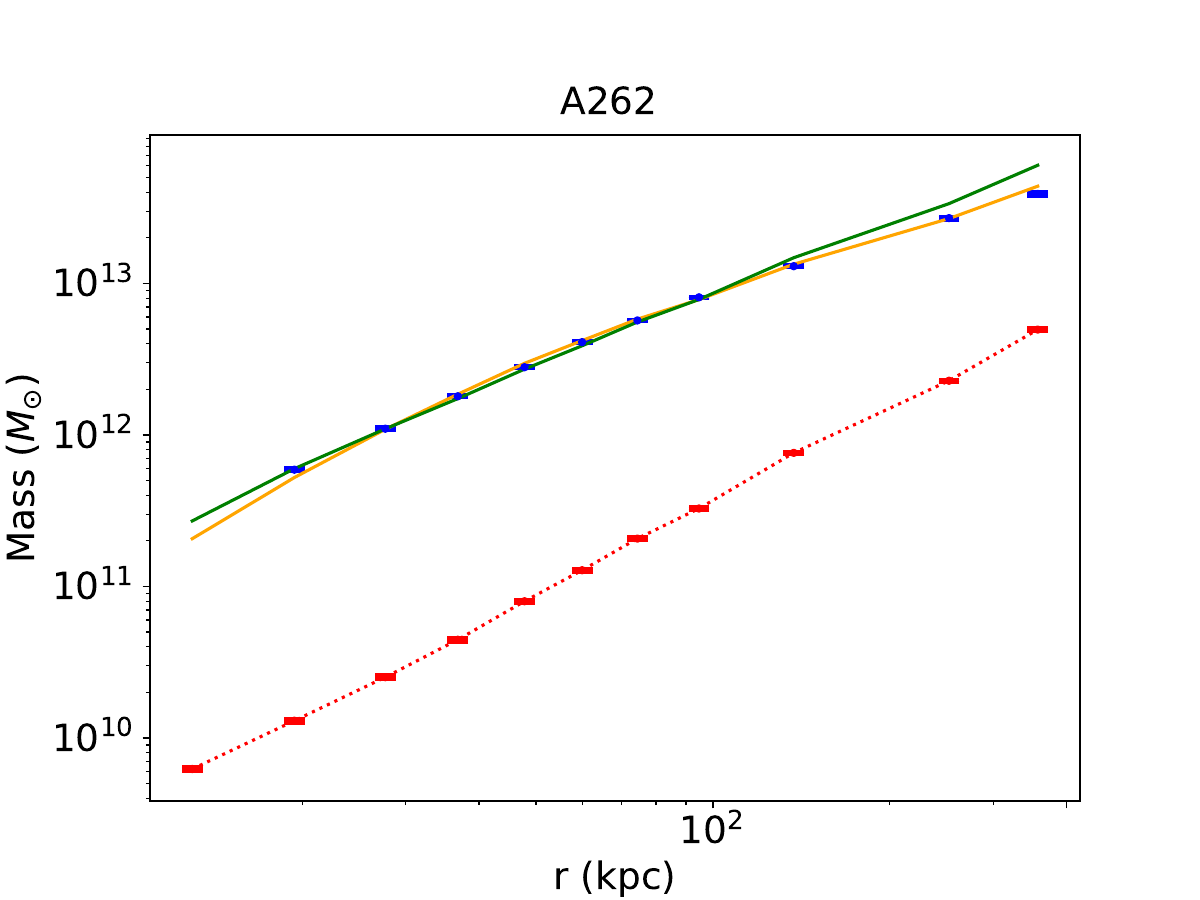}
\includegraphics[trim=0.0cm 0.0cm 1.0cm 0.0cm, clip=true, width=0.33\columnwidth]{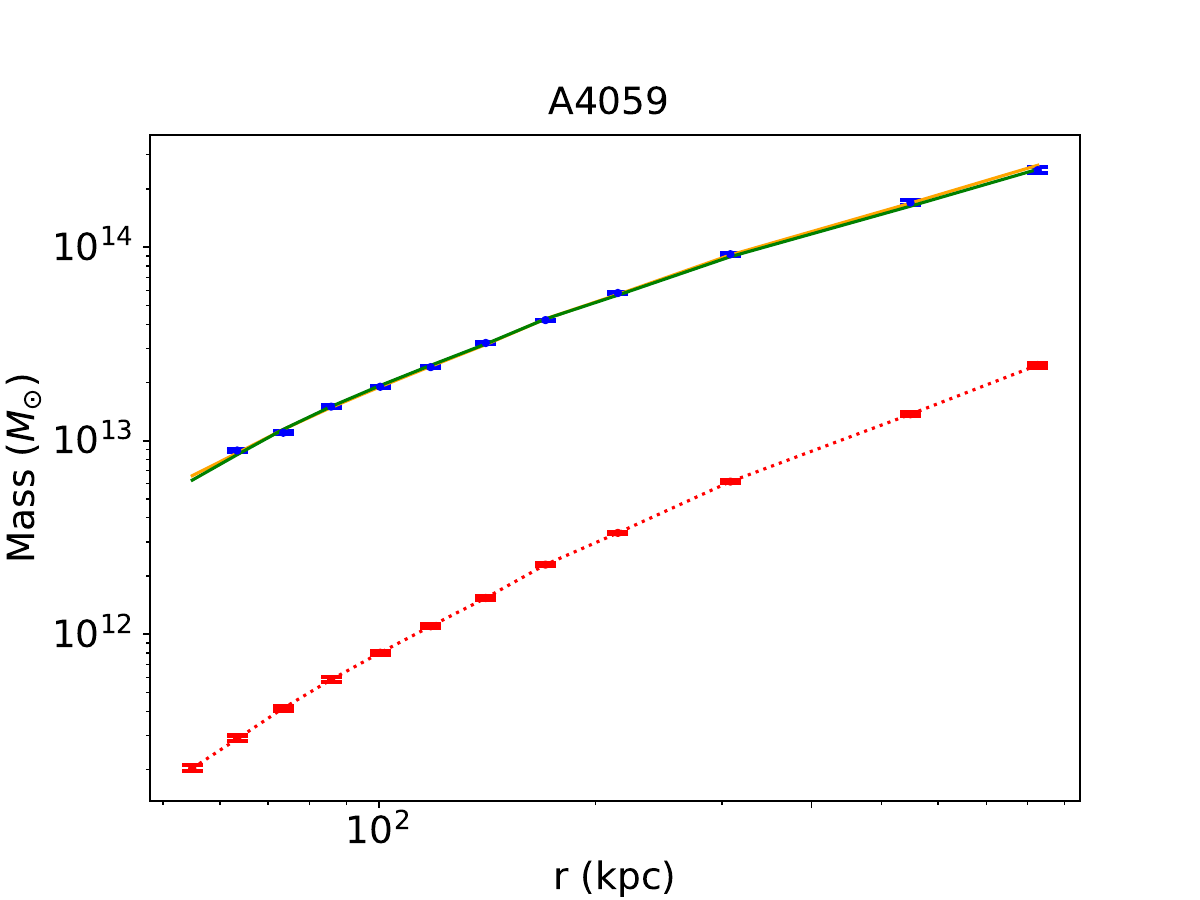}
\includegraphics[trim=0.0cm 0.0cm 1.0cm 0.0cm, clip=true, width=0.33\columnwidth]{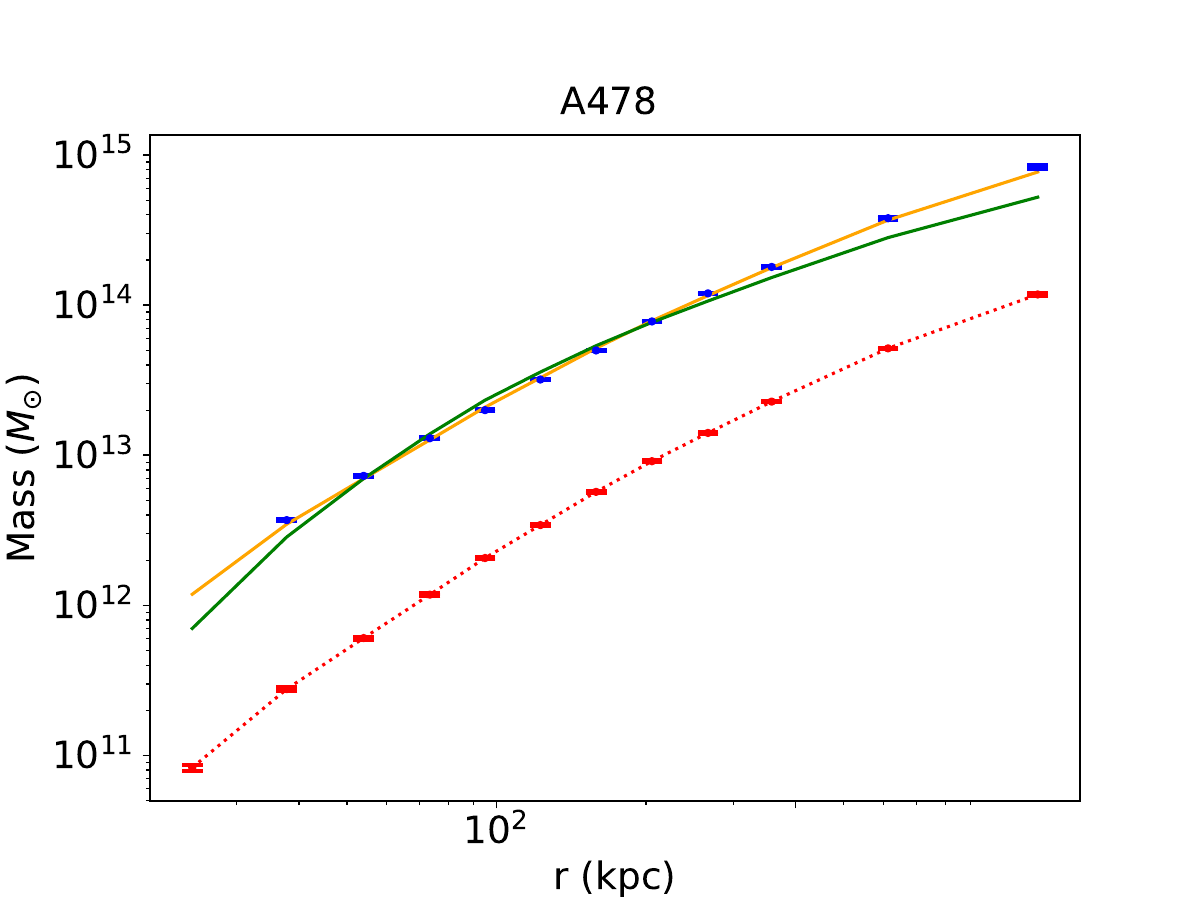}
\end{figure}
\begin{figure}
\includegraphics[trim=0.0cm 0.0cm 1.0cm 0.0cm, clip=true, width=0.33\columnwidth]{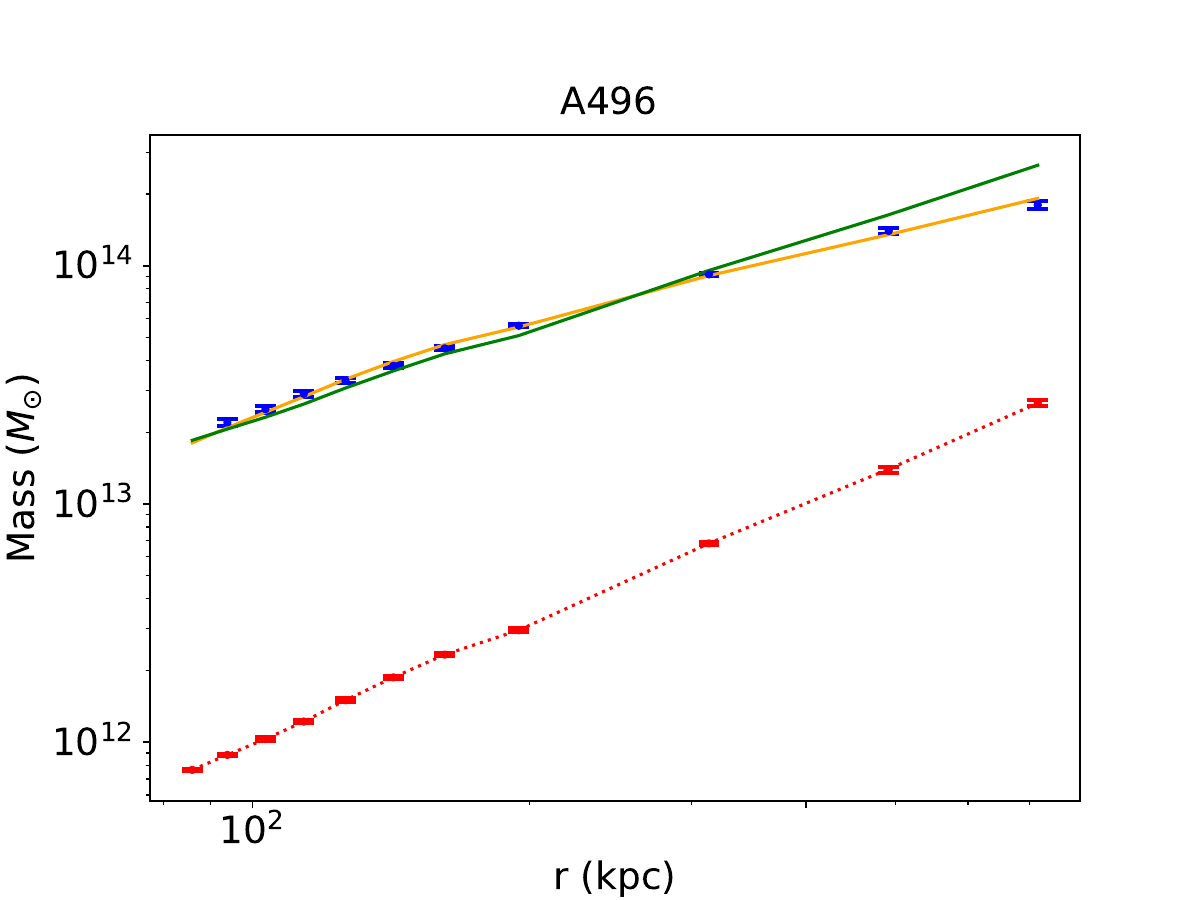}
\includegraphics[trim=0.0cm 0.0cm 1.0cm 0.0cm, clip=true, width=0.33\columnwidth]{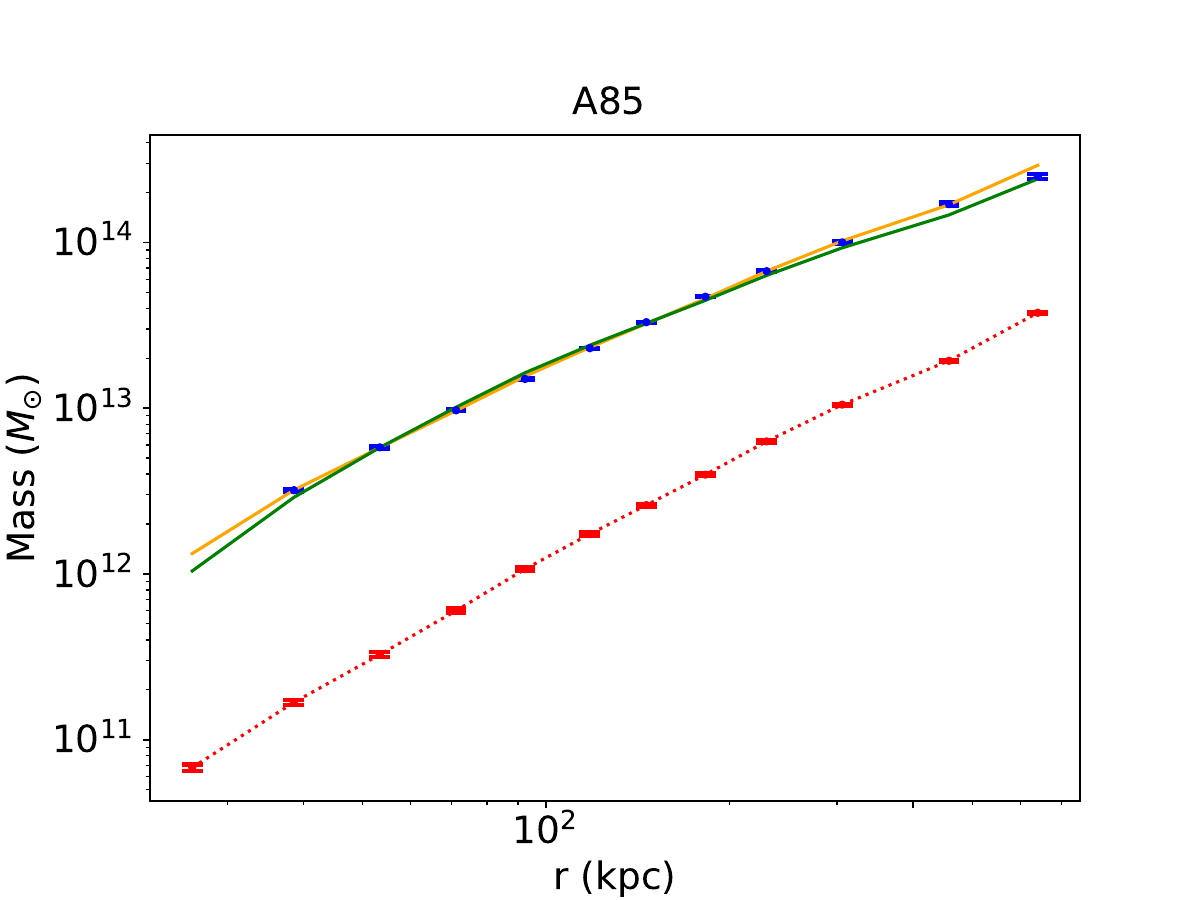}
\includegraphics[trim=0.0cm 0.0cm 1.0cm 0.0cm, clip=true, width=0.33\columnwidth]{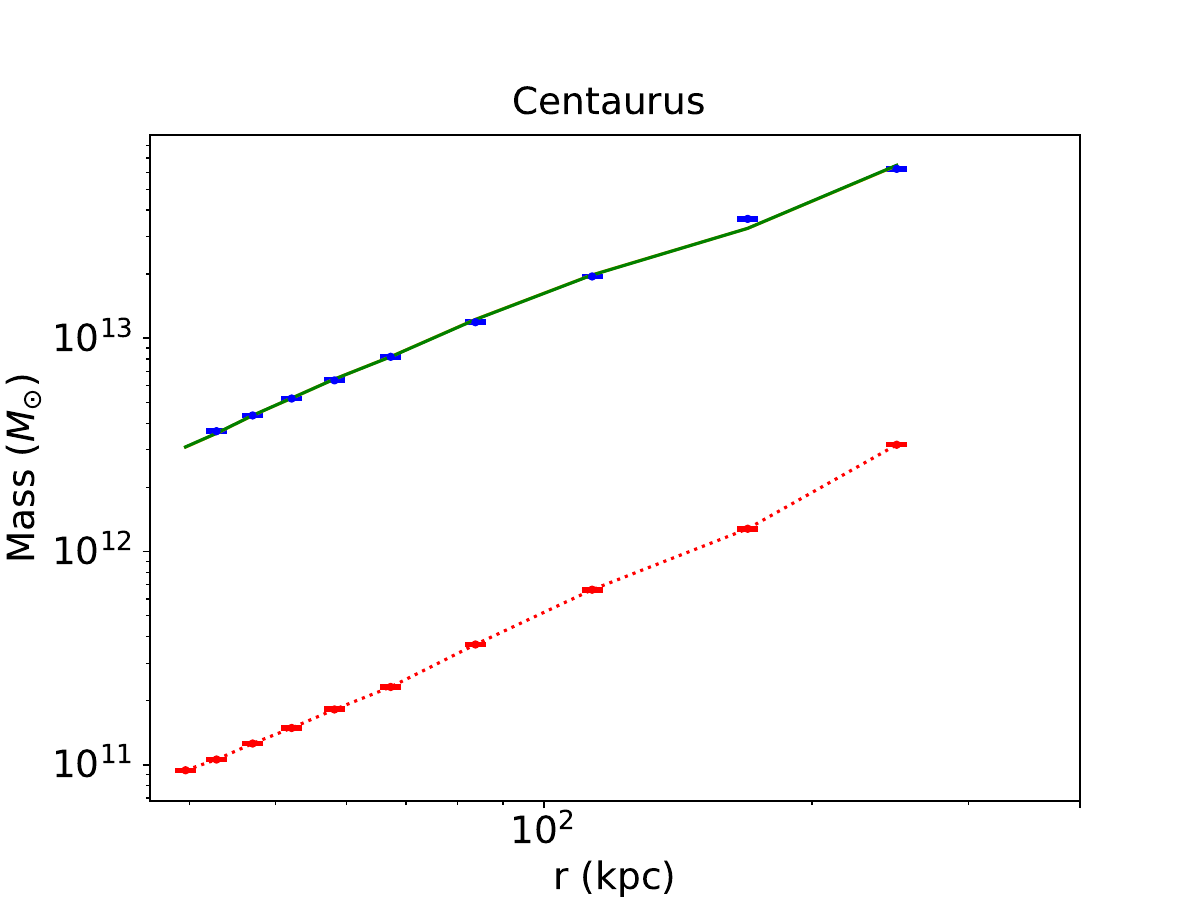}
\includegraphics[trim=0.0cm 0.0cm 1.0cm 0.0cm, clip=true, width=0.33\columnwidth]{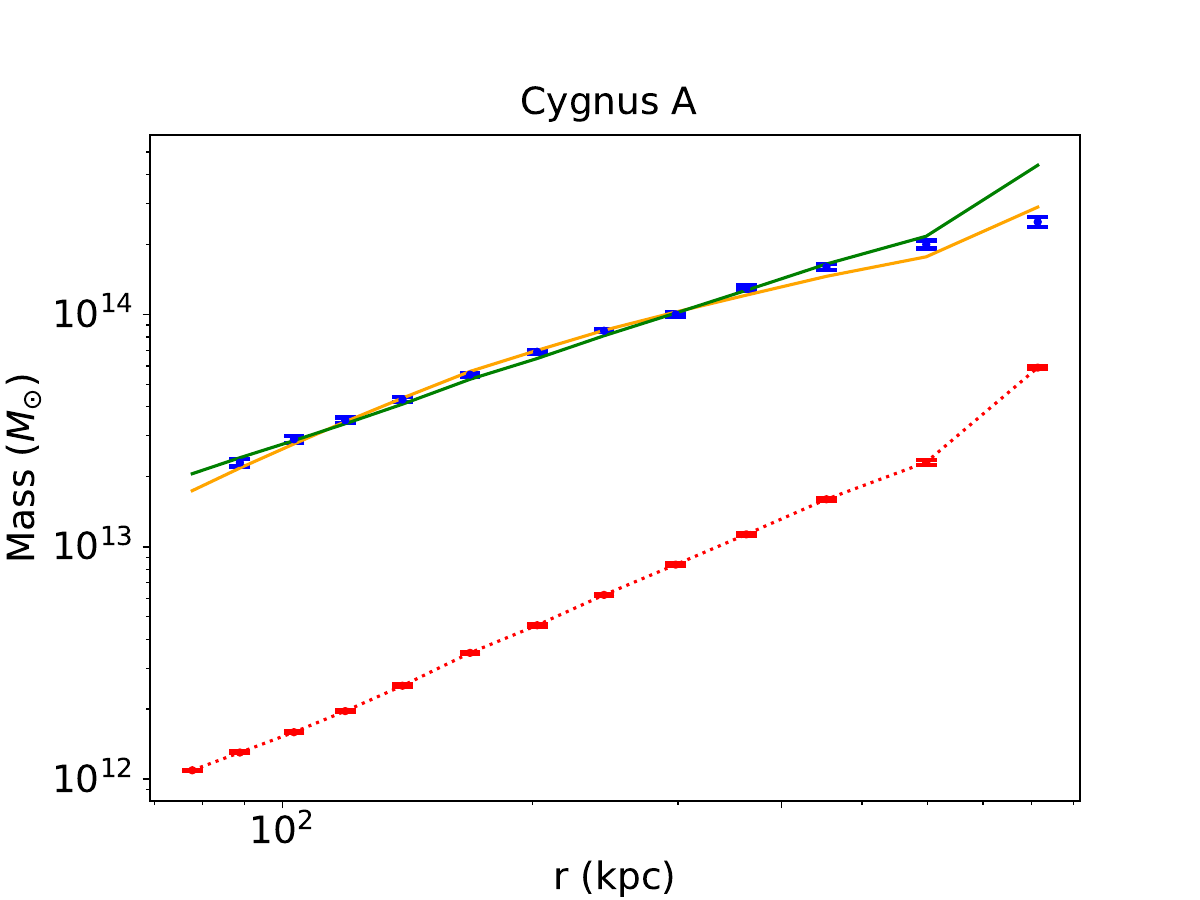}
\includegraphics[trim=0.0cm 0.0cm 1.0cm 0.0cm, clip=true, width=0.33\columnwidth]{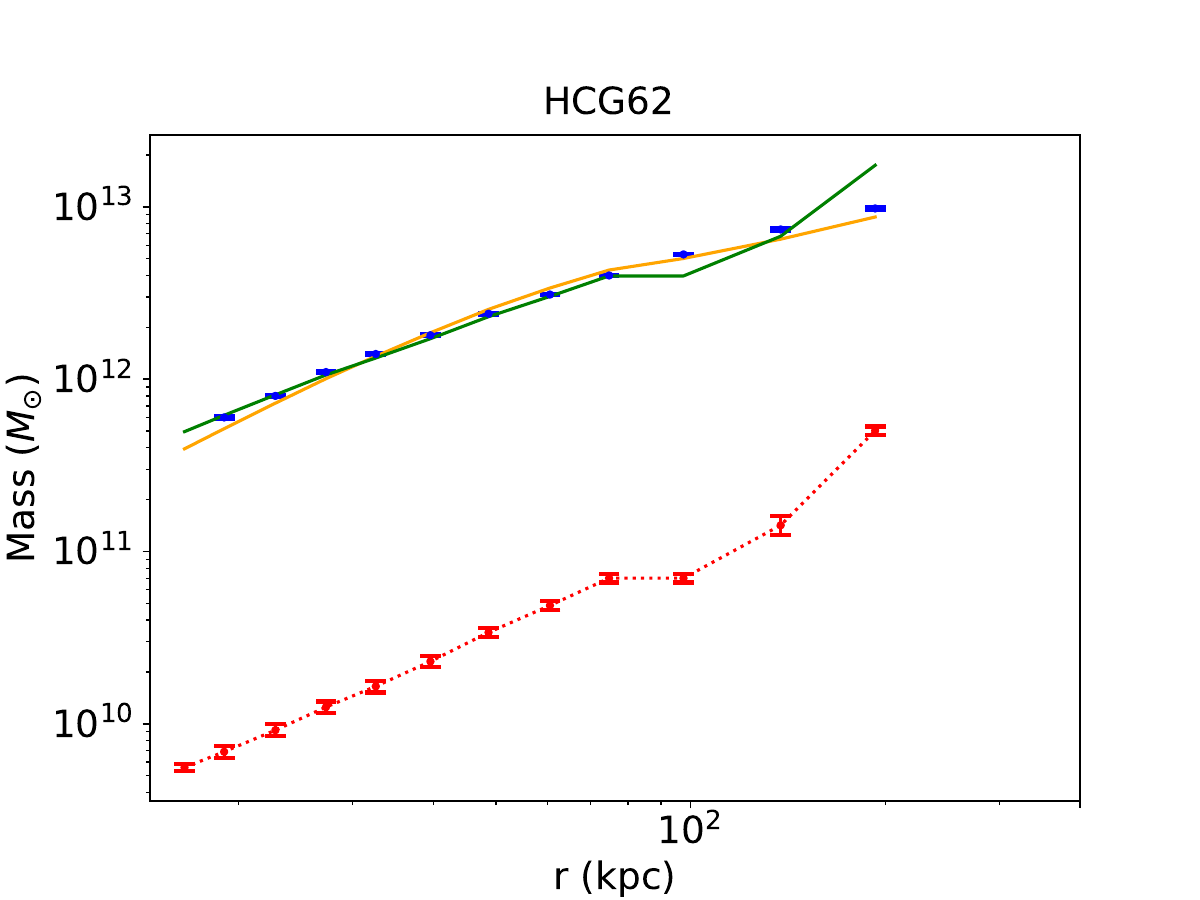}
\includegraphics[trim=0.0cm 0.0cm 1.0cm 0.0cm, clip=true, width=0.33\columnwidth]{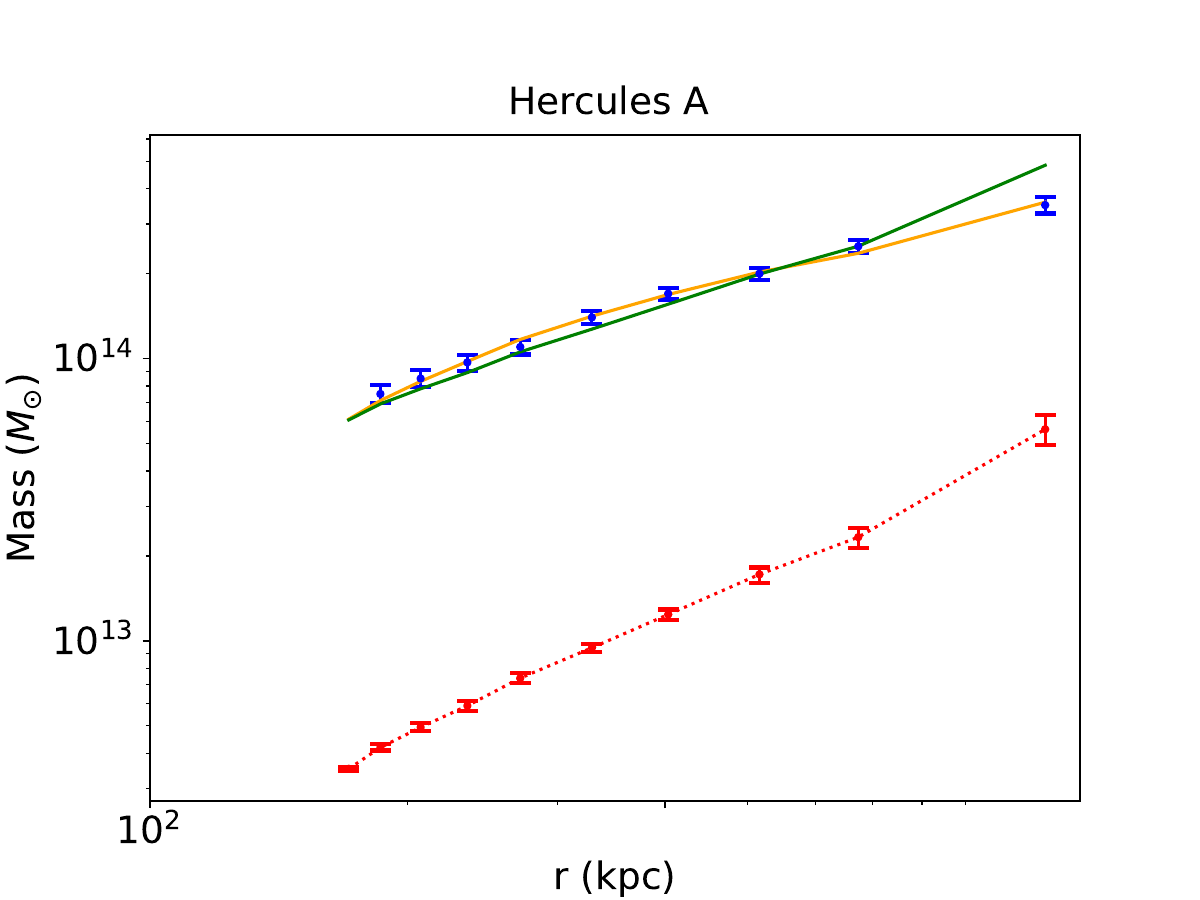}
\includegraphics[trim=0.0cm 0.0cm 1.0cm 0.0cm, clip=true, width=0.33\columnwidth]{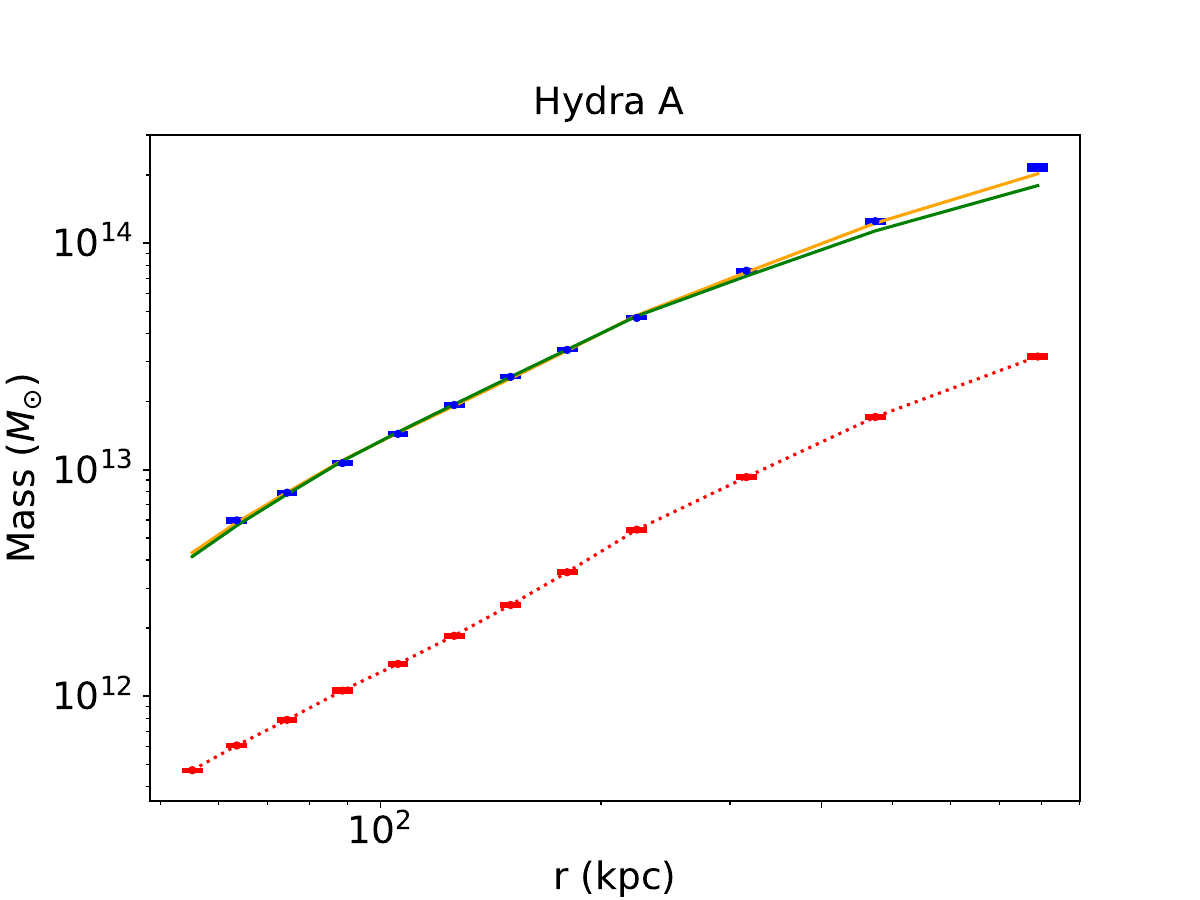}
\includegraphics[trim=0.0cm 0.0cm 1.0cm 0.0cm, clip=true, width=0.33\columnwidth]{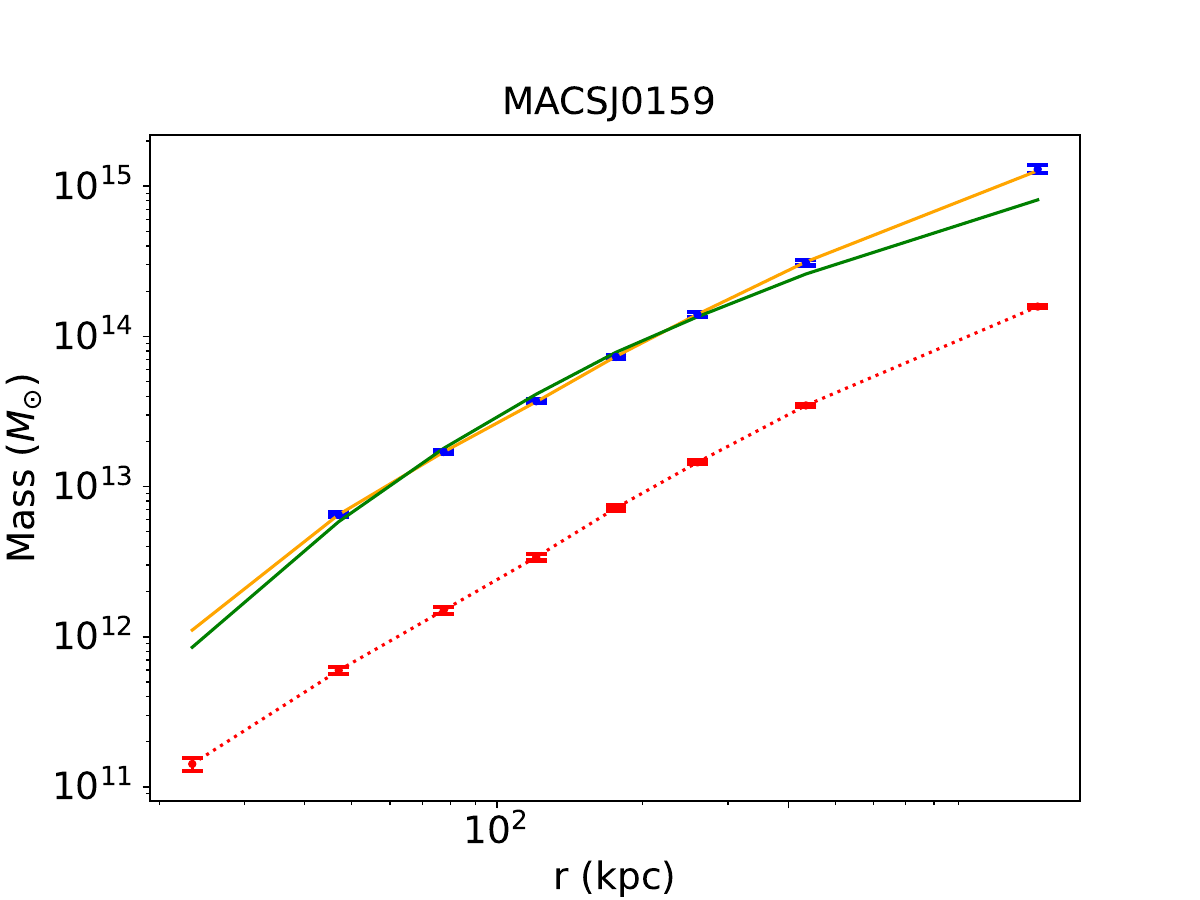}
\includegraphics[trim=0.0cm 0.0cm 1.0cm 0.0cm, clip=true, width=0.33\columnwidth]{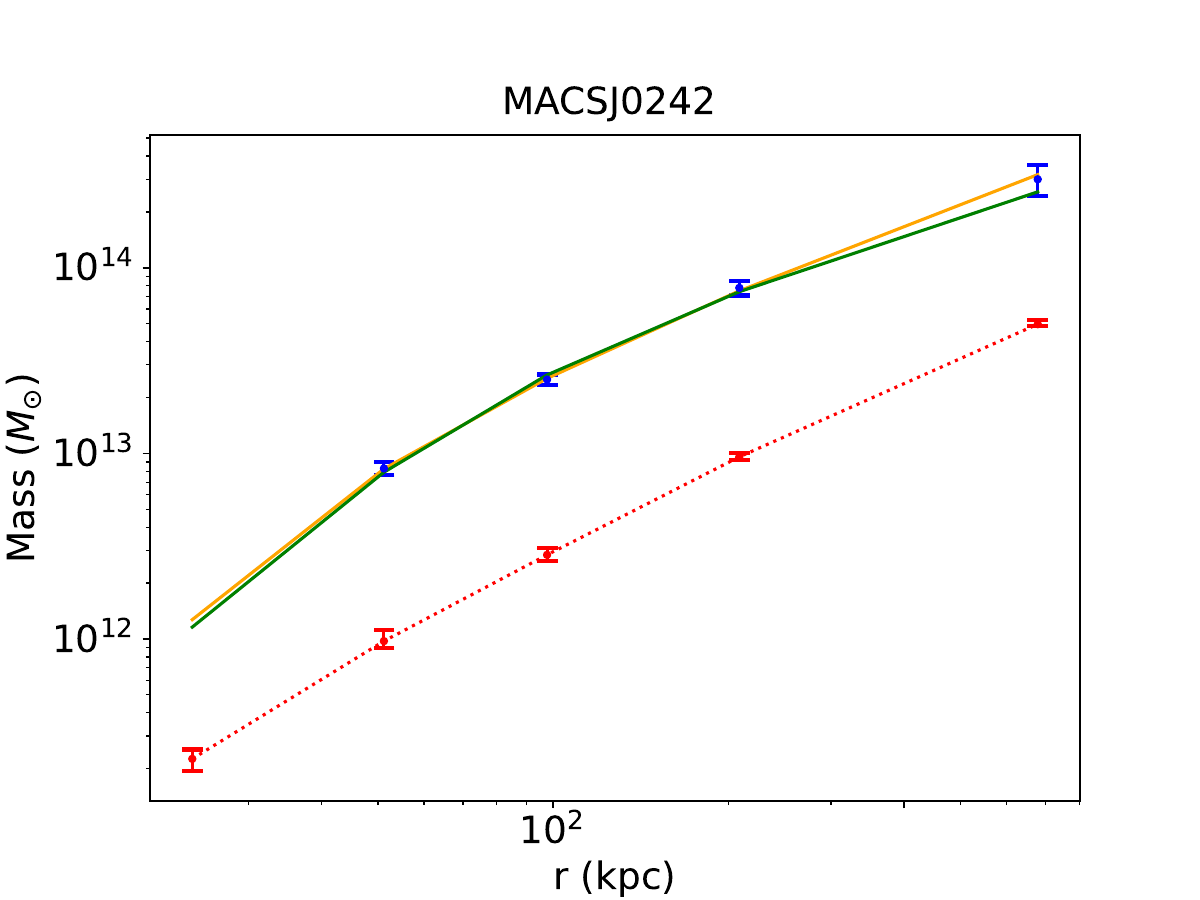}
\includegraphics[trim=0.0cm 0.0cm 1.0cm 0.0cm, clip=true, width=0.33\columnwidth]{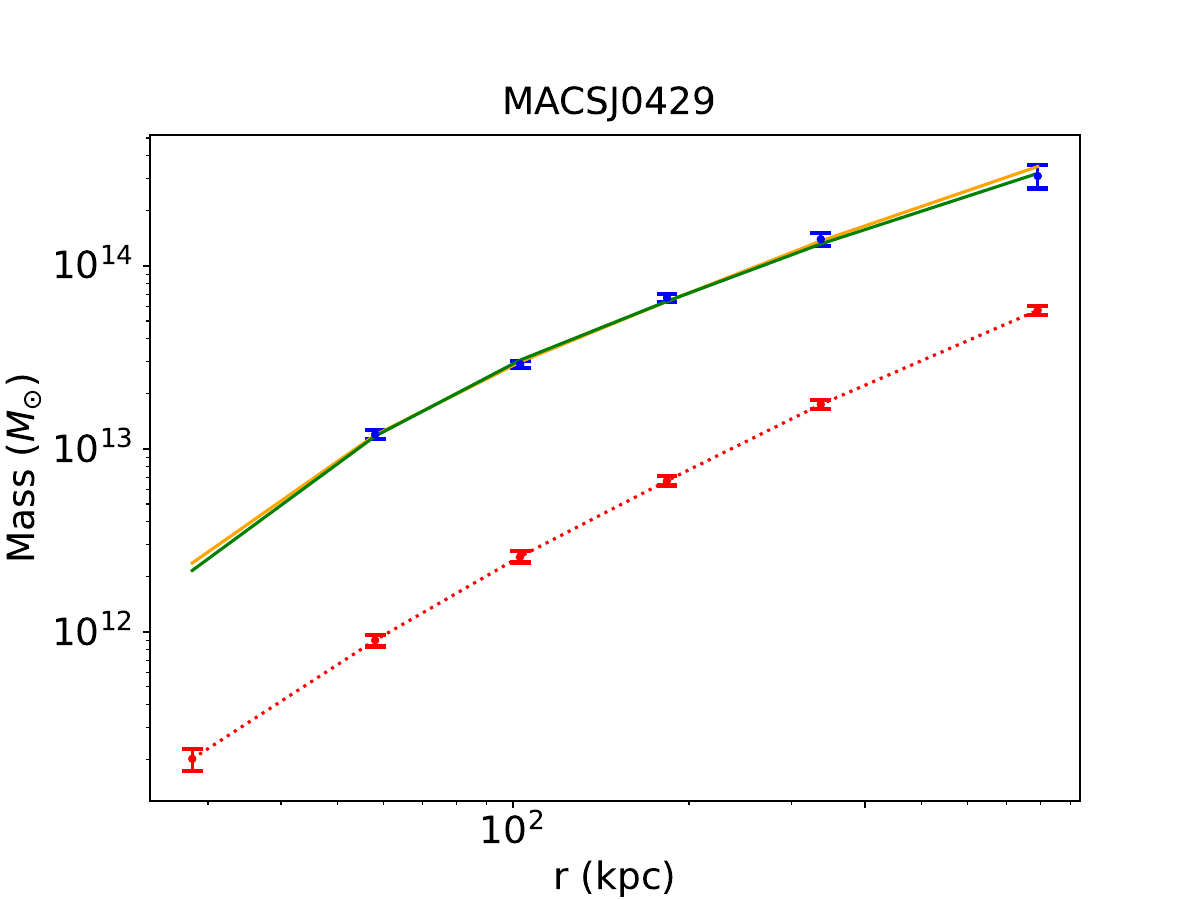}
\includegraphics[trim=0.0cm 0.0cm 1.0cm 0.0cm, clip=true, width=0.33\columnwidth]{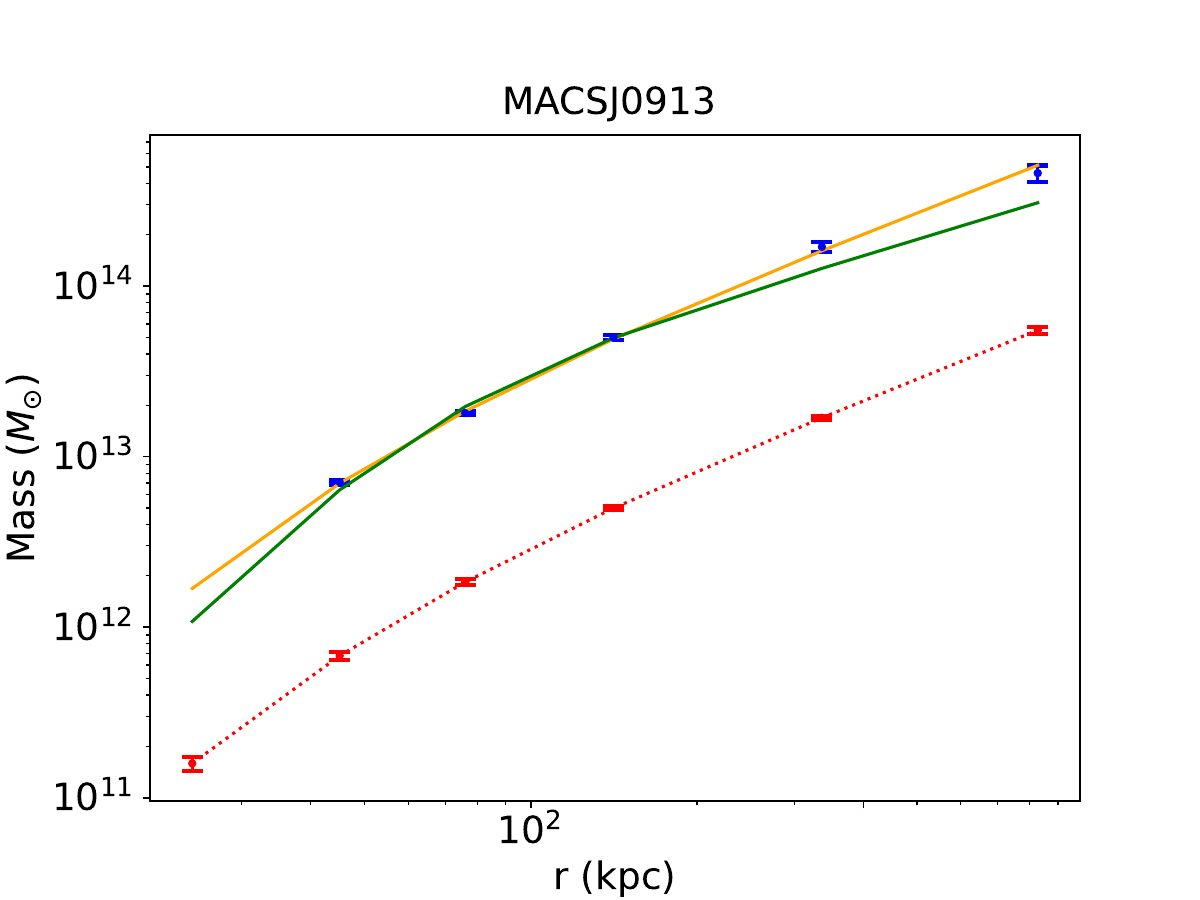}
\includegraphics[trim=0.0cm 0.0cm 1.0cm 0.0cm, clip=true, width=0.33\columnwidth]{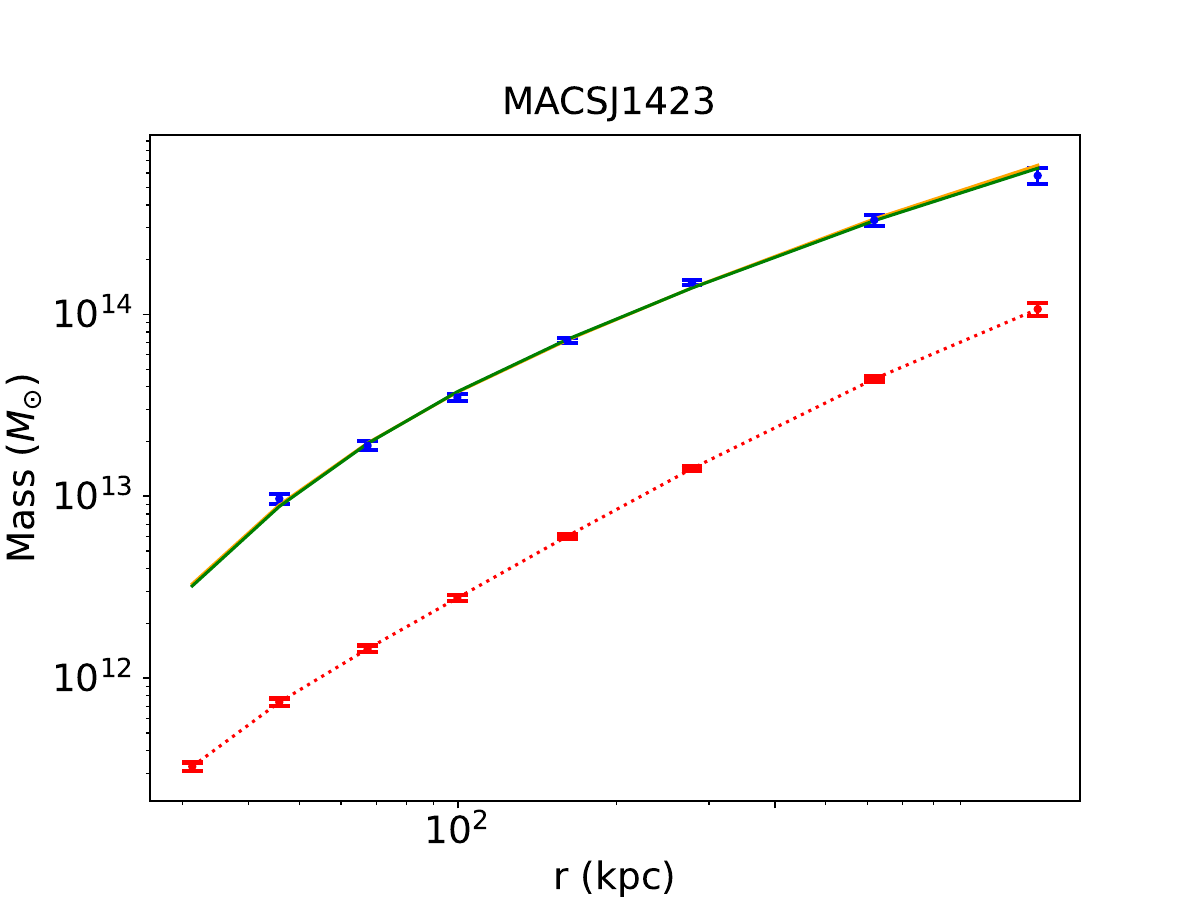}
\includegraphics[trim=0.0cm 0.0cm 1.0cm 0.0cm, clip=true, width=0.33\columnwidth]{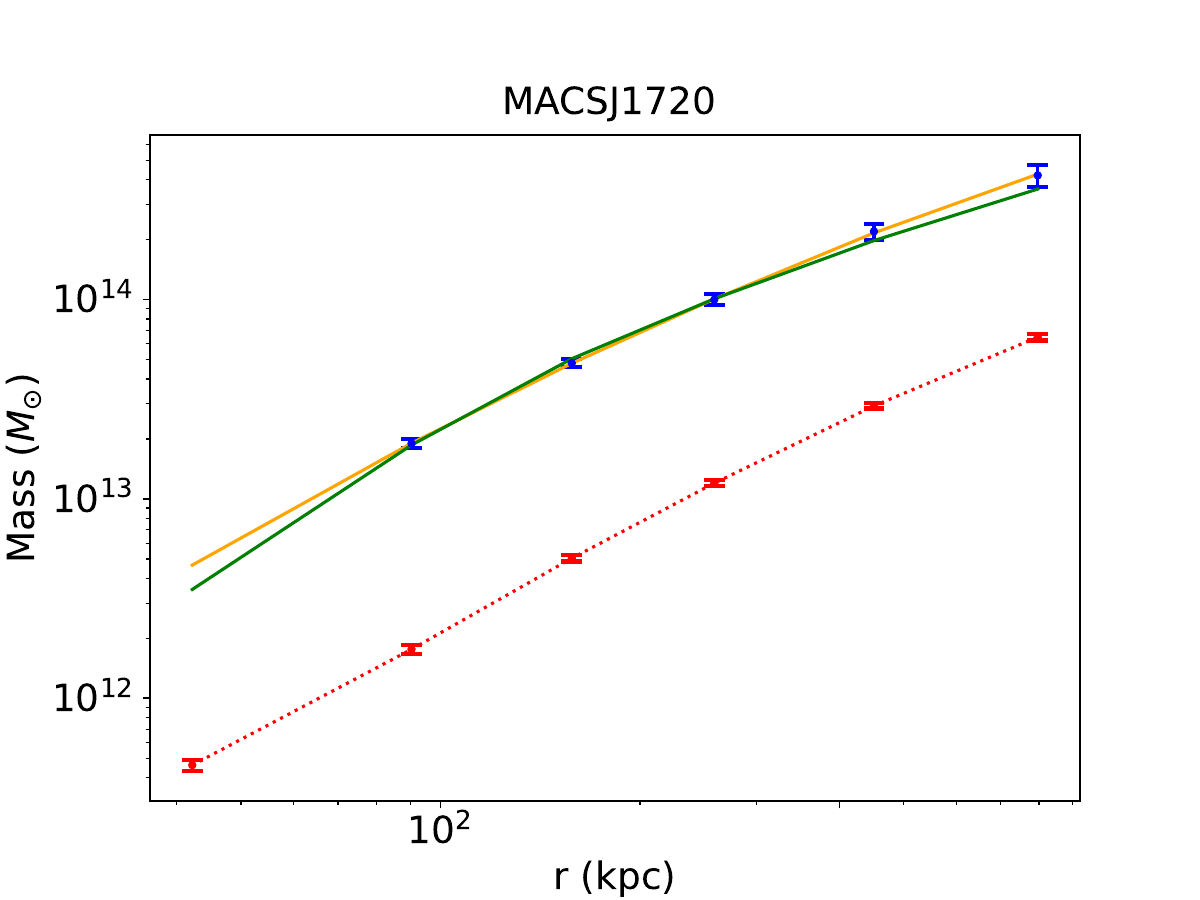}
\includegraphics[trim=0.0cm 0.0cm 1.0cm 0.0cm, clip=true, width=0.33\columnwidth]{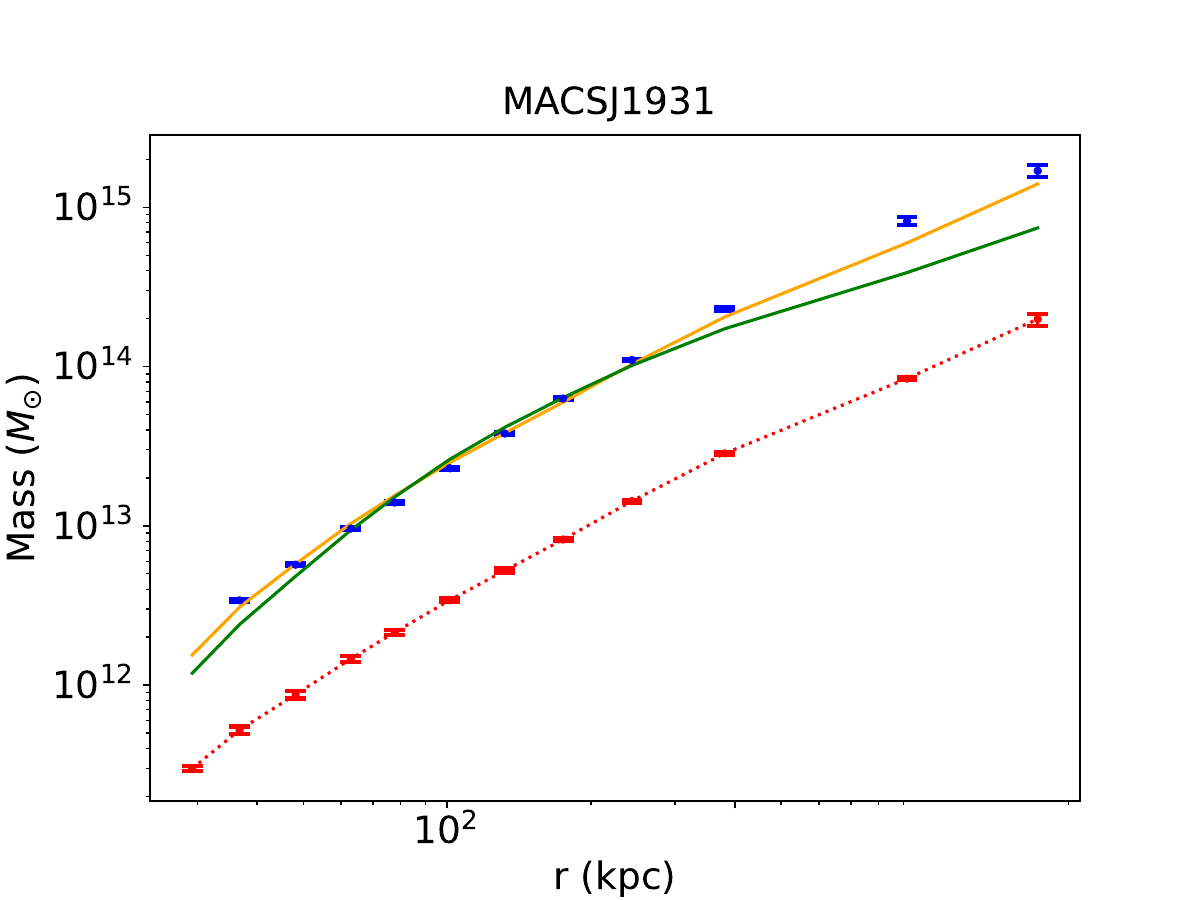}
\includegraphics[trim=0.0cm 0.0cm 1.0cm 0.0cm, clip=true, width=0.33\columnwidth]{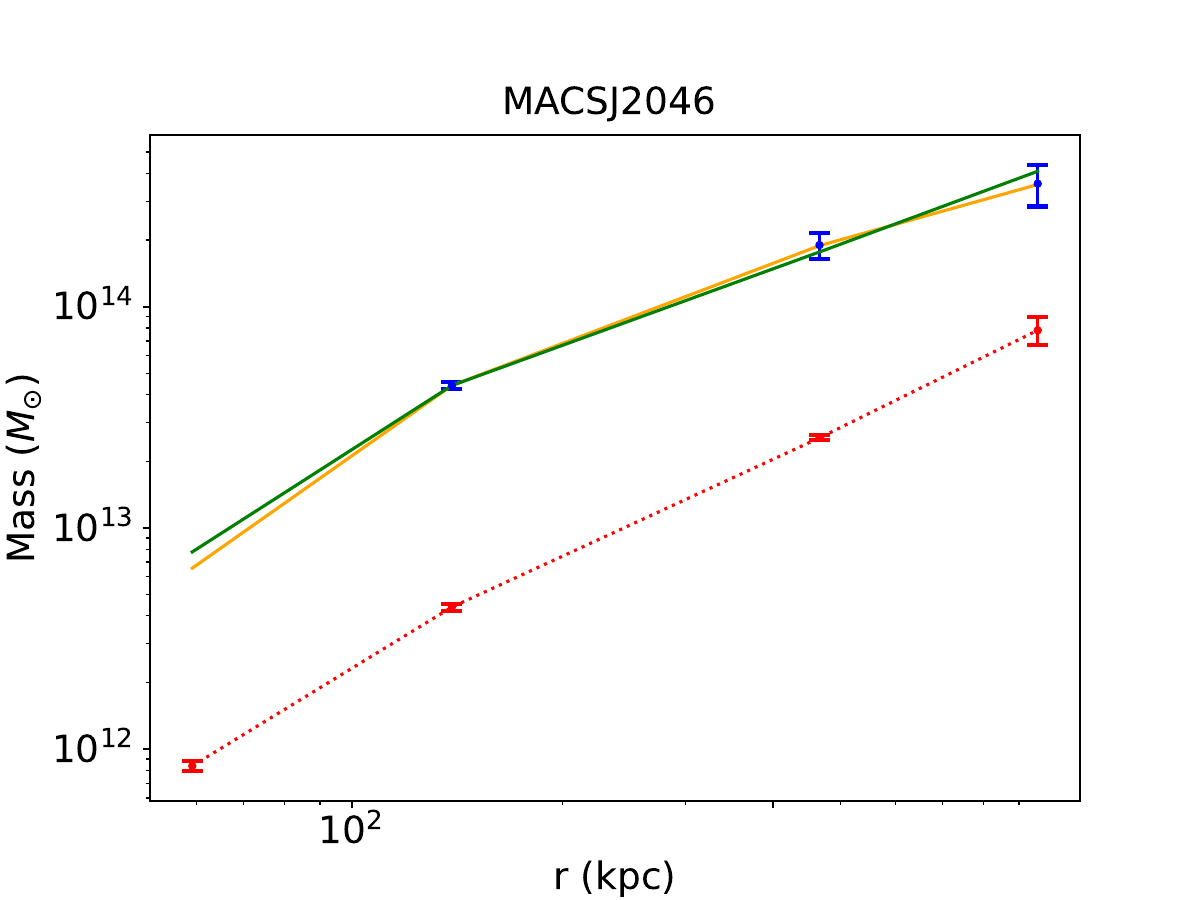}
\end{figure}
\begin{figure}
\includegraphics[trim=0.0cm 0.0cm 1.0cm 0.0cm, clip=true, width=0.33\columnwidth]{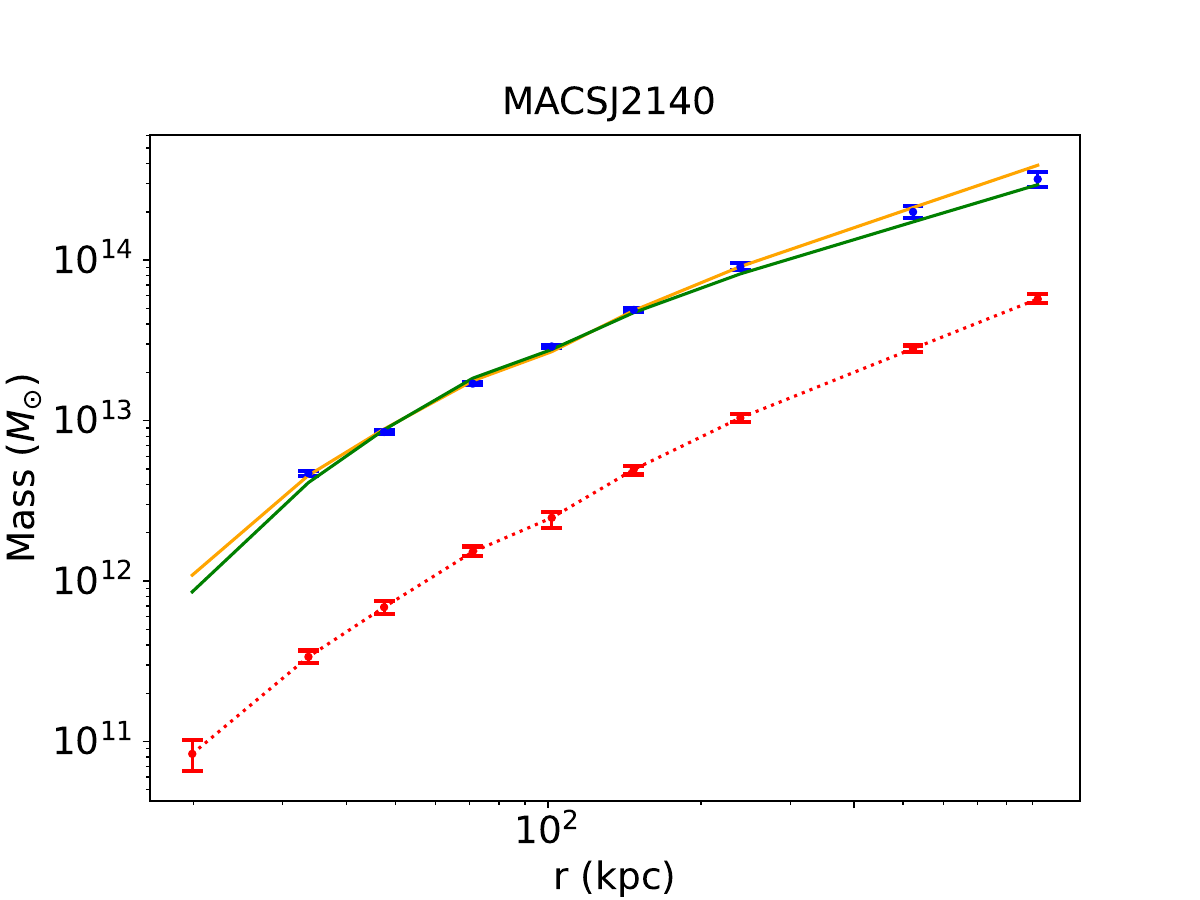}
\includegraphics[trim=0.0cm 0.0cm 1.0cm 0.0cm, clip=true, width=0.33\columnwidth]{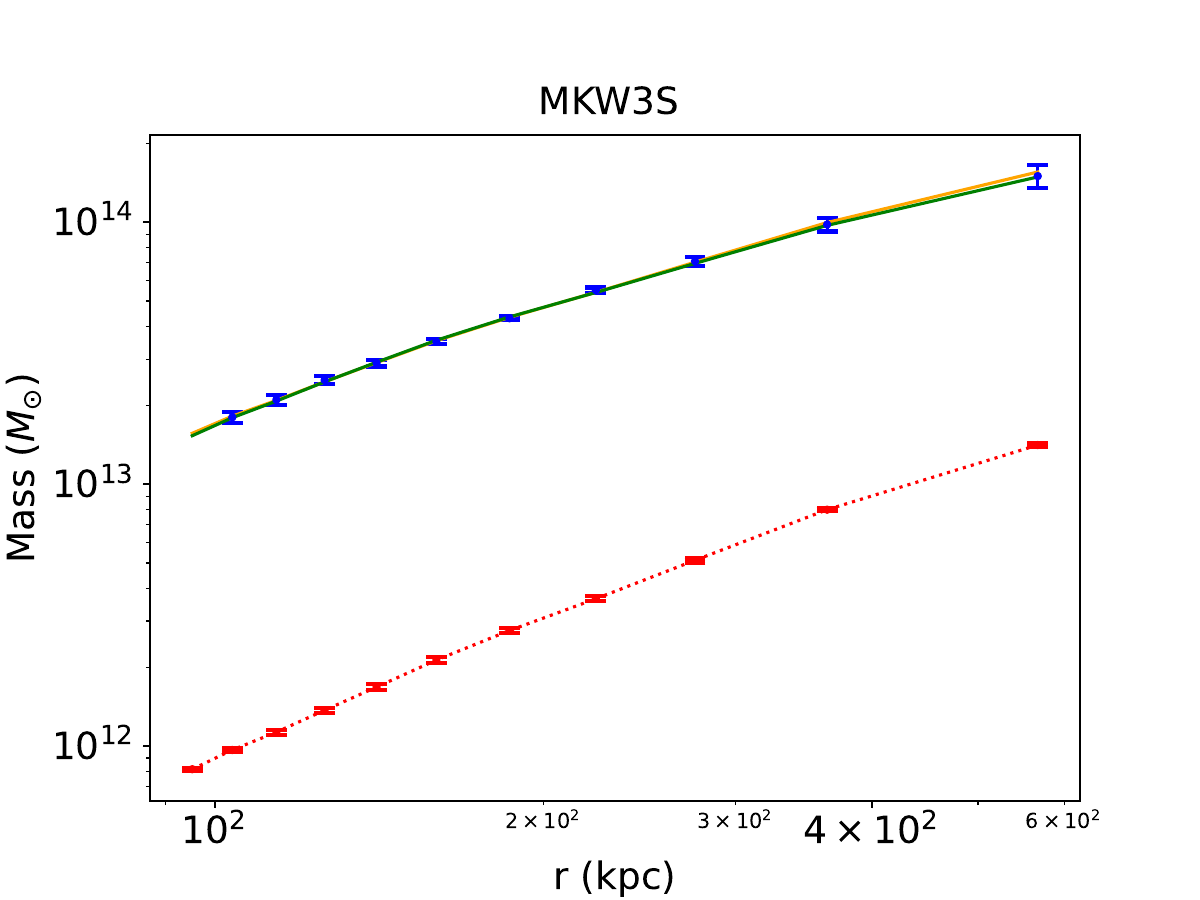}
\includegraphics[trim=0.0cm 0.0cm 1.0cm 0.0cm, clip=true, width=0.33\columnwidth]{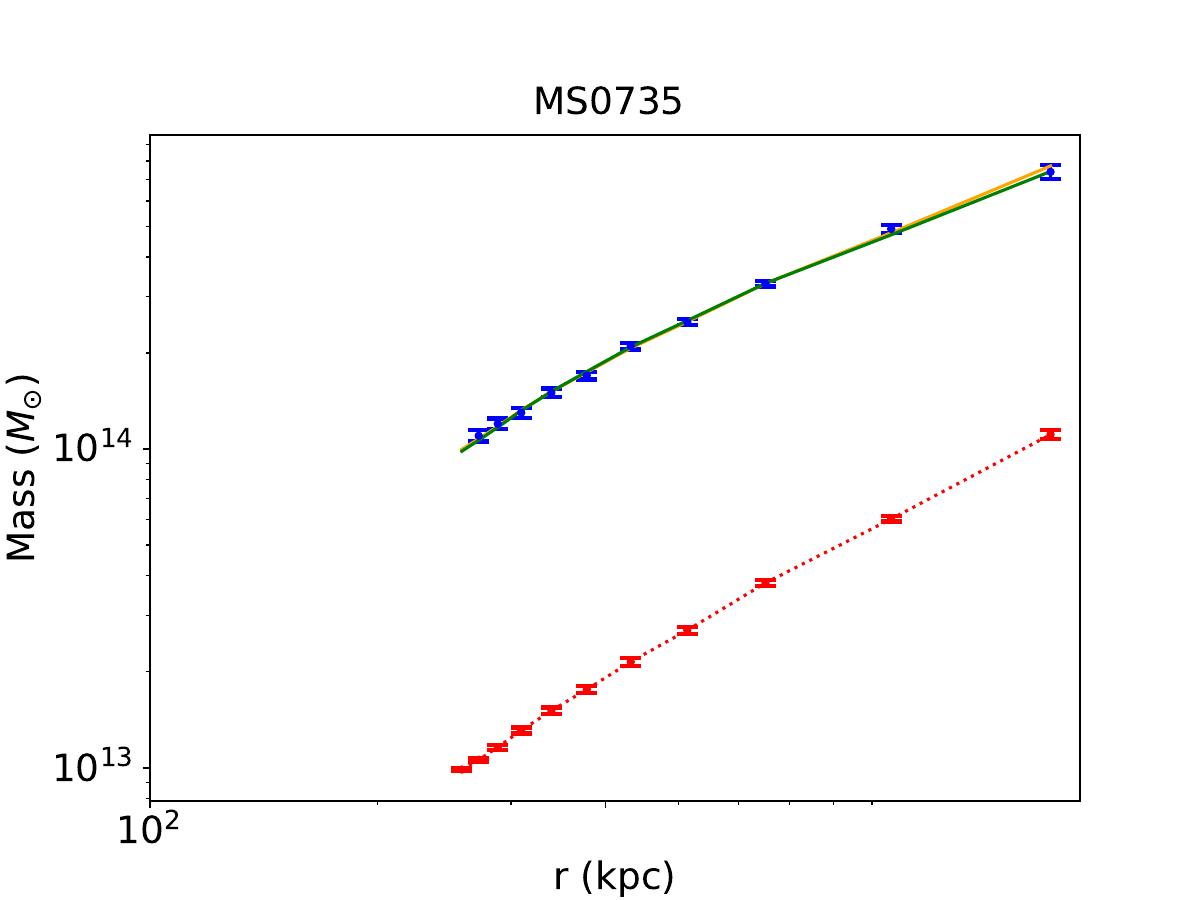}
\includegraphics[trim=0.0cm 0.0cm 1.0cm 0.0cm, clip=true, width=0.33\columnwidth]{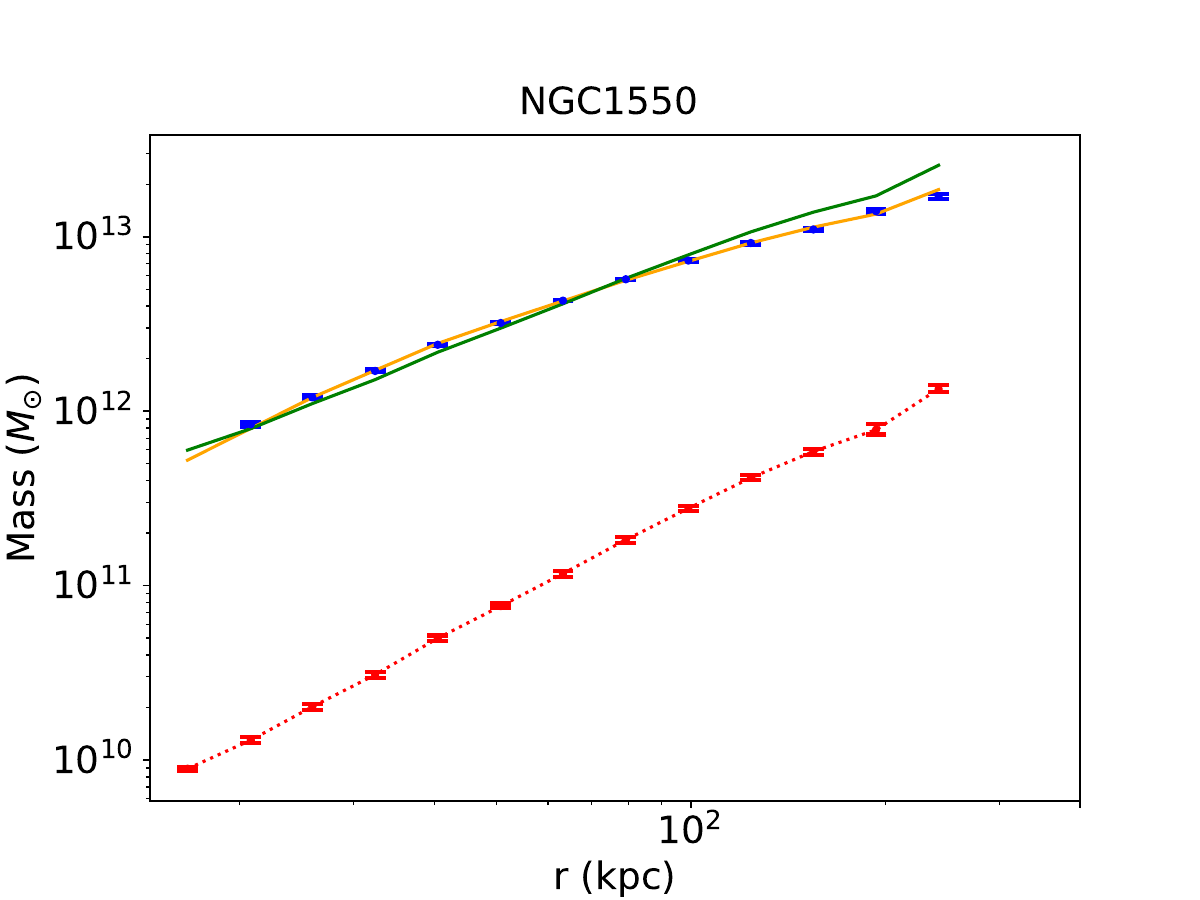}
\includegraphics[trim=0.0cm 0.0cm 1.0cm 0.0cm, clip=true, width=0.33\columnwidth]{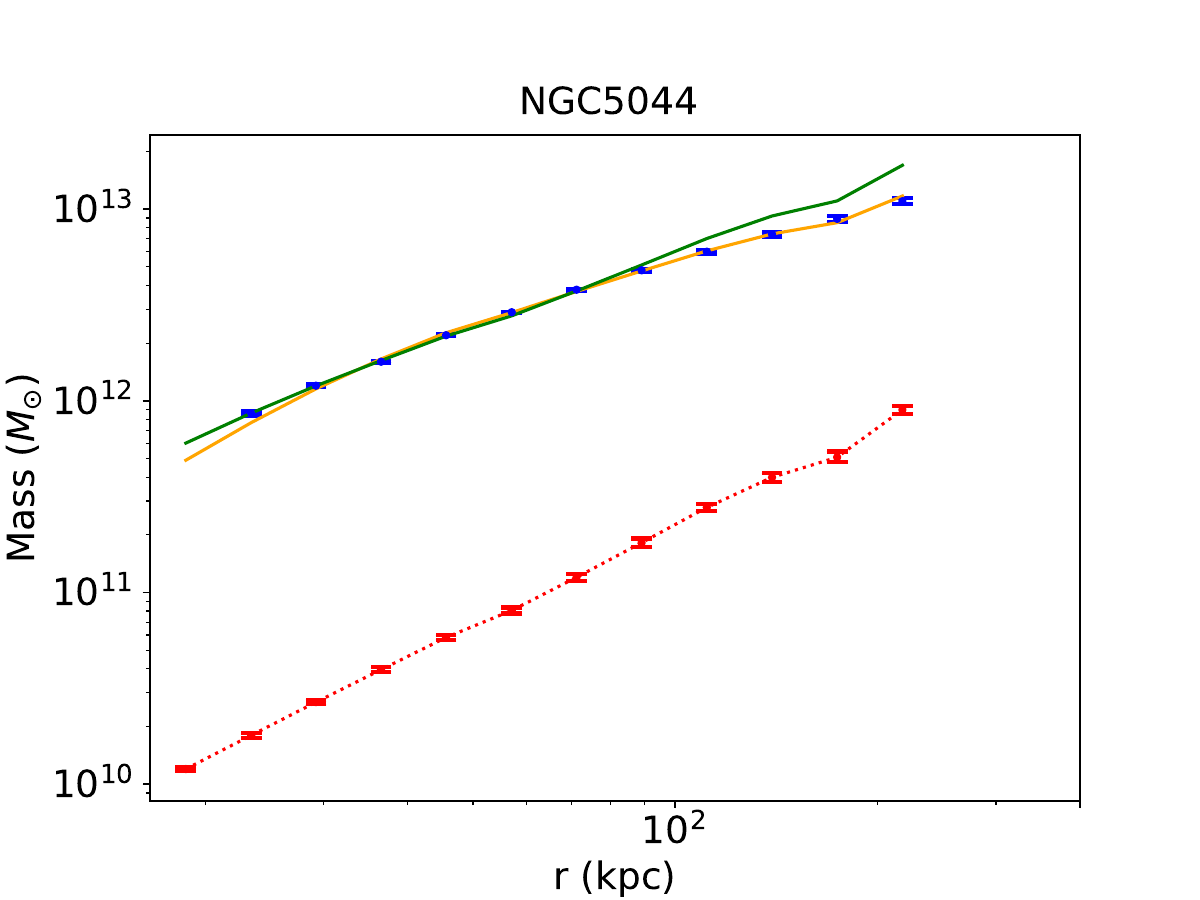}
\includegraphics[trim=0.0cm 0.0cm 1.0cm 0.0cm, clip=true, width=0.33\columnwidth]{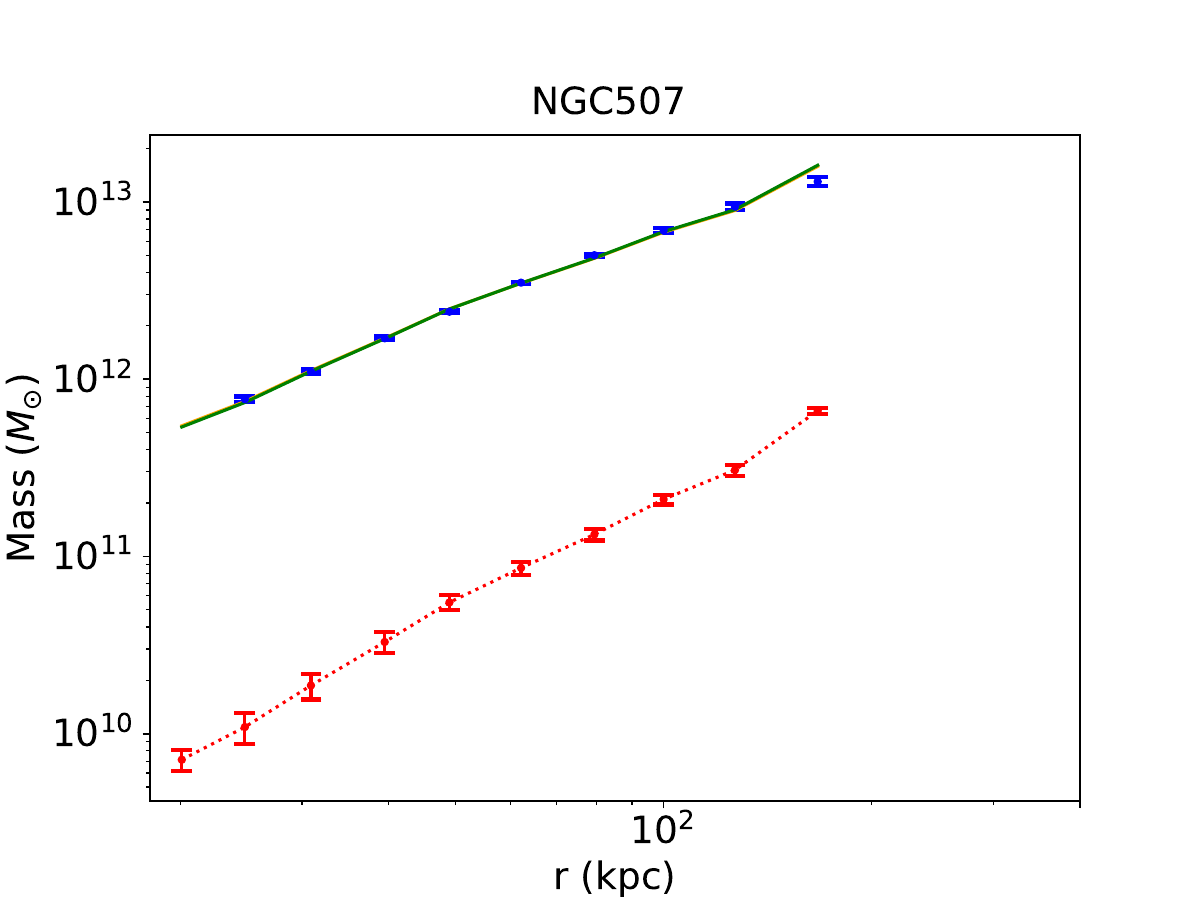}
\includegraphics[trim=0.0cm 0.0cm 1.0cm 0.0cm, clip=true, width=0.33\columnwidth]{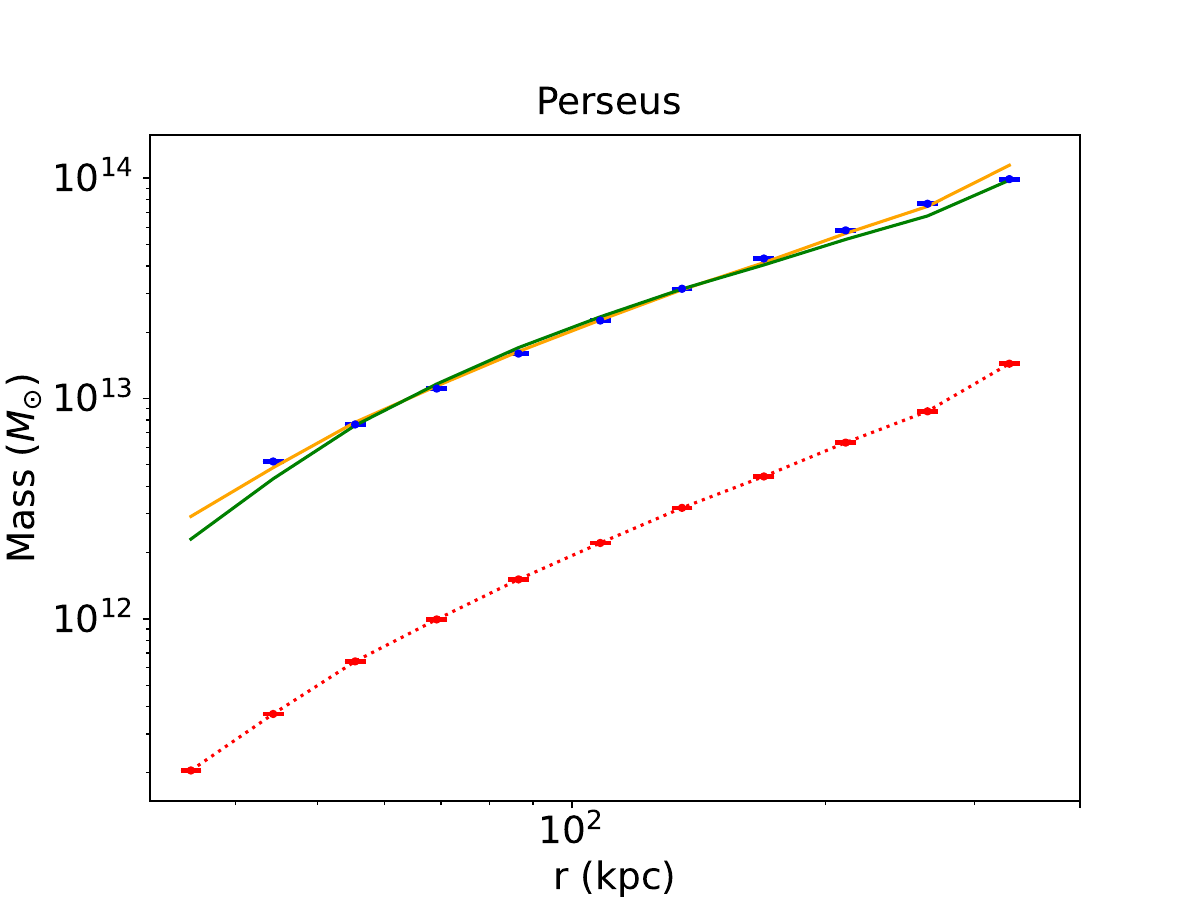}
\includegraphics[trim=0.0cm 0.0cm 1.0cm 0.0cm, clip=true, width=0.33\columnwidth]{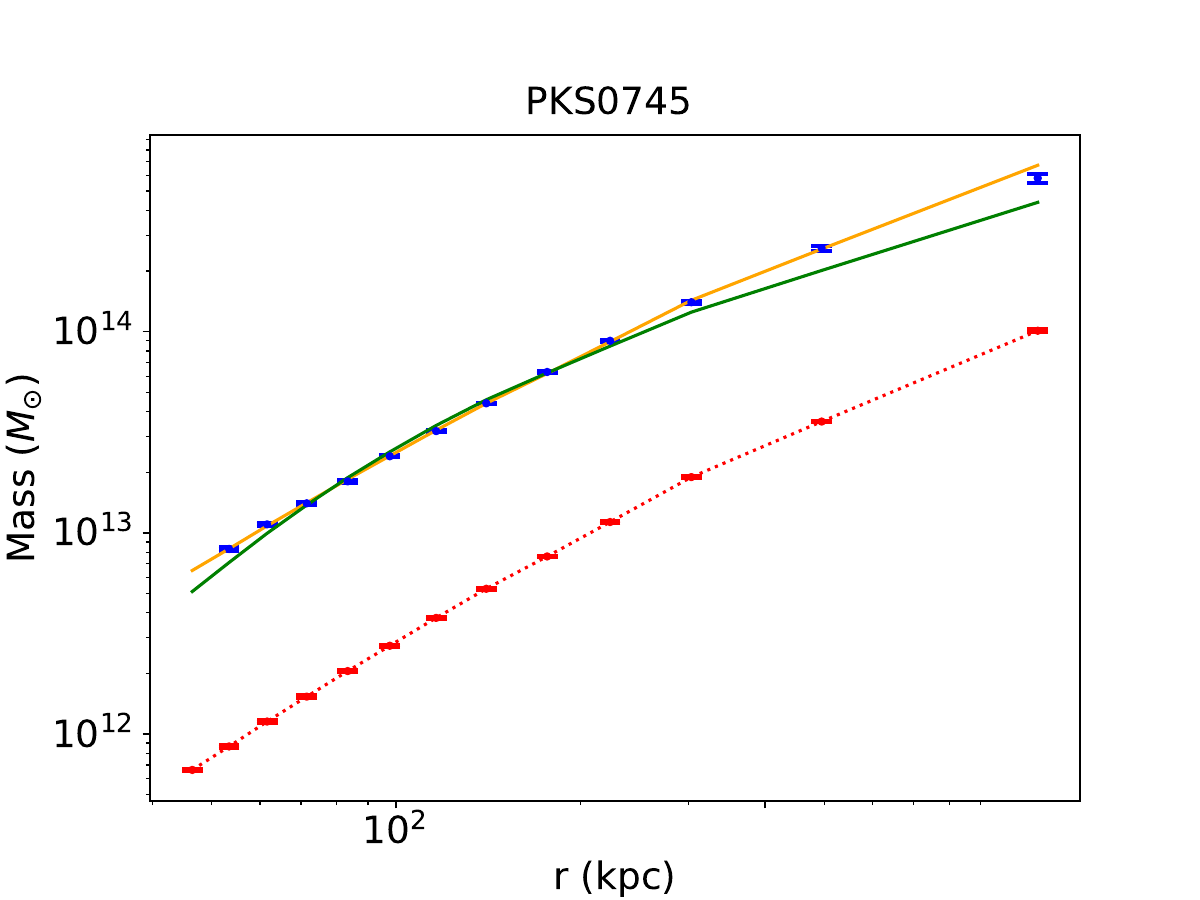}
\includegraphics[trim=0.0cm 0.0cm 1.0cm 0.0cm, clip=true, width=0.33\columnwidth]{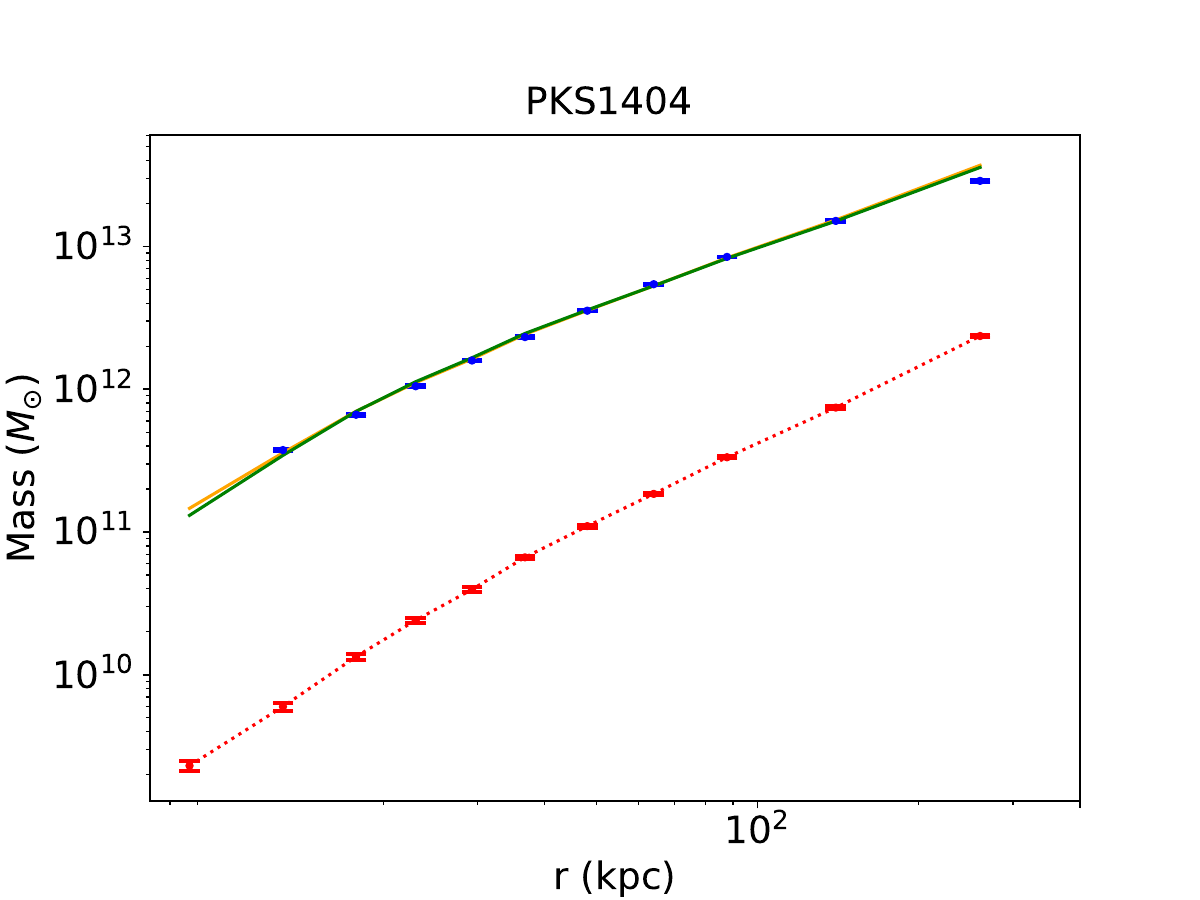}
\includegraphics[trim=0.0cm 0.0cm 1.0cm 0.0cm, clip=true, width=0.33\columnwidth]{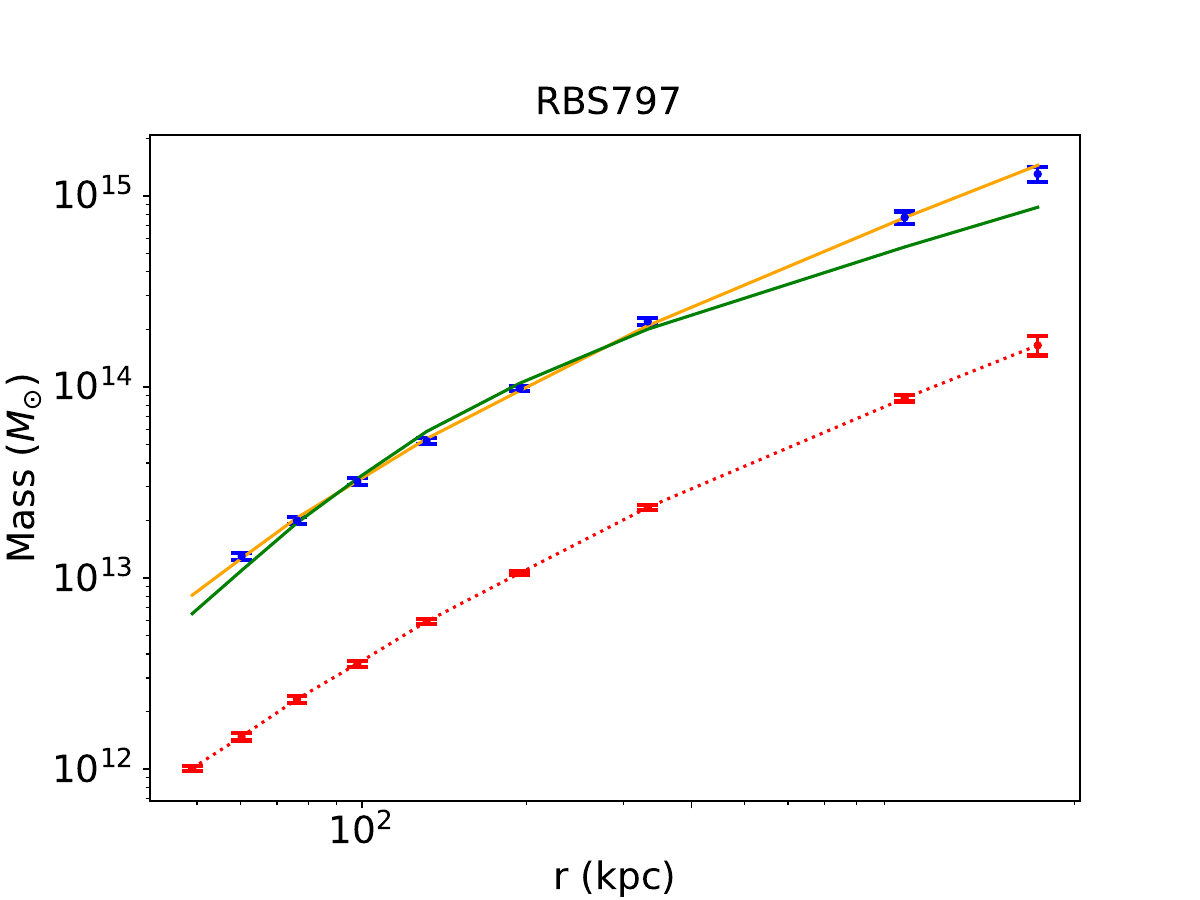}
\includegraphics[trim=0.0cm 0.0cm 1.0cm 0.0cm, clip=true, width=0.33\columnwidth]{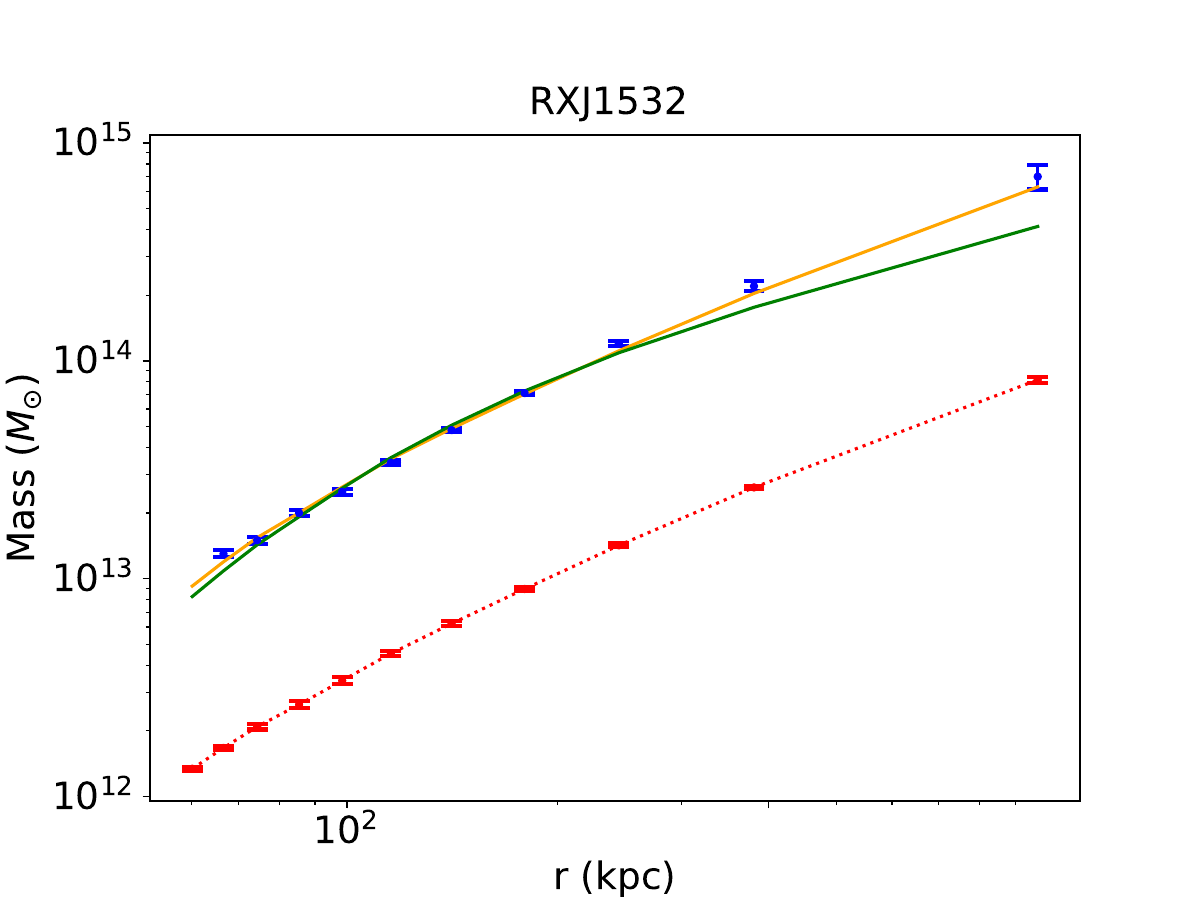}
\includegraphics[trim=0.0cm 0.0cm 1.0cm 0.0cm, clip=true, width=0.33\columnwidth]{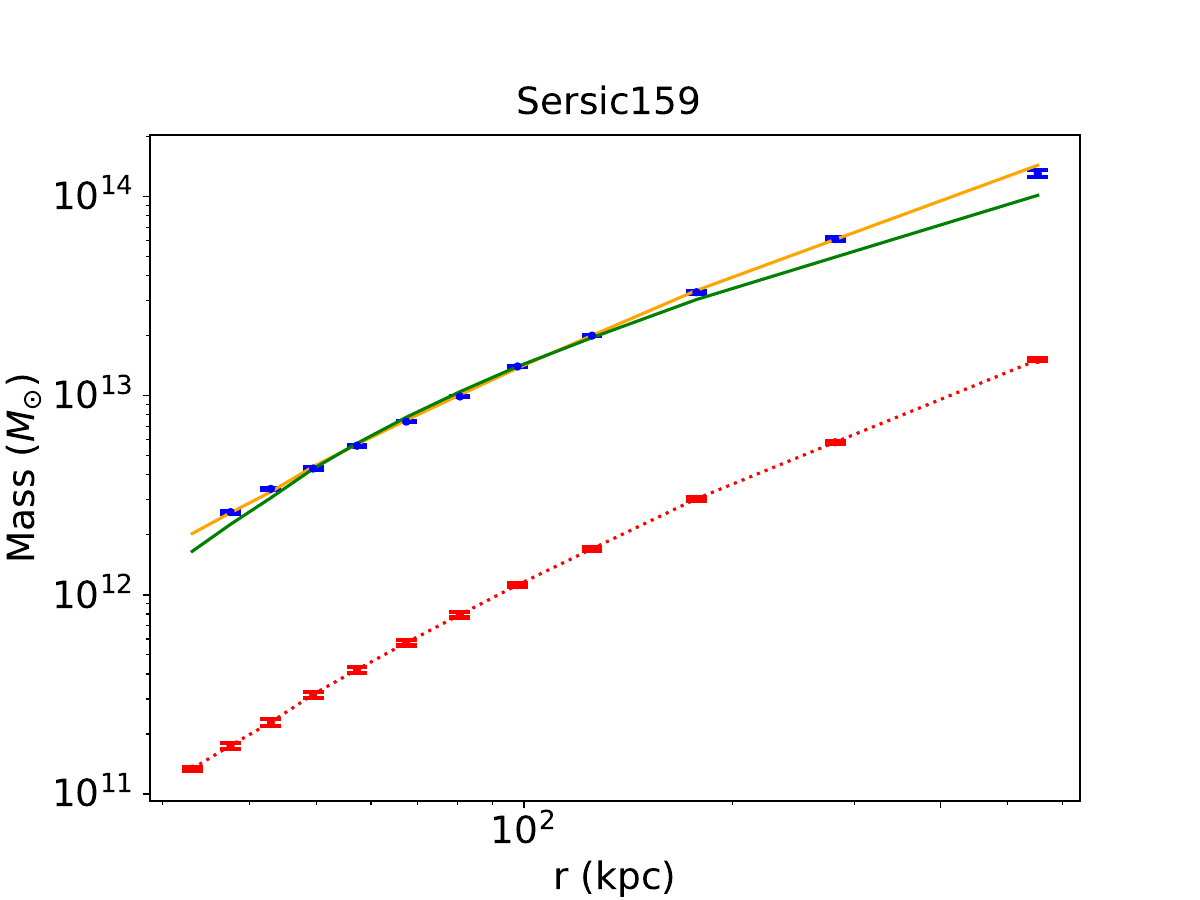}
\includegraphics[trim=0.0cm 0.0cm 1.0cm 0.0cm, clip=true, width=0.33\columnwidth]{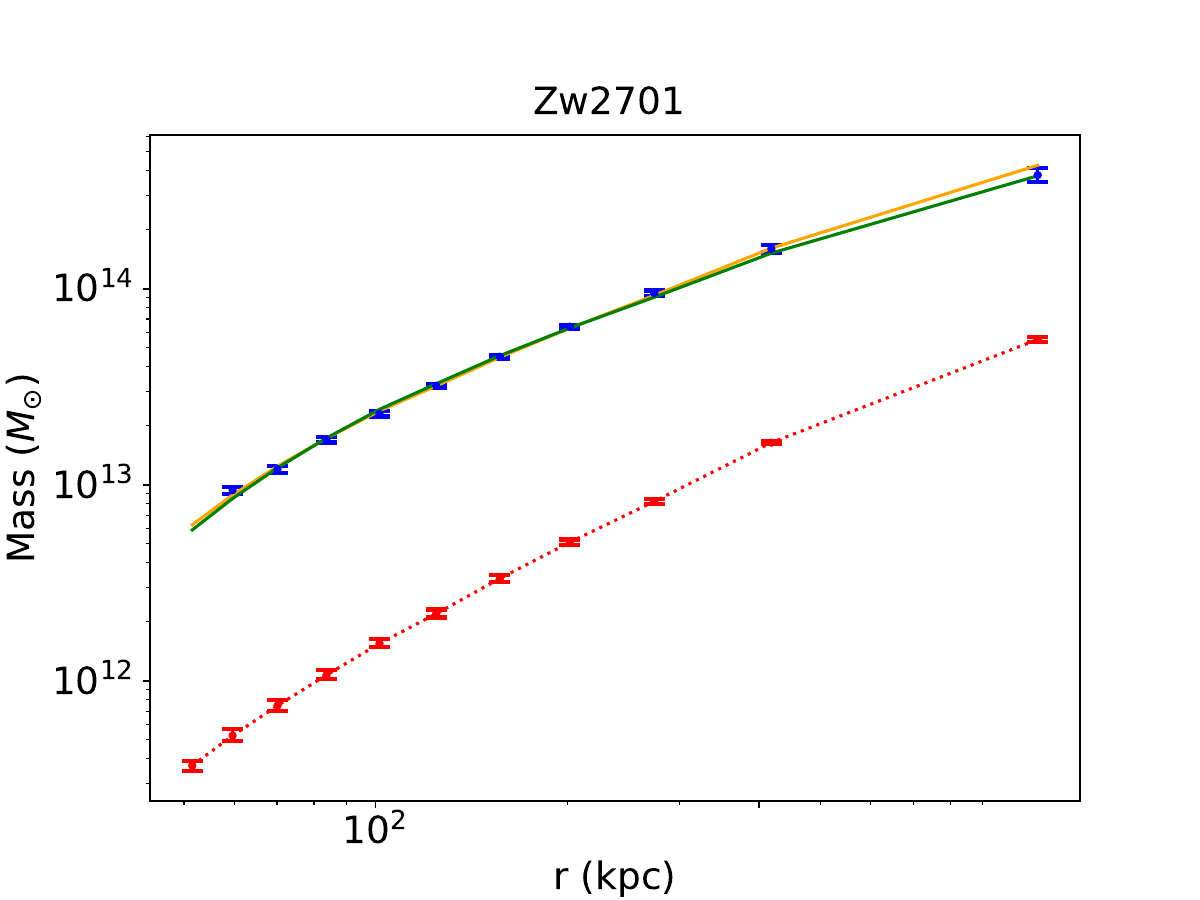}
\includegraphics[trim=0.0cm 0.0cm 1.0cm 0.0cm, clip=true, width=0.33\columnwidth]{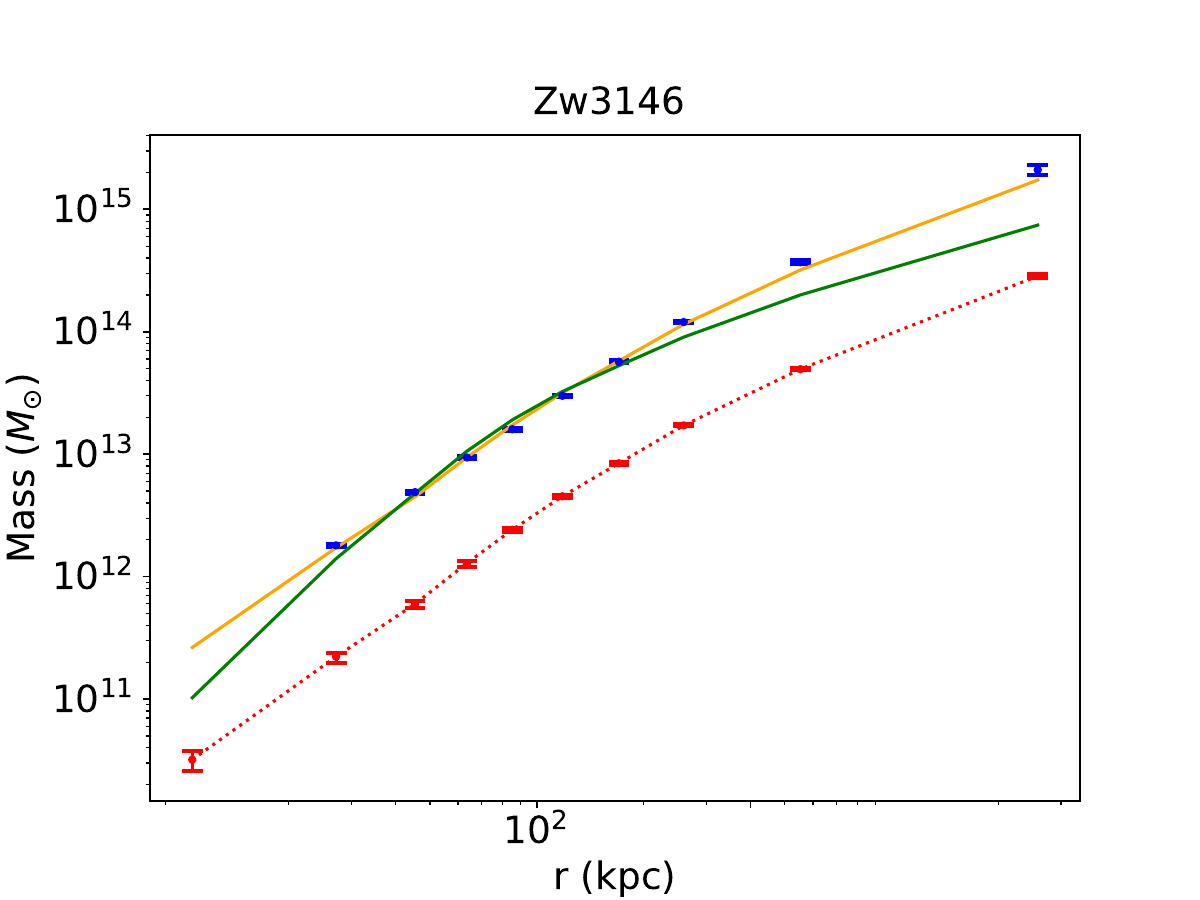}
\caption{\label{MainPaper}The radial mass profiles of various galaxy clusters are presented in this illustration. The red error bars, depicted with a dotted line, represent the physical mass of each galaxy cluster. The blue error bars indicate the Newtonian dynamic mass of the clusters. The orange line represents the best-fit line for the dynamic mass equivalent of the galaxy cluster, derived from Eq.~\ref{MachMdMbrelation}. Additionally, a green curve displays the best fit obtained when $\alpha$ is set to $0.25$. The best-fit line for the dynamic mass equivalent, derived from Machian Gravity theory, solely based on the cluster's physical gas mass, eliminates the necessity for any exotic dark matter within the clusters.}
\end{figure}
%\usepackage{geometry}
%\newgeometry{left=2cm, right=4cm}
%\addtolength{\oddsidemargin}{-.7in}

\begin{table*}
\label{clusterProperties}
\hspace*{-1.0cm}\begin{tabular}{|>{\rule[0.5cm]{0pt}{0.05cm}}p{0.09\columnwidth}|p{0.1\columnwidth}|p{0.07\columnwidth}|p{0.1\columnwidth}|p{0.1\columnwidth}|p{0.12\columnwidth}|p{0.05\columnwidth}|p{0.09\columnwidth}|p{0.08\columnwidth}|p{0.09\columnwidth}|}
\hline
Cluster & $T$ & $\rho_{0}$ & $\beta$ & $r_{c}$ & $r_{250}$ & $M_{b250}$ & $M_{d250}$ & $M_{c}$ & $\lambda^{-1}$\\
\,&  & $\times 10^{-25}$ &&  &  & $\times 10^{14}$ & $\times 10^{14}$ & $\times 10^{15}$ & \\
\,& [keV] &[$\mbox{g/cm}^{3}$] && [kpc] & [kpc] & [$M_{\odot}$] & [$M_{\odot}$] & [$M_{\odot}$] & [$r_c$]\\
\hline
$\texttt{A0085}$&$6.90^{0.40}_{-0.40}$&$0.34$&$0.532^{0.004}_{-0.004}$&$58.5^{3.3}_{-3.9}$&$2241.0^{139.0}_{-162.0}$&$1.49$&$9.02^{0.53}_{-0.53}$&$1.83^{0.05}_{-0.02}$&$0.51^{0.005}_{-0.004}$\\
$\texttt{A0119}$&$5.60^{0.30}_{-0.30}$&$0.03$&$0.675^{0.026}_{-0.023}$&$352.8^{24.7}_{-27.0}$&$1728.0^{156.0}_{-173.0}$&$0.84$&$6.88^{0.46}_{-0.44}$&$3.11^{0.06}_{-0.06}$&$0.46^{0.005}_{-0.005}$\\
$\texttt{A0133}$&$3.80^{2.00}_{-0.90}$&$0.42$&$0.530^{0.004}_{-0.004}$&$31.7^{1.9}_{-2.3}$&$1417.0^{96.0}_{-109.0}$&$0.37$&$3.13^{1.65}_{-0.74}$&$0.77^{0.09}_{-0.05}$&$0.45^{0.020}_{-0.016}$\\
$\texttt{NGC507}$&$1.26^{0.07}_{-0.07}$&$0.23$&$0.444^{0.005}_{-0.005}$&$13.4^{0.9}_{-1.0}$&$783.0^{64.0}_{-70.0}$&$0.05$&$0.48^{0.03}_{-0.03}$&$0.21^{0.08}_{-0.00}$&$0.45^{0.004}_{-0.004}$\\
$\texttt{A0262}$&$2.15^{0.06}_{-0.06}$&$0.16$&$0.443^{0.018}_{-0.017}$&$29.6^{8.5}_{-7.2}$&$1334.0^{432.0}_{-386.0}$&$0.27$&$1.39^{0.07}_{-0.07}$&$0.44^{0.07}_{-0.00}$&$0.59^{0.020}_{-0.017}$\\
$\texttt{A0400}$&$2.31^{0.14}_{-0.14}$&$0.04$&$0.534^{0.014}_{-0.013}$&$108.5^{7.8}_{-8.8}$&$1062.0^{97.0}_{-108.0}$&$0.15$&$1.42^{0.09}_{-0.09}$&$0.72^{0.07}_{-0.01}$&$0.44^{0.004}_{-0.004}$\\
$\texttt{A0399}$&$7.00^{0.40}_{-0.40}$&$0.04$&$0.713^{0.137}_{-0.095}$&$316.9^{93.9}_{-72.7}$&$1791.0^{683.0}_{-744.0}$&$0.86$&$9.51^{1.91}_{-1.39}$&$5.41^{0.30}_{-0.24}$&$0.49^{0.016}_{-0.013}$\\
$\texttt{A0401}$&$8.00^{0.40}_{-0.40}$&$0.11$&$0.613^{0.010}_{-0.010}$&$173.2^{10.7}_{-12.0}$&$2236.0^{167.0}_{-182.0}$&$1.63$&$11.96^{0.63}_{-0.63}$&$3.56^{0.05}_{-0.04}$&$0.51^{0.005}_{-0.004}$\\
$\texttt{A3112}$&$5.30^{0.70}_{-1.00}$&$0.54$&$0.576^{0.006}_{-0.006}$&$43.0^{2.8}_{-3.2}$&$1644.0^{124.0}_{-138.0}$&$0.63$&$5.50^{0.73}_{-1.04}$&$1.04^{0.07}_{-0.03}$&$0.47^{0.008}_{-0.007}$\\
$\texttt{FORNAX}$&$1.20^{0.04}_{-0.04}$&$0.02$&$0.804^{0.098}_{-0.084}$&$122.5^{13.0}_{-12.6}$&$387.0^{67.0}_{-74.0}$&$0.01$&$0.37^{0.05}_{-0.04}$&$0.83^{0.08}_{-0.04}$&$0.42^{0.009}_{-0.008}$\\
$\texttt{2A0335}$&$3.01^{0.07}_{-0.07}$&$1.07$&$0.575^{0.004}_{-0.003}$&$23.2^{1.2}_{-1.5}$&$1322.0^{74.0}_{-92.0}$&$0.33$&$2.51^{0.06}_{-0.06}$&$0.56^{0.07}_{-0.00}$&$0.95^{0.009}_{-0.010}$\\
$\texttt{IIIZw54}$&$2.16^{0.35}_{-0.30}$&$0.04$&$0.887^{0.320}_{-0.151}$&$203.5^{87.7}_{-52.7}$&$780.0^{389.0}_{-461.0}$&$0.09$&$1.54^{0.61}_{-0.35}$&$1.31^{0.22}_{-0.14}$&$0.48^{0.031}_{-0.023}$\\
$\texttt{A3158}$&$5.77^{0.10}_{-0.05}$&$0.08$&$0.661^{0.025}_{-0.022}$&$189.4^{16.2}_{-17.1}$&$1672.0^{189.0}_{-206.0}$&$0.76$&$6.91^{0.29}_{-0.24}$&$3.17^{0.05}_{-0.03}$&$0.57^{0.006}_{-0.006}$\\
$\texttt{A0478}$&$8.40^{0.80}_{-1.40}$&$0.50$&$0.613^{0.004}_{-0.004}$&$69.0^{3.2}_{-4.1}$&$2029.0^{105.0}_{-130.0}$&$1.29$&$11.45^{1.09}_{-1.91}$&$2.00^{0.07}_{-0.05}$&$0.47^{0.006}_{-0.006}$\\
$\texttt{NGC1550}$&$1.43^{0.04}_{-0.03}$&$0.15$&$0.554^{0.049}_{-0.037}$&$31.7^{10.6}_{-7.9}$&$632.0^{247.0}_{-231.0}$&$0.04$&$0.55^{0.05}_{-0.04}$&$0.41^{0.07}_{-0.01}$&$0.62^{0.018}_{-0.016}$\\
$\texttt{EXO0422}$&$2.90^{0.90}_{-0.60}$&$0.13$&$0.722^{0.104}_{-0.071}$&$100.0^{28.5}_{-21.9}$&$934.0^{338.0}_{-367.0}$&$0.15$&$2.12^{0.73}_{-0.49}$&$1.06^{0.11}_{-0.08}$&$0.50^{0.021}_{-0.019}$\\
$\texttt{A3266}$&$8.00^{0.50}_{-0.50}$&$0.05$&$0.796^{0.020}_{-0.019}$&$397.2^{22.4}_{-26.4}$&$1915.0^{132.0}_{-150.0}$&$1.35$&$12.83^{0.87}_{-0.86}$&$5.93^{0.12}_{-0.12}$&$0.49^{0.005}_{-0.005}$\\
$\texttt{A0496}$&$4.13^{0.08}_{-0.08}$&$0.65$&$0.484^{0.003}_{-0.003}$&$21.1^{1.1}_{-1.4}$&$1830.0^{111.0}_{-130.0}$&$0.73$&$4.01^{0.08}_{-0.08}$&$0.90^{0.06}_{-0.00}$&$0.77^{0.007}_{-0.008}$\\
$\texttt{A3376}$&$4.00^{0.40}_{-0.40}$&$0.02$&$1.054^{0.101}_{-0.083}$&$531.7^{53.5}_{-51.8}$&$1264.0^{169.0}_{-183.0}$&$0.35$&$4.97^{0.70}_{-0.65}$&$4.40^{0.28}_{-0.24}$&$0.46^{0.011}_{-0.011}$\\
$\texttt{A3391}$&$5.40^{0.60}_{-0.60}$&$0.05$&$0.579^{0.026}_{-0.024}$&$164.8^{18.3}_{-18.1}$&$1558.0^{227.0}_{-234.0}$&$0.51$&$5.29^{0.63}_{-0.63}$&$2.59^{0.08}_{-0.07}$&$0.44^{0.007}_{-0.007}$\\
$\texttt{A3395s}$&$5.00^{0.30}_{-0.30}$&$0.03$&$0.964^{0.275}_{-0.167}$&$425.4^{123.1}_{-86.5}$&$1223.0^{442.0}_{-501.0}$&$0.38$&$5.77^{1.70}_{-1.12}$&$4.97^{0.56}_{-0.44}$&$0.47^{0.021}_{-0.018}$\\
$\texttt{A0576}$&$4.02^{0.07}_{-0.07}$&$0.03$&$0.825^{0.432}_{-0.185}$&$277.5^{156.1}_{-89.4}$&$1066.0^{703.0}_{-903.0}$&$0.20$&$3.67^{1.93}_{-0.86}$&$3.43^{0.64}_{-0.44}$&$0.46^{0.035}_{-0.028}$\\
$\texttt{A0754}$&$9.50^{0.70}_{-0.40}$&$0.09$&$0.698^{0.027}_{-0.024}$&$168.3^{13.9}_{-14.7}$&$1402.0^{156.0}_{-170.0}$&$0.48$&$10.06^{0.84}_{-0.55}$&$9.04^{0.19}_{-0.17}$&$0.51^{0.006}_{-0.006}$\\
$\texttt{HYDRA-A}$&$4.30^{0.40}_{-0.40}$&$0.63$&$0.573^{0.003}_{-0.003}$&$35.2^{1.6}_{-2.1}$&$1502.0^{76.0}_{-95.0}$&$0.48$&$4.06^{0.38}_{-0.38}$&$0.79^{0.07}_{-0.01}$&$0.50^{0.005}_{-0.005}$\\
$\texttt{A1060}$&$3.24^{0.06}_{-0.06}$&$0.09$&$0.607^{0.040}_{-0.034}$&$66.2^{10.9}_{-9.9}$&$790.0^{171.0}_{-176.0}$&$0.07$&$1.69^{0.12}_{-0.10}$&$1.98^{0.06}_{-0.03}$&$0.55^{0.009}_{-0.009}$\\
$\texttt{A1367}$&$3.55^{0.08}_{-0.08}$&$0.03$&$0.695^{0.035}_{-0.032}$&$269.7^{20.4}_{-21.7}$&$1234.0^{131.0}_{-141.0}$&$0.32$&$3.19^{0.18}_{-0.16}$&$1.84^{0.06}_{-0.03}$&$0.47^{0.005}_{-0.004}$\\
$\texttt{MKW4}$&$1.71^{0.09}_{-0.09}$&$0.57$&$0.440^{0.004}_{-0.005}$&$7.7^{0.8}_{-0.8}$&$948.0^{108.0}_{-110.0}$&$0.09$&$0.78^{0.04}_{-0.04}$&$0.30^{0.07}_{-0.00}$&$0.51^{0.007}_{-0.006}$\\
$\texttt{ZwCl1215}$&$5.58^{0.89}_{-0.78}$&$0.05$&$0.819^{0.038}_{-0.034}$&$303.5^{23.5}_{-24.5}$&$1485.0^{155.0}_{-167.0}$&$0.57$&$7.15^{1.19}_{-1.04}$&$3.75^{0.19}_{-0.15}$&$0.47^{0.010}_{-0.009}$\\
$\texttt{NGC4636}$&$0.76^{0.01}_{-0.01}$&$0.33$&$0.491^{0.032}_{-0.027}$&$4.2^{2.1}_{-1.4}$&$216.0^{118.0}_{-92.0}$&$0.00$&$0.09^{0.01}_{-0.00}$&$0.33^{0.07}_{-0.01}$&$0.95^{0.038}_{-0.038}$\\
$\texttt{A3526}$&$3.68^{0.06}_{-0.06}$&$0.29$&$0.495^{0.011}_{-0.010}$&$26.1^{3.7}_{-3.2}$&$1175.0^{189.0}_{-174.0}$&$0.20$&$2.35^{0.06}_{-0.06}$&$1.36^{0.06}_{-0.01}$&$0.82^{0.012}_{-0.013}$\\
\hline
\end{tabular}
\end{table*}
\begin{table*}
\label{clusterProperties}
\hspace*{-1.0cm}\begin{tabular}{|>{\rule[0.5cm]{0pt}{0.05cm}}p{0.09\columnwidth}|p{0.1\columnwidth}|p{0.07\columnwidth}|p{0.1\columnwidth}|p{0.1\columnwidth}|p{0.12\columnwidth}|p{0.05\columnwidth}|p{0.09\columnwidth}|p{0.08\columnwidth}|p{0.09\columnwidth}|}
\hline
Cluster & $T$ & $\rho_{0}$ & $\beta$ & $r_{c}$ & $r_{250}$ & $M_{b250}$ & $M_{d250}$ & $M_{c}$ & $\lambda^{-1}$\\
\,&  & $\times 10^{-25}$ &&  &  & $\times 10^{14}$ & $\times 10^{14}$ & $\times 10^{15}$ & \\
\,& [keV] & [$\mbox{g/cm}^{3}$] && [kpc] & [kpc] & [$M_{\odot}$] & [$M_{\odot}$] & [$M_{\odot}$] & [$r_c$]\\
\hline
$\texttt{A1644}$&$4.70^{0.90}_{-0.70}$&$0.04$&$0.579^{0.111}_{-0.074}$&$211.3^{90.6}_{-65.9}$&$1830.0^{937.0}_{-958.0}$&$0.76$&$5.39^{1.46}_{-1.06}$&$2.02^{0.14}_{-0.10}$&$0.46^{0.021}_{-0.016}$\\
$\texttt{A1650}$&$6.70^{0.80}_{-0.80}$&$0.08$&$0.704^{0.131}_{-0.081}$&$197.9^{73.7}_{-51.2}$&$1600.0^{714.0}_{-758.0}$&$0.65$&$8.16^{1.80}_{-1.36}$&$4.45^{0.26}_{-0.22}$&$0.52^{0.020}_{-0.017}$\\
$\texttt{A1651}$&$6.10^{0.40}_{-0.40}$&$0.15$&$0.643^{0.014}_{-0.013}$&$127.5^{8.9}_{-10.1}$&$1725.0^{151.0}_{-168.0}$&$0.79$&$7.38^{0.51}_{-0.51}$&$2.46^{0.06}_{-0.04}$&$0.52^{0.006}_{-0.005}$\\
$\texttt{COMA}$&$8.38^{0.34}_{-0.34}$&$0.06$&$0.654^{0.019}_{-0.021}$&$242.3^{18.6}_{-20.1}$&$1954.0^{201.0}_{-202.0}$&$1.11$&$11.57^{0.58}_{-0.60}$&$5.88^{0.04}_{-0.06}$&$0.51^{0.004}_{-0.004}$\\
$\texttt{NGC5044}$&$1.07^{0.01}_{-0.01}$&$0.67$&$0.524^{0.002}_{-0.003}$&$7.7^{0.8}_{-0.8}$&$487.0^{50.0}_{-53.0}$&$0.01$&$0.30^{0.00}_{-0.00}$&$0.30^{0.09}_{-0.00}$&$1.77^{0.008}_{-0.020}$\\
$\texttt{A1736}$&$3.50^{0.40}_{-0.40}$&$0.03$&$0.542^{0.147}_{-0.092}$&$263.4^{125.8}_{-92.7}$&$1889.0^{1110.0}_{-1229.0}$&$0.99$&$3.86^{1.14}_{-0.79}$&$0.91^{0.09}_{-0.05}$&$0.44^{0.020}_{-0.017}$\\
$\texttt{A3558}$&$5.50^{0.40}_{-0.40}$&$0.09$&$0.580^{0.006}_{-0.005}$&$157.7^{7.5}_{-9.6}$&$2021.0^{106.0}_{-134.0}$&$1.20$&$7.03^{0.52}_{-0.51}$&$1.68^{0.06}_{-0.03}$&$0.46^{0.004}_{-0.004}$\\
$\texttt{A3562}$&$5.16^{0.16}_{-0.16}$&$0.11$&$0.472^{0.006}_{-0.006}$&$69.7^{4.6}_{-5.3}$&$1926.0^{151.0}_{-167.0}$&$0.83$&$5.14^{0.17}_{-0.17}$&$1.73^{0.05}_{-0.01}$&$0.49^{0.004}_{-0.004}$\\
$\texttt{A3571}$&$6.90^{0.20}_{-0.20}$&$0.14$&$0.613^{0.010}_{-0.010}$&$127.5^{7.3}_{-8.7}$&$1897.0^{137.0}_{-154.0}$&$0.99$&$8.76^{0.29}_{-0.29}$&$3.26^{0.04}_{-0.03}$&$0.58^{0.005}_{-0.005}$\\
$\texttt{A1795}$&$7.80^{1.00}_{-1.00}$&$0.50$&$0.596^{0.003}_{-0.002}$&$54.9^{2.4}_{-3.2}$&$1773.0^{81.0}_{-106.0}$&$0.83$&$9.03^{1.16}_{-1.16}$&$2.20^{0.07}_{-0.05}$&$0.48^{0.006}_{-0.006}$\\
$\texttt{A3581}$&$1.83^{0.04}_{-0.04}$&$0.31$&$0.543^{0.024}_{-0.022}$&$24.6^{3.7}_{-3.1}$&$840.0^{174.0}_{-169.0}$&$0.08$&$0.92^{0.05}_{-0.04}$&$0.43^{0.07}_{-0.00}$&$0.66^{0.012}_{-0.011}$\\
$\texttt{MKW8}$&$3.29^{0.23}_{-0.22}$&$0.05$&$0.511^{0.098}_{-0.059}$&$75.4^{49.4}_{-29.9}$&$977.0^{703.0}_{-619.0}$&$0.11$&$1.79^{0.37}_{-0.24}$&$1.69^{0.08}_{-0.06}$&$0.48^{0.021}_{-0.018}$\\
$\texttt{A2029}$&$9.10^{1.00}_{-1.00}$&$0.56$&$0.582^{0.004}_{-0.004}$&$58.5^{2.8}_{-3.6}$&$2200.0^{120.0}_{-146.0}$&$1.54$&$12.77^{1.41}_{-1.41}$&$2.44^{0.06}_{-0.05}$&$0.49^{0.006}_{-0.006}$\\
$\texttt{A2052}$&$3.03^{0.04}_{-0.04}$&$0.52$&$0.526^{0.005}_{-0.005}$&$26.1^{1.8}_{-2.0}$&$1373.0^{106.0}_{-119.0}$&$0.34$&$2.40^{0.04}_{-0.04}$&$0.75^{0.07}_{-0.00}$&$1.11^{0.008}_{-0.012}$\\
$\texttt{MKW3S}$&$3.70^{0.20}_{-0.20}$&$0.31$&$0.581^{0.008}_{-0.007}$&$46.5^{2.9}_{-3.4}$&$1257.0^{93.0}_{-108.0}$&$0.28$&$2.96^{0.17}_{-0.16}$&$0.99^{0.06}_{-0.01}$&$0.54^{0.005}_{-0.005}$\\
$\texttt{A2065}$&$5.50^{0.40}_{-0.40}$&$0.04$&$1.162^{0.734}_{-0.282}$&$485.9^{254.4}_{-133.8}$&$1302.0^{780.0}_{-1048.0}$&$0.50$&$8.01^{5.12}_{-2.27}$&$7.10^{1.80}_{-1.24}$&$0.50^{0.042}_{-0.036}$\\
$\texttt{A2063}$&$3.68^{0.11}_{-0.11}$&$0.12$&$0.561^{0.011}_{-0.011}$&$77.5^{5.9}_{-6.1}$&$1343.0^{127.0}_{-130.0}$&$0.32$&$3.03^{0.11}_{-0.11}$&$1.30^{0.06}_{-0.01}$&$0.55^{0.005}_{-0.005}$\\
$\texttt{A2142}$&$9.70^{1.50}_{-1.10}$&$0.27$&$0.591^{0.006}_{-0.006}$&$108.5^{6.2}_{-7.4}$&$2537.0^{167.0}_{-192.0}$&$2.40$&$15.93^{2.47}_{-1.81}$&$3.04^{0.09}_{-0.08}$&$0.48^{0.007}_{-0.007}$\\
$\texttt{A2147}$&$4.91^{0.28}_{-0.28}$&$0.03$&$0.444^{0.071}_{-0.046}$&$167.6^{72.9}_{-46.7}$&$2360.0^{1215.0}_{-1201.0}$&$1.26$&$5.62^{0.95}_{-0.66}$&$1.70^{0.07}_{-0.05}$&$0.43^{0.012}_{-0.011}$\\
$\texttt{A2163}$&$13.29^{0.64}_{-0.64}$&$0.10$&$0.796^{0.030}_{-0.028}$&$365.5^{26.7}_{-29.0}$&$2509.0^{253.0}_{-273.0}$&$3.00$&$28.52^{1.75}_{-1.70}$&$11.66^{0.22}_{-0.28}$&$0.57^{0.006}_{-0.007}$\\
$\texttt{A2199}$&$4.10^{0.08}_{-0.08}$&$0.16$&$0.655^{0.019}_{-0.021}$&$97.9^{8.2}_{-8.9}$&$1300.0^{153.0}_{-154.0}$&$0.35$&$3.81^{0.13}_{-0.14}$&$1.78^{0.05}_{-0.02}$&$0.66^{0.007}_{-0.007}$\\
$\texttt{A2204}$&$7.21^{0.25}_{-0.25}$&$0.99$&$0.597^{0.008}_{-0.007}$&$47.2^{2.9}_{-3.4}$&$2216.0^{169.0}_{-196.0}$&$1.65$&$10.46^{0.39}_{-0.38}$&$1.71^{0.06}_{-0.02}$&$0.75^{0.009}_{-0.009}$\\
$\texttt{A2244}$&$7.10^{5.00}_{-2.20}$&$0.23$&$0.607^{0.016}_{-0.015}$&$88.7^{8.6}_{-8.6}$&$1773.0^{216.0}_{-222.0}$&$0.82$&$8.36^{5.89}_{-2.60}$&$2.46^{0.38}_{-0.27}$&$0.47^{0.032}_{-0.025}$\\
$\texttt{A2256}$&$6.60^{0.40}_{-0.40}$&$0.05$&$0.914^{0.054}_{-0.047}$&$413.4^{33.1}_{-34.9}$&$1684.0^{189.0}_{-208.0}$&$0.93$&$10.51^{0.90}_{-0.84}$&$6.06^{0.18}_{-0.17}$&$0.52^{0.007}_{-0.007}$\\
$\texttt{A2255}$&$6.87^{0.20}_{-0.20}$&$0.03$&$0.797^{0.033}_{-0.030}$&$417.6^{30.3}_{-32.6}$&$1730.0^{160.0}_{-174.0}$&$0.79$&$9.82^{0.50}_{-0.47}$&$7.02^{0.13}_{-0.12}$&$0.50^{0.005}_{-0.005}$\\
$\texttt{A3667}$&$7.00^{0.60}_{-0.60}$&$0.07$&$0.541^{0.008}_{-0.008}$&$196.5^{10.9}_{-13.1}$&$2589.0^{175.0}_{-199.0}$&$2.36$&$10.70^{0.93}_{-0.93}$&$2.18^{0.06}_{-0.04}$&$0.44^{0.004}_{-0.004}$\\
$\texttt{S1101}$&$3.00^{1.20}_{-0.70}$&$0.55$&$0.639^{0.006}_{-0.007}$&$39.4^{2.2}_{-2.6}$&$1064.0^{70.0}_{-78.0}$&$0.20$&$2.23^{0.89}_{-0.52}$&$0.53^{0.08}_{-0.04}$&$0.49^{0.019}_{-0.016}$\\
$\texttt{A2589}$&$3.70^{2.20}_{-1.10}$&$0.12$&$0.596^{0.013}_{-0.012}$&$83.1^{6.6}_{-6.8}$&$1206.0^{116.0}_{-121.0}$&$0.25$&$2.90^{1.73}_{-0.87}$&$1.15^{0.15}_{-0.11}$&$0.44^{0.023}_{-0.019}$\\
$\texttt{A2597}$&$4.40^{0.40}_{-0.70}$&$0.71$&$0.633^{0.008}_{-0.008}$&$40.8^{2.2}_{-2.7}$&$1296.0^{91.0}_{-103.0}$&$0.35$&$3.96^{0.36}_{-0.63}$&$0.77^{0.07}_{-0.02}$&$0.49^{0.007}_{-0.007}$\\
$\texttt{A2634}$&$3.70^{0.28}_{-0.28}$&$0.02$&$0.640^{0.051}_{-0.043}$&$256.3^{32.8}_{-31.0}$&$1225.0^{208.0}_{-219.0}$&$0.23$&$3.05^{0.34}_{-0.31}$&$2.28^{0.08}_{-0.06}$&$0.43^{0.007}_{-0.007}$\\
$\texttt{A2657}$&$3.70^{0.30}_{-0.30}$&$0.10$&$0.556^{0.008}_{-0.007}$&$83.8^{5.0}_{-5.9}$&$1307.0^{90.0}_{-105.0}$&$0.30$&$2.94^{0.24}_{-0.24}$&$1.16^{0.06}_{-0.02}$&$0.45^{0.005}_{-0.004}$\\
\hline
\end{tabular}
\end{table*}
\begin{table*}
\label{clusterProperties}
\hspace*{-1.0cm}\begin{tabular}{|>{\rule[0.5cm]{0pt}{0.05cm}}p{0.09\columnwidth}|p{0.1\columnwidth}|p{0.07\columnwidth}|p{0.1\columnwidth}|p{0.1\columnwidth}|p{0.12\columnwidth}|p{0.05\columnwidth}|p{0.09\columnwidth}|p{0.08\columnwidth}|p{0.09\columnwidth}|}
\hline
Cluster & $T$ & $\rho_{0}$ & $\beta$ & $r_{c}$ & $r_{250}$ & $M_{b250}$ & $M_{d250}$ & $M_{c}$ & $\lambda^{-1}$\\
\,&  & $\times 10^{-25}$ &&  &  & $\times 10^{14}$ & $\times 10^{14}$ & $\times 10^{15}$ & \\
\,& [keV] & [$\mbox{g/cm}^{3}$] && [kpc] & [kpc] & [$M_{\odot}$] & [$M_{\odot}$] & [$M_{\odot}$] & [$r_c$]\\
\hline
$\texttt{A4038}$&$3.15^{0.03}_{-0.03}$&$0.26$&$0.541^{0.009}_{-0.008}$&$41.5^{3.3}_{-3.7}$&$1274.0^{121.0}_{-134.0}$&$0.27$&$2.38^{0.05}_{-0.04}$&$1.01^{0.06}_{-0.00}$&$0.95^{0.008}_{-0.010}$\\
$\texttt{A4059}$&$4.40^{0.30}_{-0.30}$&$0.20$&$0.582^{0.010}_{-0.010}$&$63.4^{4.4}_{-5.0}$&$1324.0^{116.0}_{-126.0}$&$0.33$&$3.71^{0.26}_{-0.26}$&$1.45^{0.06}_{-0.02}$&$0.51^{0.006}_{-0.005}$\\
$\texttt{A2734}$&$3.85^{0.62}_{-0.54}$&$0.06$&$0.624^{0.034}_{-0.029}$&$149.3^{19.4}_{-18.3}$&$1357.0^{226.0}_{-234.0}$&$0.35$&$3.53^{0.60}_{-0.52}$&$1.53^{0.08}_{-0.05}$&$0.45^{0.009}_{-0.009}$\\
$\texttt{A2877}$&$3.50^{2.20}_{-1.10}$&$0.03$&$0.566^{0.029}_{-0.025}$&$133.8^{14.5}_{-14.1}$&$943.0^{132.0}_{-139.0}$&$0.11$&$2.01^{1.27}_{-0.64}$&$1.74^{0.25}_{-0.18}$&$0.40^{0.022}_{-0.019}$\\
$\texttt{NGC499}$&$0.72^{0.03}_{-0.02}$&$0.20$&$0.722^{0.034}_{-0.030}$&$16.9^{1.6}_{-1.7}$&$196.0^{27.0}_{-30.0}$&$0.00$&$0.11^{0.01}_{-0.01}$&$0.34^{0.08}_{-0.01}$&$0.61^{0.009}_{-0.009}$\\
$\texttt{AWM7}$&$3.75^{0.09}_{-0.09}$&$0.09$&$0.671^{0.027}_{-0.025}$&$121.8^{13.7}_{-12.6}$&$1120.0^{157.0}_{-154.0}$&$0.23$&$3.06^{0.14}_{-0.14}$&$1.97^{0.05}_{-0.02}$&$0.58^{0.007}_{-0.007}$\\
$\texttt{PERSEUS}$&$6.79^{0.12}_{-0.12}$&$0.63$&$0.540^{0.006}_{-0.004}$&$45.1^{2.4}_{-2.9}$&$2414.0^{145.0}_{-189.0}$&$1.87$&$9.71^{0.20}_{-0.19}$&$1.86^{0.05}_{-0.01}$&$0.90^{0.008}_{-0.010}$\\
$\texttt{S405}$&$4.21^{0.67}_{-0.59}$&$0.02$&$0.664^{0.263}_{-0.133}$&$323.2^{185.0}_{-113.4}$&$1561.0^{1034.0}_{-1165.0}$&$0.44$&$4.59^{1.96}_{-1.14}$&$2.64^{0.27}_{-0.25}$&$0.44^{0.023}_{-0.021}$\\
$\texttt{3C129}$&$5.60^{0.70}_{-0.60}$&$0.03$&$0.601^{0.260}_{-0.131}$&$223.9^{125.7}_{-76.4}$&$1567.0^{1113.0}_{-1455.0}$&$0.46$&$5.67^{2.55}_{-1.38}$&$3.65^{0.55}_{-0.29}$&$0.44^{0.029}_{-0.019}$\\
$\texttt{A0539}$&$3.24^{0.09}_{-0.09}$&$0.06$&$0.561^{0.020}_{-0.018}$&$104.2^{10.2}_{-10.3}$&$1194.0^{150.0}_{-158.0}$&$0.22$&$2.36^{0.11}_{-0.10}$&$1.35^{0.06}_{-0.01}$&$0.51^{0.005}_{-0.005}$\\
$\texttt{S540}$&$2.40^{0.38}_{-0.34}$&$0.08$&$0.641^{0.073}_{-0.051}$&$91.5^{27.0}_{-21.1}$&$877.0^{305.0}_{-305.0}$&$0.10$&$1.46^{0.29}_{-0.24}$&$0.90^{0.07}_{-0.04}$&$0.50^{0.015}_{-0.013}$\\
$\texttt{A0548w}$&$1.20^{0.19}_{-0.17}$&$0.02$&$0.666^{0.194}_{-0.111}$&$139.4^{63.7}_{-44.4}$&$593.0^{311.0}_{-328.0}$&$0.03$&$0.49^{0.16}_{-0.11}$&$0.48^{0.09}_{-0.04}$&$0.43^{0.024}_{-0.021}$\\
$\texttt{A0548}$&$3.10^{0.10}_{-0.10}$&$0.05$&$0.480^{0.013}_{-0.013}$&$83.1^{9.2}_{-9.1}$&$1324.0^{177.0}_{-175.0}$&$0.25$&$2.15^{0.09}_{-0.09}$&$1.13^{0.06}_{-0.01}$&$0.48^{0.005}_{-0.005}$\\
$\texttt{A3395}$&$5.00^{0.30}_{-0.30}$&$0.02$&$0.981^{0.619}_{-0.244}$&$473.2^{270.5}_{-145.4}$&$1221.0^{783.0}_{-977.0}$&$0.30$&$5.71^{3.65}_{-1.69}$&$6.12^{1.78}_{-0.94}$&$0.44^{0.040}_{-0.029}$\\
$\texttt{UGC03957}$&$2.58^{0.41}_{-0.36}$&$0.09$&$0.740^{0.133}_{-0.086}$&$100.0^{32.0}_{-23.9}$&$764.0^{306.0}_{-339.0}$&$0.08$&$1.57^{0.38}_{-0.29}$&$1.33^{0.10}_{-0.07}$&$0.52^{0.019}_{-0.016}$\\
$\texttt{PKS0745}$&$7.21^{0.11}_{-0.11}$&$0.97$&$0.608^{0.006}_{-0.006}$&$50.0^{2.5}_{-3.1}$&$2169.0^{137.0}_{-159.0}$&$1.58$&$10.43^{0.19}_{-0.19}$&$2.32^{0.04}_{-0.01}$&$1.23^{0.009}_{-0.013}$\\
$\texttt{A0644}$&$7.90^{0.80}_{-0.80}$&$0.15$&$0.700^{0.011}_{-0.011}$&$143.0^{7.8}_{-9.4}$&$1557.0^{103.0}_{-119.0}$&$0.65$&$9.37^{0.96}_{-0.96}$&$4.23^{0.11}_{-0.10}$&$0.49^{0.006}_{-0.006}$\\
$\texttt{S636}$&$1.18^{0.19}_{-0.17}$&$0.01$&$0.752^{0.217}_{-0.123}$&$242.3^{92.1}_{-62.1}$&$742.0^{323.0}_{-336.0}$&$0.04$&$0.65^{0.22}_{-0.15}$&$0.64^{0.11}_{-0.07}$&$0.42^{0.023}_{-0.020}$\\
$\texttt{A1413}$&$7.32^{0.26}_{-0.24}$&$0.19$&$0.660^{0.017}_{-0.015}$&$126.1^{10.0}_{-10.5}$&$1794.0^{179.0}_{-194.0}$&$0.94$&$9.46^{0.42}_{-0.38}$&$3.71^{0.04}_{-0.04}$&$0.62^{0.007}_{-0.007}$\\
$\texttt{M49}$&$0.95^{0.02}_{-0.01}$&$0.26$&$0.592^{0.007}_{-0.007}$&$7.7^{0.8}_{-0.8}$&$177.0^{19.0}_{-20.0}$&$0.00$&$0.11^{0.00}_{-0.00}$&$0.74^{0.07}_{-0.01}$&$1.04^{0.012}_{-0.014}$\\
$\texttt{A3528n}$&$3.40^{1.66}_{-0.64}$&$0.07$&$0.621^{0.034}_{-0.030}$&$125.4^{13.1}_{-13.3}$&$1181.0^{179.0}_{-193.0}$&$0.25$&$2.71^{1.33}_{-0.53}$&$1.16^{0.15}_{-0.10}$&$0.44^{0.021}_{-0.017}$\\
$\texttt{A3528s}$&$3.15^{0.89}_{-0.59}$&$0.09$&$0.463^{0.013}_{-0.012}$&$71.1^{7.0}_{-6.9}$&$1872.0^{244.0}_{-250.0}$&$0.72$&$2.99^{0.85}_{-0.57}$&$0.61^{0.07}_{-0.02}$&$0.41^{0.009}_{-0.009}$\\
$\texttt{A3530}$&$3.89^{0.27}_{-0.25}$&$0.03$&$0.773^{0.114}_{-0.085}$&$296.5^{54.3}_{-46.1}$&$1150.0^{282.0}_{-308.0}$&$0.28$&$3.56^{0.58}_{-0.46}$&$2.50^{0.14}_{-0.11}$&$0.46^{0.011}_{-0.010}$\\
$\texttt{A3532}$&$4.58^{0.19}_{-0.17}$&$0.05$&$0.653^{0.034}_{-0.029}$&$198.6^{20.8}_{-20.3}$&$1372.0^{188.0}_{-199.0}$&$0.42$&$4.41^{0.29}_{-0.26}$&$2.42^{0.05}_{-0.05}$&$0.50^{0.006}_{-0.006}$\\
$\texttt{A1689}$&$9.23^{0.28}_{-0.28}$&$0.33$&$0.690^{0.011}_{-0.011}$&$114.8^{6.9}_{-7.7}$&$1898.0^{143.0}_{-154.0}$&$1.21$&$13.21^{0.45}_{-0.45}$&$4.87^{0.05}_{-0.05}$&$0.69^{0.007}_{-0.007}$\\
$\texttt{A3560}$&$3.16^{0.51}_{-0.44}$&$0.03$&$0.566^{0.033}_{-0.029}$&$180.3^{22.5}_{-21.6}$&$1402.0^{230.0}_{-240.0}$&$0.32$&$2.71^{0.46}_{-0.40}$&$1.16^{0.07}_{-0.04}$&$0.42^{0.008}_{-0.008}$\\
$\texttt{A1775}$&$3.69^{0.20}_{-0.11}$&$0.06$&$0.673^{0.026}_{-0.023}$&$183.1^{15.5}_{-16.3}$&$1391.0^{153.0}_{-167.0}$&$0.41$&$3.73^{0.25}_{-0.17}$&$1.66^{0.06}_{-0.03}$&$0.52^{0.006}_{-0.005}$\\
$\texttt{A1800}$&$4.02^{0.64}_{-0.56}$&$0.04$&$0.766^{0.308}_{-0.139}$&$276.1^{157.5}_{-94.2}$&$1284.0^{825.0}_{-949.0}$&$0.38$&$4.15^{1.80}_{-0.97}$&$2.30^{0.32}_{-0.22}$&$0.48^{0.032}_{-0.024}$\\
$\texttt{A1914}$&$10.53^{0.51}_{-0.50}$&$0.22$&$0.751^{0.018}_{-0.017}$&$162.7^{10.4}_{-11.6}$&$1768.0^{148.0}_{-162.0}$&$1.09$&$15.21^{0.82}_{-0.80}$&$7.28^{0.10}_{-0.10}$&$0.60^{0.006}_{-0.006}$\\
$\texttt{NGC5813}$&$0.52^{0.08}_{-0.07}$&$0.18$&$0.766^{0.179}_{-0.103}$&$17.6^{6.4}_{-4.3}$&$166.0^{79.0}_{-97.0}$&$0.00$&$0.07^{0.02}_{-0.01}$&$0.19^{0.08}_{-0.01}$&$0.54^{0.025}_{-0.022}$\\
\hline
\end{tabular}
\end{table*}

%\addtolength{\textwidth}{100.75in}
%\addtolength{\oddsidemargin}{-.7in}
\begin{table}
\label{clusterProperties}
\hspace*{-1.0cm}\begin{tabular}{|>{\rule[0.5cm]{0pt}{0.05cm}}p{0.09\columnwidth}|p{0.1\columnwidth}|p{0.07\columnwidth}|p{0.1\columnwidth}|p{0.1\columnwidth}|p{0.12\columnwidth}|p{0.05\columnwidth}|p{0.09\columnwidth}|p{0.08\columnwidth}|p{0.09\columnwidth}|}
\hline
Cluster & $T$ & $\rho_{0}$ & $\beta$ & $r_{c}$ & $r_{250}$ & $M_{b250}$ & $M_{d250}$ & $M_{c}$ & $\lambda^{-1}$\\
\,&  & $\times 10^{-25}$ &&  &  & $\times 10^{14}$ & $\times 10^{14}$ & $\times 10^{15}$ & \\
\,& [keV] & [$\mbox{g/cm}^{3}$] && [kpc] & [kpc] & [$M_{\odot}$] & [$M_{\odot}$] & [$M_{\odot}$] & [$r_c$]\\
\hline
$\texttt{NGC5846}$&$0.82^{0.01}_{-0.01}$&$0.47$&$0.599^{0.016}_{-0.015}$&$4.9^{0.7}_{-0.8}$&$152.0^{26.0}_{-27.0}$&$0.00$&$0.08^{0.00}_{-0.00}$&$0.60^{0.07}_{-0.01}$&$1.35^{0.018}_{-0.022}$\\
$\texttt{A2151w}$&$2.40^{0.06}_{-0.06}$&$0.16$&$0.564^{0.014}_{-0.013}$&$47.9^{4.1}_{-4.4}$&$957.0^{105.0}_{-114.0}$&$0.12$&$1.42^{0.05}_{-0.05}$&$0.75^{0.07}_{-0.01}$&$0.61^{0.007}_{-0.006}$\\
$\texttt{A3627}$&$6.02^{0.08}_{-0.08}$&$0.04$&$0.555^{0.056}_{-0.044}$&$210.6^{40.4}_{-36.5}$&$1830.0^{474.0}_{-515.0}$&$0.84$&$6.62^{0.67}_{-0.53}$&$3.01^{0.08}_{-0.07}$&$0.47^{0.009}_{-0.008}$\\
$\texttt{TRIANGUL}$&$9.60^{0.60}_{-0.60}$&$0.10$&$0.610^{0.010}_{-0.010}$&$196.5^{11.4}_{-13.1}$&$2385.0^{169.0}_{-187.0}$&$2.04$&$15.22^{0.98}_{-0.98}$&$4.56^{0.07}_{-0.06}$&$0.49^{0.005}_{-0.004}$\\
$\texttt{OPHIUCHU}$&$10.26^{0.32}_{-0.32}$&$0.13$&$0.747^{0.035}_{-0.032}$&$196.5^{18.2}_{-19.0}$&$1701.0^{224.0}_{-240.0}$&$0.93$&$14.12^{0.80}_{-0.75}$&$9.17^{0.15}_{-0.14}$&$0.59^{0.007}_{-0.006}$\\
$\texttt{ZwC174}$&$5.23^{0.84}_{-0.73}$&$0.10$&$0.717^{0.073}_{-0.053}$&$163.4^{33.1}_{-28.3}$&$1354.0^{349.0}_{-378.0}$&$0.44$&$5.49^{1.04}_{-0.87}$&$2.67^{0.14}_{-0.12}$&$0.51^{0.014}_{-0.013}$\\
$\texttt{A2319}$&$8.80^{0.50}_{-0.50}$&$0.10$&$0.591^{0.013}_{-0.012}$&$200.7^{13.5}_{-15.0}$&$2657.0^{228.0}_{-250.0}$&$2.70$&$15.08^{0.92}_{-0.91}$&$3.54^{0.05}_{-0.05}$&$0.50^{0.005}_{-0.005}$\\
$\texttt{A3695}$&$5.29^{0.85}_{-0.74}$&$0.04$&$0.642^{0.259}_{-0.117}$&$281.0^{179.3}_{-106.1}$&$1887.0^{1379.0}_{-1652.0}$&$0.96$&$6.88^{2.99}_{-1.59}$&$2.50^{0.27}_{-0.21}$&$0.47^{0.029}_{-0.022}$\\
$\texttt{IIZw108}$&$3.44^{0.55}_{-0.48}$&$0.03$&$0.662^{0.167}_{-0.097}$&$257.0^{112.5}_{-75.3}$&$1327.0^{670.0}_{-695.0}$&$0.36$&$3.20^{0.96}_{-0.65}$&$1.53^{0.13}_{-0.10}$&$0.45^{0.019}_{-0.017}$\\
$\texttt{A3822}$&$4.90^{0.78}_{-0.69}$&$0.04$&$0.639^{0.150}_{-0.093}$&$247.2^{113.2}_{-79.4}$&$1675.0^{904.0}_{-942.0}$&$0.67$&$5.63^{1.60}_{-1.15}$&$2.45^{0.19}_{-0.15}$&$0.47^{0.021}_{-0.019}$\\
$\texttt{A3827}$&$7.08^{1.13}_{-0.99}$&$0.05$&$0.989^{0.410}_{-0.192}$&$417.6^{175.5}_{-106.5}$&$1515.0^{762.0}_{-977.0}$&$0.72$&$10.82^{4.82}_{-2.67}$&$7.99^{2.08}_{-0.86}$&$0.51^{0.045}_{-0.024}$\\
$\texttt{A3888}$&$8.84^{1.41}_{-1.24}$&$0.10$&$0.928^{0.084}_{-0.066}$&$282.4^{34.5}_{-32.4}$&$1455.0^{252.0}_{-281.0}$&$0.70$&$12.62^{2.32}_{-1.99}$&$9.10^{0.61}_{-0.47}$&$0.53^{0.015}_{-0.012}$\\
$\texttt{A3921}$&$5.73^{0.24}_{-0.23}$&$0.07$&$0.762^{0.036}_{-0.030}$&$231.0^{20.7}_{-20.8}$&$1435.0^{167.0}_{-182.0}$&$0.54$&$6.70^{0.42}_{-0.38}$&$3.97^{0.08}_{-0.07}$&$0.54^{0.007}_{-0.006}$\\
$\texttt{HCG94}$&$3.45^{0.30}_{-0.30}$&$0.11$&$0.514^{0.007}_{-0.006}$&$60.6^{3.8}_{-4.4}$&$1237.0^{89.0}_{-104.0}$&$0.25$&$2.40^{0.21}_{-0.21}$&$1.01^{0.06}_{-0.02}$&$0.44^{0.005}_{-0.004}$\\
$\texttt{RXJ2344}$&$4.73^{0.76}_{-0.66}$&$0.07$&$0.807^{0.033}_{-0.030}$&$212.0^{16.7}_{-17.4}$&$1222.0^{127.0}_{-135.0}$&$0.33$&$4.97^{0.82}_{-0.72}$&$2.85^{0.14}_{-0.12}$&$0.48^{0.010}_{-0.009}$\\
\hline
\end{tabular}
\caption{\label{Table1}The table presents various cluster properties, including cluster temperature ($T$), central density ($\rho_0$), $\beta$ parameter, and core radius ($r_c$). These data points have been sourced from \cite{reiprich2003cosmological}. Additionally, the table includes the cluster's outer radius ($r_{out}$), defined as the radius where the surface intensity reaches 250 times the ambient intensity. It also lists the total baryonic mass within the cluster ($M_b$) and the cluster's dynamical mass ($M_d$) as calculated using Newtonian dynamics. The upper and lower standard deviations have been computed based on the standard deviations of $T$, $\beta$, and $r_c$. Furthermore, the table includes values for $M_c$ and $\lambda^{-1}$ derived from an MCMC analysis. Notably, $\lambda^{-1}$ is expressed relative to $r_c$, with the observation that $\lambda^{-1}$ tends to be approximately half the size of $r_c$. } 
\end{table}

\begin{table}
\label{clusterProperties}
\hspace*{-1.0cm}\begin{tabular}{|>{\rule[0.5cm]{0pt}{0.05cm}}p{0.12\columnwidth}|p{0.09\columnwidth}|p{0.09\columnwidth}|p{0.06\columnwidth}|p{0.1\columnwidth}|p{0.12\columnwidth}|p{0.11\columnwidth}|p{0.11\columnwidth}|p{0.12\columnwidth}|}
\hline
Cluster & $\rho_{0}$ & $r_{c}$ & $r_{2500}$ & $M_{b2500}$ & $M_{d2500}$ & $\alpha$ & $Q$ & $\lambda^{-1}$ \\
\,&  &  && $\times 10^{13}$ & $\times 10^{14}$ &  &  &  \\
\,&  & [kpc] & [kpc] & [$M_{\odot}$] & [$M_{\odot}$] &  &  & [kpc] \\
\hline
\texttt{2A0335} & $31.0^{2.6}_{-2.1}$ & $204.0^{28.0}_{-22.0}$ & $517.8$ & $1.59^{0.02}_{-0.02}$ & $1.24\pm0.05$ & $0.06\pm0.01$ & $10.51\pm0.21$ & $4.53\pm1.79$ \\
\texttt{3C295} & $50.9^{9.9}_{-6.7}$ & $236.0^{96.0}_{-64.0}$ & $875.0$ & $4.85^{0.47}_{-0.45}$ & $3.30\pm0.40$ & $0.17\pm0.03$ & $6.96\pm0.56$ & $11.97\pm5.03$ \\
\texttt{3C388} & $23.7^{2.5}_{-1.5}$ & $127.0^{38.0}_{-27.0}$ & $757.2$ & $1.73^{0.13}_{-0.13}$ & $1.20\pm0.17$ & $0.41\pm0.04$ & $7.55\pm0.66$ & $19.04\pm3.50$ \\
\texttt{3C401} & $30.3^{13.0}_{-5.8}$ & $210.0^{180.0}_{-82.0}$ & $816.2$ & $2.17^{0.32}_{-0.27}$ & $2.00\pm0.46$ & $0.32\pm0.08$ & $9.62\pm1.54$ & $19.84\pm11.7$ \\
\texttt{A85} & $51.1^{5.0}_{-4.0}$ & $381.0^{66.0}_{-53.0}$ & $640.4$ & $3.76^{0.05}_{-0.05}$ & $2.50\pm0.09$ & $0.18\pm0.01$ & $7.79\pm0.13$ & $7.65\pm1.69$ \\
\texttt{A133} & $36.4^{6.9}_{-3.6}$ & $314.0^{130.0}_{-75.0}$ & $654.1$ & $2.06^{0.06}_{-0.06}$ & $1.80\pm0.12$ & $0.24\pm0.02$ & $9.48\pm0.36$ & $24.04\pm3.14$ \\
\texttt{A1795} & $55.0^{1.6}_{-1.6}$ & $373.0^{21.0}_{-21.0}$ & $892.2$ & $5.75^{0.17}_{-0.15}$ & $3.81\pm0.05$ & $0.14\pm0.00$ & $7.23\pm0.11$ & $14.72\pm0.73$ \\
\texttt{A1835} & $95.7^{3.4}_{-3.7}$ & $439.0^{29.0}_{-31.0}$ & $1550.8$ & $19.00^{0.32}_{-0.32}$ & $11.00\pm0.49$ & $0.08\pm0.01$ & $6.65\pm0.14$ & $11.78\pm0.95$ \\
\texttt{A2029} & $68.8^{3.9}_{-3.4}$ & $304.0^{30.0}_{-26.0}$ & $969.7$ & $8.95^{0.11}_{-0.11}$ & $5.04\pm0.11$ & $0.19\pm0.00$ & $6.81\pm0.18$ & $2.58\pm0.87$ \\
\texttt{A2052} & $27.3^{0.48}_{-0.46}$ & $198.0^{9.7}_{-9.6}$ & $533.0$ & $1.39^{0.08}_{-0.08}$ & $1.12\pm0.01$ & $0.30\pm0.01$ & $8.45\pm0.22$ & $22.66\pm0.77$ \\
\texttt{A2199} & $45.6^{2.8}_{-2.1}$ & $363.0^{36.0}_{-26.0}$ & $444.2$ & $1.34^{0.01}_{-0.01}$ & $1.48\pm0.02$ & $0.17\pm0.01$ & $10.47\pm0.32$ & $1.90\pm0.53$ \\
\texttt{A2204} & $97.9^{3.6}_{-2.7}$ & $305.0^{54.0}_{-45.0}$ & $1581.2$ & $17.40^{0.56}_{-0.56}$ & $10.50\pm0.12$ & $0.27\pm0.01$ & $6.30\pm0.17$ & $61.24\pm3.00$ \\
\texttt{A2597} & $39.1^{1.3}_{-1.5}$ & $234.0^{20.0}_{-25.0}$ & $965.5$ & $3.95^{0.15}_{-0.15}$ & $2.73\pm0.05$ & $0.16\pm0.01$ & $7.65\pm0.17$ & $24.91\pm1.21$ \\
\texttt{A262} & $14.0^{0.52}_{-0.45}$ & $134.0^{11.0}_{-9.7}$ & $357.2$ & $0.50^{0.01}_{-0.01}$ & $0.39\pm0.01$ & $0.37\pm0.01$ & $8.82\pm0.24$ & $11.35\pm0.76$ \\
\texttt{A4059} & $40.2^{1.7}_{-1.5}$ & $237.0^{29.0}_{-25.0}$ & $825.9$ & $2.46^{0.07}_{-0.07}$ & $2.50\pm0.09$ & $0.23\pm0.01$ & $10.75\pm0.22$ & $6.30\pm3.04$ \\
\texttt{A478} & $86.7^{6.6}_{-5.5}$ & $553.0^{68.0}_{-56.0}$ & $1228.5$ & $11.80^{0.28}_{-0.28}$ & $8.30\pm0.26$ & $0.11\pm0.00$ & $6.54\pm0.15$ & $3.07\pm1.17$ \\
\texttt{A496} & $37.9^{0.66}_{-0.62}$ & $131.0^{26.0}_{-21.0}$ & $714.4$ & $2.65^{0.08}_{-0.09}$ & $1.80\pm0.07$ & $0.46\pm0.02$ & $7.20\pm0.22$ & $39.45\pm3.62$ \\
\texttt{Centaurus} & $35.1^{0.27}_{-0.38}$ & $276$ & $249.0$ & $0.32^{0.00}_{-0.00}$ & $0.62\pm0.01$ & $0.25\pm0.01$ & $20.32\pm0.51$ & $17.75\pm0.89$ \\
\texttt{Cygnus A} & $45.2^{4.5}_{-2.1}$ & $145.0^{70.0}_{-43.0}$ & $814.7$ & $5.90^{0.07}_{-0.06}$ & $2.50\pm0.12$ & $0.48\pm0.02$ & $4.91\pm0.19$ & $49.07\pm3.73$ \\
\texttt{HCG62} & $7.02^{1.9}_{-0.8}$ & $48.6^{37.0}_{-19.0}$ & $192.9$ & $0.05^{0.00}_{-0.00}$ & $0.10\pm0.00$ & $0.79\pm0.05$ & $17.67\pm0.70$ & $28.95\pm2.47$ \\
\texttt{Hercules A} & $53.7^{14.0}_{-3.4}$ & $135.0^{190.0}_{-85.0}$ & $1114.7$ & $5.62^{0.69}_{-0.70}$ & $3.50\pm0.23$ & $0.50\pm0.06$ & $6.52\pm0.39$ & $75.12\pm15.3$ \\
\texttt{Hydra A} & $35.3^{1.2}_{-0.65}$ & $341.0^{25.0}_{-24.0}$ & $789.7$ & $3.16^{0.05}_{-0.05}$ & $2.16\pm0.05$ & $0.18\pm0.01$ & $6.40\pm0.15$ & $25.67\pm0.88$ \\
\texttt{MACSJ0159} & $125.0^{100.0}_{-39.0}$ & $700.0^{730.0}_{-270.0}$ & $1311.7$ & $15.80^{0.27}_{-0.27}$ & $13.00\pm0.80$ & $0.07\pm0.02$ & $8.10\pm0.38$ & $12.16\pm3.10$ \\
\texttt{MACSJ0242} & $57.9^{63.0}_{-12.0}$ & $258.0^{230.0}_{-81.0}$ & $679.0$ & $4.97^{0.11}_{-0.26}$ & $3.00\pm0.57$ & $0.20\pm0.18$ & $6.04\pm1.18$ & $25.15\pm21.4$ \\
\texttt{MACSJ0429} & $54.5^{14.0}_{-6.8}$ & $202.0^{100.0}_{-51.0}$ & $791.1$ & $5.72^{0.34}_{-0.34}$ & $3.10\pm0.45$ & $0.21\pm0.04$ & $6.15\pm0.55$ & $12.55\pm6.34$ \\
\texttt{MACSJ0913} & $71.7^{18.0}_{-8.6}$ & $305.0^{120.0}_{-57.0}$ & $827.3$ & $5.51^{0.27}_{-0.26}$ & $4.60\pm0.50$ & $0.03\pm0.01$ & $9.04\pm0.41$ & $5.59\pm2.71$ \\
\texttt{MACSJ1423} & $70.6^{8.1}_{-6.0}$ & $211.0^{59.0}_{-46.0}$ & $1263.4$ & $10.70^{0.87}_{-0.84}$ & $5.80\pm0.58$ & $0.24\pm0.03$ & $6.12\pm0.41$ & $23.50\pm3.97$ \\
\texttt{MACSJ1720} & $67.8^{18.0}_{-16.0}$ & $405.0^{170.0}_{-150.0}$ & $796.6$ & $6.48^{0.24}_{-0.24}$ & $4.20\pm0.52$ & $0.15\pm0.04$ & $6.52\pm0.51$ & $15.82\pm8.62$ \\
\texttt{MACSJ1931} & $126.0^{43.0}_{-22.0}$ & $851.0^{390.0}_{-200.0}$ & $1721.4$ & $19.90^{1.90}_{-1.40}$ & $17.00\pm1.40$ & $0.01\pm0.00$ & $7.02\pm0.17$ & $13.59\pm1.00$ \\
\hline
\end{tabular}
\end{table}
\begin{table}
\label{clusterProperties}
\hspace*{-1.0cm}\begin{tabular}{|>{\rule[0.5cm]{0pt}{0.05cm}}p{0.12\columnwidth}|p{0.09\columnwidth}|p{0.09\columnwidth}|p{0.06\columnwidth}|p{0.1\columnwidth}|p{0.12\columnwidth}|p{0.11\columnwidth}|p{0.11\columnwidth}|p{0.12\columnwidth}|}
\hline
Cluster & $\rho_{0}$ & $r_{c}$ & $r_{2500}$ & $M_{b2500}$ & $M_{d2500}$ & $\alpha$ & $Q$ & $\lambda^{-1}$ \\
\,&  &  && $\times 10^{13}$ & $\times 10^{14}$ &  &  &  \\
\,&  & [kpc] & [kpc] & [$M_{\odot}$] & [$M_{\odot}$] &  &  & [kpc] \\
\hline
\texttt{MACSJ2046} & $52.9^{16.0}_{-5.3}$ & $210.0^{130.0}_{-45.0}$ & $957.4$ & $7.82^{1.10}_{-1.20}$ & $3.60\pm0.76$ & $0.35\pm0.13$ & $4.93\pm0.81$ & $49.03\pm29.0$ \\
\texttt{MACSJ2140} & $51.1^{5.2}_{-2.7}$ & $176.0^{36.0}_{-30.0}$ & $920.7$ & $5.73^{0.34}_{-0.39}$ & $3.20\pm0.34$ & $0.14\pm0.02$ & $6.90\pm0.35$ & $6.24\pm2.68$ \\
\texttt{MKW3S} & $34.9^{12.0}_{-4.3}$ & $214.0^{210.0}_{-90.0}$ & $567.3$ & $1.41^{0.02}_{-0.02}$ & $1.50\pm0.15$ & $0.22\pm0.04$ & $11.01\pm0.56$ & $18.07\pm9.30$ \\
\texttt{MS0735} & $66.2^{13.0}_{-3.0}$ & $368.0^{27.0}_{-140.0}$ & $1549.8$ & $11.10^{0.35}_{-0.39}$ & $7.40\pm0.36$ & $0.20\pm0.03$ & $7.02\pm0.28$ & $61.25\pm23.2$ \\
\texttt{NGC1550} & $9.62^{0.25}_{-0.16}$ & $58.7^{5.3}_{-3.7}$ & $241.6$ & $0.13^{0.01}_{-0.01}$ & $0.17\pm0.01$ & $0.40\pm0.01$ & $13.88\pm0.31$ & $8.71\pm0.46$ \\
\texttt{NGC5044} & $6.97^{0.2}_{-0.16}$ & $45.1^{5.4}_{-4.7}$ & $217.7$ & $0.09^{0.00}_{-0.00}$ & $0.11\pm0.00$ & $0.44\pm0.01$ & $12.99\pm0.36$ & $11.99\pm0.63$ \\
\texttt{NGC507} & $10.4^{2.4}_{-1.5}$ & $115.0^{77.0}_{-34.0}$ & $167.0$ & $0.07^{0.00}_{-0.00}$ & $0.13\pm0.01$ & $0.26\pm0.01$ & $23.79\pm0.70$ & $3.12\pm1.43$ \\
\texttt{Perseus} & $39.3^{0.5}_{-0.46}$ & $205.0^{4.9}_{-4.6}$ & $329.9$ & $1.44^{0.00}_{-0.00}$ & $0.99\pm0.01$ & $0.14\pm0.00$ & $7.91\pm0.33$ & $2.66\pm0.98$ \\
\texttt{PKS0745} & $68.9^{5.3}_{-4.5}$ & $351.0^{50.0}_{-42.0}$ & $1115.8$ & $10.10^{0.15}_{-0.15}$ & $5.80\pm0.30$ & $0.08\pm0.01$ & $6.63\pm0.14$ & $5.60\pm2.58$ \\
\texttt{PKS1404} & $14.4^{0.41}_{-0.37}$ & $98.2^{6.0}_{-5.4}$ & $260.3$ & $0.24^{0.00}_{-0.00}$ & $0.29\pm0.01$ & $0.25\pm0.04$ & $15.35\pm1.35$ & $3.33\pm2.08$ \\
\texttt{RBS797} & $102.0^{29.0}_{-15.0}$ & $421.0^{230.0}_{-110.0}$ & $1715.1$ & $16.50^{1.90}_{-2.00}$ & $13.00\pm1.20$ & $0.02\pm0.01$ & $8.41\pm0.29$ & $14.22\pm4.76$ \\
\texttt{RXJ1532} & $93.5^{44.0}_{-17.0}$ & $502.0^{380.0}_{-150.0}$ & $969.2$ & $8.17^{0.24}_{-0.23}$ & $7.00\pm0.89$ & $0.02\pm0.01$ & $7.52\pm0.19$ & $18.45\pm3.07$ \\
\texttt{Sersic159} & $29.6^{2.1}_{-1.8}$ & $233.0^{29.0}_{-26.0}$ & $552.9$ & $1.51^{0.03}_{-0.03}$ & $1.30\pm0.05$ & $0.10\pm0.00$ & $9.46\pm0.20$ & $3.53\pm1.45$ \\
\texttt{Zw2701} & $48.8^{6.1}_{-3.5}$ & $248.0^{79.0}_{-49.0}$ & $1098.2$ & $5.51^{0.18}_{-0.18}$ & $3.80\pm0.31$ & $0.19\pm0.02$ & $7.83\pm0.40$ & $14.67\pm4.93$ \\
\texttt{Zw3146} & $106.0^{33.0}_{-21.0}$ & $738.0^{310.0}_{-190.0}$ & $2578.5$ & $28.50^{0.94}_{-1.20}$ & $21.00\pm1.90$ & $0.03\pm0.01$ & $6.07\pm0.19$ & $2.85\pm1.06$ \\
\hline
\end{tabular}
\caption{\label{table2}The provided table presents various cluster properties, including cluster central density ($\rho_0$), core radius ($r_c$), $r_{2500}$; the calculated quantities $M_{b2500}$, $M_{d2500}$, and the fitted quantities $\alpha$, $Q$, $\lambda^{-1}$. These data points have been sourced from \cite{main2016relationship}. The unit of $\rho_0$ is so chosen that $A = 4\pi G \rho_0 r_c^2 \mu m_H$ is in keV unit.}
\end{table}

\end{document}